\documentclass[aps,prx,twocolumn,superscriptaddress,amsmath,longbibliography]{revtex4-2}
\usepackage{dcolumn}
\usepackage{float,soul}
\usepackage{array}
\usepackage{url}
\usepackage{xcolor}
\usepackage{color}
\usepackage{graphicx}
\usepackage{amsfonts}
\usepackage{natbib}
\usepackage{amsmath,amssymb}
\usepackage{bbold}
\usepackage{physics}
\usepackage{bbm,bm}
\PassOptionsToPackage{normalem}{ulem}
\usepackage{ulem}
\usepackage[english]{babel}
\usepackage{hyperref}
\hypersetup{colorlinks=true,citecolor=blue,linkcolor=blue, urlcolor=blue, breaklinks=true}

\newcommand{\tunn}[0]{\mathtt{t}}
\newcommand{\alatt}[0]{\mathtt{a}}
\newcommand{\Kconst}[0]{\mathtt{K}}
\newcommand{\opa}{\hat{a}}

\newcommand{\Id}{\hat{\mathbbm{1}}}
\newcommand{\T}{\hat{\mathcal{T}}}

\newcommand{\mysymb}[2]{\mathord{\vcenter{\hbox{\includegraphics[height=#2 ex]{#1}}}}}
\newcommand{\myket}[1]{\lvert #1 \rangle }

\begin{document}

\preprint{APS/123-QED}

\title{Diagnosing electronic phases of matter using photonic correlation functions}
\author{Gautam Nambiar}
\email{nambiar@terpmail.umd.edu}
\affiliation{Joint Quantum Institute, University of Maryland,
College Park, Maryland 20742, USA}
\author{Andrey Grankin}
\affiliation{Joint Quantum Institute, University of Maryland,
College Park, Maryland 20742, USA}
\author{Mohammad Hafezi}
\affiliation{Joint Quantum Institute, University of Maryland,
College Park, Maryland 20742, USA}

\begin{abstract}
In the past couple of decades, there have been significant advances in measuring quantum properties of light, such as quadratures of squeezed light and single-photon counting. Here, we explore whether such tools can be leveraged to probe electronic correlations in the many-body quantum regime. Specifically, we show that it is possible to probe certain spin, charge, and topological orders in an electronic system by measuring the correlation functions of scattered photons. We construct a mapping from the correlators of the scattered photons to those of a correlated insulator, particularly for Mott insulators described by a single-band Fermi-Hubbard model at half-filling. We show that frequency filtering before photodetection plays a crucial role in determining this mapping.  We find that if the ground state of the insulator is a gapped spin liquid, a photon-pair correlation function, i.e.,  $G^{(2)}$, can detect the presence of anyonic excitations with fractional mutual statistics. Moreover, we show that correlations between electromagnetic quadratures can be used to detect expectation values of static spin chirality operators on both the kagome and triangular lattices, thus being useful in detecting chiral spin liquids. More generally, we show that a series of hitherto unmeasured spin-spin and spin-charge correlation functions of the material can be extracted from photonic correlations. This work opens up access to probe correlated materials, beyond the linear response paradigm, by detecting quantum properties of scattered light.
\end{abstract}

\maketitle

\tableofcontents

\section{Introduction}\label{sec:intro}
    Strongly interacting quantum many-body systems can host a variety of exotic phases of matter. However, there exists a gap between theoretical models and experimental observations in terms of accessible physical observables. A hallmark example is topologically ordered phases, beginning with the experimental observation of the Fractional Quantum Hall effect, whose excitations exhibit fractional statistics \cite{arovas1984fractional,wen1990ground,wen1990anyons,wen1991topological,nakamura2020direct,bartolomei2020fractional}. There are strong theoretical reasons ~\cite{hastings2004lieb,oshikawa2000commensurability,lieb1961two} to expect that spin systems dubbed ``quantum spin liquids" also host topological order and other exotic gapless field theories \cite{kitaev2006anyons, knolle2019field, savary2016quantum, lee2007high, lee2006doping,broholm2020quantum,clark2021quantum}. However, experimental verification of such claims has been extremely challenging. At the same time, other unconventional phases, including high-temperature superconductors and correlated insulators, have been experimentally observed in cuprate-like strongly correlated materials \cite{keimer2015quantum}, and more recently in moir\'e materials \cite{andrei2021marvels,mak2022semiconductor}. While there is a thriving theoretical effort to explain many of these phases, conclusively matching theory to experiment is generally difficult. The main challenge stems from the fact that the nontrivial nature of many of these phases is encoded in correlation functions that are difficult to measure experimentally. Conventionally, probes for accessing electronic correlation functions work within the linear response paradigm. For example, Raman scattering has been employed to study potential spin liquid candidates ~\cite{devereaux2007inelastic, wulferding2019raman,shastry1990theory,shastry1991raman,ko2010raman,nasu2016fermionic,wang2020range,nambiar2023monopole,perkins2013raman}. However, given how difficult it is to characterize exotic phases, it is important to (1) develop novel experimental protocols to measure a wider class of correlation functions and (2) understand how these new correlation functions can assist in diagnosing the phase of matter under study.
\begin{figure}[t]
  \includegraphics[width=0.48\textwidth]{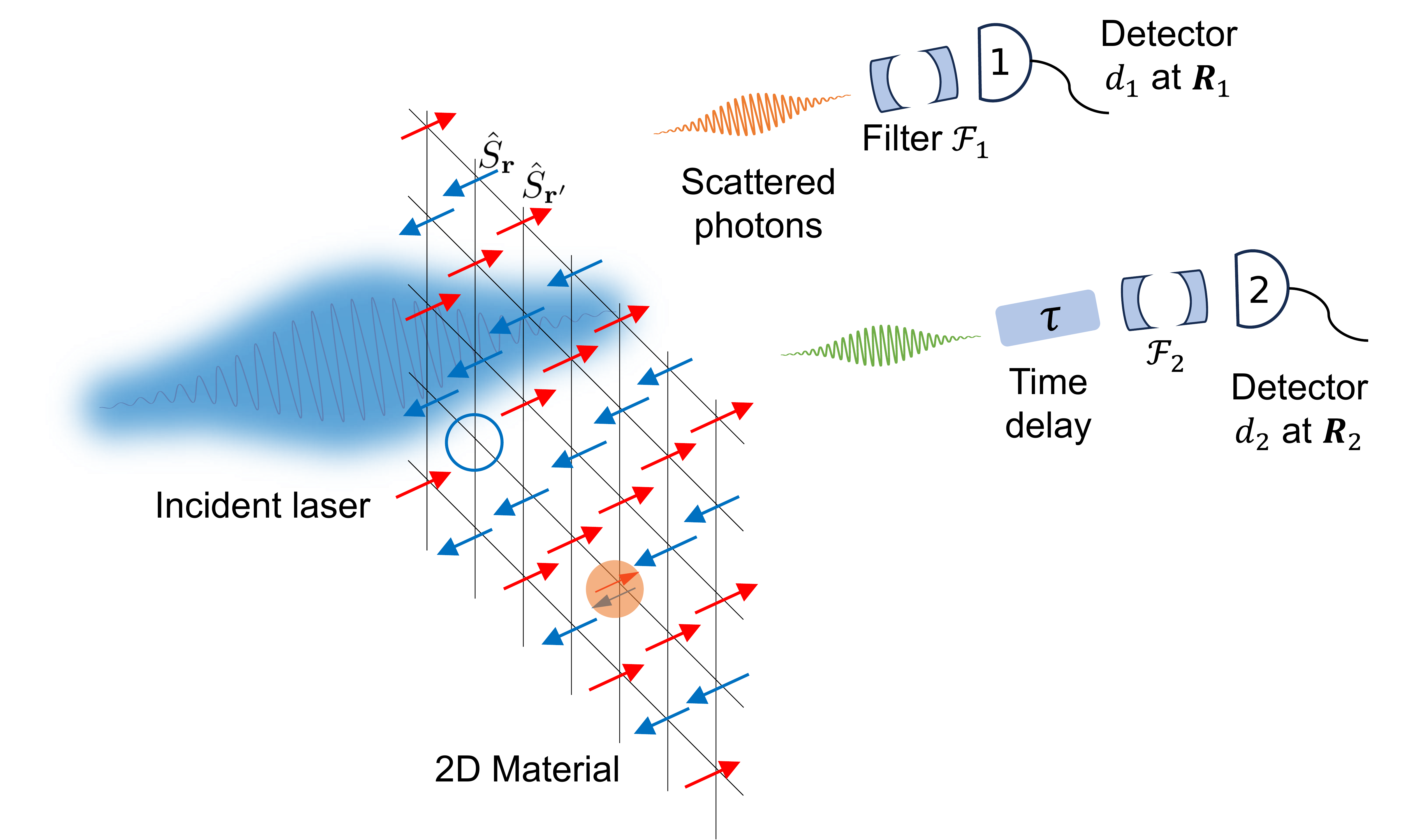}
  \caption{ A system of itinerant electrons is irradiated with a laser. Conventionally, photodetectors measure the intensity of the scattered photons, and the correlations are ignored. In this paper, we propose a Hanbury Brown-Twiss-type setup to measure correlations between pairs of photons. We allow for frequency filters, $\mathcal{F}_1$ and $\mathcal{F}_2$, before detection, and a time delay, $\tau$, between detection events.}
  \label{fig:schem1}
\end{figure}

 A promising approach is to move beyond the linear-response paradigm by studying the nonlinear response to an external electromagnetic drive \cite{wan2019resolving, mahmood2021observation,fava2021hydrodynamic,choi2020theory, nandkishore2021spectroscopic,hart2023extracting,konig2020spin,li2021photon,sim2023nonlinear,sim2023microscopic,negahdari2023nonlinear,gao2023two,potts2024exploiting,zhang2024terahertz,parameswaran2020asymptotically,gerken2022unique,zhang2024disentangling, krupnitska2023finite, watanabe2024revealing,alpichshev2017origin, devereaux2023angle, mcginley2024signatures, mcginley2024anomalous,tai2023quantum}. One example is two-dimensional coherent spectroscopy, an extension of pump-probe spectroscopy \cite{shen1984principles, gelzinis2019two,smallwood2018multidimensional}, in which two pump pulses, separated by a fixed time delay, are applied to the sample, followed by a probe measurement at a fixed delay relative to the last pulse. A recent work predicted that a setup like this can be used to detect anyonic excitations \cite{mcginley2024anomalous,mcginley2024signatures}. It is also possible to go beyond the linear-response paradigm by studying higher-order correlations within the output signal generated by external stimuli. A classic example is shot noise in electric current \cite{kobayashi2021shot} that has been used to detect fractionally charged excitations \cite{saminadayar1997observation, de1998direct}, Cooper pairing in normal state \cite{bastiaans2021direct,zhou2019electron} and absence of quasiparticles \cite{chen2023shot}.  

Thus, it is intriguing to ask whether quantum correlations between scattered photons could carry useful information beyond what can be obtained from both linear response and pump-probe techniques. It is known in quantum optics that when two or more photons are scattered off an optically nonlinear medium, they can become correlated \cite{chang2014quantum}. Our work demonstrates that when this medium is a strongly correlated electronic system, its nontrivial correlations are inherited by the scattered photons. Therefore, analyzing photonic correlations in the outcome of spectroscopy experiments reveals many-body observables that are not accessible through current linear and nonlinear probes.

  The paradigmatic setup for the detection of the photonic correlations consists of a Hanbury Brown-Twiss interferometer where the scattered photons are divided into two separate paths \cite{brown1956correlation} (See Fig.~\ref{fig:schem1}). Then, the photons are detected in each path separately. In this work, we also allow for the possibility of frequency and polarization filtering before detection. This scheme allows for the measurement of a four-point correlation function $G_{d_1,d_2}^{(2)}(\tau)=\langle{\hat{a}_{d_1}^{\dagger}(0)\hat{a}_{d_2}^{\dagger}(\tau)\hat{a}_{d_2}(\tau)\hat{a}_{d_1}(0)}\rangle$, where $\hat{a}_{d_j}$ are the annihilation operators of the filtered photonic mode in the corresponding ($j$-th) interferometer arm. $\tau$ denotes the time delay between the detection events \footnote{The correction due to retardation is clarified in Eq.~\eqref{eq:taudef} in Sec.~\ref{sec:photonicdefs}.}. We note that in addition to $G^{(2)}$, the described setup allows for the measurement of correlation functions of electromagnetic quadrature including $\expval{\hat{a}_{d_1}(0)}$ and $\expval{\hat{a}_{d_2}(\tau)\hat{a}_{d_1}(0)}$. This is done by mixing the scattered photons with a strong reference coherent field having a fixed phase. 
    
  In this work, we first develop a systematic procedure to map these photonic correlators to dynamical correlation functions of the material in its undriven state. As a specific example, we provide the mapping for a single-band Fermi-Hubbard model at half-filling in a Mott insulator state, where in the limit of strong on-site repulsion, the many-body energy spectrum splits into distinct sectors corresponding to charge and spin degrees of freedom, well-separated by the Mott gap \footnote{Strictly speaking, only the many-body states connected through application of local operators on the ground state split into distinct charge and spin sectors.}. Furthermore, we demonstrate that the frequency filtering of scattered photons plays a crucial role in our scheme. In particular, it enables the measurement of matter correlations restricted to a particular energy sector of interest, i.e. a pure spin or a mixed spin-charge sector. 
    
   We then present several salient applications: (1) We show that by measuring the first-order quadrature of the scattered light and its fluctuations, one can determine the static expectation value of the spin chirality operator on the kagome and triangular lattices, respectively, which is currently inaccessible using current techniques. These operators acquire nonzero expectation values in chiral spin liquids \cite{kalmeyer1987equivalence,wen1989chiral,gong2014emergent,he2014chiral,hickey2016haldane,wietek2017chiral,cookmeyer2021four}, and play the role of chiral mass terms in $U(1)$ Dirac spin liquids \cite{lee1992gauge, song2019unifying}.  (2)   We demonstrate that the correlations of scattered photons can be used to probe the mixed spin-charge correlators, providing insights into the dynamics of a hole and a doubly-occupied site (called a doublon) in a Mott insulator. (3) We derive the contributions to the $G^{(2)}$ correlation functions from non-interacting magnons in a magnetically ordered system.  (4) We establish that  $G^{(2)}$ can be used to diagnose whether the state in the spin sector is a spin liquid with excitations carrying fractional mutual statistics. For this, we follow the semiclassical argument in Refs.~\cite{mcginley2024anomalous,mcginley2024signatures} for pump-probe spectroscopy, and show that it also applies to $G^{(2)}$ spectroscopy. 

    The rest of the paper is organized as follows. In Sec.~\ref{sec:Summary}, we provide an overview of the physical setting and the main results of the paper, along with sufficient background and context to allow each of the remaining sections to be read independently. In Sec.~\ref{sec:formalism}, we provide details of the $\T$-matrix formalism that we use for describing different  scenarios of photon scattering.  Sec.~\ref{sec:photonicdefs}  presents a mathematical definition of the photonic correlators studied in this work. Technical details for experimental measurement schemes are provided in Appendix~\ref{app:homodyne}.
    Next, in Sec.~\ref{sec:Hubbardsetup}, we apply the scattering formalism to the Fermi-Hubbard model at half-filling. Given the separation of energy scales discussed in Sec.~\ref{sec:scales}, we show in Sec.~\ref{sec:Tmatrixsimp} that correlation functions of photonic operators map to those of pure matter operators denoted as $\hat{R}^{(1)}$ and $\hat{R}^{(2)}$. We study the microscopic structure of these operators in Sec.~\ref{sec:operators}, with the technical details provided in Appendix~\ref{sec:explicitmicro}.  In Sec.~\ref{sec:temporal}, we discuss the temporal structure of the matter correlation functions, with the emphasis on the role of frequency filters. Starting from Sec.~\ref{sec:spinchirality}, we provide a detailed discussion of the proposed salient applications. First, in Sec.~\ref{sec:spinchirality}, we show that by measuring the squeezing spectrum of the scattered photons it is possible to determine the expectation values of the scalar spin chirality operators. Next, in Sec.~\ref{sec:magnons}, we calculate $G^{(2)}$ of scattered photons for the case when the spin sector has non-interacting bosonic excitations such as magnons. In Sec.~\ref{sec:fracstat}, we discuss the conditions under which their contributions can be filtered out, enabling the measurement of more exotic properties of the matter. In particular, we demonstrate that the presence of anyonic excitations in a spin sector can be established using both conditional $G^{(1)}(\tau)$ and  $G^{(2)}$ measurements. In Sec.~\ref{sec:exprealization}, we comment on experimental feasibility. Finally, in Sec.~\ref{sec:conclusions}, we present an outlook.

    \section{Summary of the paper}\label{sec:Summary}
\subsection{The setting}\label{sec:setupsummary}
We now consider the problem of photon scattering off a 2-dimensional strongly correlated electronic system, as shown in Fig.~\ref{fig:schem1}. The material, initially prepared in a thermal equilibrium state, is irradiated by a monochromatic laser having polarization $\vb{e}_L$ and frequency $\omega_L$. The laser driving is weak, in the sense that $g_L\equiv \sqrt{2\pi I_L \alpha}\alatt/\omega_L \ll 1$, where $I_L$ is the laser intensity, $\alpha$ is the fine-structure constant and $\alatt$ is the lattice spacing of the electronic system.
Throughout this work, we assume the initial state of the electromagnetic field to be in either a Fock state  $|\mathcal{N}_L\rangle=(\mathcal{N}_L!)^{-1/2}(\hat{a}^{\dagger}_L)^{\mathcal{N}_L}|0\rangle$  or in a coherent state $e^{\phi_L \hat{a}_L^{\dagger}-\phi_L^*\hat{a}_L}\ket{0}$, where $\ket{0}$ is the vacuum, $\hat{a}_L$ is the annihilation operator of the laser mode (see Appendix~\ref{app:Tmatrixreview} for a more careful treatment of the laser as a wavepacket), and $\phi_L\equiv e^{-i\theta_L}\abs{\phi_L}$. Since the driving is weak, we can restrict our consideration to the subspace containing at most 2 scattered photons \cite{hafezi2011photonic,gorshkov2011photon}. These  photons, before being detected, are separated into the two arms of a Hanbury Brown-Twiss interferometer, each containing frequency and polarization filters. We define the filter function in the $j$-th interferometer arm as $\mathcal{F}_j(\omega)$ and assume that it is centered around frequency $\omega_{j}$ with possibly some spread. Furthermore, we impose the  causality condition, i.e. $\tilde{\mathcal{F}}_j(t)\equiv \int_{-\infty}^{\infty}\frac{\dd{\omega}}{2\pi}\mathcal{F}_j(\omega)e^{-i\omega t}$ is $0$ for $t<0$. Then the effective photon annihilation operator corresponding to the detection of a photon in $j$-th interferometer arm takes the form $\hat{a}_{d_j}\sim i \sum_{\vb{k}}\mathcal{F}_j(\omega_{\vb{k}}) \hat{a}_{\vb{k},\vb{e}_j}$, i.e., a superposition of normal modes of free space, $\hat{a}_{\vb{k},\vb{e}_j}$ labeled by momenta $\vb{k}$ and polarization $\vb{e}_j$ (here, the subscript $d_j$ stands for detector $j$). For a precise definition of $\hat{a}_{d_j}$, see Eq.~(\ref{eq:addef}, \ref{eq:filterredef}).  Finally, all measurements are considered to be made in the asymptotic future  (formally defined in Sec.~\ref{sec:formalism}).

For the matter side, the general form of our results applies to any correlated insulator with an optical gap. For concreteness, we consider a 2-dimensional system of strongly interacting itinerant electrons in a Mott-insulator state, which we model by the single-band Fermi-Hubbard model at half-filling~\cite{arovas2022hubbard,qin2022hubbard}. In the limit of strong on-site interaction $U\gg \tunn$, where $\tunn$ is the tunneling amplitude between nearest sites, the many-body energy spectrum splits into sectors having a definite number of doubly occupied sites (``doublon excitations"). The lowest energy manifold (``spin sector") consists of pure spin excitations described by the Heisenberg model. Depending on the lattice geometry, and strengths of tunnelings between next-nearest-neighbors and beyond, there are several possibilities for the ground state of the spin sector. It could be a magnetically ordered state such as a N\'eel or a $120^\circ$ antiferromagnet, in which case its excitations are magnons. It could also be a quantum spin liquid whose excitations are charged under an emergent gauge group, or have fractional statistics or both. 
 
\begin{figure*}[t]
  \centering
  \includegraphics[width=0.90\textwidth]{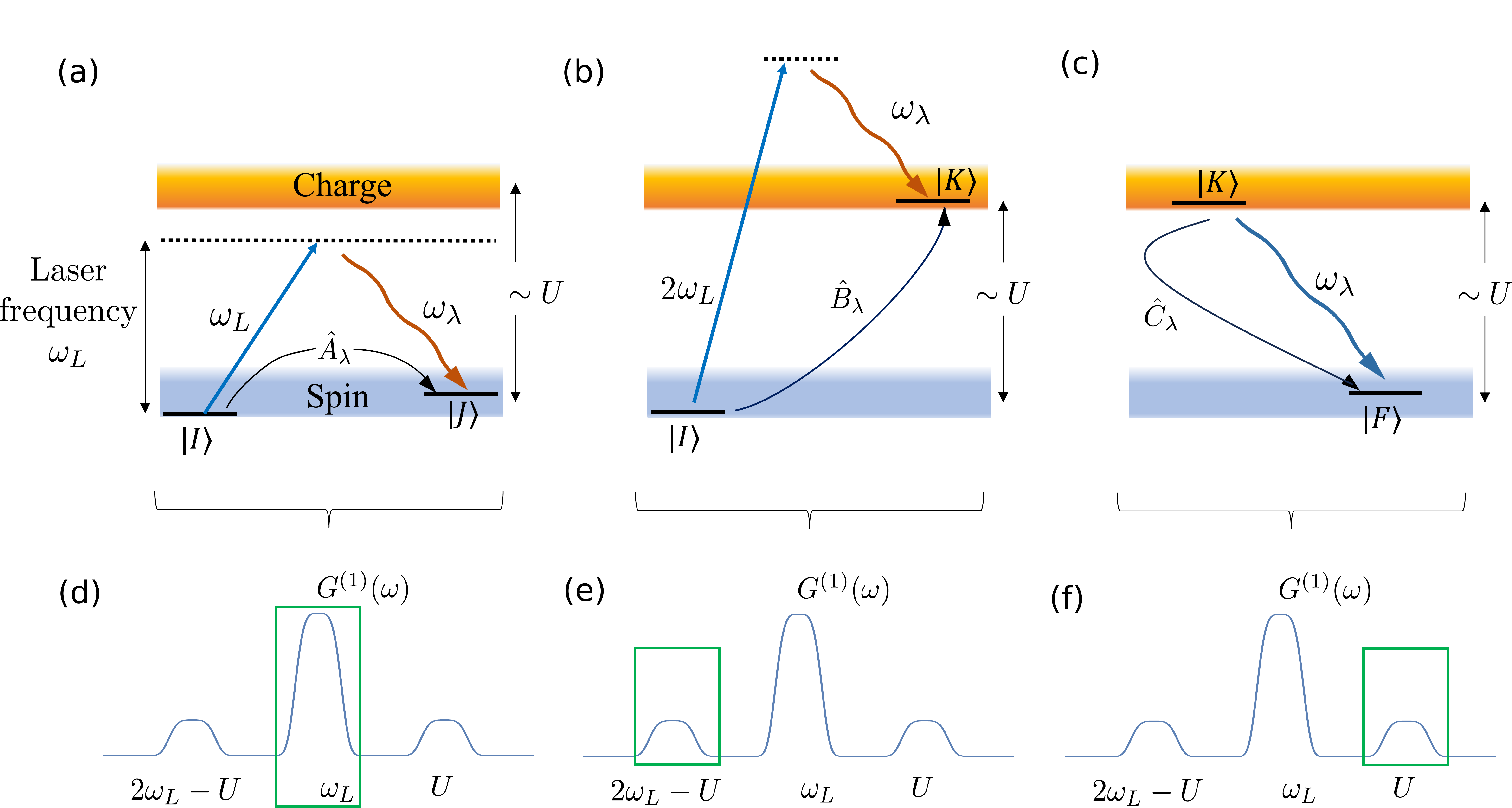}
  \caption{Schematic illustration of different photon scattering
processes. \textbf{(a-c)} The
blue- and orange-shaded regions represent the lower and higher energy
sectors, respectively, separated by an optical gap of order $U$.
For the Fermi-Hubbard model at half-filling, $U$ corresponds to the
 on-site repulsion, with the low- and high-energy sectors identified as spin
and charge sectors, respectively. More generally, the presented formalism
applies to any insulator with an optical gap. The laser frequency
$\omega_{L}$ is assumed to be detuned from $U$. The three terms
in $\hat{M}_{\lambda}$, defined in Eq.~\eqref{eq:Meffint} correspond
to different pathways leading to emission of a photon. These three
pathways are depicted schematically as follows: \textbf{(a)}:
Raman process -- absorption of a laser photon followed by the emission
of a photon with a frequency near $\omega_{L}$. This process is governed
by the effective matter operator $\hat{A}_{\lambda}$, which mediates
transitions from the state $\ket{I}$ to $\ket{J}$ within the same
(lower energy) sector. \textbf{(b)}:
Absorption of two photons followed by emission of a photon of frequency
near $2\omega_{L}-U$. This process involves the effective matter
operator $\hat{B}_{\lambda}$, which transitions the state $\ket{I}$
in the lower energy sector to $\ket{K}$ in the higher energy sector.
\textbf{(c)}: Emission of a photon
of frequency near $U$ originating from a state $\ket{K}$ in the
higher energy sector that was previously accessed via process (b).
This emission is mediated by the effective matter operator $\hat{C}_{\lambda}$, which transitions the state $\ket{K}$ to $\ket{F}$ in the lower
energy sector. Panels (d-f) illustrate the corresponding intensity
versus frequency profiles. Notably, emission into sideband (e) is
necessarily accompanied by simultaneous emission into sideband (f).}
  \label{fig:sectorsbasic}
\end{figure*}

Throughout this work, we assume the incident laser is detuned from the optical gap such that  $\tunn \ll \abs{\omega_L-U}\ll U$. The absorption and emission of photons couples the spin and charge sectors of the system as shown in Fig.~\ref{fig:sectorsbasic}. Furthermore, the photons can be scattered inelastically with their amplitudes being dependent on the finer structure of the spin and charge sectors. We note that in contrast to the conventional Raman spectroscopy, we are interested in the scattering of multiple photons, which, as we demonstrate, provides additional information on the state of the electronic system. The relation between the inelastic photon scattering amplitudes and the matter correlations constitutes the main focus of this work and will be reviewed in the section below. 

\subsection{Mapping between photonic and matter correlators (Input-output relations)}\label{sec:iomap}
We first summarize the main assumptions in this work:
\begin{enumerate}
    \item The material is initially in equilibrium at a temperature much smaller than $U$, the optical gap.
    \item $\tunn \ll \abs{\omega_L -U}\ll U$. More generally, going beyond the single-band Fermi-Hubbard model, this assumption means that the detuning of the drive frequency from the optical gap is much smaller than the optical gap. In addition, the intrinsic coupling scale between different optical sectors is much smaller than the detuning.
    \item Weak driving: $g_L\equiv \sqrt{2\pi I_L \alpha}\alatt/\omega_L \ll 1$. This implies that light-matter scattering processes can be treated order by order in the number of photons involved.
    \item The frequencies of detected photons are postselected to be of the order of the drive frequency $\omega_L$ up to corrections of order $\abs{U-\omega_L}$.   As discussed later (Sec.~\ref{sec:Tmatrixsimp}), this requirement, together with Assumption 2, implies that light scattering processes conserve the total number of photons.
    \item Dipole approximation, i.e., momentum exchange between light and matter is neglected (discussed below).
    \item We consider a single electronic orbital per lattice site and per spin, and assume that the light-matter coupling in this truncated Hilbert space is given by Peierls' substitution.
\end{enumerate}
Later, in Sec.~\ref{sec:operators}, we also assume the absence of spin-orbit coupling for simplicity. But this assumption can be dropped straightforwardly by starting with a spin-orbit coupled microscopic tunneling Hamiltonian as done in Ref.~\cite{yang2021non}. 
We start by providing a qualitative picture of the laser-induced processes. For simplicity we assume the system is initially prepared in the lowest energy sector (for a Mott insulator, this is the spin sector).  
Emission of a photon by the driven material can result from one of the three pathways shown in Fig.~\ref{fig:sectorsbasic}(a-c). Fig.~\ref{fig:sectorsbasic}(a) depicts the conventional Raman scattering process, where the off-resonant absorption of a single photon is followed by emission at a different frequency,
$\omega_{\lambda}$. This process is accompanied by the  material transitioning from a state $\ket{I}$ to $\ket{J}$ belonging to the same energy sector. Energy conservation implies that $\omega_{\lambda}=\omega_{L}+E_I-E_J$, where $E_{\ldots}$ denotes the energy of the state $|\ldots\rangle$. Under our approximations, the emitted photon has a frequency close to $\omega_L$. 

At weak driving, Raman scattering involving emission of a single photon represents the leading contribution to photonic observables. However, the Hanbury Brown-Twiss measurement scheme post-selects the scattering amplitudes that necessarily involve two photons. We identify the following main contributions to these amplitudes. (1) Two Raman processes, occurring sequentially, such that both emitted photons have a frequency $\sim\omega_L$ (see Fig.~\ref{fig:sectorsbasic}(d)), and (2) the process shown in Figs.~\ref{fig:sectorsbasic}(b, c) involving the real excitation of an intermediate state $|K\rangle$ within the charge sector (upper Hubbard band).

In the second scenario, after the emission of the first photon, the material transitions from the state $\ket{I}$
in the spin sector to 
$\ket{K}$
 in the charge sector. The frequency of the emitted photon is given by $\omega_{\lambda}=2\omega_{L}+E_I-E_K\sim2\omega_L-U$.  The material can then return to the lowest energy sector by emitting a photon with frequency $E_K-E_F\sim U$. Consequently, the intensity spectrum of scattered light exhibits sidebands at the frequencies $\sim U$ and $2\omega_L-U$, as schematically shown in Figs.~\ref{fig:sectorsbasic}(e, f). At weak driving, these sidebands have a smaller spectral weight compared to the central peak at $\sim\omega_L$ because they involve two-photon emission processes. The microscopic details of these processes in the correlated insulators described by the Fermi-Hubbard model at half-filling are provided in Sec.~\ref{sec:operators}.

 Let us now assume that the  evolution of light and matter in the absence of light-matter interactions is described by the Hamiltonian $\hat{H}_0$. In the remainder of this Subsection, we provide a quick way to arrive at the mapping between correlation functions of scattered photons and those of matter operators in equilibrium. A more explicit treatment by quantizing the vector potential and using the $\T$-matrix formalism to compute scattering amplitudes is provided in Secs.~\ref{sec:Tmatrixsimp}, \ref{sec:photoncorrsimp}, and \ref{sec:temporal}. To motivate our results here, we directly write down the following effective interaction Hamiltonian that captures all the photon emission processes in Fig.~\ref{fig:sectorsbasic}: 
 \begin{table*}
    \centering
    \begin{tabular}{|p{6.4cm} | p{10.7cm}|}
        \hline
         \textbf{Photonic correlator} & \textbf{Matter correlator} \\\hline
          \vspace{0.0001in}First-order quadrature: $e^{i\theta}\expval{\hat{a}_{d_j}(0)}_{\text{out}}+\text{c.c.}$& \vspace{0.0001in} $e^{i\theta}\mathcal{F}_j (\omega_L)\expval{\hat{M}_j(0)}_0 +\text{c.c.}$ \\[2ex]
         \hline
         \vspace{1mm}Intensity: $G^{(1)}_{d_j}(0)=\langle\hat{a}_{d_j}^{\dagger}(0)\hat{a}_{d_j}(0)\rangle_{\text{out}}$& \vspace{1mm}$\iint_{0}^{\infty}\dd{t}\dd{t'}\Tilde{\mathcal{F}}_j(t)\left[\Tilde{\mathcal{F}}_j(t')\right]^* \expval{\left[\hat{M}_{j}(-t')\right]^{\dagger}\hat{M}_{j}(-t)}_0$ \\[2ex]
         \hline
         \vspace{0.0001in}$G^{(1)}_{d_j}(\tau)= \langle\hat{a}^{\dagger}_{d_j}(\tau)\hat{a}_{d_j}(0)\rangle_{\text{out}}+\text{c.c.}$ & \vspace{0.0001in} $\left\{\iint_{0}^{\infty}\dd{t}\dd{t'}\tilde{\mathcal{F}}_j(t)\left[\tilde{\mathcal{F}}_j(t')\right]^*\expval{\hat{M}^{\dagger}_j(\tau-t')\hat{M}_j(-t)}_0\right\}+\text{c.c.}$\\[2ex]
         \hline
         \vspace{0.0001in}Phase-dependent part of the quadrature fluctuations: $e^{2i\theta}\expval{\hat{a}_{d_2}(\tau)\hat{a}_{d_1}(0)}_{\text{out}}+\text{c.c.}$&\vspace{0.0001in}$   e^{2i\theta}\iint_{0}^{\infty}\dd{t_1}\dd{t_2}\tilde{\mathcal{F}}_1(t_1)\tilde{\mathcal{F}}_2(t_2) \expval{{\mathbb{T}\left[\hat{M}_2(\tau-t_2)\hat{M}_1(-t_1)\right]}}_0$+\text{c.c.} \\ [2ex]
         \hline
         \vspace{0.001in}Photon-pair correlation function: & \vspace{0.001in}$\iiiint_{0}^{\infty}\dd{t_1}\dd{t_2}\dd{t'_1}\dd{t'_2}\tilde{\mathcal{F}}_1(t_1)\tilde{\mathcal{F}}_2(t_2)[\tilde{\mathcal{F}}_1(t'_1)]^*[\tilde{\mathcal{F}}_2(t'_2)]^*$\\ $G^{(2)}_{d_1,d_2}(\tau)=\expval{\hat{a}^{\dagger}_{d_1}(0)\hat{a}_{d_2}^{\dagger}(\tau)\hat{a}_{d_2}(\tau)\hat{a}_{d_1}(0)}_{\text{out}}$ & $\quad \quad \times\left\langle\bar{\mathbb{T}}\left[\hat{M}^{\dagger}_1(-t'_1)\hat{M}^{\dagger}_2(\tau-t'_2)\right]\mathbb{T}\left[\hat{M}_2(\tau-t_2)\hat{M}_1(-t_1)\right]\right\rangle_0$
        \\[2ex]
         \hline 
        \vspace{0.0001in}Conditional $G^{(1)}$ defined as $H_{d_1, d_2}(t,\tau)$& \vspace{0.001in}$\Bigl\{\iiiint_{0}^{\infty}\dd{t_1}\dd{t_2}\dd{t'_1}\dd{t'_2}\tilde{\mathcal{F}}_1(t_1)\tilde{\mathcal{F}}_2(t_2)[\tilde{\mathcal{F}}_1(t'_1)]^*[\tilde{\mathcal{F}}_2(t'_2)]^*$\\
         $= \langle\hat{a}^{\dagger}_{d_1}(0)\hat{a}^{\dagger}_{d_2}(t+\tau)\hat{a}_{d_2}(t)\hat{a}_{d_1}(0)\rangle_{\text{out}}+\text{c.c.}$ & $\quad \quad \times\left\langle\bar{\mathbb{T}}\left[\hat{M}^{\dagger}_1(-t'_1)\hat{M}^{\dagger}_2(t+\tau-t'_2)\right]\mathbb{T}\left[\hat{M}_2(t-t_2)\hat{M}_1(-t_1)\right]\right\rangle_0\Bigr\}+\text{c.c.}$\\[2ex]
         \hline
    \end{tabular}
    \caption{Correspondence between the correlation functions of scattered photons (left column) and the matter correlation functions in equilibrium (right column). In the left column, $\expval{.}_{\text{out}}$ indicates that the expectation value $\mel{\text{out}}{.}{\text{out}}$ is taken in the full light-matter post-scattering state $\ket{\text{out}}$. In the right column, $\expval{.}_0$ indicates that the expectation value is taken in a pure matter energy eigenstate, or more generally in any mixed state diagonal in the energy eigenstate basis. The matter operators $\hat{M}_j$ are defined in Eq.~\eqref{eq:Meffint} and \eqref{eq:MjtoABC}. $\mathcal{F}_j(\omega_j)$ is the effective filter function, and $\tilde{\mathcal{F}}_j(t)$ is its Fourier transform. Here $\mathbb{T}[\ ]$ and $\bar{\mathbb{T}}[\ ]$ denote time- and anti-time-ordering respectively. Note that $G^{(1)}_{d_j}(0)$ is the same as intensity of Raman-scattered photons~\cite{shastry1990theory,shastry1991raman}. For measuring the first-order quadrature and two-mode squeezing, the output photons are interfered with a strong local oscillator whose frequency is equal to the drive frequency, $\omega_L$, and whose phase \textit{relative} to the drive is $\theta$. We provide a detailed discussion on the above photonic correlators and the experimental schemes to measure them in Sec.~\ref{sec:photonicdefs} and Appendix~\ref{app:homodyne}.} 
    \label{tab:dictionary1}
\end{table*}
 \begin{equation}\label{eq:Heffint}
     \begin{aligned}
     \hat{H}^I_{\text{eff}}(t)&=\sum_{\lambda}\hat{M}_{\lambda}(t)\hat{a}_{\lambda}^{\dagger}e^{i\omega_{\lambda}t}+\text{h.c.,}
     \end{aligned}
 \end{equation}
where $\hat{X}(t)\equiv e^{i\hat{H}_0 t}\hat{X}e^{-i\hat{H}_0 t}$. The operator $\hat{a}^{\dagger}_{\lambda}$ denotes the photon creation operator in the mode $\lambda$, which is a composite index for momentum and polarization. $\hat{M}_{\lambda}$ is an effective pure matter operator. For now, $\hat{M}_{\lambda}$ can be thought of as the operator implementing the transitions shown in Fig.~\ref{fig:sectorsbasic}, but we will define $\hat{M}_{\lambda}$ precisely in Sec.~\ref{sec:operators} and provide intuition in the remainder of this Section.  For energy scales considered in this work, the momentum transfer between transverse electromagnetic radiation and electrons is a negligible fraction of the reciprocal lattice momenta. So, we have used the dipole approximation and dropped the spatial dependence of the photon operators --  an approximation discussed further in Sec.~\ref{sec:Ramanrev} and Sec.~\ref{sec:conclusions}. The matter operator $\hat{M}_j$ does depend on the polarization of the absorbed and emitted photons. Through this polarization dependence, $\hat{M}_j$ encodes symmetry properties of light matter interactions, as elaborated later in this Section, and in Sec.~\ref{sec:spinchirality}.  For the purpose of this section, photon absorption from the laser is treated classically, while emission is treated quantum mechanically. A full quantum mechanical treatment is provided in Sec.~\ref{sec:photoncorrsimp}. Under the assumption about energy scales and post-selection discussed above, we can write the matter operator $\hat{M}_{\lambda}$ as a sum of three terms corresponding to the photon-emission processes in Fig.~\ref{fig:sectorsbasic} (a), (b) and (c) respectively:  \begin{equation}\label{eq:Meffint}
     \begin{aligned}
     \hat{M}_{\lambda}(t)=&\left[e^{-i(\omega_L t+\theta_L)}\hat{A}_{\lambda}(t)+ e^{-2i(\omega_Lt+\theta_L)}\hat{B}_{\lambda}(t)\right.\\
     &\left.\quad \quad +\hat{C}_{\lambda}(t)\right]+\text{h.c.},
     \end{aligned}
 \end{equation}
where $\theta_L$ is the phase of the drive laser, which we set to zero without loss of generality. Here, the first term corresponds to emission of a photon after absorption of a photon from the laser (hence the factor $e^{-i\omega_Lt}$). As shown in Fig.~\ref{fig:sectorsbasic} (a), the operator $\hat{A}_{\lambda}$  couples the states $\ket{I}$ and $\ket{J}$ within the spin sector and corresponds to Raman scattering of photons. Operators $\hat{B}_{\lambda}$ and $\hat{C}_{\lambda}$ induce transitions between the states belonging to different energy sectors $|I\rangle\rightarrow|K\rangle$ and $|K\rangle\rightarrow|F\rangle$ (see Figs.~\ref{fig:sectorsbasic}(b, c)). The operator $\hat{B}_{\lambda}$ is multiplied by $e^{-2i\omega_Lt}$ because as shown in Fig.~\ref{fig:sectorsbasic} (b), a photon is emitted after absorbing two photons from the laser. Operator $\hat{C}_{\lambda}$, as shown in Fig.~\ref{fig:sectorsbasic} (c) corresponds to spontaneous emission from the higher energy sector. As a result of the dipole approximation, the matter operators contributing to $\hat{M}_{\lambda}$ depend only on the polarization of the emitted photons, encoded in $\lambda$. For the detection of these photons, their polarization must match that of the filter in the corresponding arm of the interferometer. Accordingly, in the following, we adopt a notation $\hat{M}_{j}$, where $j=1,2$ denotes the polarizations of the detectors in our measurement scheme.

To determine the mapping between photonic and matter correlators, let us consider time evolution under the effective Hamiltonian, Eq.~\eqref{eq:Heffint}. From the term $\hat{M}_{\lambda}(t)\hat{a}^{\dagger}_{\lambda} (t)$, we see that the emission of a photon into mode $\lambda$ is accompanied by application of the matter operator $\hat{M}_{\lambda}(t)$ on the matter state. This is analogous to photon emission by an oscillating dipole \cite{dalibard1983correlation}  with $\hat{M}_\lambda$ playing the role of an effective dipolar transition operator. Following this analogy, one can also expect the creation/annihilation operators of the scattered photons to be related to that of the matter via the input-output relation:

\begin{equation}\label{eq:atoMmap}
\hat{a}_{d_{j}}(\tau)\mapsto\int_{0}^{\infty}\dd{t}\tilde{\mathcal{F}}_{j}(t)\hat{M}_{j}(\tau-t).
\end{equation}
In this formula we have also taken into account the modification due to frequency filtering. Physically it means that for a photon to be detected at time $\tau$, it must have been emitted at some earlier time, e.g., $\tau - t$ for $t > 0$, spending the remaining time in the filter. The greater the frequency resolution in the filter, the greater the uncertainty in time $t$.
The complete mapping is summarized in Table~\ref{tab:dictionary1}. The left column lists photonic correlators that can be experimentally measured. The subscript `$\text{out}$' indicates that the measurement is made post scattering in the asymptotic future (as explained in Sec.~\ref{sec:formalism}). When mode $\hat{a}_{d_j}$ corresponds to an inelastically scattered photon, $G^{(1)}_{d_j}$ simply denotes Raman intensity. In Sec.~\ref{sec:photonicdefs}, we explain how the remaining correlators in the left column of Table~\ref{tab:dictionary1} can be measured using standard quantum optics tools, such as homodyne detection and the Hanbury-Brown Twiss setup. The right column maps these photonic correlators to matter correlators in the thermal equilibrium state (in the absence of the laser drive).

To gain some intuition for these matter correlators, let us assume the matter operator $\hat{M}_j(t)$ associated with photon emission creates an excitation. Then two-photon correlations  probe the dynamics of pairs of such excitations. One might na\"ively think that the retarded delay $\tau$ between the detection of the two photons is the same as the delay between the two matter excitation events. However, this is not the case when frequency filtering occurs before detection. The higher the frequency resolution, the greater the importance of interference between amplitudes for excitations created at different times, and hence, more the uncertainty in time. This motivates the convolution between $\hat{M}_j$ and the Fourier transform of the detector's filter function, i.e., $\tilde{\mathcal{F}}_j$ that appears in Eq.~\eqref{eq:atoMmap}. Now let us understand the mapping in Table~\ref{tab:dictionary1} and take $G^{(2)}_{d_1,d_2}(\tau)$ as an example. If we na\"ively substitute Eq.~\eqref{eq:atoMmap} into the photonic correlator, we almost recover the formula in the right column except for the temporal ordering. Physically, this temporal ordering reflects causality of light-matter interaction -- the excitation created at the time of the second-photon emission happens after the first.  We provide a detailed derivation in Sec.~\ref{sec:temporal}. Related treatments of theory of time-resolved spectroscopy for intensities were given in Ref.~\cite{freericks2009theoretical,wang2020time}.

Now let us summarize the microscopic structure of $\hat{M}_j$ coming from Eq.~\eqref{eq:Meffint}. For a detailed explanation, see Sec.~\ref{sec:operators}. As shown earlier in Fig.~\ref{fig:sectorsbasic}, the frequency of emitted photons is either around the central peak $\omega_L$, or located in pairs of sidebands, near $2\omega_L-U$ and $U$. The corresponding matter operators $\hat{A}_j$, $\hat{B}_j$ and $\hat{C}_j$ are studied in detail for the single-band Fermi-Hubbard model in Sec.~\ref{sec:operators}. Here, we summarize their key features. The operator $\hat{A}_{j}$ is a sum of spin singlet terms modulated by the polarizations of the incoming laser and the scattered photon. We define  the spin of an electron at site $\vb{r}$ as $\hat{\vb{S}}_{\vb{r}}$.  To leading order in $\tunn/\abs{\omega_L-U}$, $\hat{A}_j\sim \frac{\tunn^2}{\omega_L-U}\sum_{\vb{r},\bm{\mu}}(\bm{\mu}\cdot \vb{e}^*_j)(\bm{\mu}\cdot \vb{e}_L)\left(4\hat{\vb{S}}_{\vb{r}}\cdot \hat{\vb{S}}_{\vb{r}+\bm{\mu}}-1\right)$, where $\bm{\mu}$  labels the directions available for electron-tunneling. This is the Loudon-Fleury operator~\cite{fleury1968scattering,shastry1991raman,shastry1990theory}, which is defined in detail in Eq.~\eqref{eq:AiExp}). Operator $\hat{A}_j$ can be seen as a sum of $2\times 2$ tensors in the polarization directions, i.e., $(e_j^*)^{u}(e_L)^{v}$ for $u, v \in \{x,y\}$. It can thus be decomposed into channels that are irreducible representations of the crystalline point group of the lattice. One can check that at order $\tunn^2/\abs{\omega_L - U}$, the component of $\hat{A}_j$ in the channel $(e_j^*)^{x}(e_L)^{y}-(e_j^*)^{y}(e_L)^{x}$ (invariant under spatial rotations but odd under reflection) is zero for all lattices. Ref.~\cite{ko2010raman} showed that on the kagome lattice, the leading nonzero term for $\hat{A}_j$ in this channel, appearing at order $\tunn^4/(\omega_L-U)^3$ is a sum of spin chirality operators $\sim \hat{\vb{S}}_{\vb{r}}\cdot \left(\hat{\vb{S}}_{\vb{r}'} \times \hat{\vb{S}}_{\vb{r}''}\right)$. 

Next, the operator $\hat{B}_{j}$ is a mixed spin-charge operator that involves both a spin operator and electron tunneling creating a doublon-hole pair (shown in Fig.~\ref{fig:Kmufig}, \ref{fig:SandC}, and defined precisely in Eq.~\eqref{eq:BiExp}).  Finally, $\hat{C}_{j}$ is a sum of electron tunneling operators and is proportional to the total electric current (See Eq.~\eqref{eq:CiExp}). Given that $\hat{A}_j$ acts within a single energy sector, while $\hat{B}_j$ and $\hat{C}_j$ couple different ones, it is convenient to treat them separately. Therefore, we focus on the case where the filter functions have enough resolution to distinguish the three peaks in Fig.~\ref{fig:sectorsbasic}, while within each peak, the filter may be either broad or narrow. Then,
\begin{equation}\label{eq:MjtoABC}
    \hat{M}_j(t)=\begin{cases}e^{-i\omega_L t}\hat{A}_{j}(t) &\text{ if detector }d_j \text{ detects near $\omega_L$,}\\
    e^{-2i\omega_L t}\hat{B}_{j}(t) &\text{ if detector }d_j \\&\quad \text{  detects near $2\omega_L-U$,}\\
    \hat{C}_{j}(t) &\text{ if detector }d_j \text{ detects near $U$.}
    \end{cases}
\end{equation}
Spectral resolution thus enables selection of the energy window of excitations (see Eq.~\eqref{eq:MjtoABC} above). We thus see that even though light-matter interaction occurred well before photodetection, the frequency filtering during detection can drastically affect the information obtained about the matter. In fact, this is true even for two-photon scattering from a simple two-level system ~\cite{dalibard1983correlation,masters2023simultaneous,orre2019interference}. For example, in the presence of strong frequency filtering, emitted photons can be ``stored'' in the filter, and therefore the signature anti-bunching may disappear.

 \subsection{Applications}
Given the access to new matter correlation functions, what do we learn that cannot be obtained from photon intensity measurements alone? Here, we summarize key applications for Mott insulators. We also show how photonic $G^{(2)}$ and quadrature correlation functions can reveal concrete signatures of spin liquids, which are generally notoriously difficult to observe otherwise.

\subsubsection{Measuring static spin chirality}
A chiral spin liquid is an equivalent of a $1/2$- bosonic fractional quantum Hall state occurring in an electrically neutral system ~\cite{kalmeyer1987equivalence}. Spin chirality operators defined earlier ($\sim \hat{\vb{S}}_{\vb{r}}\cdot (\hat{\vb{S}}_{\vb{r}'} \times \hat{\vb{S}}_{\vb{r}''})$) can spontaneously acquire a nonzero expectation value in chiral spin liquids ~\cite{wen1989chiral,gong2014emergent,he2014chiral,hickey2016haldane,wietek2017chiral,cookmeyer2021four}. So far, there are proposals to use neutron  \cite{lee2013proposal} and Raman scattering ($G^{(1)}(0)$ in our notation) \cite{ko2010raman} to measure the fluctuations in spin chirality. Here, we show that quadrature measurements can directly measure the \textit{static} expectation values of spin chirality in both kagome and triangular lattices.

According to Table~\ref{tab:dictionary1} and Eq.~\eqref{eq:MjtoABC},  when the detector frequency filter is tuned near $\omega_L$, the first-order quadrature measurement $\expval{\hat{a}_{d_j}(0)}_{\text{out}}$ directly provides the static expectation value of operator $\hat{A}_j$. As discussed in Ref.~\cite{ko2010raman}, on the kagome lattice, the operator $\hat{A}_j$ is equal to the sum of spin chirality terms in the channel $(e_j^*)^{x}(e_L)^{y}-(e_j^*)^{y}(e_L)^{x}$, termed as $A_{2g}$. Therefore, the measurement of the quadrature operator allows for the detection of chiral spin liquids.

For the triangular lattice, it was shown in Ref.~\cite{ko2010raman} that the spin chirality terms
vanish at order $\tunn^{4}/(\omega_{L}-U)^{3}$ in the same channel
($(e_{j}^{*})^{x}(e_{L})^{y}-(e_{j}^{*})^{y}(e_{L})^{x}$).
In this work, we show that the expectation
value of the static spin chirality can still be measured using a setup shown in Fig.~\ref{fig:homodyneG2}, and described in Sec.~\ref{sec:photonicdefs} and Appendix~\ref{app:homodyne}. This setup measures the phase-sensitive part of quadrature fluctuations, i.e., two-mode squeezing between photons detected in arms $d_1$ and $d_2$. The detected photons should be linearly polarized and the frequency filters in arms $d_1$ and $d_2$ should be tuned to the different sidebands of the scattered
light, $\sim2\omega_{L}-U$ and $\sim U$ respectively. Within each sideband, the filters
need to be broadband. Further, the detection events are post-selected such that the retarded time-delay $\tau=0$. Specifically, we show in Sec.~\ref{sec:spinchirality} that the two-mode squeezing $\Im\left[\expval{\hat{a}_{d_{2}}(0)\hat{a}_{d_{1}}(0)}_{{\rm {out}}}\right]$ yields $\sim\sum_{{\bf r},{\bf r}',{\bf r}''\in\bigtriangleup}\hat{\vb{S}}_{\vb{r}}\cdot(\hat{\vb{S}}_{\vb{r}'}\times\hat{\vb{S}}_{\vb{r}''})$. For measuring this, one should take linear combinations of the experimental data, so as to extract the component along the polarization channels specified in Sec.~\ref{sec:spinchirality} (Eqs.~(\ref{eq:channelAa}-\ref{eq:channelAb})) that are invariant under rotation and odd under reflection.  We note that although this type of measurement
is accompanied by the real excitation of the charge sector, by post-selecting
detection events having no time delay, one ensures that the system
leaves this sector immediately, and hence the charge sector does not get entangled with
light.

Here, we considered the case $\tau=0$. Next, we discuss the more general case (when $\tau \neq 0$), in which case, the matter operators whose correlation functions are probed, lie in the charged sector.
 \subsubsection{Charged sector dynamics}
By detecting a photon pair in the sidebands, one can extract a mixed spin-charge correlation function. For example, for a Mott insulator, described by the Fermi-Hubbard model at half-filling, the emission of one photon can be accompanied by the creation of a doublon-hole pair at a bond, followed by application of specific spin operators on sites neighboring this bond. These operators include spin singlet projectors. The doublon-hole pair can then propagate till it is forced to recombine by emitting the second photon. The resulting correlator can potentially capture information about possible bound states in the charge sector called Mott excitons~\cite{gallagher1997excitons,essler2001excitons,wrobel2002excitons,tohyama2002resonant,maeda2004third,matsueda2005excitonic,ono2005direct,tohyama2006symmetry,gossling2008mott,novelli2012ultrafast,kim2014excitonic,zhou2014doublon,huang2023spin,mehio2023hubbard} and their dependence on the spin background. We show the general form of such mixed spin-charge operators in Sec.~\ref{sec:R2micro}. However, making quantitative predictions for the charged sector of correlated insulators is the subject of future research.

\subsubsection{$G^{(2)}$ from magnon excitations}
In Sec.~\ref{sec:magnons}, we study the contribution of non-interacting magnon excitations in a spin system to the connected photonic correlator $\mathcal{G}^{(2)}_{d_1,d_2}(\tau)\equiv G^{(2)}_{d_1,d_2}(\tau)-G^{(1)}_{d_1}(0)G^{(1)}_{d_2}(0)$. We demonstrate that $\mathcal{G}^{(2)}_{d_1,d_2}(\tau)$ is zero when the excitations are non-interacting bosons, provided four conditions are met: (1) both the detectors have sharp frequency filters around $\omega_1$ and $\omega_2$ respectively, (2) $\omega_j\neq\omega_L$, i.e., no elastic scattering, (3) $\omega_1\neq\omega_2$, and (4) the polarization symmetry channels (discussed in Sec.~\ref{sec:iomap}) for the two detectors are distinct. In this limit
of sharp spectral resolution, the result is independent of the time
delay $\tau$ due to energy-time uncertainty. Therefore any nonzero contribution to $\mathcal{G}^{(2)}$ should arise from interactions between magnons or from topological properties of excitations.

 \subsubsection{Anyonic excitations -- detecting fractional statistics in spin liquid candidates}
 If the ground state in the spin sector is topologically ordered, its excitations can have fractional statistics. In this case, the emission
of each photon can be accompanied by the creation of an anyon pair.
Anyons, created by different photon emissions, can braid nontrivially
during their propagation. Nontrivial mutual statistics during such braiding can lead to a nonzero contribution to the connected $\mathcal{G}_{d_{1},d_{2}}^{(2)}$. Using the semiclassical
argument developed in \citep{mcginley2024anomalous,mcginley2024signatures},
in Sec.~\ref{sec:subfracstat}, we demonstrate that nontrivial braiding
leads to a universal singularity in the connected part of the $\mathcal{G}_{d_{1},d_{2}}^{(2)}$
function, depending on the photon filter frequencies. More specifically,
we find $G_{d_{1},d_{2}}^{(2)}\sim\theta(\Omega_{1}-\Delta_{1})\theta(\Omega_{2}-\Delta_{2})\left[K_{2}(\Omega_{2})(\Omega_{1}-\Delta_{1})^{-3/2}+(1\leftrightarrow2)\right]$,
where $\Omega_{j}=\omega_{L}-\omega_{j}$ denotes the Raman shift,
$\theta(\omega)$ is the Heaviside step function, $\Delta_{1}$ and
$\Delta_{2}$ are energy thresholds for creating anyon pairs via operators
$\hat{A}_{1}$ and $\hat{A}_{2}$ respectively, and $K_{j}(\Omega_{j})$
are system-specific functions. This singularity is a sharp signature
of the existence of fractional mutual statistics in a spin liquid candidate.
\section{The formalism and definitions in detail}\label{sec:setting}
\begin{figure*}[t]
  \centering
  \includegraphics[width=0.99\textwidth]{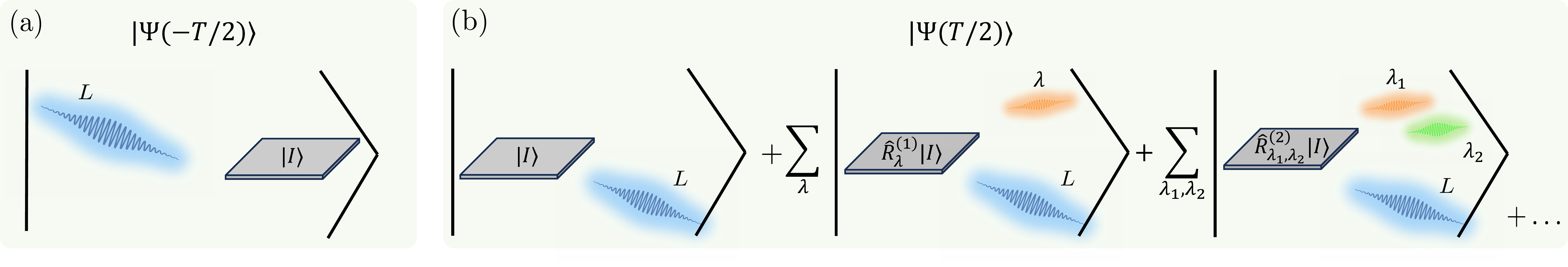}
  \caption{Schematic representation of the scattering process.
\textbf{(a)} The initial state
in the asymptotic past (at time $t=-T/2$, in the limit where it approaches
$-\infty$), $\ket{\Psi(-T/2)}$ consists of the electromagnetic
field in a laser-produced wavepacket state far away from the material.
The material is assumed to be in an energy eigenstate
$\ket{I}$. Around time $t=0$, the wavepacket spatially overlaps
with the material and interacts with it for a duration proportional
to the length of the wavepacket, which we assume to be much larger
than its central wavelength \textbf{(b)}
At $t=T/2$, in the asymptotic future, light and the material are
entangled with each other and the resulting superposition is schematically
depicted in the figure. The expansion depicted here is in the number
of photon modes in the final state. The first term corresponds to
the elastic scattering of light. The second set of terms corresponds
to the emission of a photon in mode $\lambda$, leaving the material
in a state $\hat{R}_{\lambda}^{(1)}\ket{I}$, where $\hat{R}_{\lambda}^{(1)}$
is an operator acting purely in the matter sector. For brevity, we have left out energy-conserving $\delta$-functions in the above schematic (for a more careful treatment, see Eq.~\eqref{eq:outforcoherent}). The third set corresponds to the
emission of a pair of photons in modes $\lambda_{1}$ and $\lambda_{2}$.
Therefore, correlation functions of photons map to correlation functions of matter operators like $\hat{R}^{(1)}_\lambda$ and $\hat{R}^{(2)}_{\lambda_1,\lambda_2}$. In this paper, we present a formalism
to derive expressions for these matter correlation functions. }   
  \label{fig:schem2}
\end{figure*}
In this section, we present the $\T$-matrix formalism that we use to model the scattering process. Then we describe in more detail the photonic correlators studied in this work. 
\subsection{Review of formalism}\label{sec:formalism}
 We model the experimental scenario as a scattering process under a time-independent Hamiltonian $\hat{H}\equiv \hat{H}_0 + \hat{V}$, where $\hat{H}_0$ is the Hamiltonian for light and matter, excluding light-matter interactions, and $\hat{V}$ is the light-matter interaction. The initial state $\ket{\Psi(-T/2)}$ at time $t=-T/2$ (in the limit $T\to \infty$) has light and matter decoupled -- with light being in a laser-produced wavepacket state spatially far away from the material (see Fig.~\ref{fig:schem2}(a)). We assume the initial state of matter is an energy eigenstate $\ket{I}$ of energy $E_I$. This situation can be readily extended to any mixed initial state diagonal in the eigenstate basis.

Around the time $t=0$ for a duration proportional to the length of
the laser wavepacket (which we later take to $\infty$), the wavepacket
spatially overlaps with the material and interacts with it. At $t=T/2$,
which is in the asymptotic future, the electromagnetic part of the
state is once again composed of wavepackets spatially far away from
the material. During the scattering, the electromagnetic
field becomes entangled with the matter, as schematically shown in
Fig.~\ref{fig:schem2}(b). Optical measurements are assumed to be
made at $t=T/2$ in the state $\ket{\Psi(T/2)}\equiv e^{-i\hat{H}T}\ket{\Psi(-T/2)}$.

It is convenient to define $\ket{\text{out}}\equiv e^{i\hat{H}_{0}T/2}\ket{\Psi(T/2)}$,
$\ket{\text{in}}\equiv e^{-i\hat{H}_{0}T/2}\ket{\Psi(-T/2)}$ and
the expectation value of energy in the initial state $E_{\text{in}}^{0}\equiv\mel{\text{in}}{\hat{H}_{0}}{\text{in}}$.
If the experiment duration is sufficiently long, i.e., when the uncertainty
in energy $(\delta E_{\text{in}})\ll1/T$, we can use the $\T$-matrix
formalism \cite{goldberger2004collision} to approximate $\ket{\text{out}}$.
By defining an energy eigenstate (belonging to the full light-matter Hilbert
space) of $\hat{H}_{0}$ with energy $E_{j}^{0}$ as $\ket{\Psi_{j}^{0}}$
we get (see Appendix~\ref{app:Tmatrixreview}):  
\begin{equation}\label{eq:Tmatresrepeat}
\begin{aligned}
    &\ket{\text{out}}\\&=\ket{\text{in}}-\sum_{j,k} 2\pi i \delta(E^0_{j}-E^0_{k})\ket{\Psi^0_k}\bra{\Psi^0_k}\T \ket{\Psi^0_j}\bra{\Psi^0_j}\ket{\text{in}},
    \end{aligned}
\end{equation}
where the $\T$-matrix can be represented as a Dyson series:
\begin{equation}\label{eq:tmatentry}
\begin{aligned}
    \T&=\hat{V} + \hat{V}\frac{1}{E^0_{\text{in}}-\hat{H}_0 - \hat{V}+i0^+} \hat{V}\\
    &=\hat{V}+\hat{V}\hat{\mathbb{G}}_0\hat{V}+\hat{V}\hat{\mathbb{G}}_0\hat{V}\hat{\mathbb{G}}_0\hat{V}+\hat{V}\hat{\mathbb{G}}_0\hat{V}\hat{\mathbb{G}}_0\hat{V}\hat{\mathbb{G}}_0\hat{V}+\ldots,
\end{aligned}
\end{equation}
with the free propagator denoted as $\hat{\mathbb{G}}_0\equiv \left(E^0_{\text{in}}-\hat{H}_0+i0^+\right)^{-1}$. Eq.~\eqref{eq:Tmatresrepeat} is a generalization of Fermi's Golden rule to all orders in $\hat{V}$. In Appendix~\ref{app:Tmatrixreview}, we provide a review of the $\T$-matrix formalism along with a derivation of Eq.~\eqref{eq:Tmatresrepeat}.

\subsection{Definition and measurement of photonic correlation functions}\label{sec:photonicdefs}

We now consider a wide class of photonic correlation
functions (to be introduced in this section) which can be measured
in a quantum optical setting. As a specific example, we focus on the
setup shown in Fig.~\ref{fig:schem1}. We assume
the detectors $d_{1}$ and $d_{2}$ are located at $\vb{R}_{1}$ and
$\vb{R}_{2}$ respectively (Fig.~\ref{fig:schem1}). The
spatial origin is chosen to be at the center of mass of the material.
The positive-frequency part of the filtered electric field operator
(in the Interaction Picture) along the direction of the detected polarization
$\vb{e}_{j}$, seen by the detector $d_{j}$ at location $\vb{R}_{j}$
(for $j\in\{1,2\}$), is given by 
\begin{equation}
\begin{aligned} & \vb{e}_{j}^{*}\cdot\hat{\vb{E}}^{(+)}\left(\vb{R}_{j},\tfrac{T}{2}\right)\\
 & =i\sum_{\vb{k}}\sqrt{\frac{\omega_{\vb{k}}}{2\varepsilon\mathcal{V}}}\hat{a}_{\vb{k},\vb{e}_{j}}e^{i\left(\vb{k}\cdot\vb{R}_{j}-\omega_{\vb{k}}\tfrac{T}{2}\right)}f_{j}(\omega_{\vb{k}}),\text{ and}\\
 & \hat{\vb{E}}^{(-)}(\vb{R}_{j})\equiv\left[\hat{\vb{E}}^{(+)}(\vb{R}_{j})\right]^{\dagger},
\end{aligned}
\label{eq:det1}
\end{equation}
where $\hat{a}_{\vb{k},\vb{e}_{j}}$ is a photon
annihilation operator for each orthogonal plane-wave mode of momentum
$\vb{k}$ and polarization $\vb{e}_{j}$, and $\omega_{\vb{k}}=c\abs{\vb{k}}$.
Here, $\varepsilon$ is the dielectric constant and $\mathcal{V}$
is the mode-volume of free-space (which can be taken to infinity at
a suitable point). Each detector $j$ could have its own sensitivity
profile for modes of different frequencies determined by $f_{j}(\omega)$,
a dimensionless filter function. Causality implies that the poles
of $f_{j}(\omega)$, when seen as a function of complex frequency
$\omega$, can only lie in the lower half-plane.

We now assume $\abs{\vb{R}_{1}}=cT/2$ for simplicity.
Furthermore, we assume the two detectors $j=1,2$ are sensitive to
the photons having momenta $\vb{k}$ within the thin solid angles
$\delta S_{j}$ around the directions $\vb{R}_{j}$. We also allow
for the time delay between the detection events $\tilde{\tau}$. It
is also convenient to introduce the retarded time $\tau$ as 
\begin{equation}
\tau\equiv\tilde{\tau}-\frac{\abs{\vb{R}_{2}}-\abs{\vb{R}_{1}}}{c}.\label{eq:taudef}
\end{equation}
We can now define an effective annihilation operator (which
is related to the local electric field)
that describes the photodetection by the detector $d_{j}$, as follows:
\begin{equation}
\hat{a}_{d_{j}}\equiv i\sum'_{\vb{k}}\sqrt{\frac{c\omega_{\vb{k}}}{2\mathcal{V}}}f_{j}(\omega_{\vb{k}})\hat{a}_{\vb{k},\vb{e}_{j}},\label{eq:addef}
\end{equation}
where $\sum'_{\vb{k}}$ restricts the direction of
$\vb{k}$ to be within a solid angle $(\delta S)_{j}$ around the
direction $\vb{R}_{j}$. The sum over $\vb{k}$ can be converted into an integral over frequencies
$\omega_{\vb{k}}$: 
\begin{equation}
\sum'_{\vb{k}}\sqrt{\frac{c\omega_{\vb{k}}}{2\mathcal{V}}}\longrightarrow(\delta S)_{j}\int\dd{\omega_{\vb{k}}}\sqrt{\frac{c\omega_{\vb{k}}\mathcal{V}}{2}}\rho(\omega_{\vb{k}}).\label{eq:sumtoint}
\end{equation}
Here, $\rho(\omega)$ represents the density of states
of light modes per unit volume. In free space, $\rho(\omega)=\omega^{2}/(2\pi c)^{3}$.
More generally, this density can be altered, for example, by the presence
of a cavity. The conversion from a summation over wavevectors ${\bf k}$
to an integral over frequencies $\omega$ is valid because all matrix
elements involving $\hat{a}_{\vb{k},\vb{e}_{j}}$ depend on ${\bf k}$
only through $\omega_{{\bf k}}$. We also define an effective filter
function (for $j\in\{1,2\}$): 
\begin{equation}\label{eq:filterredef}
    \mathcal{F}_j(\omega)\equiv 2\pi c \rho(\omega)(\delta S)_j f_j(\omega).
\end{equation}
When the filter is represented by a Fabry-Pérot cavity, we get:
\begin{equation}\label{eq:Lorentzdef}
    \mathcal{F}_{j,\text{ Lorentzian}}(\omega)=\frac{i\Kconst_j\Gamma_j}{\omega - \omega_j+i\Gamma_j},
\end{equation}
where $\omega_j$ denotes the resonant frequency and 
\begin{equation}
\label{eq:Kdef}
    \Kconst_j \equiv  2\pi c\rho(\omega_j)(\delta S)_j.
\end{equation} 
In deriving this expression, we assumed that the photonic density of states remains constant within the frequency window (filter resolution) defined by \(\Gamma_j\), centered around \(\omega_j\). Similarly, other causal effective filter functions with different selectivity profiles could be considered~\cite{ngaha2023frequency}. 

With this setup in mind, we can provide an explicit mathematical definition of the photonic correlation functions.

\subsubsection{Photon intensity $G^{(1)}(0)$}
The intensity of light measured by the detector $d_j$ of polarization $\vb{e}_j$, located at point $\vb{R}$ is 
\begin{equation}
\begin{aligned}
    &G^{(1)}_{d_j}(0)\equiv c\varepsilon\\ 
    &\times \bra{\Psi(T/2)}\left(\vb{e}_j \cdot \hat{\vb{E}}^{(-)}(\vb{R})\right)\left(\vb{e}^*_j\cdot\hat{\vb{E}}^{(+)}(\vb{R})\right)\ket{\Psi(T/2)}.
\end{aligned} 
\end{equation}
Here, the states and operators are represented in the Schrödinger picture. Transitioning to the interaction picture, using the definitions of \(\ket{\text{in}}\), \(\ket{\text{out}}\), and the mode \(\hat{a}_{d_j}\) detected by the detector in \eqref{eq:addef}, we can simplify the definition of \(G^{(1)}\) to:
\begin{equation}\label{eq:G1def}
     G^{(1)}_{d_j}(0)= \bra{\text{out}}\hat{a}_{d_j}^{\dagger}(0)\hat{a}_{d_j}(0)\ket{\text{out}}\equiv \expval{\hat{a}_{d_j}^{\dagger}(0)\hat{a}_{d_j}(0)}_{\text{out}}.
\end{equation}
Here, we have used the notation $\expval{.}_{\text{out}}$ to denote an expectation value taken in state $\ket{\text{out}}$.
\subsubsection{Photon-pair correlation function $G^{(2)}(\tau)$}
$G^{(2)}(\tau)$ characterizes the coincidence detection rate of two photons with a time delay \(\tilde{\tau}\) (or \(\tau\), as defined in terms of the retarded time in Eq.~\eqref{eq:taudef}). Let the polarization states selected by the two detectors be \(\mathbf{e}_{1}\) and \(\mathbf{e}_{2}\), respectively. The photon-pair correlation function can then be defined as:
\begin{equation}\label{eq:G2def1}
\begin{aligned}
    &G^{(2)}_{d_1,d_2}(\tilde{\tau})=c^2 \varepsilon^2\\
    &\times \bra{\Psi(T/2)}\left[\vb{e}_1 \cdot \hat{\vb{E}}^{(-)}(\vb{R}_1)\right]e^{i\hat{H}\tilde{\tau}}\left[\vb{e}_2 \cdot \hat{\vb{E}}^{(-)}(\vb{R}_2)\right]\\
    &\quad \times \left[\vb{e}^*_2 \cdot \hat{\vb{E}}^{(+)}(\vb{R}_2)\right]e^{-i\hat{H}\tilde{\tau}}\left[\vb{e}^*_1 \cdot \hat{\vb{E}}^{(+)}(\vb{R}_1)\right]\ket{\Psi(T/2)}.
\end{aligned}
\end{equation}
The exponentials $e^{-i\hat{H}\tilde{\tau}}$ in this expression represent the fact that the whole system continues to evolve between the detection of the first and second photons. Let us suppose that the detection is in the far-field limit, i.e., the distance from the material $L=cT/2$ to the detector is much larger than the wavelengths of incident and scattered light, as well as the correlation length of the sample. Moreover, the time delay between detection events $\tau$ should be much smaller than $1/\gamma$, where $\gamma$ is a relaxation rate of matter excitations, given by the maximum of various contributions such as spontaneous emission and phonon-mediated relaxation.   Under these conditions, (see Fig.~\ref{fig:schem2}(b)), the light-matter interaction $\hat{V}$ has almost zero expectation value in the final state. This is because the light-matter interaction is $\sim \hat{\vb{A}}_T(\vb{x})\cdot \hat{\bm{\mathcal{J}}}(\vb{x})$ (where $\hat{\vb{A}}_T(\vb{x})$ is the transverse part of the electromagnetic field and $\bm{\mathcal{J}}(\vb{x})$ is the electric current), and the support of $\hat{\vb{A}}_T(\vb{x})$ and that of $\hat{\mathcal{\bm{J}}}(\vb{x})$ are spatially separated. Therefore, $\hat{V}$ does not have any effect after $t=T/2$. One can thus replace $e^{-i\hat{H}\tilde{\tau}}$ with the free evolution $e^{-i\hat{H}_0\tilde{\tau}}$ in Eq.~\ref{eq:G2def1}. This equation can be even further simplified in the interaction picture, by using the retarded time $\tau$ (see Eq.~\ref{eq:taudef}), and using Eq.~\eqref{eq:addef} to obtain: 
\begin{equation}\label{eq:G2def2}
    G^{(2)}_{d_1,d_2}(\tau)= \expval{\hat{a}^{\dagger}_{d_1}(0)\hat{a}_{d_2}^{\dagger}(\tau)\hat{a}_{d_2}(\tau)\hat{a}_{d_1}(0)}_\text{out}.
\end{equation}
The ``connected" component of this correlation function can be defined as:
\begin{equation}\label{eq:G2connected}
    \begin{aligned}
        \mathcal{G}^{(2)}_{d_1,d_2}(\tau)\equiv G^{(2)}_{d_1,d_2}(\tau)-G^{(1)}_{d_1}(0)G^{(1)}_{d_2}(0).
    \end{aligned}
\end{equation}
\subsubsection{Quadrature measurements}\label{sec:quadraturemeasurements}
\begin{figure*}
  \centering
  \includegraphics[width=0.93\textwidth]{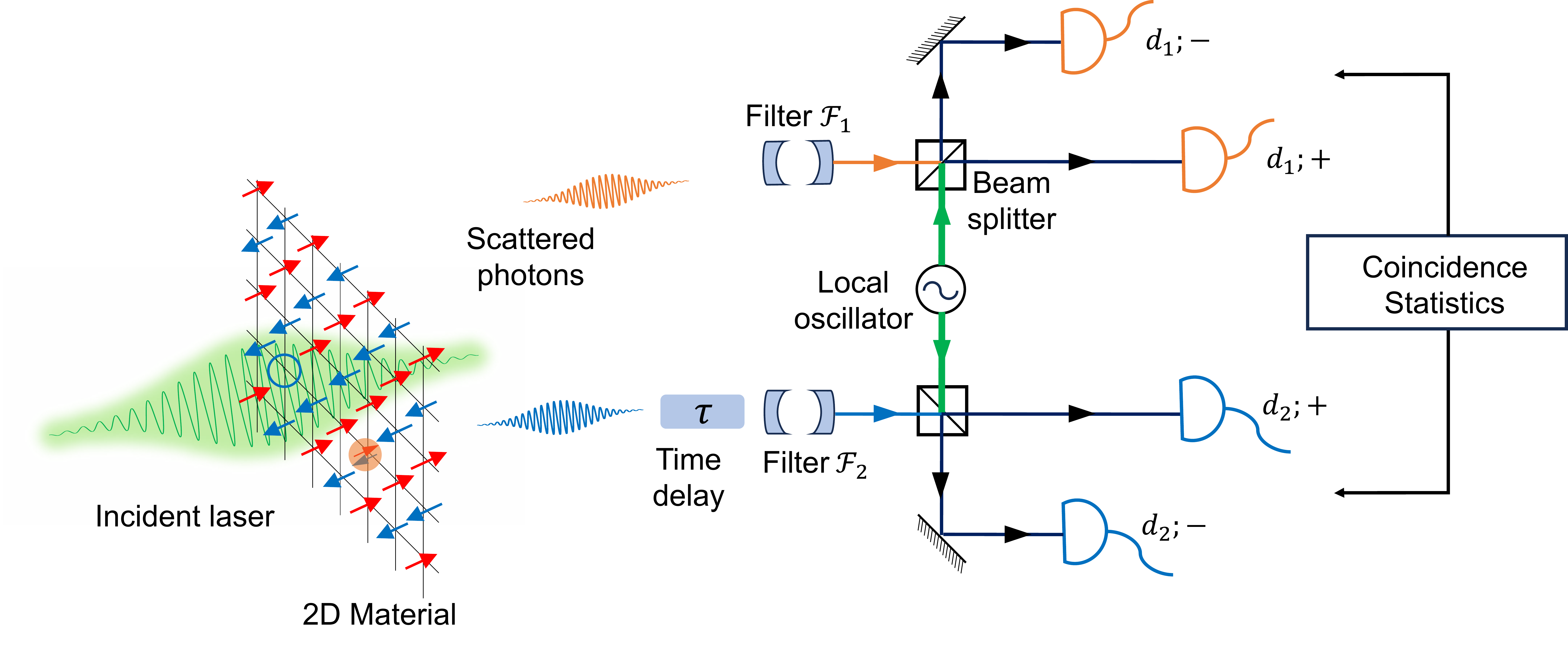}
  \caption{Optical scheme for the measurement of the phase-sensitive second-order quadrature correlations between a pair of photons scattered off the material Eq.~(\ref{eq:X2def}). One of the photons (shown as blue) is subjected to an additional (retarded) time delay $\tau$. After both photons pass through the respective frequency filters $\mathcal{F}_{j}$, each photon is mixed with a strong field of a local oscillator (annihilation operator denoted as $\hat{a}_{\text{L.O.}}$) using a beamsplitter. In our work, we consider the frequency of the local oscillator to be equal to the drive frequency $\omega_L$. The phase difference of the local oscillator with respect to the drive laser can be tuned. If $\hat{a}_{\lambda_j}$ is a scattered photonic mode, then for each of the two beam-splitters, the mode through each of the two output arms, $+1$ and $-1$ is $\tfrac{1}{\sqrt{2}}\left(\hat{a}_{\lambda_j}\pm \hat{a}_{\text{L.O.}} \right)$ respectively. First, let us consider the output from just one of the filters $\mathcal{F}_j$. The difference $G^{(1)}_{d_j;+}-G^{(1)}_{d_j;-}$ between the two arms of the beam splitter is proportional to a quadrature measurement $\expval{\opa_{d_j}}e^{i\theta}+\text{c.c.}$ Now, let us consider the output from both the filters $\mathcal{F}_1$ and $\mathcal{F}_2$. We show that by measuring $G^{(2)}$ correlations between the output arms of the beam-splitter and by taking a suitable linear combination [Eq.~\eqref{eq:homodynetrick}], one can measure phase-sensitive second-order quadrature correlations between the scattered photons $\expval{\opa_{d_2}(\tau)\opa_{d_1}(0)}$.}
  \label{fig:homodyneG2}
\end{figure*}
So far, we considered the photonic correlation functions $G^{(1)}(0)$ and $G^{(2)}(\tau)$, which are related to photon \textit{number} and its \textit{fluctuations} respectively, and lack information about the phase of the electric field. The phase-sensitive information can be measured by using a homodyne detection scheme where the scattered photons are mixed with a strong field of a local oscillator (see for example, Chapter 4 of Ref.~\cite{scully1997quantum} and Chapter 7 of Ref.~\cite{walls2008quantum}). We review below how one can measure an arbitrary ``quadrature" of the electric field $\hat{a}_{\lambda}e^{i\theta}+\text{h.c.}$, for any $\theta$. The setup is shown in Fig.~\ref{fig:homodyneG2}. 

Consider a strong local oscillator $\hat{a}_{\text{L.O.}}$ with the same frequency $\omega_L$ as the incoming laser. Here, we will replace $\sqrt{\tfrac{\omega_L c}{\mathcal{V}}}\hat{a}_{\text{L.O.}}(t)$ by its classical expectation value $-i\sqrt{I_{\text{L.O.}}} e^{-i(\omega_L t + \theta+\theta_L)}$. Here, $I_{\text{L.O.}}\equiv \tfrac{\langle\hat{a}^{\dagger}_{\text{L.O.}} \hat{a}_{\text{L.O.}}\rangle\omega_L c}{\mathcal{V}}$ is the intensity of the local oscillator, $\mathcal{V}$ is its mode volume, and $\theta$ is the tunable \textit{relative} phase difference between the local oscillator and the drive laser (whose phase is $\theta_L$).

In our proposed setup, each scattered photon after passing through the frequency filter is admixed with the local oscillator using a beamsplitter. If the input modes going into a beamsplitter are $\hat{a}_{\lambda}$ and $\hat{a}_{\text{L.O.}}$, the output modes are $\tfrac{1}{\sqrt{2}}\left(\hat{a}_{\lambda}\pm\hat{a}_{\text{L.O.}}\right)$. Under filtering followed by admixture with the local oscillator (as shown in Fig.~\ref{fig:homodyneG2}), the annihilation operator $\hat{a}_{d_j}$ corresponding to detector $d_j$ (defined in Eq.~\eqref{eq:addef}) gets transformed as
\begin{equation}\label{eq:homodynetransform}
     \hat{\mathbb{a}}_{d_j;s}(t)\equiv \frac{\hat{a}_{d_j}(t)}{\sqrt{2}}\pm \frac{\sqrt{I_{\text{L.O.}}}}{2} e^{-i\left(\omega_L t+\theta+\theta_L\right)},
 \end{equation}
  depending on $s$, the output arm of the beamsplitter, where $s$ can be $+$ or $-$. 
  
First, one measures $G^{(1)}(0)$ of the transformed modes: 
\begin{equation}\label{eq:G1homodyne}
     G^{(1)}_{d_j;s}(\theta;0;t)=\expval{\hat{\mathbb{a}}_{d_j;s}^{\dagger}(t)\hat{\mathbb{a}}_{d_j;s}(t)}_{\text{out}}.
 \end{equation}
Here, in writing $G^{(1)}_{d_j;s}(0;t)$, the $\theta$ in the argument refers to the phase of the local oscillator with respect to the drive, the $0$ implies that the creation and annihilation operators are at equal time, while $t$ denotes the time at which the measurement is done. The $t$-dependence arises because of the time-dependence of the local oscillator.

 By substituting Eq.~\eqref{eq:homodynetransform} in Eq.~\eqref{eq:G1homodyne}, one can show that
 \begin{equation}\label{eq:quadraturetrick}
 \begin{aligned}
    &G^{(1)}_{d_j;+}(\theta;0;t)-G^{(1)}_{d_j;-}(\theta;0;t)\\
    &=\sqrt{\frac{I_{\text{L.O.}}}{2}}\left(\expval{\hat{a}_{d_j}(t)}_{\text{out}}e^{i(\omega_L t + \theta)}+ \text{c.c.}\right).
    \end{aligned}
 \end{equation}
 We show in Appendix~\ref{app:homodyne} that upon time-averaging the above measurement, one can extract the $t$-independent contribution $\expval{\opa_{d_j}(0)e^{i\theta}}+\text{c.c.}$ reported in Table~\ref{tab:dictionary1}. By taking appropriate linear combinations of this quantity, one can thus measure: 
 \begin{equation}\label{eq:X1def}
    X^{+}_{d_j}(0)= \expval{\hat{a}_{d_j}(0)}_\text{out}.
\end{equation}
\subsubsection{Phase-sensitive second-order quadrature measurements}
 Furthermore, we show below that the setup in Fig.~\ref{fig:homodyneG2} can be used to obtain phase-sensitive second order quadrature correlations: 
 \begin{equation}\label{eq:X2def}
     X^{++}_{d_1,d_2}(\tau)= \expval{\hat{a}_{d_2}(\tau)\hat{a}_{d_1}(0)}_\text{out}.
\end{equation}
 To obtain the above, we consider measurements of $G^{(2)}$ between different arms of the beamsplitters:\begin{equation}\label{eq:G2homodyne}
\begin{aligned}
     &G^{(2)}_{d_1,d_2;s_1,s_2}(\theta;\tau;t)\\
     &=\expval{\hat{\mathbb{a}}_{d_1;s_1}^{\dagger}(t)\hat{\mathbb{a}}_{d_2;s_2}^{\dagger}(t+\tau)\hat{\mathbb{a}}_{d_2;s_2}(t+\tau)\hat{\mathbb{a}}_{d_1;s_1}(t)}_{\text{out}}.
\end{aligned}
 \end{equation}
By substituting Eq.~\eqref{eq:homodynetransform} in Eq.~\eqref{eq:G2homodyne}, one can show that (we suppress the indices $d_1,d_2$ for simplicity):
 \begin{equation}\label{eq:homodynetrick}
 \begin{aligned}
     &\left(G^{(2)}_{+,+}+G^{(2)}_{-,-}-G^{(2)}_{+,-}-G^{(2)}_{-,+}\right)(\theta;\tau;t)\\
     &=\frac{I_{\text{L.O.}}}{2}\left[e^{2i(\omega_L t +\theta)+i\omega_L \tau}\expval{\hat{a}_{d_2}(t+\tau)\hat{a}_{d_1}(t)}_{\text{out}}+\text{c.c.}\right.\\
     &\quad \quad \quad \quad +\left. e^{-i\omega_L \tau}\langle\hat{a}_{d_2}^{\dagger}(t+\tau)\hat{a}_{d_1}(t) \rangle_{\text{out}} + \text{c.c.} \right].
\end{aligned}
 \end{equation}
The above correlation function comprises both a phase-dependent contribution $e^{2i(\omega_L t +\theta)+i\omega_L \tau}\expval{\hat{a}_{d_2}(t+\tau)\hat{a}_{d_1}(t)}_{\text{out}}$ as well as a phase-independent contribution $e^{-i\omega_L \tau}\langle\hat{a}_{d_2}^{\dagger}(t+\tau)\hat{a}_{d_1}(t) \rangle_{\text{out}}$. These two terms can be experimentally separated out by measuring Eq.~\eqref{eq:homodynetrick} for different values of $\theta$, because the former depends on $\theta$ (the phase of the local oscillator), but the latter does not. In other words, we consider the following difference for different values of $\theta_A$ and $\theta_B$:
\begin{equation}\label{eq:homodynetrick2}
 \begin{aligned}
     &\left(G^{(2)}_{+,+}+G^{(2)}_{-,-}-G^{(2)}_{+,-}-G^{(2)}_{-,+}\right)(\theta_A;\tau;t)\\
     -&\left(G^{(2)}_{+,+}+G^{(2)}_{-,-}-G^{(2)}_{+,-}-G^{(2)}_{-,+}\right)(\theta_B;\tau;t)\\
     =&\frac{I_{\text{L.O.}}}{2}\left[\left(e^{2i\theta_A}-e^{2i\theta_B}\right)e^{i\omega_L( 2t + \tau)}\expval{\hat{a}_{d_2}(t+\tau)\hat{a}_{d_1}(t)}_{\text{out}}\right.\\
     &\quad \quad \quad \quad \left.+ \text{ c.c.} \right].
\end{aligned}
 \end{equation}
Once again, the above equation has a $t$-dependence. We show in Appendix~\ref{app:homodyne} that by time-averaging, one can extract the phase-sensitive second order quadrature correlator presented in Table~\ref{tab:dictionary1} and Eq.~\eqref{eq:X2def}.

We note that the above phase-sensitive second order quadrature measurement shown in Fig.~\ref{fig:homodyneG2} is related to two-mode squeezing that has been measured experimentally \cite{boyer2008entangled,de2024characterizing}.
\begin{figure}
  \centering
  \includegraphics[width=0.48\textwidth]{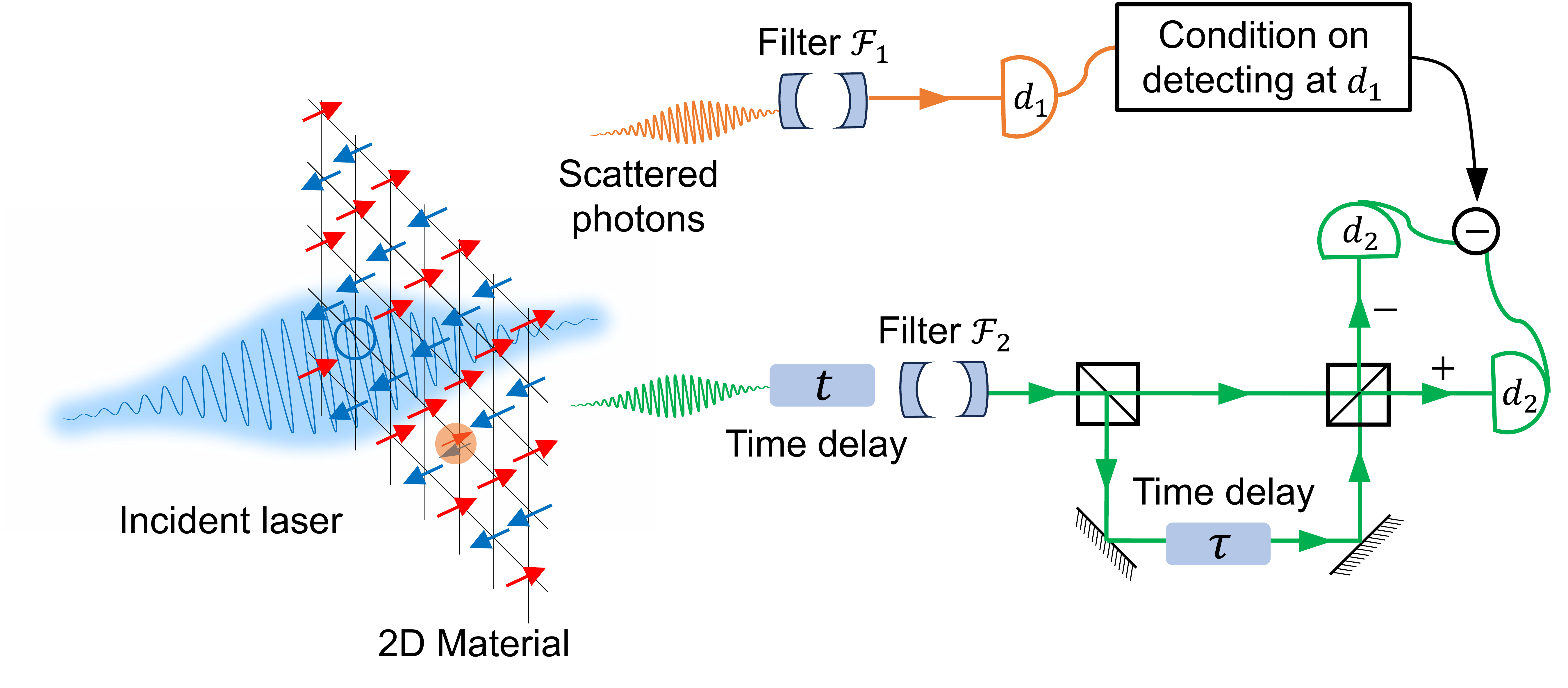}
  \caption{Measurement scheme for $H_{d_1,d_2}(t,\tau)\equiv \langle\hat{a}^{\dagger}_{d_1}(0)\hat{a}^{\dagger}_{d_2}(t+\tau)\hat{a}_{d_2}(t)\hat{a}_{d_1}(0)\rangle_{\text{out}}+\text{c.c.}$, which can be thought of as a $G^{(1)}(\tau)$ measurement at detector $d_2$ conditioned on detecting a photon at $d_1$. The mode shown in green here is split into two paths using a beamsplitter, and one of the paths is given a time delay $\tau$ with respect to the other. The two paths are made to interfere, and two $G^{(2)}$ measurements are taken between $d_1$ and each arm of $d_2$. We show in Appendix~\ref{app:homodyne} that the desired correlator can be obtained this way.}
  \label{fig:ConditionalG2}
\end{figure}
\subsubsection{$G^{(1)}(\tau)$ and Conditional $G^{(1)}(\tau)$}
By beating a mode with its own time-delayed counterpart, one can also measure the following correlator:
\begin{equation}\label{eq:X2minusdef}
     G^{(1)}_{d_j}(\tau)= \langle\hat{a}^{\dagger}_{d_j}(\tau)\hat{a}_{d_j}(0)\rangle_\text{out}+\text{c.c.}.
\end{equation}
In this work, we also consider the following correlation function, obtained by measuring $G^{(1)}(\tau)$ by detector $d_2$ conditioned on detecting a photon at $d_1$. 
\begin{equation}\label{eq:ConditionalG1}
     H_{d_1,d_2}(t,\tau)= \langle\hat{a}^{\dagger}_{d_1}(0)\hat{a}^{\dagger}_{d_2}(t+\tau)\hat{a}_{d_2}(t)\hat{a}_{d_1}(0)\rangle_\text{out}+\text{c.c.}
\end{equation}
The corresponding measurement scheme is shown in Fig.~\ref{fig:ConditionalG2} and described in Appendix~\ref{app:homodyne}.

Having defined the photonic correlators, we now turn to the matter sector in the next section. To establish a connection from the photonic correlators in Eq.~\eqref{eq:G1def}, Eq.~\eqref{eq:G2def2}, and Eq.~(\ref{eq:X1def}-\ref{eq:ConditionalG1}), to the corresponding matter correlation functions, it is necessary to know the form of the $\T$-matrix that acts on $\ket{\text{in}}$. This requires knowledge of the material's Hamiltonian $\hat{H}_0$ and light-matter interaction $\hat{V}$.
\section{Light scattering off a single-band Fermi-Hubbard model at half-filling}\label{sec:Hubbardsetup}
In this section, we describe the microscopic matter Hamiltonian, light-matter interaction, and the relevant energy scales. We explicitly work out the mapping between photonic and matter correlations when $\hat{H}_0$ is a single-band Fermi-Hubbard model of spin $1/2$ electrons at half-filling, and the light-matter interaction $\hat{V}$ is obtained by Peierls' substitution. Our procedure can then be readily adapted to any other system as long as its ground state is an insulator, i.e., the electrically charged degrees of freedom are gapped. Let us now consider the following Hamiltonian,
\begin{equation}\label{eq:1bhm}
\begin{aligned}
    \hat{H}_0=& \sum_{(\vb{r},\bm{\mu}),\sigma}\left[-\tunn_{\vb{r},\vb{r}+\bm{\mu}}\hat{c}^{\dagger}_{\vb{r}+\bm{\mu}, \sigma}\hat{c}_{\vb{r}, \sigma}\right]+\text{h.c.}+U\sum_{\vb{r}}\hat{n}_{\vb{r},\uparrow}\hat{n}_{\vb{r},\downarrow}\\
&+\sum_{\vb{k},\vb{e}_{\vb{k}}} \omega_{\vb{k}}\left(\hat{a}^{\dagger}_{\vb{k},\vb{e}_{\vb{k}}}\hat{a}_{\vb{k},\vb{e}_{\vb{k}}}+\frac{1}{2}\right),
\end{aligned}
\end{equation}
where $\tunn_{\vb{r},\vb{r}+\bm{\mu}}$ is the tunneling coefficient between sites $\vb{r}$ and $\vb{r}+\bm{\mu}$. We denote the nearest-neighbor tunneling by $\tunn$. For example, for a square lattice with nearest-neighbor tunneling, $\bm{\mu}$ is $\alatt( 1,0)$ or $\alatt(0, 1)$. However, in general, $\bm{\mu}$ can connect site $\vb{r}$ to an arbitrary site $\vb{r}+\bm{\mu}$, if $\tunn_{\vb{r},\vb{r}+\bm{\mu}}\neq 0$, and we say there is a \textit{bond} $\left(\vb{r},\bm{\mu}\right)$ regardless of whether it is nearest-neighbor. To avoid double counting, $\bm{\mu}$ runs through half the set of bonds.  Operator $\hat{c}^{\dagger}_{\vb{r},\sigma}$ creates an electron of spin $\sigma=\uparrow$ or $\downarrow$  in a Wannier orbital localized at lattice site $\vb{r}$. The electronic spin operator is defined as:
\begin{equation}\label{eq:Stofermions}
    \hat{\vb{S}}_{\vb{r}}=\frac{1}{2}\sum_{\alpha,\beta}\hat{c}_{\vb{r},\alpha}^{\dagger}\bm{\sigma}\hat{c}_{\vb{r},\beta},
\end{equation}
where $\bm{\sigma}$ is a vector formed by the three Pauli matrices.

 Under the Peierls' substitution, the full light-matter Hamiltonian $\hat{H}$ is obtained from $\hat{H}_0$ by substituting $\tunn_{\vb{r},\vb{r}+\bm{\mu}}$  with
\begin{equation}\label{eq:peierls}
    \tunn_{\vb{r},\vb{r}+\bm{\mu}} \to \tunn_{\vb{r},\vb{r}+\bm{\mu}} e^{iq_e\int_{\vb{r}}^{\vb{r}+\bm{\mu}} \dd{\vb{x}}\cdot \hat{\vb{A}}(\vb{x})},
\end{equation}
where $q_e$ is the charge of the electron,  $\hat{\vb{A}}(\vb{r})$ is the vector potential of the electromagnetic field. By Helmholtz decomposition, any vector field can be decomposed uniquely as a sum of transverse (divergence-free) and longitudinal (curl-free) components. The transverse component of $\hat{\vb{A}}(\vb{r})$ is invariant under an arbitrary gauge transformation $\vb{A}(\vb{r})\to \vb{A}(\vb{r}) +\grad\theta(\vb{r})$ for some scalar field $\theta(\vb{r})$. The transverse component, called $\hat{\vb{A}}_T(\vb{r})$ can be expanded in terms of the normal modes of radiation in free space, $\hat{a}_{\vb{k},\vb{e}_{\vb{k}}}$ (introduced in Eq.~\eqref{eq:det1}) as:  
\begin{equation}\label{eq:Aexp}
    \hat{\vb{A}}_T(\vb{r})=\sum_{\vb{k},\vb{e}_{\vb{k}}}\frac{1}{\sqrt{2\varepsilon \mathcal{V}\omega_{\vb{k}}}}\left(\vb{e}_{\vb{k}}\hat{a}_{\vb{k},\vb{e}_{\vb{k}}}e^{i\vb{k}\cdot \vb{r}}+\vb{e}_{\vb{k}}^*\hat{a}^{\dagger}_{\vb{k},\vb{e}_{\vb{k}}}e^{-i\vb{k}\cdot \vb{r}}\right)
\end{equation}
and $\vb{e}_{\vb{k}}$ is the mode polarization, and satisfies $\vb{k}\cdot \vb{e}_{\vb{k}}=0$. We note that quantizing the transverse part of the vector potential is a gauge-invariant notion. For simplicity, we also choose the Coulomb gauge, i.e.,

\begin{equation}
    \vb{A}(\vb{r})=\vb{A}_T(\vb{r}).
\end{equation}
In the Coulomb gauge, longitudinal electric fields are obtained entirely from the scalar potential $\phi$ as $\vb{E}_L(\vb{r})=-\grad \phi$. Gauss's law implies that $-\grad^2 \phi(\vb{x}) = q_e \rho(\vb{x})$, where $\rho(\vb{x})$ is the electron density. Integrating out $\phi$ results in the usual density-density Coulomb interaction found in all electronic systems. For the purpose of this work, this Coulomb interaction can be thought to have been screened leaving behind the on-site Hubbard repulsion in the Hamiltonian Eq.~\eqref{eq:1bhm}. Thus, longitudinal electric fields arise purely from matter density fluctuations and not from electromagnetic radiation. Therefore, we can consider light-matter interactions in the Coulomb gauge without loss of generality.

Thus, the light-matter interaction $\hat{V}$ (defined such that the full Hamiltonian $\hat{H}=\hat{H}_0 +\hat{V}$), resulting from Peierls' substitution in Eq.~\eqref{eq:peierls} is given by:

\begin{equation}\label{eq:lmpeierls}   
    \begin{aligned}
        &\hat{V}=\sum_{(\vb{r},\bm{\mu}),\sigma}\left\{-i\left(\tunn_{\vb{r},\vb{r}+\bm{\mu}}\hat{c}^{\dagger}_{\vb{r}+\bm{\mu}, \sigma}\hat{c}_{\vb{r}, \sigma}-\text{h.c.}\right)\sin\left(q_e \hat{A}_{\vb{r},\vb{r}+\bm{\mu}}\right)\right.\\
        &\quad \left.  +\left(\tunn_{\vb{r},\vb{r}+\bm{\mu}}\hat{c}^{\dagger}_{\vb{r}+\bm{\mu}, \sigma}\hat{c}_{\vb{r}, \sigma}+\text{h.c.}\right)\left[1-\cos\left(q_e \hat{A}_{\vb{r},\vb{r}+\bm{\mu}}\right)\right] \right\},
    \end{aligned}
\end{equation}
where $\hat{A}_{\vb{r},\vb{r}+\bm{\mu}}\equiv \int_{\vb{r}}^{\vb{r}+\bm{\mu}} \dd{\vb{x}}\cdot \hat{\vb{A}}(\vb{x})\approx  \hat{\vb{A}}\left(\vb{r}+\bm{\mu}/2\right)\cdot \bm{\mu}$.

A remark is due here regarding Peierls' substitution. The single-band Fermi-Hubbard model should be viewed as the projection of a continuum model into the lowest energy Wannier orbital.  Determining the true light-matter coupling term in such projected models is not trivial. The interaction terms could generically be modified \cite{li2020electromagnetic,dmytruk2021gauge}, for instance, by pair-hopping terms modulated by coupling to the gauge field. In this work, however, following Ref.~\cite{shastry1990theory,shastry1991raman,ko2010raman}, we restrict ourselves to the toy problem where the light-matter coupling is obtained by Peierls' substitution, which is manifestly gauge-invariant.
\subsection{Overview of energy scales and sectors}\label{sec:scales} 
\begin{figure}
  \centering
  \includegraphics[width=0.44\textwidth]{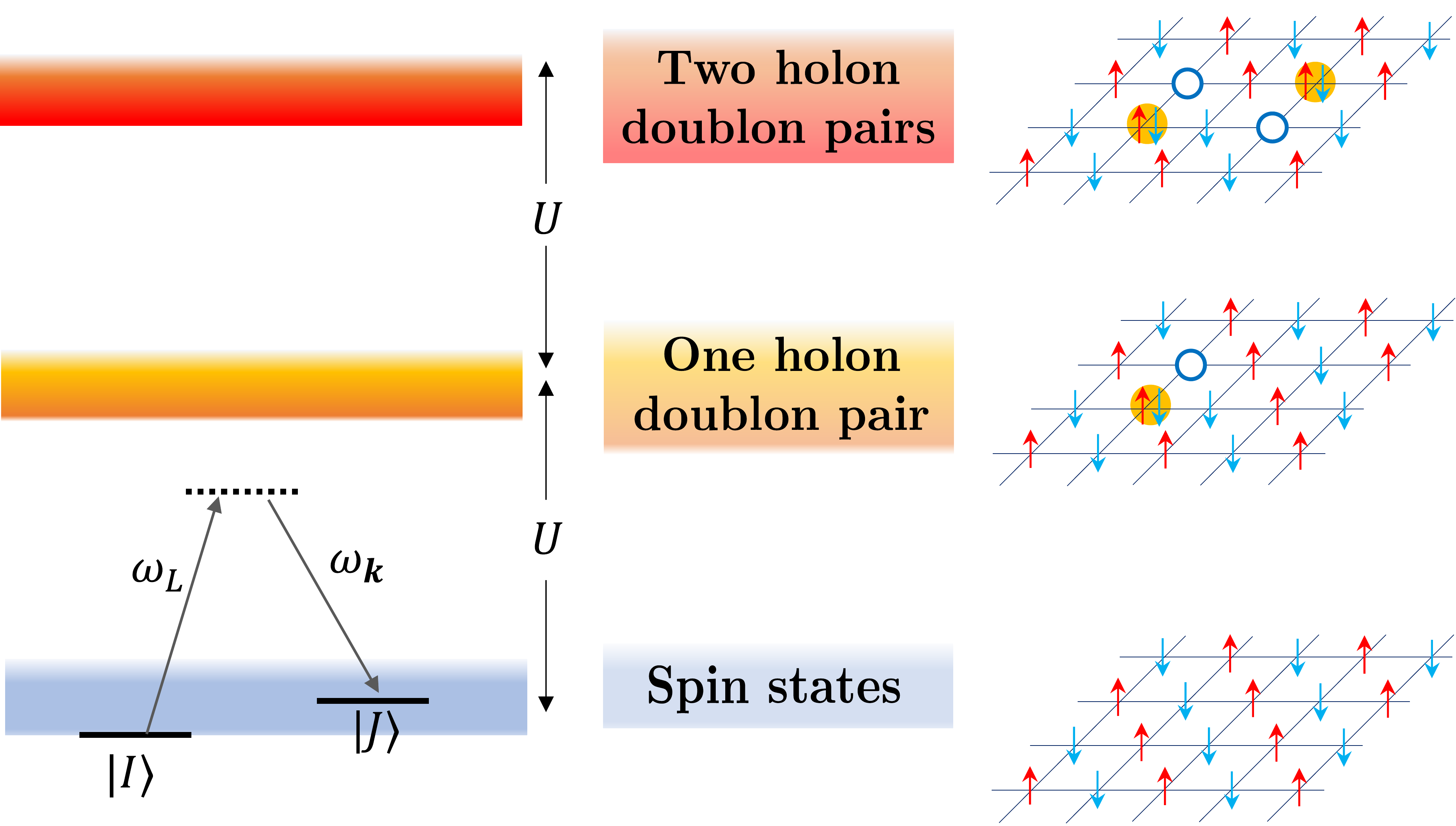}
  \caption{Schematic representation of the energy eigenstates accessible by applying local operators to the ground state of the Fermi-Hubbard model at half-filling. The eigenstates are divided into distinct sectors: the lowest-energy sector (light blue) consists of spin states with frozen charge degrees of freedom, while the next two sectors (orange and red) are separated by approximately \(U\) and correspond to states with one and two doublon-hole pairs, respectively. A Raman process is illustrated: starting from a state \(\ket{I}\) in the spin sector, photon absorption leads to virtual occupation of a state in the one-doublon-hole sector, followed by photon emission into mode \(\mathbf{k}\), resulting in the material returning to a potentially different state \(\ket{J}\) in the spin sector. Here we use dotted lines to depict virtually occupied states. The dotted line here is placed at the energy $E_I+\omega_L$, even though there is no state at that energy in the matter Hamiltonian.}
  \label{fig:sectorsraman}
\end{figure}

Let us now look at the energy scales in the problem at half-filling (i.e., the number of electrons equals the number of lattice sites) in the limit $\tunn\ll U$, where the ground state is a Mott-insulator. In this limit, the manifold of energy eigenstates \textit{that can be accessed by applying local operators on the ground state} split up into sectors as shown in Fig.~\ref{fig:sectorsraman}.

In the lowest energy sector (shaded blue), called the spin sector, or lower Hubbard band, the charge degree of freedom is frozen and the excitations lie purely in the spin sector. If the ground state is a conventionally ordered state, these excitations are magnons and if the ground state is a quantum spin liquid, then these states can be composites of fractionalized excitations. The bandwidth of this sector is of the order of $J\approx \tunn^2/U$. The sector shaded yellow, roughly separated by an energy $U$ from the spin sector, called the charge sector or upper Hubbard band, consists of states where one site is doubly occupied (called a doublon) and one site is empty (called a hole). The states in this sector include bound-states of doublons and holes, called Mott excitons \cite{gallagher1997excitons,essler2001excitons,wrobel2002excitons,tohyama2002resonant,maeda2004third,matsueda2005excitonic,ono2005direct,tohyama2006symmetry,gossling2008mott,novelli2012ultrafast,kim2014excitonic,zhou2014doublon,huang2023spin,mehio2023hubbard}, as well as their scattering states. We expect the bandwidth of this sector to be of order $\tunn$. The sector shaded red consists of states with two doublon-hole pairs and has an energy of order $2U$ relative to the spin sector. 

Similar  to Ref.~\cite{shastry1990theory,shastry1991raman,ko2010raman}, we assume that the laser frequency $\omega_L$ satisfies
\begin{equation}
    \tunn \ll \abs{U-\omega_L} \ll U.
\end{equation}
In other words, the laser is detuned from the Fermi-Hubbard repulsion $U$, and the detuning, though small compared to $U$, is large compared to $\tunn$. This assumption allows us to do perturbation theory in both $\tunn/\abs{U-\omega_L}$ and $\tunn/U$.

The second small parameter is the laser-matter coupling that is involved
during each photon absorption. Expanding the light-matter interaction from Peierls' substitution in Eq.~\eqref{eq:peierls} to leading order in $q_e$, one gets the paramagnetic interaction $q_e\hat{\vb{A}}\cdot \hat{\bm{\mathcal{J}}}$, where $\hat{\bm{\mathcal{J}}}$ is the electric current. Further, expanding $\hat{\vb{A}}$ using Eq.~\eqref{eq:Aexp}, we get an interaction term $\sim\tunn\frac{q_{e}\alatt}{\sqrt{2\varepsilon\mathcal{V}\omega_{L}}}\hat{a}_{L}\hat{c}_{\vb{r}}^{\dagger}\hat{c}_{\vb{r}'}$.
 If the input state (of the laser field) is a Fock state, the operator $\hat{a}_{L}$ can be replaced by $\sqrt{\mathcal{N}_{L}}$,
where $\mathcal{N}_{L}$ is the average number of photons in the laser mode.
For a coherent state input $e^{\phi_{L}a_{L}^{\dagger}-\phi_{L}^{*}a_{L}}\left|0\right\rangle $
having the amplitude $\phi_{L}$, $N_{L}=|\phi_{L}|^{2}$. Denoting
the intensity of the laser as $I_{L}=\frac{\mathcal{N}_{L}\omega_{L}c}{\mathcal{V}}$
and $I_{L}=\frac{\left|\phi_{L}\right|^{2}\omega_{L}c}{\mathcal{V}}$
for the Fock-state and coherent inputs, respectively, we can define a dimensionless light-matter coupling:
\begin{equation}\label{eq:gLdef}
    g_L=\frac{q_e \alatt\sqrt{\mathcal{N}_L}}{\sqrt{2\varepsilon \mathcal{V}\omega_L}}\equiv\frac{\sqrt{2\pi I_L \alpha}\ \alatt}{\omega_L},
\end{equation}
where $\alatt$ is the lattice spacing and where $\alpha=\frac{q_e^2}{4\pi \varepsilon c}$ is the fine-structure constant. We can also interpret  $g_L$ as the ratio of the effective Rabi frequency of the laser drive to the laser frequency $\omega_L$. In this work, we will assume that $g_L \ll 1$. 

Similarly, the interaction term corresponding to a photon emission into an unoccupied mode $(\vb{k},\vb{e}_j)$ is $\sim \tunn \frac{q_e \alatt}{\sqrt{2\varepsilon \mathcal{V}\omega_{\vb{k}}}}\hat{a}^{\dagger}_{\vb{k},\vb{e}_{j}} \hat{c}^{\dagger}_{\vb{r}}\hat{c}_{\vb{r}'}$. This suggests another dimensionless coupling in the problem: $\tfrac{q_e \alatt}{\sqrt{2\varepsilon \mathcal{V}\omega_{\vb{k}}}}$, which we rewrite as $\sqrt{\pi \alpha} \alatt\sqrt{\tfrac{2c}{\omega_{\vb{k}}\mathcal{V}}}$. We see from Eq.~\eqref{eq:addef},\eqref{eq:sumtoint},\eqref{eq:filterredef}  that the factor $\sqrt{\tfrac{2c}{\omega_{\vb{k}}\mathcal{V}}}$  cancels in the final expressions for $G^{(1)}$ and $G^{(2)}$. Hence we define the following dimensionful light-matter coupling
\begin{equation}
    g\equiv \sqrt{\pi\alpha}\alatt,
\end{equation}
with the understanding that the dimensionless small parameter corresponding to $g$ is essentially $\sqrt{\alpha}$.

In the following sections, we present expressions for \(G^{(1)}\), \(G^{(2)}\), and quadrature correlations, calculated to leading order in the small parameters \(\tunn/U\), \(\tunn/|U - \omega_L|\), \(g_L\), and $\sqrt{\alpha}$.

\subsection{Form of relevant terms in the $\T$-matrix}\label{sec:Tmatrixsimp}
According to the expansion of the $\T$-matrix, the amplitude of the emission of each additional photon is accompanied by a multiplicative factor of \(g\), (and therefore a factor of the small parameter $\sqrt{\alpha}$). Therefore, we can restrict our analysis to processes that result in the emission of at most two photons, which is the minimal number required for a nonzero \(G^{(2)}\).

With this in mind, let us examine the relevant terms in the expression $\T=\hat{V}+\hat{V}\hat{\mathbb{G}}_0\hat{V}+\hat{V}\hat{\mathbb{G}}_0\hat{V}\hat{\mathbb{G}}_0\hat{V}+\hat{V}\hat{\mathbb{G}}_0\hat{V}\hat{\mathbb{G}}_0\hat{V}\hat{\mathbb{G}}_0\hat{V}+\ldots$, where $\hat{V}$ is given by Eq.~\eqref{eq:lmpeierls}. We will do a small $\sqrt{\alpha}$ expansion, expanding Eq.~\eqref{eq:lmpeierls} order by order in $q_e \hat{\vb{A}}$. Later, in Sec.~\ref{sec:operators}, we provide a discussion about the order at which this expansion can be truncated. But for now, let us make some general conclusions. The (transverse part of) vector potential $\hat{\vb{A}}$ can be expanded in terms of $\sim \hat{a}_{\lambda}+\hat{a}_{\lambda}^{\dagger}$, where $\hat{a}^{\dagger}_{\lambda}$ is a photon creation operator in mode $\lambda$. Thus, each factor of $q_e \hat{\vb{A}}$ can lead to either absorption (via $\hat{a}_{\lambda}$) or emission ($\hat{a}^{\dagger}_{\lambda}$) of a photon. In general, one can have arbitrary numbers of photon absorptions and emissions. However, using the energy scales discussed in Sec.~\ref{sec:scales}, i.e., $\tunn\ll\abs{U-\omega_L}\ll U$, we can substantially reduce the number of possibilities. Further, we make the approximation that photons can be only absorbed from the laser. Once a photon is emitted into a different mode $\lambda$, it never gets absorbed.

With these simplifications, let us consider the part of $\T$ leading to the emission of one photon, labeled $\T^{(1)}$. As shown in Fig.~\ref{fig:sectorsraman}, starting from a state in the spin sector, the emission of a photon requires the absorption of a photon from the laser. Consequently, to leading order:
\begin{equation}\label{eq:VGVexp}
\T^{(1)}=\sum_{\vb{k}}\hat{R}_{(\vb{k},\vb{e}_j)}^{(1)}\otimes \hat{a}^{\dagger}_{\vb{k},\vb{e}_j}\hat{a}_L\sqrt{\frac{2c}{\mathcal{V}\omega_{\vb{k}}\mathcal{N}_L}}+\ldots ,
\end{equation}
where $\vb{e}_j$ is the polarization selected by the detector. The normalization factor \(\sqrt{\frac{2c}{\mathcal{V}\omega_{\mathbf{k}}\mathcal{N}_L}}\) is chosen to simplify subsequent expressions.

We now turn to processes that result in the emission of two photons. In this work, we assume that the detector selectively filters photons, allowing only those with frequencies within the range \(\omega_L \pm O(|\omega_L - U|)\). For example, as a result, photons with a frequency of \(2\omega_L - U\) are detected, while those with a frequency of \(\omega_L / 2\) are not. Under these conditions, we can see that: (1) it is sufficient to consider processes where two photons are absorbed, and (2) after absorbing two photons and emitting two photons, the final state of the material is in the spin sector. 

Thus, the term in $\T$ that contributes to the emission of two photons, which we label as $\T^{(2)}$, takes the following form (we use the shorthand notation $\lambda_j$ for the mode $(\vb{k},\vb{e}_{\lambda_j})$ of frequency $\omega_{\lambda_j}$, where $\vb{e}_{\lambda_j}$ is the polarization detected by detector $j$):
\begin{equation}\label{eq:Tsimp}    \T^{(2)}=\sum_{\lambda_1,\lambda_2}\hat{R}^{(2)}_{\lambda_1, \lambda_2}\otimes \frac{\hat{a}^{\dagger}_{\lambda_1}\hat{a}^{\dagger}_{\lambda_2}\left(\hat{a}_L\right)^2}{2\sqrt{\mathcal{N}_L (\mathcal{N}_L-1)}}\frac{2c}{\mathcal{V}\sqrt{\omega_{\lambda_1}{\omega_{\lambda_2}}}} + \ldots
\end{equation}
where $\hat{R}^{(2)}_{\lambda_1, \lambda_2}$ is a pure matter operator and will be calculated in the next section. $\hat{R}^{(2)}_{\lambda_1, \lambda_2}$ is symmetric in $\lambda_1$ and $\lambda_2$. Here, $\hat{a}^{\dagger}_{\lambda_1}\hat{a}^{\dagger}_{\lambda_2}\left(\hat{a}_L\right)^2$ signifies that two photons are absorbed from the laser and two photons are emitted into modes $\lambda_1$ and $\lambda_2$. The remaining factors are due to our normalization convention. We note that while summing over modes $\lambda_1$ and $\lambda_2$, the polarizations can be fixed to correspond to those selected by the detectors, reducing the summation to the associated momenta only. The full $\T$-matrix is the sum $\T=\T^{(1)}+\T^{{(2)}}+\ldots$.
\subsection{Expressions for photonic correlation functions in terms of matter operators $\hat{R}^{(1)}$ and $\hat{R}^{(2)}$}\label{sec:photoncorrsimp}
Recall that the photonic correlations defined in Eqs.~\eqref{eq:G1def}, \eqref{eq:G2def2}, \eqref{eq:X1def}, \eqref{eq:X2def} and \eqref{eq:X2minusdef} are expectation values in the $\ket{\text{out}}$ state, which can be determined from the $\ket{\text{in}}$ state using the $\T$-matrix. 
We now use the form of the $\T$-matrix in Eqs.~\eqref{eq:VGVexp} and \eqref{eq:Tsimp} to express the photonic correlation functions in terms of matter operators $\hat{R}^{(1)}$ and $\hat{R}^{(2)}$. 
The result of this subsection should be seen as an intermediate step, which will later be used along with more input from microscopic details to derive the results in Table~\ref{tab:dictionary1}.

To begin, let us assume that the \(\ket{\text{in}}\) state is a product state between the matter and light degrees of freedom.
\begin{equation}\label{eq:instatedef}
    \ket{\text{in}}=\ket{I}_M \otimes  \myket{\psi^{(0)}_L}_R,
\end{equation}
where $\ket{I}_M$ is an energy eigenstate of the matter part of $\hat{H}_0$ (hence subscript $M$) with energy $E_I$. $\myket{\psi^{(0)}_L}_R$ is a state in the electromagnetic sector with the laser mode occupied (we use subscript $L$ to denote ``laser" and subscript $R$ to denote the ``radiation" sector). Since Eq.~\eqref{eq:Tmatresrepeat} is easiest to apply for an $\ket{\text{in}}$ state that is an eigenstate of $\hat{H}_0$, we can consider $\myket{\psi^{(0)}_L}_R$ to be a Fock state with $\mathcal{N}_L$ photons in laser mode $L$, a mode with well-defined wavevector and polarization, i.e., 
\begin{equation}\label{eq:fockstate}
    \myket{\psi^{(0)}_L}_R=\frac{1}{\sqrt{\mathcal{N}_L!}}\left(\hat{a}^{\dagger}_L\right)^{\mathcal{N}_L}\ket{0}_R.
\end{equation}

 In Appendix~\ref{app:coherentcase}, we show that we can also have the laser mode $L$ to be populated with a coherent state and still be able to calculate $\ket{\text{out}}$ to use Eq.~\eqref{eq:Tmatresrepeat}. In that case,
 \begin{equation}\label{eq:coherentstate}
    \myket{\psi^{(0)}_L}_R=e^{\phi_L \hat{a}_L^{\dagger}-\phi_L^*\hat{a}_L}\ket{0}_R.
\end{equation}
 Here, $\phi_L\equiv \abs{\phi_L}e^{-i\theta_L}$. Using Eq.~\eqref{eq:VGVexp}, \eqref{eq:Tsimp} and Eq.~\eqref{eq:Tmatresrepeat}, to leading order in $g$, we get the $\ket{\text{out}}$ state to be a superposition of the unscattered $\ket{\text{in}}$ state, and states resulting from one and two-photon scattering: 
\begin{equation}\label{eq:outforcoherent}
    \begin{aligned}
        \ket{\text{out}}&=\ket{I}\otimes \myket{\psi^{(0)}_L}\\ &-2\pi i \sum_{\lambda}\sum_{F}\delta(E_{FI}+\omega_{\lambda}-\omega_L)\ket{F}\\
        & \quad \quad \quad\times \sqrt{\frac{2c}{\mathcal{V}\omega_{\lambda}}} e^{-i\theta_L}\bra{F}\hat{R}_{\lambda}^{(1)}\ket{I}\otimes \hat{a}^{\dagger}_{\lambda}\myket{\psi^{(1)}_L}\\
        &-2\pi i \sum_{\lambda_1,\lambda_2}\sum_{F}\delta(E_{FI}+\omega_{\lambda_1}+\omega_{\lambda_2}-2\omega_L)\ket{F}\\
        & \quad \quad \quad\times \frac{2c e^{-2i\theta_L}}{\mathcal{V}\sqrt{\omega_{\lambda_1}{\omega_{\lambda_2}}}}\bra{F}\hat{R}_{\lambda_1,\lambda_2}^{(2)}\ket{I}\otimes \frac{\hat{a}^{\dagger}_{\lambda_1}\hat{a}^{\dagger}_{\lambda_2}}{2}\myket{\psi^{(2)}_L}.
    \end{aligned}
\end{equation}
We used the notation $E_{FI}\equiv E_F - E_I$. $\myket{\psi^{(n)}_L}$ is defined as $(\hat{a}_L)^n \myket{\psi^{(0)}_L}$, normalized so that $\langle\psi^{(n)}_L\myket{\psi^{(n)}_L}=1$. For example, if $\myket{\psi^{(0)}_L}$ is a Fock state as in Eq.~\eqref{eq:fockstate}, then $\myket{\psi^{(1)}_L}$ and $\myket{\psi^{(2)}_L}$ are the corresponding Fock states with $\mathcal{N}_L-1$ and $\mathcal{N}_L-2$ photons respectively in mode $L$. On the other hand, if $\myket{\psi^{(0)}_L}$ is a coherent state as in Eq.~\eqref{eq:coherentstate}, then both $\myket{\psi^{(1)}_L}$ and $\myket{\psi^{(2)}_L}$ are equal to $\myket{\psi^{(0)}_L}$.

It is important to note that the application of photonic annihilation operators to the $|\text{out}\rangle$ state in the phase-sensitive correlation functions \(\expval{\hat{a}_{d_j}}_{\text{out}}\) and \(\expval{\hat{a}_{d_2}(\tau)\hat{a}_{d_1}(0)}_{\text{out}}\) changes the total number of photons. In contrast, the \(\T\)-matrix is photon-number preserving, as long as the approximations outlined in Sec.~\ref{sec:scales} and Sec.~\ref{sec:Tmatrixsimp} are applied. Therefore, if the initial state of the laser, i.e., $\myket{\psi^{(0)}_L}$ is a Fock state, then the states $\ket{\text{out}}$ and $\hat{a}_{d_j}\ket{\text{out}}$ are orthogonal. In contrast, they have a finite overlap if $\myket{\psi^{(0)}_L}$ is a coherent superposition of different photon number sectors. Therefore, for quadrature correlators, we assume that the input laser state $\myket{\psi^{(0)}_L}$ is necessarily a coherent state (see Eq.~\eqref{eq:coherentstate}). On the other hand, the photonic operators in $G^{(1)}(0)$ [Eq.~\eqref{eq:G1def}], $G^{(2)}(\tau)$ [Eq.~\eqref{eq:G2def2}], $G^{(1)}(\tau)$ [Eq.~\eqref{eq:X2minusdef}], and $H_{d_1,d_2}(t,\tau)$ [Eq.~\eqref{eq:ConditionalG1}] do conserve total photon number, so both a coherent state and a Fock state input can lead to a nonzero measurement result. 

With this in mind, we can now obtain expressions for photonic correlators in terms of the matter operators $\hat{R}^{(1)}_{\lambda}$ and $\hat{R}^{(2)}_{\lambda_1,\lambda_2}$. Recall that $G^{(1)}_{d_j}=\langle\text{out}\lvert \hat{a}^{\dagger}_{d_j}(0)\hat{a}_{d_j}(0)\rvert\text{out}\rangle$. Into this, we substitute $\ket{\text{out}}$ [given in Eq.~\eqref{eq:outforcoherent}], use the definition of operator $\hat{a}_{d_j}$ in Eq.~\eqref{eq:addef}, the definition of the effective filter function $\mathcal{F}_j(\omega_{\lambda_j})$ [Eq.~\eqref{eq:filterredef}], and use Eq.~\eqref{eq:sumtoint} (for converting sums to integrals) to obtain:
\begin{equation}\label{eq:G1simp}
\begin{aligned}
     G^{(1)}_{d_j}(0)\approx & \sum_{J} \biggl|\int_{-\infty}^{\infty} \dd{\omega_{\lambda_j}}\mathcal{F}_j(\omega_{\lambda_j})  \\
   &\quad \times \delta \left(E_{JI} + \omega_{\lambda_j} - \omega_L \right) \mel{J}{\hat{R}^{(1)}_{\lambda_j}(\omega_{\lambda_j})}{I}\biggr|^2.
\end{aligned}
\end{equation}
Here, the polarization of mode $\lambda_j$ is $\vb{e}_j$, as set by detector $d_j$. We also assumed that the dependence of the matter operators $\hat{R}^{(1)}_{\lambda}$ and $\hat{R}^{(2)}_{\lambda_1,\lambda_2}$ on the photonic mode $\lambda$ is only through its frequency $\omega_{\lambda}$ and polarization $\vb{e}_{\lambda}$, an assumption that we will justify in Sec.~\ref{sec:Ramanrev}.

For a  nonzero time delay, the $G^{(1)}_{d_j}$ function takes the following form:
\begin{equation}\label{eq:X2minussimp}
    \begin{aligned}
        &G^{(1)}_{d_j}(\tau)=\expval{\hat{a}^{\dagger}_{d_j}(\tau)\hat{a}_{d_j}(0)}_{\text{out}}\\
        &\approx \sum_J \int_{-\infty}^{\infty} \dd{\omega_{\lambda_j}}\biggl[\mathcal{F}_j(\omega_{\lambda_j})\mathcal{F}^{*}_j(\omega_{\lambda_j})e^{i\omega_{\lambda_j}\tau}\\ & \quad \times\delta\left(E_{JI}+\omega_{\lambda_j}-\omega_L\right)\left.\bra{I}\left[\hat{R}^{(1)}_{\lambda_j}\right]^{\dagger}\ketbra{J}\hat{R}^{(1)}_{\lambda_j}\ket{I}\right].
    \end{aligned}
\end{equation}
Analogously for the $G^{(2)}$ function we get:
\begin{equation}\label{eq:G2simp2}
\begin{aligned}
    &G^{(2)}_{d_1,d_2}(\tau)\equiv \expval{\hat{a}^{\dagger}_{d_1}(0)\hat{a}^{\dagger}_{d_2}(\tau)\hat{a}_{d_2}(\tau)\hat{a}_{d_1}(0)}_{\text{out}}\\
    \approx \sum_F &\Biggl|\iint_{-\infty}^{\infty} \frac{\dd{\omega_{\lambda_1}}\dd{\omega_{\lambda_2}}}{(2\pi)}\Bigl[\mathcal{F}_1(\omega_{\lambda_1})\mathcal{F}_2(\omega_{\lambda_2})\\ \times &  \delta\left(E_{FI}+\omega_{\lambda_1}+\omega_{\lambda_2}-2\omega_L\right)\\
    \times & \left.e^{i\omega_{\lambda_1}\tau} \mel{F}{\hat{R}^{(2)}_{\lambda_1,\lambda_2}(\omega_{\lambda_1},\omega_{\lambda_2})}{I}\right]\Biggr|^2.
\end{aligned}
\end{equation}
The photon in mode $\lambda_2$ is detected after time $\tau$ following the photon detection in mode $\lambda_1$, leading to the phase $e^{i\omega_{\lambda_1}\tau}$ coming from time- evolution in between. \footnote{It may appear from the definition of $G^{(2)}(\tau)$ that this phase is $e^{-i\left(\omega_{\lambda_2}+E_F+(\mathcal{N}_L-2)\omega_L\right)\tau}$. But this reduces to $e^{i\omega_{\lambda_1}\tau}$ after using the $\delta$-function that imposes energy conservation, and after getting rid of overall phases independent of indices $F,\lambda_1,\lambda_2$.}

Finally, we consider phase-sensitive quadrature measurements for a coherent state input. Such correlators can be measured using an interferometric scheme, explained in Sec.~\ref{sec:quadraturemeasurements}. The first-order and second-order quadrature correlators are respectively:
\begin{equation}\label{eq:X1simp}
 \begin{aligned}   e^{i(\theta+\theta_L)}\expval{\hat{a}_{d_j}(0)}_{\text{out}}+\text{c.c.}\approx &\mathcal{F}_j(\omega_L) e^{i\theta} 
  \mel{I}{\hat{R}^{(1)}_{\lambda_j}(\omega_{L})}{I}\\
  &+\text{c.c.}, \text{ and}
  \end{aligned}
\end{equation}
\begin{equation}\label{eq:X2plussimp}
\begin{aligned}
    &e^{2i(\theta+\theta_L)}\expval{\hat{a}_{d_2}(\tau)\hat{a}_{d_1}(0)}_{\text{out}}+\text{c.c.}\\
    \approx& i \iint_{-\infty}^{\infty} \frac{\dd{\omega_{\lambda_1}}\dd{\omega_{\lambda_2}}}{(2\pi)}\Bigl[e^{-i\omega_{\lambda_2}\tau}\mathcal{F}_1(\omega_{\lambda_1})\mathcal{F}_2(\omega_{\lambda_2})\\
    &\quad\times  \left. e^{2i\theta}\delta\left(\omega_{\lambda_1}+\omega_{\lambda_2}-2\omega_L\right)\mel{I}{\hat{R}^{(2)}_{\lambda_1,\lambda_2}(\omega_{\lambda_1},\omega_{\lambda_2})}{I}\right]. 
  \end{aligned}
\end{equation}
As expected, the final expressions only depend on the \textit{relative} phase difference between the local oscillator (used for measuring the quadrature) and the drive. Further details of the above two correlation functions are provided in Appendix~\ref{app:homodyne}.
\section{Microscopic structure of matter operators $\hat{R}^{(1)}$ and $\hat{R}^{(2)}$ }\label{sec:operators}
In this section, we provide the explicit expressions for the matter operators $\hat{R}^{(1)}_{\lambda_1}$ and $\hat{R}^{(2)}_{\lambda_1,\lambda_2}$  for the single-band Fermi-Hubbard model at half-filling defined in Sec.~\ref{sec:Hubbardsetup}.

We expand $\T=\hat{V}+\hat{V}\hat{\mathbb{G}}_0\hat{V}+\hat{V}\hat{\mathbb{G}}_0\hat{V}\hat{\mathbb{G}}_0\hat{V}+\hat{V}\hat{\mathbb{G}}_0\hat{V}\hat{\mathbb{G}}_0\hat{V}\hat{\mathbb{G}}_0\hat{V}+\ldots$, treating all terms at the same order in $q_e \hat{A}_{\vb{r},\vb{r}'}$ on equal footing. Recall from the discussion in Sec.~\ref{sec:Tmatrixsimp} that it suffices to consider processes involving at most two photons being absorbed and at most two photons emitted. So, we should expand the light-matter interaction $\hat{V}$ given in Eq.~\eqref{eq:lmpeierls}  up to fourth order in $q_e$. The fourth order term $\left(\tunn_{\vb{r},\vb{r}+\bm{\mu}}\hat{c}^{\dagger}_{\vb{r}+\bm{\mu}, \sigma}\hat{c}_{\vb{r}, \sigma}+\text{h.c.}\right)\left(q_e \hat{A}_{\vb{r},\vb{r}+\bm{\mu}}\right)^4$ can however be dropped. The reason is that this term can contribute to processes involving two-photon absorption and two-photon emission at most at linear order in $\hat{V}$. However, the factor $\left(\tunn_{\vb{r},\vb{r}+\bm{\mu}}\hat{c}^{\dagger}_{\vb{r}+\bm{\mu}, \sigma}\hat{c}_{\vb{r}, \sigma}+\text{h.c.}\right)$ results in the material transitioning out of the spin sector (lower Hubbard band), which we forbid by postselecting the frequencies of detected photons. Therefore, in powers of $q_e$, the part of $\hat{V}$ [Eq.~\eqref{eq:lmpeierls}] containing linear or``paramagnetic" ($\hat{V}_P$), quadratic or ``diamagnetic" ($\hat{V}_D$), and cubic ($\hat{V}_C$) terms  is:
\begin{equation}\label{eq:lminteraction}   
    \begin{aligned}
        &\hat{V}=\sum_{(\vb{r},\bm{\mu}),\sigma}\Biggl\{-i\left(\tunn_{\vb{r},\vb{r}+\bm{\mu}}\hat{c}^{\dagger}_{\vb{r}+\bm{\mu}, \sigma}\hat{c}_{\vb{r}, \sigma}-\text{h.c.}\right)\\
        & \quad \quad \quad \quad \quad \times\left[q_e \hat{A}_{\vb{r},\vb{r}+\bm{\mu}} -\frac{\left(q_e \hat{A}_{\vb{r},\vb{r}+\bm{\mu}}\right)^3}{6} \right]\\
        &\quad \quad \quad     +\left(\tunn_{\vb{r},\vb{r}+\bm{\mu}}\hat{c}^{\dagger}_{\vb{r}+\bm{\mu}, \sigma}\hat{c}_{\vb{r}, \sigma}+\text{h.c.}\right)\frac{\left(q_e \hat{A}_{\vb{r},\vb{r}+\bm{\mu}}\right)^2}{2} \Biggr\}\\&\equiv \hat{V}_P +\hat{V}_C+\hat{V}_D.
    \end{aligned}
\end{equation}
Thus, according to Eq.~\eqref{eq:lminteraction}, the paramagnetic term $\hat{V}_P$ involves an electron current between a pair of sites by absorbing or emitting a photon. The cubic term $\hat{V}_C$ consists of the same electronic process as $\hat{V}_P$, but involves three photons in total. In this formulation  of light-matter interaction, the diamagnetic term $\hat{V}_D$ couples to the point-split local density $\frac{1}{2}\sum_{\sigma}\left(\tunn_{\vb{r},\vb{r}+\bm{\mu}}\hat{c}^{\dagger}_{\vb{r}+\bm{\mu}, \sigma}\hat{c}_{\vb{r}, \sigma}+\text{h.c.}\right)$ unlike just the local density in a free-electron gas. $\hat{V}_D$ thus involves hopping of an electron by either absorbing two photons, emitting two photons, or by absorbing one photon along with emitting another. 

We now separately study the outcomes where one and two photons are emitted respectively.
\subsection{Matter operator $\hat{R}_{\lambda}^{(1)}$: Review of Raman scattering}\label{sec:Ramanrev}
To leading order in $g$, we consider processes leading to the emission of one photon. Processes involving absorption and emission of two or more photons are of higher order.
 
 At half-filling, electron tunneling  results in double occupancy at a site and this costs energy $U$. Since $\tunn \ll \abs{U-\omega_L}$, just one insertion of $\hat{V}$ alone cannot result in a photon absorption. But two insertions of $\hat{V}$ via the term $\hat{V}_P\hat{\mathbb{G}}_0\hat{V}_P$ can result in absorption and     emission of a photon. More specifically, this term represents that a laser photon can be absorbed off-resonantly via the paramagnetic term (first arrow from the left in Fig.~\ref{fig:sectorsraman}). This should then be followed by an electron in the doubly occupied site returning to its empty neighbor by emitting a photon, thus leaving the material in a possibly different state in the spin sector (second arrow from the left in Fig.~\ref{fig:sectorsraman}). This process can be seen as a photon-assisted superexchange. Therefore, to leading order in light-matter interaction $g$ and $\tunn/\abs{U-\omega_L}$, the main contribution to $\hat{R}_{\lambda}^{(1)}$ is from such a Raman process \cite{fleury1968scattering,shastry1990theory}. By simplifying the term $\hat{V}_P\hat{\mathbb{G}}_0\hat{V}_P$, we obtain the Fleury-Loudon operator, a sum of projectors to spin singlets, modulated by the polarizations of the incoming laser and the detected photon~\cite{fleury1968scattering,shastry1990theory,shastry1991raman,ko2010raman}:
\begin{equation}\label{eq:R1exp}
    \begin{aligned}
    \hat{R}_{(\vb{k},\vb{e}_j)}^{(1)}\equiv \hat{A}_j=&\sum_{\left(\vb{r},\bm{\mu}\right)}\frac{\abs{\tunn_{\vb{r},\vb{r}+\bm{\mu}}}^2} g_Lg {\omega_L-U}\left(4\hat{\vb{S}}_{\vb{r}}\cdot \hat{\vb{S}}_{\vb{r}+\bm{\mu}}-1\right)\\
    &\quad \quad \quad \quad  \times \left(\vb{e}^*_{j} \cdot \bar{\bm{\mu}}\right)\left(\vb{e}_L\cdot \bar{\bm{\mu}}\right)+\ldots 
    \end{aligned}
\end{equation}
We define $\bar{\bm{\mu}}\equiv \bm{\mu}/\alatt$. In our summation convention, $\sum_{\left(\vb{r},\bm{\mu}\right)}$ runs through each bond exactly once. The ``$\ldots$" above denotes terms of order $\tunn^3/\abs{\omega_L-U}^2$ and higher.
 
In Eq.~\eqref{eq:R1exp} we ignored the momentum transfer between electrons and photons by omitting the $e^{i\left(\vb{k}_{L}-\vb{k}\right)\cdot \left(\vb{r}+\bm{\mu}/2\right)}$ factor. This is because $\omega_L$ and $\omega_{\lambda}$ are of order $U$, the Hubbard interaction. For typical materials, the corresponding wavelength is  equal to several thousand lattice spacings. Hence, throughout this paper, we will ignore the spatial variation of the laser field. Therefore, $\hat{R}^{(1)}_{(\vb{k},\vb{e}_j)}$ depends on the emitted mode $(\vb{k},\vb{e}_j)$, only through its polarization $\vb{e}_j$ and, in general, the frequency $\omega_{\vb{k}}$. In the near future, in systems with much larger lattice spacings, such as moir\'e materials, significant momentum transfer to electrons may become possible via optical excitation. In this case, our formalism can be adjusted  to include finite-momentum matter operators.

\subsection{Processes leading to the emission of two photons}\label{sec:processes}
\begin{figure*}[t]
  \centering
  \includegraphics[width=0.99\textwidth]{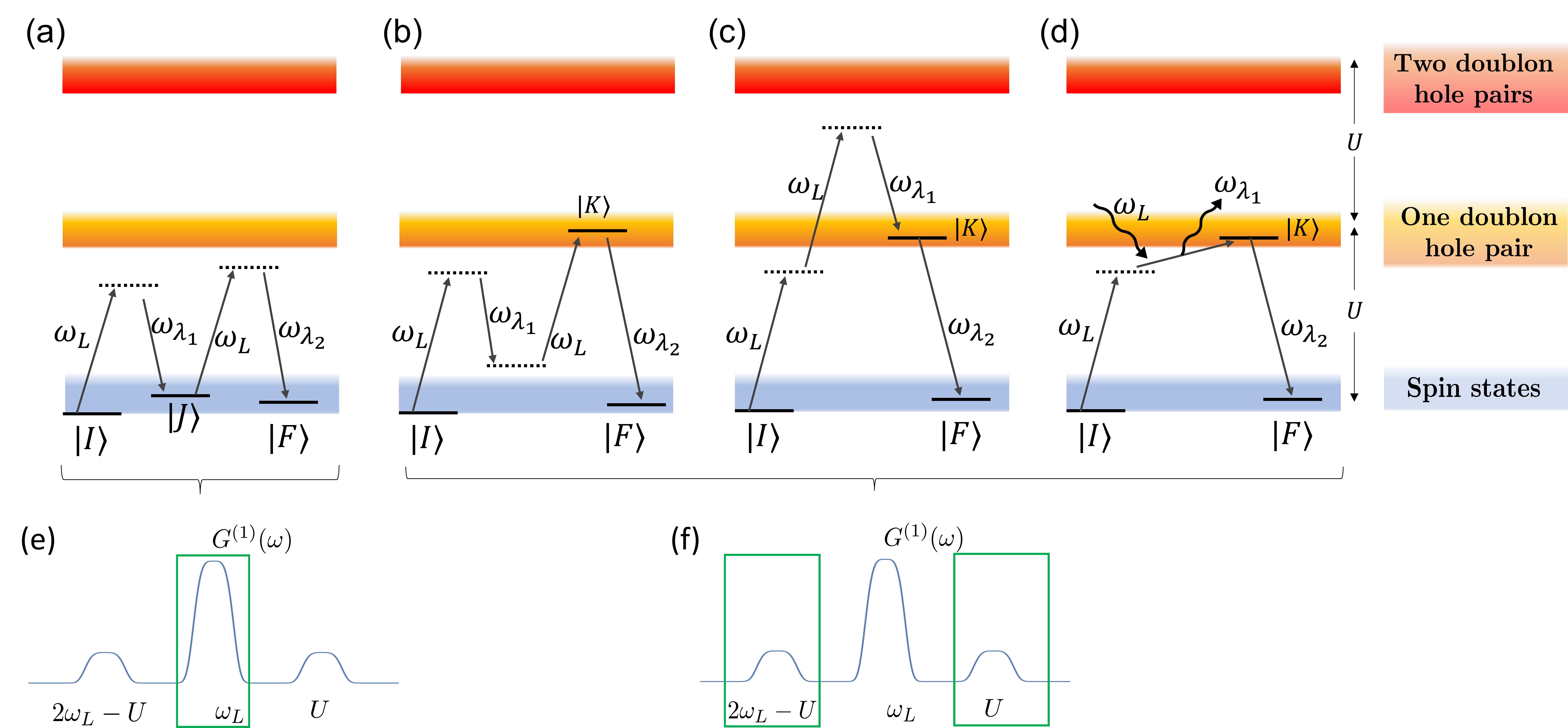}
  \caption{(a)-(d) illustrate the four processes contributing to $G^{(2)}(\tau)$ to leading order in $g_L$ and $\tunn/\abs{\omega_L-U}$. (e) and (f) show a cartoon intensity profile of the emitted photons. Process (a) contributes to the central peak highlighted in (e) and processes (b-d) contribute to the sidebands highlighted in (f). \textbf{(a)}: Raman process occurring twice, with the intermediate state $\ket{J}$ being in the spin sector. \textbf{(b)}: Scattering of two photons, accompanied with the real, i.e., resonant excitation of a state $|K\rangle$ in the charge sector. This is reminiscent of the sideband process in Ref.~\cite{dalibard1983correlation}. \textbf{(c)}: This process corresponds to successive absorption of two photons resulting in virtual occupation of the two-doublon-hole sector, followed by emission of two photons into the sidebands in (f). \textbf{(d)}: This process differs from (a), (b) and (c) in that it involves the diamagnetic term. After a photon absorption and virtual occupation of the single doublon-hole sector, the diamagnetic term results in the scattering of a laser photon (wavy line labelled $\omega_L$) into an emitted mode of frequency $\omega_{\lambda_1}$, which lies in the sideband near $2\omega_L-U$. This results in excitation of a matter state $\ket{K}$ in the single doublon-hole sector. Finally, a photon of frequency $\omega_{\lambda_2}$ is emitted into the sideband around $U$. \textbf{(e)}: The central peak is highlighted, corresponding to photons of frequency around $\omega_L$.  
  \textbf{(f)}: The sidebands corresponding to pairs of photons of frequency around $2\omega_L-U$ and $U$ are highlighted. 
  }
  \label{fig:sectors}
\end{figure*}

We now describe the processes that contribute to $\hat{R}^{(2)}$, i.e., lead to the emission of two photons. Unlike the case for $\hat{R}^{(1)}_{\lambda}$, the diamagnetic and cubic terms can also contribute to the emission of two photons. We show that $\hat{R}^{(2)}_{\lambda_1,\lambda_2}$ can be extracted from Eq.~\eqref{eq:Tsimp} by expanding $\hat{V}_P\hat{\mathbb{G}}_0\hat{V}_P\hat{\mathbb{G}}_0\hat{V}_P\hat{\mathbb{G}}_0\hat{V}_P+\hat{V}_P\hat{\mathbb{G}}_0\hat{V}_D\hat{\mathbb{G}}_0\hat{V}_P$$+\hat{V}_C\hat{\mathbb{G}}_0\hat{V}_P+\hat{V}_P\hat{\mathbb{G}}_0 \hat{V}_C$. First, we provide a qualitative overview. We then discuss the result for $\hat{R}^{(2)}_{\lambda_1,\lambda_2}$ for the case of the single band Fermi-Hubbard model at half-filling. Explicit details of the derivation are provided in Appendix~\ref{sec:explicitmicro}.

In Fig.~\ref{fig:sectors}, we schematically show the two-photon scattering processes on the energy diagram. Their microscopic counterparts are provided in Fig.~\ref{fig:description2} for the Fermi-Hubbard model at half-filling.  As we explicitly demonstrate in this section below, the photon intensity \(G^{(1)}(0)\) as a function of frequency qualitatively resembles the profile in Fig.~\ref{fig:sectors}(e-f). Specifically, it features a central peak around \(\omega_L\) with a width of \(\sim \tunn^2 / U\) (Fig.~\ref{fig:sectors}(e)), along with two sidebands near \(2\omega_L - U\) and \(U\) (Fig.~\ref{fig:sectors}(f)). The central peak can be anticipated from our previous discussion of Raman scattering,  where the emitted photon's frequency deviates from the laser's frequency by an amount of the order of the spin excitation energy scale. Furthermore, in this section, we demonstrate  that the sidebands are generated by the processes in Fig.~\ref{fig:sectors}(b-d) leading to pairs of photons around frequencies $2\omega_L-U$ and $U$. Let us now study all the processes in Fig.~\ref{fig:sectors} one by one.
\begin{figure*}
  \centering
  \includegraphics[width=1.0\textwidth]{description2.pdf}
  \caption{ Microscopic processes corresponding to Fig.~\ref{fig:sectors}: We show the square lattice for concreteness, but our results are general. In all the subfigures, a curved blue arrow indicates that an electron \textit{tunneled} from the tail to the head of the arrow, and the configuration shown in a subfigure is a consequence of the hop shown in the \textit{same} subfigure. The legend is provided at the bottom. We have organized the subfigures into rows on the basis of the subfigures of Fig.~\ref{fig:sectors}. ($\mathbf{a_1}$-$\mathbf{a_6}$): This row corresponds to Fig.~\ref{fig:sectors}(a). The system absorbs a photon via the paramagnetic term, virtually creating a doublon-hole pair (a$_2$). An electron then tunnels back emitting a photon. This results in applying the spin exchange operator $4\vb{S}_A\cdot \vb{S}_B-1$ along the bond, say, $AB$ colored blue in (a$_3$). A similar process then repeats along a different bond via absorption of a second laser photon, resulting in the emission of a second photon. ($\mathbf{b_1}$-$\mathbf{b_6}$): This row corresponds to Fig.~\ref{fig:sectors}(b). Here, the photons are emitted into the sidebands (see the discussion below Eq.~\eqref{eq:R2ab_bigexp1}). Next, the process in Fig.~\ref{fig:sectors}(c) amounts to two successive photon absorptions followed by the emission of two photons. This can occur in two distinct ways -- (c$_1$-c$_6$) and ($c'_1$-$c'_6$). ($\mathbf{c_1}$-$\mathbf{c_5}$): Two doublon-hole pairs are created, and the same ones are annihilated. Note that during the time between emission of the two photons, the doublon-hole pair can move around ($c_5$). ($\bm{c'_1}$-$\bm{c'_5}$): Two doublon-hole pairs are created. A hole from one pair and a doublon from a different pair recombine ($c'_4$). ($\mathbf{d_1}$-$\mathbf{d_5}$): This row corresponds to Fig.~\ref{fig:sectors}(d). A doublon-hole pair is virtually created by absorbing a photon via the paramagnetic term. Then, either a doublon or a hole scatters to a different site by both absorbing and emitting a photon via the diamagnetic term (d$_3$). After time evolution, the doublon-hole pair recombines by emitting a photon via the paramagnetic term.}
  \label{fig:description2}
\end{figure*}
\begin{enumerate}
\item Fig.~\ref{fig:sectors}(a) shows the Raman process occurring twice in succession. First, a doublon-hole pair is virtually created along a bond by a photon absorption via the paramagnetic term $\hat{V}_P$ (Fig.~\ref{fig:description2}(a$_1$-a$_2$)). This pair then recombines via a photon emission, with the system returning to the spin sector -- in a possibly excited state. This effectively results in the application of a spin operator on the matter state, which to leading order in $\tunn/\abs{\omega_L -U}$ is the Fleury-Loudon sum of spin singlet projectors~\cite{fleury1968scattering}, $\hat{A}_1$ defined in Eqs.~(\ref{eq:R1exp}, \ref{eq:AiExp}). The system then undergoes a time-evolution until a second Raman process occurs (Fig.~\ref{fig:description2}(a$_5$)), resulting in the application of a second Fleury-Loudon operator $\hat{A}_2$ (See Appendix~\ref{sec:4a4b} for technical details). 

Both the photons emitted have frequencies close to $\omega_L$ (the central peak in Fig.~\ref{fig:sectors}(e)), and their difference from $\omega_L$ is of the order of $J\sim \tunn^2/U$, i.e., the energy scale of the spin sector. 
\item The microscopic process corresponding to Fig.~\ref{fig:sectors}(b) is shown in Fig.~\ref{fig:description2}(b$_1$-b$_6$). It starts similarly to the process in Fig.~\ref{fig:sectors}(a) with the difference being that the two emitted photons have frequencies of order $2\omega_L - U$ and $U$ respectively (belonging to sidebands in Fig.~\ref{fig:sectors}(e,f)). See also Appendix~\ref{sec:4a4b} for technical details. 
\item 
Figure~\ref{fig:sectors}(c) illustrates a process in which two photons (\(\omega_L\)) are absorbed sequentially, resulting in the creation of two virtual doublon-hole pairs at different locations (see Fig.~\ref{fig:description2}(c\(_2\)-c\(_3\), \(c'_2\)-\(c'_3\))). Subsequently, one of these doublon-hole pairs recombines via the emission of a photon. The recombining pair may consist of one of the initially created pairs (Fig.~\ref{fig:description2}(c\(_4\))), in which case the two pairs could have been arbitrarily separated. Alternatively, the recombining pair may involve a doublon from one pair and a hole from the other pair (Fig.~\ref{fig:description2}(\(c'_4\))). This latter scenario requires the bonds associated with the two pairs to be connected via a third bond (Fig.~\ref{fig:description2}(\(c'_3\))). The emitted photon in this process has a frequency \(\omega_{\lambda_1}\) approximately equal to \(2\omega_L - U\).

Following the emission of the first photon, the system transitions to the single-doublon-hole pair sector, where it undergoes time evolution (Fig.~\ref{fig:description2}(c\(_5\), \(c'_5\))) until a second photon is emitted. This second photon, with a frequency on the order of \(U\), arises from the recombination of the doublon-hole pair, which may have moved to a distant bond (Fig.~\ref{fig:description2}(c\(_6\), \(c'_6\))). This sequence of events also corresponds to the sidebands depicted in Fig.~\ref{fig:sectors}(f).
  
The intermediate state $\ket{K}$, involved in processes
shown in Fig.~\ref{fig:sectors}(b, c), belongs to the single doublon-hole
sector. This implies that the scattering of photons in this case,
is accompanied by the action of matter operators that induce transitions
between energy sectors. We also notice that the elementary processes
in Fig.~\ref{fig:description2}(b$_{4}$-b$_{6}$) and \ref{fig:description2}(c$_{4}$-c$_{6}$)
look identical. As a result the contributions to $\hat{R}_{\lambda_{1},\lambda_{2}}^{(2)}$
originating from the scattering amplitudes corresponding to rows (b)
and (c) in Fig.~\ref{fig:description2} have opposite signs (arising
from opposite signs of two-photon detuning) and nearly cancel each
other. The incomplete cancellation can be explained as follows. As
shown in Figs.~\ref{fig:description2}(b$_{2}$) and \ref{fig:description2}(b$_{4}$),
the two bonds along which light-assisted tunneling occurs, can be
arbitrary, and, in principle, can even share a site or coincide. But for two doublon-hole pairs to be created, the bonds in Figs.~\ref{fig:description2}(c$_{2}$) and \ref{fig:description2}(c$_{3}$)
must not share sites. The associated effect on the matter state  can be mathematically expressed as an application of a sum of local terms. These terms involve a spin-singlet projection along a bond, e.g. \((\vb{r}_{1}, \bm{\mu}_{1})\), followed by electron tunneling along   bonds touching the bond \((\vb{r}_{1}, \bm{\mu}_{1})\).

Similarly, the elementary processes depicted in row \(c'\) of Fig.~\ref{fig:description2}, up to the emission of the first photon, can be described mathematically as the application of a sum of local terms, consisting of a spin operator supported on a bond followed by an electron tunneling operator supported near that bond.

Finally, the recombination of the doublon-hole pair, accompanied by the emission of the second photon (see Fig.~\ref{fig:description2}(b\(_6\), c\(_6\), \(c'_6\)) and Fig.~\ref{fig:description2}(d\(_5\))), is governed by an electron tunneling operator that is proportional to the global electric current.

So far, the considered processes correspond to the action of $\hat{V}_P \hat{\mathbb{G}}_0 \hat{V}_P \hat{\mathbb{G}}_0 \hat{V}_P \hat{\mathbb{G}}_0 \hat{V}_P$ in Eq.~\eqref{eq:Tsimp}, and do not involve the diamagnetic term.
\item The process depicted in Fig.~\ref{fig:sectors}(d) involves a contribution from the diamagnetic term and represents the leading-order term in \(\tunn / |U - \omega_L|\). The corresponding elementary steps are schematically illustrated in Fig.~\ref{fig:description2}(d\(_1\)-d\(_5\)). In this sequence, an electron first tunnels by absorbing a photon of frequency \(\omega_L\) via the paramagnetic term. Subsequently, through the diamagnetic term, another photon of frequency \(\omega_L\) is absorbed, and a photon with frequency \(\omega_{\lambda_1} \sim 2\omega_L - U\) is emitted. Finally, via the paramagnetic term, a photon with frequency \(\omega_{\lambda_2} \sim U\) is emitted, returning the material to the spin sector. This process corresponds to the term \(\hat{V}_P \hat{\mathbb{G}}_0 \hat{V}_D \hat{\mathbb{G}}_0 \hat{V}_P\) in Eq.~\eqref{eq:Tsimp} and contributes to the sidebands shown in Fig.~\ref{fig:sectors}(f). 
\end{enumerate}
Finally, the processes involving the cubic term $\hat{V}_C$ are explained in Appendix~\ref{sec:cubicappendix}.

In summary, all processes contributing to the emission into the sidebands (Fig.~\ref{fig:sectors}(b, c, d)) involve an intermediate step where the material undergoes a resonant transition to a state \(\ket{K}\) in the single doublon-hole sector. In contrast, the process responsible for emission into the central peak (Fig.~\ref{fig:sectors}(a)) involves only off-resonant excitation of this sector. The physical implication is that, in the cases shown in Fig.~\ref{fig:sectors}(b, c, d), the scattering of light is governed by mixed spin-charge operators. Meanwhile, the process in Fig.~\ref{fig:sectors}(a) corresponds to the action of operators confined entirely to the spin sector. This distinction is formally demonstrated in Appendix~\ref{sec:explicitmicro}.

\subsection{Microscopic expression for $\hat{R}^{(2)}$}\label{sec:R2micro}
Following the discussion above,
in this subsection, we provide the final expression for the operator
$\hat{R}_{\lambda_{1},\lambda_{2}}^{(2)}$ (see Appendix~\ref{sec:explicitmicro}
for a complete derivation using the $\T$-matrix formalism). We start
by providing several intermediate definitions.
\begin{figure*}
  \centering
  \includegraphics[width=0.7\textwidth]{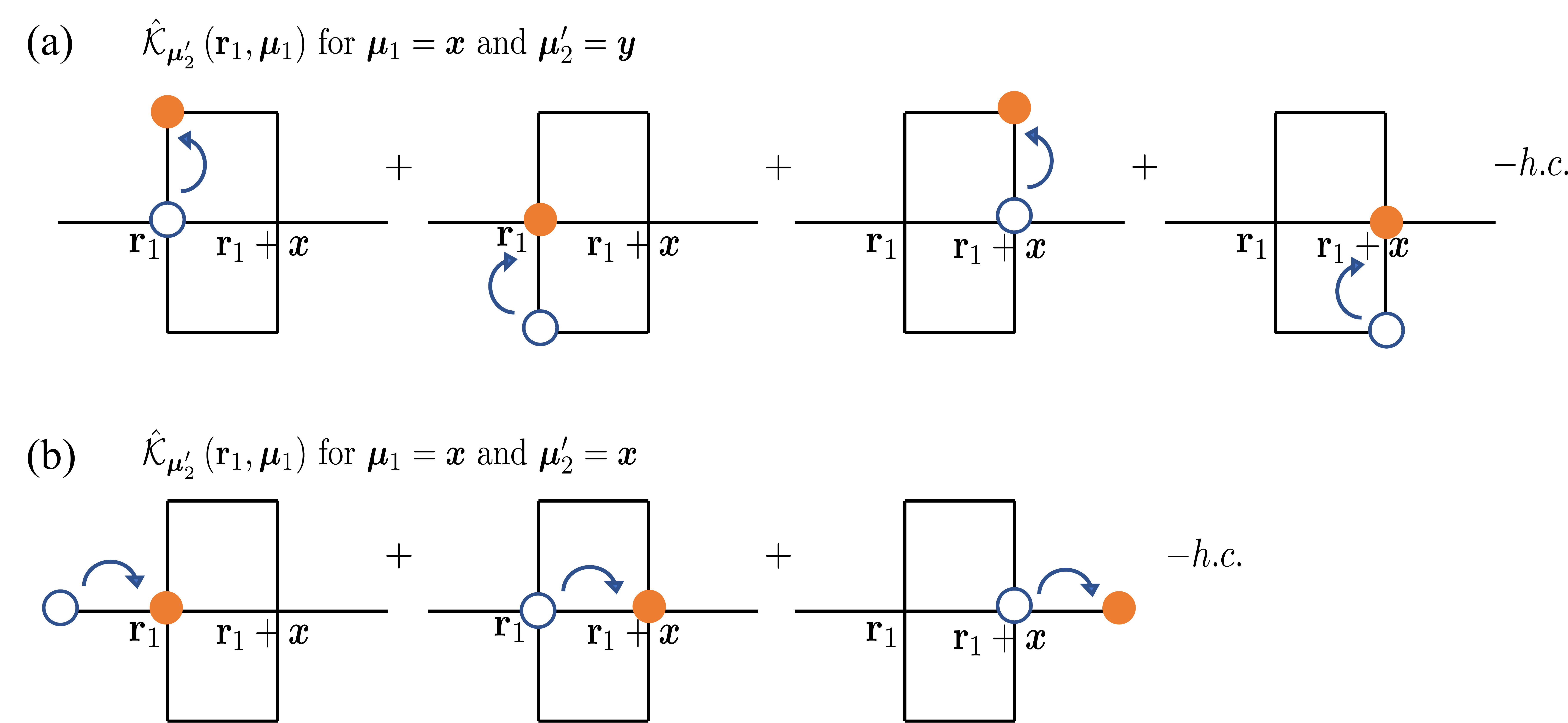}
  \caption{A visualization of the definition of the operator $\hat{\mathcal{K}}_{\bm{\mu}'_2}\left(\vb{r}_1,\bm{\mu}_1\right)$ (see Eq.~\eqref{eq:defKmu}) on the square lattice. This is an operator that creates a doublon-hole pair when acting on the spin sector. \textbf{(a)}: $\bm{\mu}_1=\bm{x}$ (lattice vector in the $x$-direction) and $\bm{\mu}'_2=\bm{y}$. \textbf{(b)}: $\bm{\mu}_1=\bm{x}$ and $\bm{\mu}'_2=\bm{x}$.}
  \label{fig:Kmufig}
\end{figure*}
\begin{figure*}
  \centering
  \includegraphics[width=0.94\textwidth]{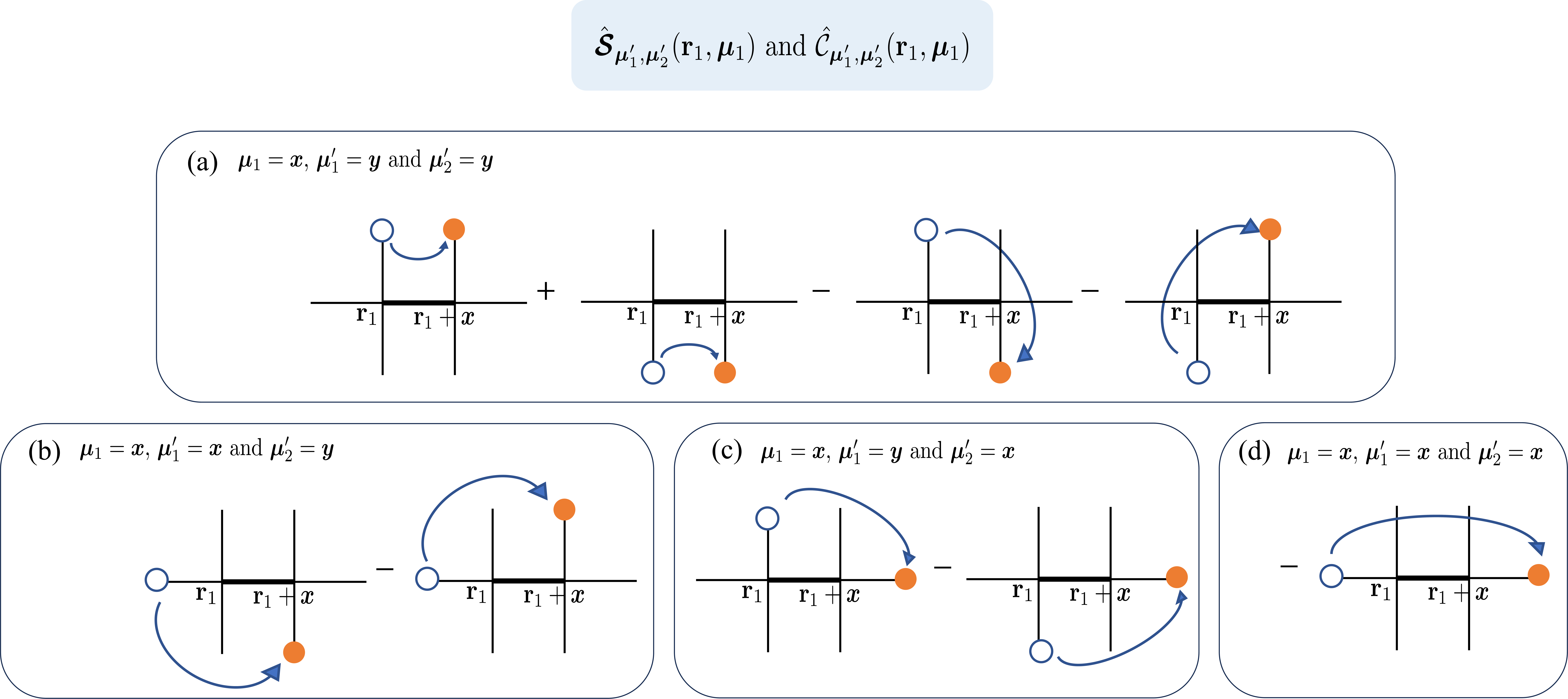}
  \caption{A visualization of the definition of the operators $\hat{\bm{\mathcal{S}}}_{\bm{\mu}'_1,\bm{\mu}'_2}(\vb{r}_1,\bm{\mu}_1)$ and $\hat{\mathcal{C}}_{\bm{\mu}'_1,\bm{\mu}'_2}(\vb{r}_1,\bm{\mu}_1)$ (see Eq.~\eqref{eq:defmathcalS} and Eq.~\eqref{eq:defmathcalC}). Both the operators create a doublon-hole pair when acting on a state in the spin sector. $\hat{\bm{\mathcal{S}}}$ is a spin triplet operator, while $\hat{\mathcal{C}}$ is a spin singlet operator. Since this is the only difference between the two, we use the same figure to denote both $\hat{\bm{\mathcal{S}}}$ and $\hat{\mathcal{C}}$. The operators are specified by a bond $(\vb{r}_1,\bm{\mu}_1)$ and two additional lattice directions $\bm{\mu}'_1$ and $\bm{\mu}'_2$. The four subfigures show the definition for different choices of these directions.}
  \label{fig:SandC}
\end{figure*}

\subsubsection{Defining operators $\hat{A}_j$, $\hat{B}_j$, and $\hat{C}_j$}\label{sec:ABCdefs}
We emphasize that throughout this section, bonds $(\vb{r},\vb{r}+\bm{\mu})$ are \textit{not} restricted to be nearest-neighbors. First, let us define 
\begin{equation}\label{eq:defJmu}
    \hat{\mathcal{J}}_{\bm{\mu}}\equiv i\sum_{\vb{r},\sigma}\left(\tunn_{\vb{r},\vb{r}+\bm{\mu}}\hat{c}^{\dagger}_{\vb{r}+\bm{\mu}, \sigma}\hat{c}_{\vb{r}, \sigma}-\tunn^*_{\vb{r},\vb{r}+\bm{\mu}}\hat{c}^{\dagger}_{\vb{r}, \sigma}\hat{c}_{\vb{r}+\bm{\mu}, \sigma}\right).
\end{equation}
 $\hat{\mathcal{J}}_{\bm{\mu}}$ is related to the global electric current in the direction $\bm{\mu}$.

Next, we define an operator $\hat{\mathcal{K}}_{\bm{\mu}'_{2}}\left(\vb{r}_{1},\bm{\mu}_{1}\right)$
which is a sum of operators that induce electron tunneling along
the lattice vector $\bm{\mu}'_{2}$ with the following additional
constraint. As shown in Fig.~\ref{fig:Kmufig}, the constituent terms
belong to the vicinity of the bond $(\vb{r}_{1},\bm{\mu}_{1})$, such
that each term in $\hat{\mathcal{K}}_{\bm{\mu}'_{2}}\left(\vb{r}_{1},\bm{\mu}_{1}\right)$
includes the sites $\vb{r}_{1}$ or $\vb{r}_{1}+\bm{\mu}_{1}$ as
a starting or the ending point of the tunneling. The formal definition
is 
\begin{equation}\label{eq:defKmu}
 \begin{aligned}   &\hat{\mathcal{K}}_{\bm{\mu}'_2}\left(\vb{r}_1,\bm{\mu}_1\right)\\
 &\equiv i\sum_{\vb{r}'_2,\sigma'_2}\eta^{(\vb{r}'_2,\bm{\mu}'_2)}_{(\vb{r}_1,\bm{\mu}_1)}\left(\tunn_{\vb{r}'_2,\vb{r}'_2+\bm{\mu}'_2}\hat{c}^{\dagger}_{\vb{r}'_2+ \bm{\mu}'_2,\sigma'_2}\hat{c}_{\vb{r}'_2\sigma'_2}-\text{h.c.}\right),
 \end{aligned}
 \end{equation}
where we have introduced a symbol $\eta_{(\vb{r}_1,\bm{\mu}_1)}^{(\vb{r}'_2,\bm{\mu}'_2)}$ that is a function of two bonds $(\vb{r}_1,\bm{\mu}_1)$ and $(\vb{r}'_2,\bm{\mu}'_2)$, and is symmetric in its two arguments. It is defined as
\begin{equation}\label{eq:defeta}
    \eta_{(\vb{r}_1,\bm{\mu}_1)}^{(\vb{r}'_2,\bm{\mu}'_2)}=\begin{cases}
        1,\text{ if bonds }(\vb{r}'_2,\bm{\mu}'_2)\text{ and }(\vb{r}_1,\bm{\mu}_1)\text{ have}\\ \ \quad  \text{at least one site in common.}\\
        0, \text{ otherwise.}
    \end{cases}
\end{equation}

Now, we define another set of tunneling operators
--- a spin triplet operator $\hat{\bm{\mathcal{S}}}_{\bm{\mu}'_{1},\bm{\mu}'_{2}}(\vb{r}_{1},\bm{\mu}_{1})$
and a spin singlet operator $\hat{\mathcal{C}}_{\bm{\mu}'_{1},\bm{\mu}'_{2}}(\vb{r}_{1},\bm{\mu}_{1})$,
shown in Fig.~\ref{fig:SandC}. They induce an electron tunneling
between the two distinct sites adjacent to the bond $(\vb{r}_{1},\bm{\mu}_{1})$
(see Fig.~\ref{fig:SandC}). The locations of these two sites can be
parametrized as $\vb{r}_{1}\pm\bm{\mu}'_{1}$ and $\vb{r}_{1}+\bm{\mu}_{1}\pm\bm{\mu}'_{2}$. The constraint that these two sites together with $\vb{r}_1$ and $\vb{r}_1+\bm{\mu}_1$ constitute four distinct sites is enforced using the following notation (for nonzero lattice vectors $\bm{\mu}$,
$\bm{\nu}$ and $\bm{\rho}$):
\begin{equation}\label{eq:defhfunction}
    h\left(\bm{\mu},\bm{\nu},\bm{\rho}\right)=\begin{cases}
        &0,\text{ if }\bm{\mu}+\bm{\rho}=\vb{0},\\ &\quad \ \text{or if }\bm{\nu}=\bm{\mu}\\ &\quad \ \text{or if }\bm{\mu}+\bm{\rho}=\bm{\nu}.\\
        &1, \text{ otherwise}.
    \end{cases}
\end{equation}
Let us also define:
\begin{equation}
    \begin{aligned}
&\tunn^3_{\text{eff}}\left(\bm{\mu}'_1,\bm{\mu}'_2,\vb{r}_1,\bm{\mu}_1\right)\\
&\equiv \tunn_{\vb{r}_1+s'_1\bm{\mu}'_1,\vb{r}_1}\tunn_{\vb{r}_1,\vb{r}_1+\bm{\mu}_1}\tunn_{\vb{r}_1+\bm{\mu}_1,\vb{r}_1+\bm{\mu}_1+s'_2\bm{\mu}'_2}.
    \end{aligned}
\end{equation}
With this notation, $\hat{\bm{\mathcal{S}}}_{\bm{\mu}'_1,\bm{\mu}'_2}(\vb{r}_1,\bm{\mu}_1)$ and $\hat{\mathcal{C}}_{\bm{\mu}'_1,\bm{\mu}'_2}(\vb{r}_1,\bm{\mu}_1)$ are defined as
\begin{equation}\label{eq:defmathcalS}
\begin{aligned}
    &\hat{\bm{\mathcal{S}}}_{\bm{\mu}'_1,\bm{\mu}'_2}(\vb{r}_1,\bm{\mu}_1)=\frac{1}{2}\sum_{\substack{s'_1\in\{\pm 1\}\\s'_2\in\{\pm 1\}}}s'_1 s'_2\Biggl\{\Biggl[ h\left(\bm{\mu}_1,s'_1\bm{\mu}'_1,s'_2\bm{\mu}'_2\right)\\
    & \times \sum_{\alpha,\beta}\left(\tunn^3_{\text{eff}}\left(\bm{\mu}'_1,\bm{\mu}'_2,\vb{r}_1,\bm{\mu}_1\right)\hat{c}^{\dagger}_{\vb{r}_1+\bm{\mu}_1+s'_2\bm{\mu}'_2,\alpha}\bm{\sigma}_{\alpha \beta}\hat{c}_{\vb{r}_1+s'_1\bm{\mu}'_1,\beta}\right.\\
    &\left.\quad \quad\quad \quad +\text{h.c.}\right)\Biggr]+ \left[\bm{\mu}'_1 \leftrightarrow \bm{\mu}'_2\right]\Biggr\}, \text{ and}
    \end{aligned}
\end{equation}
\begin{equation}\label{eq:defmathcalC}
\begin{aligned}
    &\hat{\mathcal{C}}_{\bm{\mu}'_1,\bm{\mu}'_2}(\vb{r}_1,\bm{\mu}_1)=\frac{i}{2}\sum_{\substack{s'_1\in\{\pm 1\}\\s'_2\in\{\pm 1\},\alpha}} s'_1 s'_2\Biggl\{\Biggl[ h\left(\bm{\mu}_1,s'_1\bm{\mu}'_1,s'_2\bm{\mu}'_2\right)\\
    &\times\left(\tunn^3_{\text{eff}}\left(\bm{\mu}'_1,\bm{\mu}'_2,\vb{r}_1,\bm{\mu}_1\right)\hat{c}^{\dagger}_{\vb{r}_1+\bm{\mu}_1+s'_2\bm{\mu}'_2,\alpha}\hat{c}_{\vb{r}_1+s'_1\bm{\mu}'_1,\alpha}-\text{h.c.}\right)\Biggr]\\
    &\quad \quad \quad+ \left[\bm{\mu}'_1 \leftrightarrow \bm{\mu}'_2\right]\Biggr\}.
    \end{aligned}
\end{equation}
Next, we define a symmetric tunneling operator 
\begin{equation}\label{eq:defHrr}
    \begin{aligned}
    &\hat{\mathcal{H}}_{\vb{r},s\bm{\mu},s'\bm{\mu}'}\\
    &=\sum_{\alpha}\tunn^*_{\vb{r},\vb{r}+s\bm{\mu}}\tunn_{\vb{r},\vb{r}+s'\bm{\mu}'}\hat{c}^{\dagger}_{\vb{r}+s'\bm{\mu}',\alpha}\hat{c}_{\vb{r}+s\bm{\mu},\alpha}+\text{h.c.}
    \end{aligned}
\end{equation}
and a local spin current
\begin{equation}\label{eq:defJSrr}
\begin{aligned}
    &\hat{\bm{\mathcal{J}}}^S_{\vb{r},s\bm{\mu},s'\bm{\mu}'}\\
    &=i\sum_{\alpha}\left(\tunn^*_{\vb{r},\vb{r}+s\bm{\mu}}\tunn_{\vb{r},\vb{r}+s'\bm{\mu}'}\hat{c}^{\dagger}_{\vb{r}+s'\bm{\mu}',\alpha}\bm{\sigma}_{\alpha \beta}\hat{c}_{\vb{r}+s\bm{\mu},\beta}-\text{h.c.}\right).
\end{aligned}    
\end{equation}

Using these local operators, we now define the following
global (zero-momentum) matter operators $\hat{A}_{j}$, $\hat{B}_{j}$
and $\hat{C}_{j}$ that describe photoemission with the polarization
matching that of the $j$-th detector $j\in\{1,2\}$. To leading order
in laser-matter coupling $g_{L}$ and $\tunn/\abs{\omega_{L}-U}$,
the operators are:
\begin{equation}\label{eq:AiExp}
\begin{aligned}
   &\hat{A}_j=\frac{g_L g}{\omega_L - U}\\
   &\times\sum_{(\vb{r},\bm{\mu})}\abs{\tunn_{\vb{r},\vb{r}+\bm{\mu}}}^2\left(4\hat{\vb{S}}_{\vb{r}}\cdot \hat{\vb{S}}_{\vb{r} + \bm{\mu}}-1\right)\left(\bar{\bm{\mu}}\cdot \vb{e}_L\right)\left(\bar{\bm{\mu}}\cdot\vb{e}^*_j\right),
\end{aligned}
\end{equation}
The operator $\hat{A}_{j}$ is therefore equivalent
to the Fleury-Loudon operator that describes the conventional Raman
scattering (Eq.~\eqref{eq:R1exp}).

In contrast, operator $\hat{B}_{j}$ is a mixed spin-charge
operator that creates a doublon-hole pair and spin excitations. It
is defined as (where we make use of the definitions in Eq.~\eqref{eq:defHrr},
\eqref{eq:defJSrr}, \eqref{eq:defKmu}, \eqref{eq:defmathcalS} and \eqref{eq:defmathcalC}):
\begin{widetext}
\begin{equation}\label{eq:BiExp}
    \begin{aligned}
        &\hat{B}_j = g_L^2  g \left\{ \frac{1}{\omega_L-U}\sum_{\vb{r},\bm{\mu}',\bm{\mu}}\sum_{\substack{s, s' =\pm  1\\s\bm{\mu} \neq s' \bm{\mu}'}}\biggl[\left(s'\bar{\bm{\mu}}'\cdot \vb{e}_L\right)\left(s\bar{\bm{\mu}}\cdot \vb{e}_L\right)\vb{e}^*_j\cdot \Bigl[ \left(s\bar{\bm{\mu}}-s'\bar{\bm{\mu}}'\right)\hat{\bm{\mathcal{J}}}^S_{\vb{r},s\bm{\mu},s'\bm{\mu}'}\cdot\hat{\vb{S}}_{\vb{r}}+\frac{i}{2}\left(s\bar{\bm{\mu}}+s'\bar{\bm{\mu}}'\right)\hat{\mathcal{H}}_{{\vb{r}},s\bm{\mu},s'\bm{\mu}'}\Bigr]\biggr]\right. \\
         &\quad \quad\quad +\frac{1}{(\omega_L-U)^2}\sum_{(\vb{r},\bm{\mu}),\bm{\mu}'}\biggl[\left(\bar{\bm{\mu}}'\cdot \vb{e}_L\right)\left(\bar{\bm{\mu}}\cdot \vb{e}_L\right)\left(\bar{\bm{\mu}}\cdot\vb{e}^*_j\right)\hat{\mathcal{K}}_{\bm{\mu}'}\left(\vb{r},\bm{\mu}\right)\abs{\tunn_{\vb{r},\vb{r}+\bm{\mu}}}^2\left(4\hat{\vb{S}}_{\vb{r}}\cdot \hat{\vb{S}}_{\vb{r} + \bm{\mu}}-1\right)\biggr]\\ 
          &\quad\quad \quad +\frac{1}{\left(\omega_L-U\right)^2}\sum_{\substack{(\vb{r},\bm{\mu}),\\ \bm{\mu}'_1,\bm{\mu}'_2}}\left[\left(\bar{\bm{\mu}}'_1\cdot \vb{e}_L\right)\left(\bar{\bm{\mu}}'_2\cdot \vb{e}_L\right)\left(\bar{\bm{\mu}}\cdot\vb{e}^*_j\right) \left[i\hat{\bm{\mathcal{S}}}_{\bm{\mu}'_1,\bm{\mu}'_2}(\vb{r},\bm{\mu})\cdot\biggl(\frac{\hat{\vb{S}}_{\vb{r}}-\hat{\vb{S}}_{\vb{r}+\bm{\mu}}}{2}-i\hat{\vb{S}}_{\vb{r}}\times \hat{\vb{S}}_{\vb{r}+\bm{\mu}}\biggr) \right.\right.\\
    &\quad\quad\quad\quad \quad\quad\quad \quad\quad \quad\left.   +\hat{\mathcal{C}}_{\bm{\mu}'_1,\bm{\mu}'_2}(\vb{r},\bm{\mu})\left(\hat{\vb{S}}_{\vb{r}}\cdot \hat{\vb{S}}_{\vb{r} + \bm{\mu}}-\frac{1}{4}\right)\Biggr]\Biggr]\vphantom{\sum_{\bm{\mu}',\bm{\mu}}\sum_{\substack{s, s' =\pm  1\\s\bm{\mu} \neq s' \bm{\mu}'}}}-\frac{1}{2}\sum_{\bm{\mu}}\left(\bar{\bm{\mu}}\cdot\vb{e}_L \right)^2\left(\bar{\bm{\mu}}\cdot\vb{e}^*_j \right)\hat{\mathcal{J}}_{\bm{\mu}}\right\}.
    \end{aligned}
\end{equation}
The first line in Eq.~\eqref{eq:BiExp} comes from
the process shown in Fig.~\ref{fig:description2}(d$_{1}$-d$_{5}$),
and involves scattering described by the diamagnetic term. The second
line above, involving $\hat{\mathcal{K}}_{\bm{\mu}'}(\vb{r},\bm{\mu})$
results from the incomplete cancellation of the processes in Fig.~\ref{fig:description2}(b$_{1}$-b$_{6}$)
and Fig.~\ref{fig:description2}(c$_{1}$-c$_{6}$) (see Sec.~\ref{sec:processes}
for a discussion). The last two lines involving $\hat{\bm{\mathcal{S}}}_{\bm{\mu}'_{1},\bm{\mu}'_{2}}(\vb{r},\bm{\mu})$
and $\hat{\mathcal{C}}_{\bm{\mu}'_{1},\bm{\mu}'_{2}}(\vb{r},\bm{\mu})$
originate from the process shown in Fig.~\ref{fig:description2}($c'_{1}$-$c'_{6}$).

Finally, the operators $\hat{C}_{j}$ that also couple
the spin and charge sectors are defined as: 
\begin{equation}\label{eq:CiExp}
\begin{aligned}
   \hat{C}_j=&ig\sum_{(\vb{r},\bm{\mu}),\sigma}\left(\bar{\bm{\mu}}\cdot\vb{e}^*_j\right)\left(\tunn_{\vb{r},\vb{r}+\bm{\mu}}\hat{c}^{\dagger}_{\vb{r}+\bm{\mu}, \sigma}\hat{c}_{\vb{r}, \sigma}-\tunn^*_{\vb{r},\vb{r}+\bm{\mu}}\hat{c}^{\dagger}_{\vb{r}, \sigma}\hat{c}_{\vb{r}+\bm{\mu}, \sigma}\right)\\
   \equiv&g\sum_{\bm{\mu}}\left(\bar{\bm{\mu}}\cdot \vb{e}^*_j\right)\hat{\mathcal{J}}_{\bm{\mu}} .
\end{aligned}
\end{equation}
$\hat{C}_j$ is proportional to the component of the global electron current along the direction of $\vb{e}^*_j$. 
\end{widetext}
We can run a sanity check on our expressions Eqs.~(\ref{eq:AiExp}-\ref{eq:CiExp})
by performing a symmetry analysis. Specifically, because of the absence of spin-orbit coupling in the microscopic Hamiltonian, all
terms should transform as a singlet under spin rotation. One can easily check
that this holds for the operators $\hat{A}_{j}$ and $\hat{C}_{j}$.
This also holds for $\hat{B}_j$. For example, the tunneling operator $\hat{\bm{\mathcal{S}}}_{\bm{\mu}'_1,\bm{\mu}'_2}(\vb{r},\bm{\mu})$ transforms as a triplet, but after taking the scalar product with $\left((\hat{\vb{S}}_{\vb{r}}-\hat{\vb{S}}_{\vb{r}+\bm{\mu}})/2-i\hat{\vb{S}}_{\vb{r}}\times \hat{\vb{S}}_{\vb{r}+\bm{\mu}}\right)$, the result transforms as a spin singlet. Next, if the input light and the detectors are linearly polarized, one can check that each term in $\hat{B}_j$ and $\hat{C}_j$ is odd under time-reversal transformation. This is also expected from  the microscopic form of the light-matter coupling  $\sim\hat{\vb{A}}\cdot \hat{\bm{\mathcal{J}}}$  and the fact that the electric current $\hat{\bm{\mathcal{J}}}$ is odd under time-reversal transformation. Therefore, for linearly polarized input and output, since $\hat{B}_j$ arises from a three-photon process (absorption of two and emission of one), and $\hat{C}_j$ arises from a single-photon process (emission of one), they should both be odd under time-reversal.

We note that in Eq.~\eqref{eq:BiExp}, the leading-order terms at face value (in $\tunn/\abs{\omega_L-U}$), are those in the first line, involving the diamagnetic term. However, after taking the expectation value in a matter state, we expect the contribution from this set of terms to be small. We can directly see that this is the case when the time delay between application of $\hat{B}_1$ and $\hat{C}_2$  on the matter state is 0. This is because when acting on a state in the spin sector,
\begin{align}
    \left(\hat{c}^{\dagger}_{\vb{r}_2,\alpha}\hat{c}_{\vb{r}_1,\beta}-\text{h.c.}\right)\left(\hat{c}^{\dagger}_{\vb{r}_2,\beta}\bm{\sigma}_{\beta\gamma}\hat{c}_{\vb{r}_1,\gamma}-\text{h.c.}\right)=0,\text{ and}\label{eq:diamzero1}\\
    \left(\hat{c}^{\dagger}_{\vb{r}_2,\alpha}\hat{c}_{\vb{r}_1,\beta}-\text{h.c.}\right)\left(\hat{c}^{\dagger}_{\vb{r}_2,\beta}\hat{c}_{\vb{r}_1,\beta}+\text{h.c.}\right)=0\label{eq:diamzero2}.
\end{align}
However, if the time-delay is nonzero, the contribution from the diamagnetic term may be nonzero.
\subsubsection{Matrix element of $\hat{R}^{(2)}$ in terms of $\hat{A}_j$, $\hat{B}_j$ and $\hat{C}_j$}
Using operators Eqs.~(\ref{eq:AiExp}-\ref{eq:CiExp})
we now consider the full operator $\hat{R}^{(2)}$ describing the amplitude to absorb two photons and emit two photons. The matrix elements of $\hat{R}^{(2)}$
between matter energy states $\ket{I}$ and $\ket{F}$ (both in the
spin sector) have the following form (see Appendix~\ref{sec:explicitmicro} for further details):
\begin{align}\label{eq:R2final}
    &\mel{F}{\hat{R}^{(2)}_{\lambda_1,\lambda_2}}{I}\nonumber \\=&-\sum_J\left[ \frac{\mel{F}{\hat{A}_2}{J}\mel{J}{\hat{A}_1}{I}}{\omega_{\lambda_1} - \left(\omega_L-E_{JI}+i0^+\right)}+\left(1\leftrightarrow 2\right)\right]\nonumber \\
    &-\sum_K\left[ \frac{\mel{F}{\hat{C}_2}{K}\mel{K}{\hat{B}_1}{I}}{\omega_{\lambda_1} - \left(2\omega_L-E_{KI}+i0^+\right)}+\left(1\leftrightarrow 2\right)\right],
\end{align}
where $\ket{J}$ and $\ket{K}$ are the intermediate
many-body eigenstates, belonging to the spin and charge sectors respectively.
$E_{JI}$, the energy difference between states $\left|J\right\rangle $
and $\left|I\right\rangle $ is of order $\tunn^{2}/U$, while $E_{KI}$
is of order $U\pm\text{order}(\tunn)$. Therefore, from Eq.~\eqref{eq:R2final}
we explicitly see that the generation of photon pair with frequencies near $\omega_{L}$ (central peak peak in Fig.~\ref{fig:sectors}(e))
is accompanied by the action of pure spin operators $\hat{A}_{j}$
on the matter state. In contrast, the sideband emission with the frequencies
$2\omega_{L}-U$ and $U$(Fig.~\ref{fig:sectors}(f)), is associated
with the action of operators $\hat{B}_{j}$ and current operator $\hat{C}_{j}$
respectively.

The energy dependent factors in Eq.~\eqref{eq:R2final} can be absorbed into the Heisenberg evolution of the matter operators (see Appendix~\ref{sec:explicitmicro} for a derivation), and Eq.~\eqref{eq:R2final} can be written as

\begin{equation}\label{eq:R2finaltime}
    \begin{aligned}
        &\mel{F}{\hat{R}^{(2)}_{\lambda_1,\lambda_2}}{I}\\
        &=-i\int_{-\infty}^{\infty}\dd{t}e^{-i(\omega_{\lambda_1}-\omega_L)t}\mel{F}{\mathbb{T}\left[\hat{A}_2(0) \hat{A}_1(-t)\right]}{I}\\
        & -i\int_{-\infty}^{\infty} \dd{t}\bra{F}\left[\theta(t)e^{-i\left(\omega_{\lambda_1}- 2\omega_L\right) t}\hat{C}_2(0)\hat{B}_1(-t)\right.\\
        & \left. \quad \quad \quad \quad \quad \quad +\theta(-t)e^{-i\omega_{\lambda_1} t} \hat{C}_1(-t)\hat{B}_2(0)\right]\ket{I},
    \end{aligned}
\end{equation}
where $\mathbb{T}[\ ]$ denotes time-ordering of operators inside $[\ ]$.

Note that the formula for $G^{(2)}(\tau)$ in Eq.~\eqref{eq:G2simp2} involves an integral over $\omega_{\lambda_1}$ and $\omega_{\lambda_2}$, i.e., coherent superpositions of the different $\hat{R}^{(2)}_{\lambda_1,\lambda_2}$'s. Therefore, the frequency filter functions $\mathcal{F}_i(\omega)$ of the detectors play a crucial role in determining $G^{(2)}(\tau)$. With this in mind, in the following section, we investigate the temporal structure of the matter correlation functions inferred from measurement of \(G^{(2)}(\tau)\), with an emphasis on their dependence on the frequency filter functions of the detectors.
\section{Temporal structure of correlation functions}\label{sec:temporal}
In this section, we combine
the results of Sec.~\ref{sec:R2micro} (expressions for matter operators $\hat{R}^{(1)}$ and $\hat{R}^{(2)}$) and Sec.~\ref{sec:Hubbardsetup} (Eqs.~(\ref{eq:G1simp}-\ref{eq:X2plussimp}) to derive the central
relations of this work summarized in Table~\ref{tab:dictionary1}.

It is convenient to work with the Fourier transform
of the filter function defined as $\tilde{\mathcal{F}}_{j}(t)=\int_{-\infty}^{\infty}\frac{\dd{\omega}}{2\pi}\mathcal{F}_{j}(\omega)e^{-i\omega t}$.
In order to simplify the expressions, we absorb phases such as $e^{-iE_{JI}t}$
into Heisenberg time-evolution of the operators involved in the correlation
functions. 
 \subsection{Intensity $G^{(1)}(0
 )$}
 The expression for $G^{(1)}_{d_j}(0)=\bra{\text{out}}\hat{a}_{d_j}^{\dagger}(0)\hat{a}_{d_j}(0)\ket{\text{out}}$ in Eq.~\eqref{eq:G1simp} can be simplified to
\begin{equation}\label{eq:G1simp2}
    \begin{aligned}
        G^{(1)}_{d_j}(0)\approx &\iint_{-\infty}^{\infty}\dd{t}\dd{t'}\Tilde{\mathcal{F}}_j(t)\left[\Tilde{\mathcal{F}}_j(t')\right]^*e^{i\omega_L (t-t')}\\ 
        &\quad \quad \quad \times \mel{I}{\left[\hat{A}_j(-t')\right]^{\dagger}\hat{A}_j(-t)}{I},
    \end{aligned}
\end{equation}
where  $\hat{A}_{j}$ is defined in Eq.~\eqref{eq:R1exp}.
The operators in Eq.~\eqref{eq:G1simp2} are represented in the Heisenberg
picture such that $\hat{A}(t)\equiv e^{i\hat{H}_{0}t}\hat{A}e^{-i\hat{H}_{0}t}$.
To connect to the known results~\cite{shastry1990theory,shastry1991raman,ko2010raman},
let us consider the case of a Lorentzian effective filter function,
defined in Eq.~\eqref{eq:Lorentzdef} that is peaked in frequency
around $\omega_{j}$ with a width $\Gamma_{j}$. In terms of temporal variables, we get:
\begin{equation}\label{eq:Lorentztildedef}
    \tilde{\mathcal{F}}_{j,\text{ Lorentzian}}(t)=\Kconst_j\Gamma_j \theta(t)e^{-i\omega_j t}  e^{-\Gamma_j t},
\end{equation}
where $\Kconst$ is a constant and was defined in Eq.~\eqref{eq:Kdef}. In temporal variables, Eq.~\eqref{eq:G1simp2} reads:
\begin{equation}\label{eq:G1LorentzFilter}
 \begin{aligned}
    &G^{(1)}(\omega_j)\approx \abs{\Kconst_j}^2 \Gamma_j/2 \\
    & \quad \times\int_{-\infty}^{\infty}\dd{t} e^{-\Gamma_j\abs{t}}e^{i(\omega_L - \omega_j)t} \expval{\left[\hat{A}_j(t)\right]^{\dagger}\hat{A}_j(0)}_0,
\end{aligned}   
\end{equation}
As in Eq.~\eqref{eq:G1simp2}, the expectation value
is taken in a matter eigenstate $\ket{I}$ in the spin sector. It
can be straightforwardly generalized to any state or a density
matrix describing thermal equilibrium (henceforth denoted with a subscript
$0$ as in Eq.~\eqref{eq:G1LorentzFilter}) within the spin sector.
Eq.~\eqref{eq:G1simp2} is equivalent to the results of Ref.~\cite{shastry1990theory,shastry1991raman,ko2010raman}.
Thus, $G^{(1)}$ measures the dynamical fluctuations of spin singlet
projection operators. In the special case of $A_{2g}$ channel for
the kagome lattice, it measures the dynamical fluctuations of spin
chirality operators. 
\subsection{First-order quadrature correlator $X^{+}$}
Using Eqs.~(\ref{eq:X1simp},\ref{eq:R1exp}),
we get the first-order quadrature operator expectation value: 
\begin{equation}\label{eq:X1matter}
    \begin{aligned}
       \expval{\hat{a}_{d_j}(0)}_{\text{out}}=\mathcal{F}_j(\omega_L) \expval{\hat{A}_j(0)}_0.
    \end{aligned}
\end{equation}
It follows that, first, the signal for the quadrature correlator \(X^{+}\) is sharply peaked at the laser frequency \(\omega_L\), corresponding to the \textit{elastic} scattering of photons. Second, this correlator provides a direct measurement of the static expectation value of the operator \(\hat{A}_j\), which, in most cases, is a sum of spin-singlet projection operators. However, on the kagome lattice in $A_{2g}$ ($\left(e^x_j\right)^*e^y_L-\left(e^y_j\right)^* e^x_L$) channel, to leading order in $\tunn/\abs{\omega_L-U}$, this operator is a linear combination of spin chirality operators \cite{ko2010raman}. While fluctuations of spin chirality have been proposed to be measured via neutron  \cite{lee2013proposal} and Raman scattering \cite{ko2010raman}, the first-order photonic quadrature correlator introduced in this work enables a direct measurement of static spin chirality. It is important to note that a nonzero signal in this channel is only possible if the ground state spontaneously breaks reflection and time-reversal symmetries. 
\subsection{Phase-sensitive second order quadrature correlation $X^{++}(\tau)$}
Analyzing Eqs.~(\ref{eq:X2plussimp}, \ref{eq:G2simp2}),
we find that to derive the matter correlators measured by $X_{d_{1},d_{2}}^{++}(\tau)=\expval{\hat{a}_{d_{2}}(\tau)\hat{a}_{d_{1}}(0)}_{\text{out}}$
and $G_{d_{1},d_{2}}^{(2)}(\tau)=\langle\hat{a}_{d_{1}}^{\dagger}(0)\hat{a}_{d_{2}}^{\dagger}(\tau)\hat{a}_{d_{2}}(\tau)\hat{a}_{d_{1}}(0)\rangle_{\text{out}}$,
we need to express the following term using the expressions for $\hat{A}_{j}$,
$\hat{B}_{j}$ and $\hat{C}_{j}$:
\begin{align}\label{eq:interamp}
    &i \int_{-\infty}^{\infty} \frac{\dd{\omega_{\lambda_1}}}{2\pi}\mathcal{F}_{1}(\omega_{\lambda_1})\mathcal{F}_{2}(2\omega_L-\omega_{\lambda_1}-E_{FI})  \nonumber \\& \quad \quad \quad \times e^{i\omega_{\lambda_1}\tau} \mel{F}{\hat{R}^{(2)}_{\lambda_1,\lambda_2}(\omega_{\lambda_1},\omega_{\lambda_2})}{I}. 
\end{align}

In the following, we assume that the filters $\mathcal{F}_{i}(\omega)$
are sensitive enough to distinguish frequencies around $\omega_{L}$
from those around $2\omega_{L}-U$ and $U$. Therefore, we ignore
any interference between these sets of amplitudes. Without loss of
generality, we assume that $\tau>0$, i.e., detector $2$ clicks after
detector $1$.

Substituting the matrix element $\mel{F}{\hat{R}^{(2)}_{\lambda_1,\lambda_2}(\omega_{\lambda_1},\omega_{\lambda_2})}{I}$ obtained in Eq.~\eqref{eq:R2finaltime} into the expression in Eq.~\eqref{eq:interamp}, and only keeping the term corresponding to the frequency range of interest, we get:
\begin{widetext}
    \begin{equation}\label{eq:transampfull}
    \begin{aligned}
        &i\int_{-\infty}^{\infty} \frac{\dd{\omega_{\lambda_1}}}{2\pi}\mathcal{F}_{1}(\omega_{\lambda_1})\mathcal{F}_{2}(2\omega_L-\omega_{\lambda_1}-E_{FI})  e^{i\omega_{\lambda_1}\tau}  \mel{F}{\hat{R}^{(2)}_{\lambda_1,\lambda_2}}{I}\\
        & \quad \quad \quad= e^{i(2\omega_L-E_{FI})\tau} \int\limits_0\limits^{\infty}\int\limits_0\limits^{\infty}\dd{t_1}\dd{t_2}\tilde{\mathcal{F}}_1(t_1)\tilde{\mathcal{F}}_2(t_2)\mel{F}{\hat{M}^{(2)}_{d_1,d_2}(\tau-t_2,-t_1)}{I}, \text{ where}\\
        &\hat{M}^{(2)}_{d_1,d_2}(\tau-t_2,-t_1)=\begin{cases} e^{i\omega_L\left( t_1+ t_2 - \tau\right)}\mathbb{T}\left[\hat{A}_2(\tau-t_2)\hat{A}_1(-t_1)\right]
        \\
        \quad  \text{if both detectors detect near $\omega_L$,}\\
        \\
        e^{2i\omega_L t_1}\theta(t_1+\tau-t_2)\hat{C}_2(\tau-t_2)\hat{B}_1(-t_1)\\
        \quad  \text{if $d_1$ detects near $2\omega_L-U$ and $d_2$ detects near $U$,}\\
        \\
        e^{2i \omega_L( t_2 -\tau)}\theta(t_2-t_1-\tau)\hat{C}_1(-t_1)\hat{B}_2(\tau-t_2)\\
        \quad  \text{if  $d_1$ detects near $U$ and  $d_2$ detects near $2\omega_L-U$}.
        \end{cases}
    \end{aligned}
    \end{equation}
    \end{widetext}
    The above operator is directly related to the operator $\hat{M}_{j}(t)$ introduced in Sec.~\ref{sec:iomap} [see Eq.~\eqref{eq:Meffint} and Eq.~\eqref{eq:MjtoABC}] as:
    \begin{equation}
        \hat{M}^{(2)}_{d_1,d_2}(\tau-t_2,-t_1)=\mathbb{T}\left[\hat{M}_2(\tau-t_2)\hat{M}_1(-t_1)\right].
    \end{equation}
We now substitute Eq.~\eqref{eq:transampfull} in Eq.~\eqref{eq:X2plussimp} and obtain:
\begin{equation}\label{eq:X2matter}
    \begin{aligned}
        \expval{\hat{a}_{d_2}(\tau)\hat{a}_{d_1}(0)}_{\text{out}}\approx  &\iint_{-\infty}^{\infty}\dd{t_1}\dd{t_2}\tilde{\mathcal{F}}_1(t_1)\tilde{\mathcal{F}}_2(t_2)\\ &\times\expval{\hat{M}^{(2)}_{d_1,d_2}(\tau-t_2,-t_1)}_0,
    \end{aligned}
    \end{equation}
where $\hat{M}^{(2)}_{d_1,d_2}(\tau-t_2,-t_1)$ is a time-ordered product of operators, as defined in Eq.~\eqref{eq:transampfull}. 

Assuming  Lorentzian  filter functions defined in Eq.~\eqref{eq:Lorentztildedef}, and taking the large $\Gamma_1, \Gamma_2$ limit, i.e., the limit of frequency selectivity being broad, the above equation becomes
\begin{equation}\label{eq:quad2broad}
    \begin{aligned}
    \expval{\hat{a}_{d_2}(\tau)\hat{a}_{d_1}(0)}_{\text{out}}\Big|_{\Gamma_1,\Gamma_2\to \infty}\approx  \Kconst_1\Kconst_2\expval{\hat{M}^{(2)}_{d_1,d_2}(0,-\tau)}_0.
    \end{aligned}
\end{equation}
We remind the reader that measurement of the above correlator requires a phase-sensitive quadrature measurement in which the input drive is in a coherent state. Such a scheme is described in Sec.~\ref{sec:quadraturemeasurements}.  
\subsection{Photon pair correlation function $G^{(2)}(\tau)$}
Now, we are ready to derive the matter correlator measured by $G^{(2)}_{d_1,d_2}(\tau)=\langle\hat{a}^{\dagger}_{d_1}(0)\hat{a}_{d_2}^{\dagger}(\tau)\hat{a}_{d_2}(\tau)\hat{a}_{d_1}(0)\rangle_{\text{out}}$. We substitute Eq.~\eqref{eq:transampfull} into Eq.~\eqref{eq:G2simp2} and sum over final states $\ket{F}$ to obtain: 
    \begin{widetext}
\begin{equation}\label{eq:G2matterfull}
    \begin{aligned}
        &G^{(2)}_{d_1,d_2}(\tau)\approx \iint_{0}^{\infty}\iint_{0}^{\infty}\dd{t_1}\dd{t_2}\dd{t'_1}\dd{t'_2}\tilde{\mathcal{F}}_1(t_1)\tilde{\mathcal{F}}_2(t_2)[\tilde{\mathcal{F}}_1(t'_1)]^*[\tilde{\mathcal{F}}_2(t'_2)]^* \ \mathcal{C}^{(2)}_{d_1,d_2}(-t'_1,\tau-t'_2;\tau-t_2,-t_1),
    \end{aligned}
    \end{equation} 
    where 
 \begin{equation}\label{eq:C2matterfull}
    \begin{aligned}
        &\mathcal{C}^{(2)}_{d_1,d_2}(-t'_1,\tau-t'_2;\tau-t_2,-t_1)\\
        & \ =\begin{cases}e^{i\omega_L\left(t_1-t'_1 + t_2-t'_2\right)}\expval{{\bar{\mathbb{T}}\left[\hat{A}^{\dagger}_1(-t'_1)\hat{A}^{\dagger}_2(\tau-t'_2)\right]\mathbb{T}\left[\hat{A}_2(\tau-t_2)\hat{A}_1(-t_1)\right]}}_0
        \\
        \quad  \text{if both detectors detect near $\omega_L$,}\\
        \\ \theta(t'_1+\tau-t'_2)\theta(t_1+\tau-t_2)e^{2i \omega_L\left(t_1-t'_1\right)}\expval{\hat{B}^{\dagger}_1(-t'_1)\hat{C}^{\dagger}_2(\tau-t'_2)\hat{C}_2(\tau-t_2)\hat{B}_1(-t_1)}_0\\
        \quad  \text{if $d_1$ detects near $2\omega_L-U$ and  $d_2$ detects near $U$,}\\
        \\
         \theta(t_2-t_1-\tau)\theta(t'_2-t'_1-\tau)e^{2i\omega_L\left(t_2-t'_2\right)}\expval{\hat{B}^{\dagger}_2(\tau-t'_2)\hat{C}^{\dagger}_1(-t'_1)\hat{C}_1(-t_1)\hat{B}_2(\tau-t_2)}_0\\
        \quad  \text{if  $d_1$ detects near $U$ and  $d_2$ detects near $2\omega_L-U$}.
        \end{cases}
    \end{aligned}
    \end{equation}   
\end{widetext}
Here, the matter operators $\hat{A}_i$, $\hat{B}_j$ and $\hat{C}_i$ are the same as  in Eqs.~(\ref{eq:AiExp}, \ref{eq:BiExp}, \ref{eq:CiExp}), and $\tilde{\mathcal{F}}_j(t_j)$ for detector $j$ is the Fourier transform of the frequency filter function  $\mathcal{F}_j(\omega)$.
\begin{figure}
  \centering
  \includegraphics[width=0.46\textwidth]{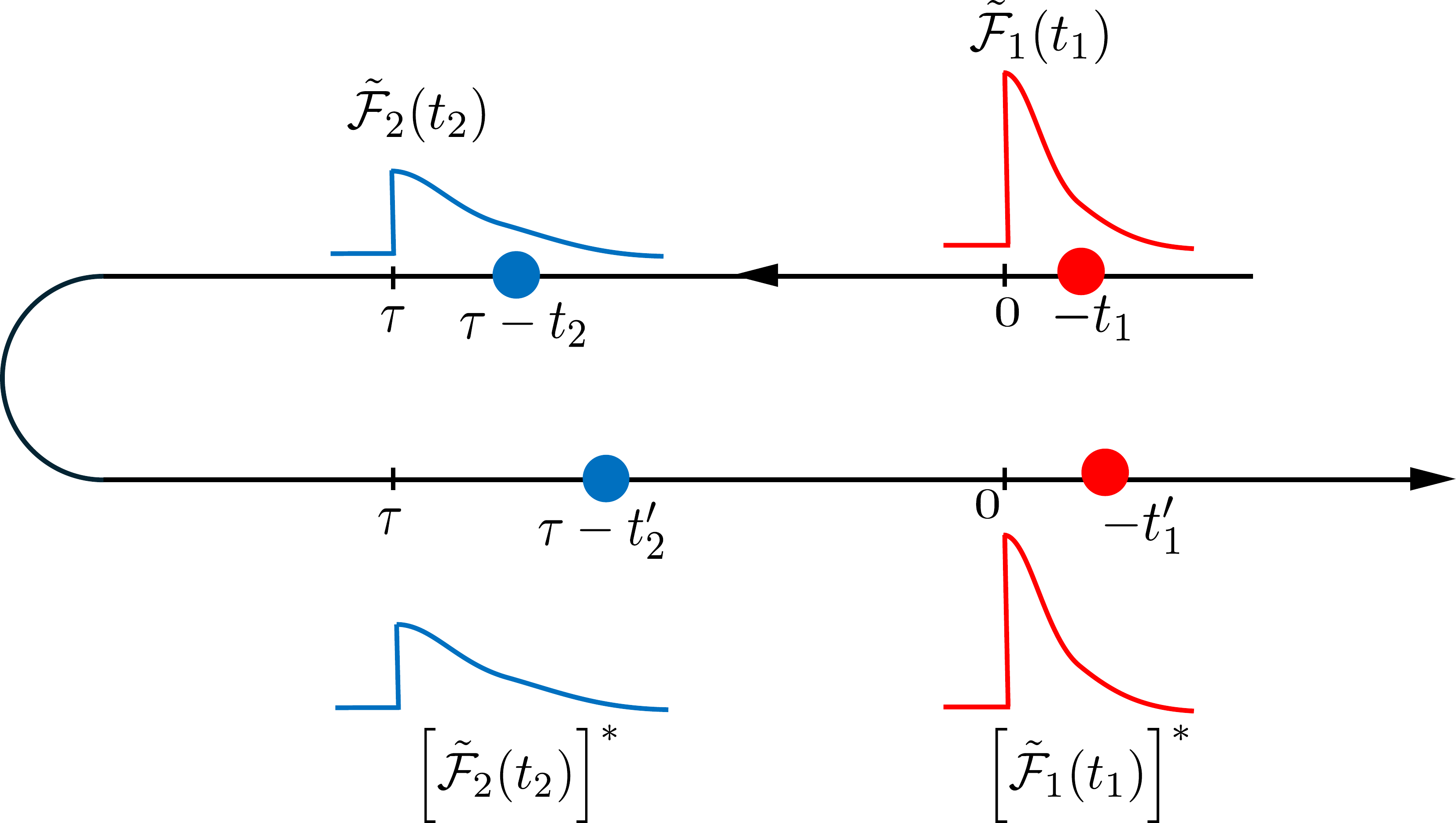}
  \caption{Time contour representation of the matter correlation function $G^{(2)}_{d_1,d_2}(\tau)$, given in Eq.~(\ref{eq:G2matterfull}, \ref{eq:C2matterfull}). Here, time flows forward from right to left. The red and blue circles denote operators that couple to the photon detected first and second, respectively. The forward time segment of the contour (top) is time-ordered, while the backward time segment (bottom) is anti-time-ordered. The profiles of the Fourier transformed \textit{causal} filter functions $\tilde{\mathcal{F}_j}(t_j)$ are shown schematically.}
  \label{fig:ContourG2}
\end{figure}

The temporal structure is illustrated in Fig.~\ref{fig:ContourG2}. Given that the photonic correlator is of the form $\sim\expval{\hat{a}_1^{\dagger}(0)\hat{a}_2^{\dagger}(\tau)\hat{a}_2(\tau)\hat{a}_1(0)}$, one might expect the matter correlator to be of the form $\sim\expval{\hat{A}^{\dagger}_1(0)\hat{A}^{\dagger}_2(\tau)\hat{A}_2(\tau)\hat{A}_1(0)}$. This structure is indeed correct if the filter functions are broad in frequency, and hence narrow in the time domain. However, in a more general scenario, the photon can spend time in the causal filter, say a cavity, before being detected. Consequently, the time delay between the detection events is not necessarily equal to the time interval between the photon emission events. In fact, if the filter's frequency selectivity is narrower than \(1/\tau\), its Fourier transform can become sufficiently broad such that the first emitted photon may be detected after the second. Thus, the matter operators must be convolved with the Fourier-transformed filter functions, as expressed in Eqs.~(\ref{eq:G2matterfull}, \ref{eq:C2matterfull}). 

Let us now consider Lorentzian effective filter functions as defined in Eq.~\eqref{eq:Lorentzdef}, i.e., $\mathcal{F}_j(\omega)=i\Kconst\Gamma_j/\left(\omega - \omega_{j}+i\Gamma_j\right)$, for detectors $j=1$ and $2$ respectively. Substituting its Fourier transform, Eq.~\eqref{eq:Lorentztildedef} into Eq.~\eqref{eq:G2matterfull}, we get:
\begin{equation}\label{eq:filteredG2}
    \begin{aligned}
        &G^{(2)}_{d_{1},d_{2}}(\tau)=\abs{\Kconst_1 \Kconst_2 \Gamma_1\Gamma_2}^2\int\limits_{0}\limits^{\infty}\int\limits_{0}\limits^{\infty}\int\limits_{0}\limits^{\infty}\int\limits_{0}\limits^{\infty}\dd{t_1}\dd{t_2}\dd{t'_1}\dd{t'_2}\biggl[\\
        &\times e^{-\left[\Gamma_1\left(t_1+t'_1\right)+\Gamma_2\left(t_2+t'_2\right)\right]}\mathcal{C}^{(2)}_{d_{1},d_{2}}(-t'_1,\tau-t'_2;\tau-t_2,-t_1)\\
        &\times e^{i\omega_1(t'_1-t_1)+i\omega_2 (t'_2-t_2)}\biggl]
    \end{aligned}
\end{equation}
Let us now look at the different limiting cases of the frequency selectivity being broad (large $\Gamma_j$) and narrow (small $\Gamma_j$). In the large $\Gamma_j$ limit, we approximate $\theta(t)\Gamma_j e^{-\Gamma_j t}\approx \delta(t)$ \footnote{For Eq.~(\ref{eq:G2matterfull}, \ref{eq:C2matterfull}) to hold, we want $\Gamma_j <\abs{\omega_L-U}$, so that the spin operators ($\hat{A}_j$) do not interfere with the charge operators ($\hat{B}_j$, $\hat{C}_j$). Therefore, the approximation $\theta(t)\Gamma_j e^{-\Gamma_j t}\approx \delta(t)$ is justified, only if $\abs{\omega_L - U}$ is much greater than the bandwidth of matter excitations. Since this bandwidth can be at most the tunneling $\tunn$, the approximation is justified in our setting.}.

\subsubsection{Detector $d_{1}$ is broad in frequency\label{subsec:Detector--is}}

\label{sec:broad1only} First, let us take the limit
$\Gamma_{1}\to\infty$, while assuming $\Gamma_{2}$ is finite. Then,
Eq.~\eqref{eq:filteredG2} yields: 
\begin{equation}
\begin{aligned}G_{d_{1},d_{2}}^{(2)}(\tau)\Big|_{\Gamma_{1}\to\infty}\approx & \abs{\Kconst_{1}\Kconst_{2}}^{2}\Gamma_{2}^{2}\int\limits _{0}^{\infty}\dd{t_{2}}\int\limits _{0}^{\infty}\dd{t'_{2}}e^{-\Gamma_{2}(t_{2}+t'_{2})}\\
 & \times e^{i\omega_{2}(t'_{2}-t_{2})}\mathcal{C}_{d_{1},d_{2}}^{(2)}(0,\tau-t'_{2};\tau-t_{2},0).
\end{aligned}
\end{equation}
Here, the matter operators corresponding to the photon
detected first (red dot in Fig.~\ref{fig:ContourG2}) are taken at
a fixed time $0$, while those, corresponding to the second photon
(blue dot in Fig.~\ref{fig:ContourG2}) can contribute at times earlier
than $\tau$. 

\subsubsection{Detector $d_{2}$ is broad in frequency}

\label{sec:broad2only} Next, let us take the limit
$\Gamma_{2}\to\infty$, while assuming $\Gamma_{1}$ is finite. Then,
Eq.~\eqref{eq:filteredG2} yields :
\begin{equation}
\begin{aligned}G_{d_{1},d_{2}}^{(2)}(\tau)\Big|_{\Gamma_{2}\to\infty}\approx & \abs{\Kconst_{1}\Kconst_{2}}^{2}\Gamma_{1}^{2}\int\limits _{0}^{\infty}\dd{t_{1}}\int\limits _{0}^{\infty}\dd{t'_{1}}e^{-\Gamma_{1}(t_{1}+t'_{1})}\\
 & \times e^{i\omega_{1}(t'_{1}-t_{1})}\mathcal{C}_{d_{1},d_{2}}^{(2)}(-t'_{1},\tau;\tau,-t_{1}).
\end{aligned}
\label{eq:broad1only}
\end{equation}
Here, the matter operators corresponding to the photon
detected second (blue dots in Fig.~\ref{fig:ContourG2}) are taken
at a fixed time $\tau$, while those, corresponding to the first photon
can contribute earlier than time $0$. 
  \subsubsection{Both detectors are broad in frequency}

\label{sec:broadboth} Next, let us take the limit
where both $\Gamma_{1}\to\infty$ and $\Gamma_{2}\to\infty$. Then,
Eq.~\eqref{eq:filteredG2} yields: 
\begin{equation}
\begin{aligned}G_{d_{1},d_{2}}^{(2)}(\tau)\Big|_{\Gamma_{1},\Gamma_{2}\to\infty}\approx & \abs{\Kconst_{1}\Kconst_{2}}^{2}\mathcal{C}_{d_{1},d_{2}}^{(2)}(0,\tau;\tau,0).\end{aligned}
\label{eq:broad2only}
\end{equation}
In this limit the photonic correlation function exactly
reflects that of the matter. 

 \subsubsection{Both detectors are narrow in frequency}\label{sec:narrowboth}
Let us consider the opposite limit of
sharp frequency resolution, when both $\Gamma_{1}$ and $\Gamma_{2}$
are much less than $1/\tau$. In this limit, $G_{d_{1},d_{2}}^{(2)}(\tau)$ is independent of $\tau$:
 \begin{equation}\label{eq:narrowboth}
 \begin{aligned}
    G^{(2)}_{d_1,d_2}\Big|_{\Gamma_1,\Gamma_2\to 0}\propto \iiint_{-\infty}^\infty & \dd{t_2} \dd{t'_1}\dd{t'_2}e^{i\omega_1 t'_1 +i\omega_2(t'_2-t_2)} \\
    &\times \mathcal{C}^{(2)}_{d_{1},d_{2}}(-t'_1,-t'_2;-t_2,0) .
    \end{aligned}
\end{equation}

In summary, Eqs.~\eqref{eq:G1simp2}, \eqref{eq:X1matter}, \eqref{eq:X2matter}, and \eqref{eq:G2matterfull}, establish a one-to-one correspondence between the matter and photonic correlation functions. In our derivation, the expectation value \(\expval{.}_{0}\) on the right-hand side is evaluated with respect to the unperturbed energy eigenstate \(\ket{I}\) of the matter Hamiltonian, within the spin sector. As a corollary, this expectation value can also be taken with respect to a mixed state (within the spin sector) that is diagonal in the energy eigenbasis, such as a thermal state. Consequently, our formalism remains valid at nonzero temperatures, provided that the temperature is much smaller than \(U\).

With this mapping established, we now explore several key applications that enable the characterization of phases in spin systems.
\section{Application I: Measurement of static spin chirality operators}\label{sec:spinchirality} 
In this section, we demonstrate that the static spin chirality on the triangular lattice can be probed through the phase-dependent part of fluctuations of the first-order quadrature operators of the scattered photons, i.e. \(\Im \expval{\hat{a}_{d_2}(0)\hat{a}_{d_1}(0)}_{\text{out}}\).
 For this, the system should be driven by a coherent state input. Additional conditions are that the delay time $\tau=0$, and the filters should be such that $d_1$ and $d_2$ select the sidebands near $2\omega_L-U$, and $U$ respectively, but are broad in their respective sidebands. Also, the polarization of the drive as well as that of the detectors should be linear, i.e., $\vb{e}_L$, $\vb{e}_1$, and $\vb{e}_2$ are all real.  The experimental scheme for the photonic measurement is provided in Appendix~
\ref{app:homodyne}.

 We have shown in Sec.~\ref{sec:R2micro} that the correlations between photons emitted into the sidebands probe correlations between matter operators $\hat{B}_j$ [Eq.~\eqref{eq:BiExp}] and $\hat{C}_j$ [Eq.~\eqref{eq:CiExp}], both of which couple the spin sector to the charge sector. How could we then measure a pure spin correlator via photons in the sidebands? The reason lies in the two conditions we mentioned above. The absence of filtering implies that photons do not spend any additional time in the filter after they are emitted by the material. Furthermore, $\tau=0$ implies that the doublon-hole pair formed at the time of the first photon emission should immediately (within a temporal uncertainty $\sim 1/\abs{\omega_L - U}$) recombine to emit the second photon. In other words, the time evolution step between Fig.~\ref{fig:description2}($c'_4$) and ($c'_5$) is not present anymore. Thus, the net result of the two-photon scattering on the material is the application of an operator purely in the spin sector. Further, we are able to circumvent the no-go result in Ref.~\cite{ko2010raman} (where the spin chirality term was zero on the triangular lattice) because we have access to three polarizations $\vb{e}_L$, $\vb{e}_1$ and $\vb{e}_2$. 

Using the mapping from photonic to electronic correlator in Eq.~\eqref{eq:quad2broad}, we have 
\begin{equation}\label{eq:homodyne2chiral}
    \Im \expval{\hat{a}_{d_2}(0)\hat{a}_{d_1}(0)}_{\text{out}}=\Kconst_1\Kconst_2\Im \expval{\hat{C}_2(0)\hat{B}_1(0)}_0.
\end{equation}
The expression for $\hat{B}_1$ is given in Eq.~\eqref{eq:BiExp}. Operator $\hat{C}_2$ is the global electron current along the direction $\vb{e}_2$, consisting of nearest-neighbor tunnelings. For the system to return to the spin sector after applying $\hat{C}_2$, the electron tunneling implemented by $\hat{B}_1$ should also be along a nearest-neighbor bond. 

We  discuss the anticipated  form of the spin operators measured by Eq.~\eqref{eq:homodyne2chiral} using a symmetry analysis.
First, let us examine the terms in Eq.~\eqref{eq:BiExp}. Recall from Eqs.~(\ref{eq:diamzero1}-\ref{eq:diamzero2}) in Sec.~\ref{sec:ABCdefs} that the terms in the first line of Eq.~\eqref{eq:BiExp} do not contribute to $\expval{\hat{C}_2(0)\hat{B}_1(0)}$.  
Furthermore, since $\vb{e}_1$, $\vb{e}_2$, and $\vb{e}_L$ are all real, both $\hat{B}_1(0)$ and $\hat{C}_2(0)$ are odd under time-reversal. So, $\hat{C}_2(0)\hat{B}_1(0)$ is even under time-reversal. Therefore,  $-i\left(\hat{C}_2(0)\hat{B}_1(0)-\hat{B}^{\dagger}_1(0)\hat{C}^{\dagger}_2(0)\right)/2$ is both a Hermitian operator and odd under time-reversal. It also transforms as a spin singlet under spin rotations since there is no spin-orbit coupling. Since all terms considered in this work are at most eighth order in fermionic operators, they are at most fourth-order in spin operators. The only possible spin operator compatible with these conditions is a sum of spin chirality terms of the form $\sim \sum\hat{\vb{S}}_{\vb{r}}\cdot \left(\hat{\vb{S}}_{\vb{r}'}\times \hat{\vb{S}}_{\vb{r}'''}\right)$ with real coefficients. Specifically, we are interested in scalar spin chirality, which is invariant under lattice rotation and odd under reflection. To isolate the scalar spin chirality, one must take linear combinations of experimental data for different directions of linear polarizations $\vb{e}_L$, $\vb{e}_1$, and $\vb{e}_2$, and expand the result in terms of irreducible representations of the crystalline point group of the triangular lattice.  There are two such polarization channels that transform the same way as scalar spin chirality does:
\begin{align}
    \mathcal{A}_a\equiv&\left(e^x_1 e^y_2+e^x_2 e^y_1\right)\left[(e^x_L)^2-(e^y_L)^2\right]\nonumber\\
    &-\left(e^x_1 e^x_2-e^y_1 e^y_2\right)(2e^x_Le^y_L)\text{, and}\label{eq:channelAa}\\
    \mathcal{A}_b\equiv & \left(e^x_1 e^y_2-e^x_2 e^y_1\right) \left[(e^x_L)^2+(e^y_L)^2\right].\label{eq:channelAb}
\end{align}
These two channels transform identically as the channel termed $A_{2g}$ in Ref.~\cite{ko2010raman}). Let us now study Eq.~\eqref{eq:homodyne2chiral} in each of these channels separately.

\begin{figure}
  \centering
  \includegraphics[width=0.48\textwidth]{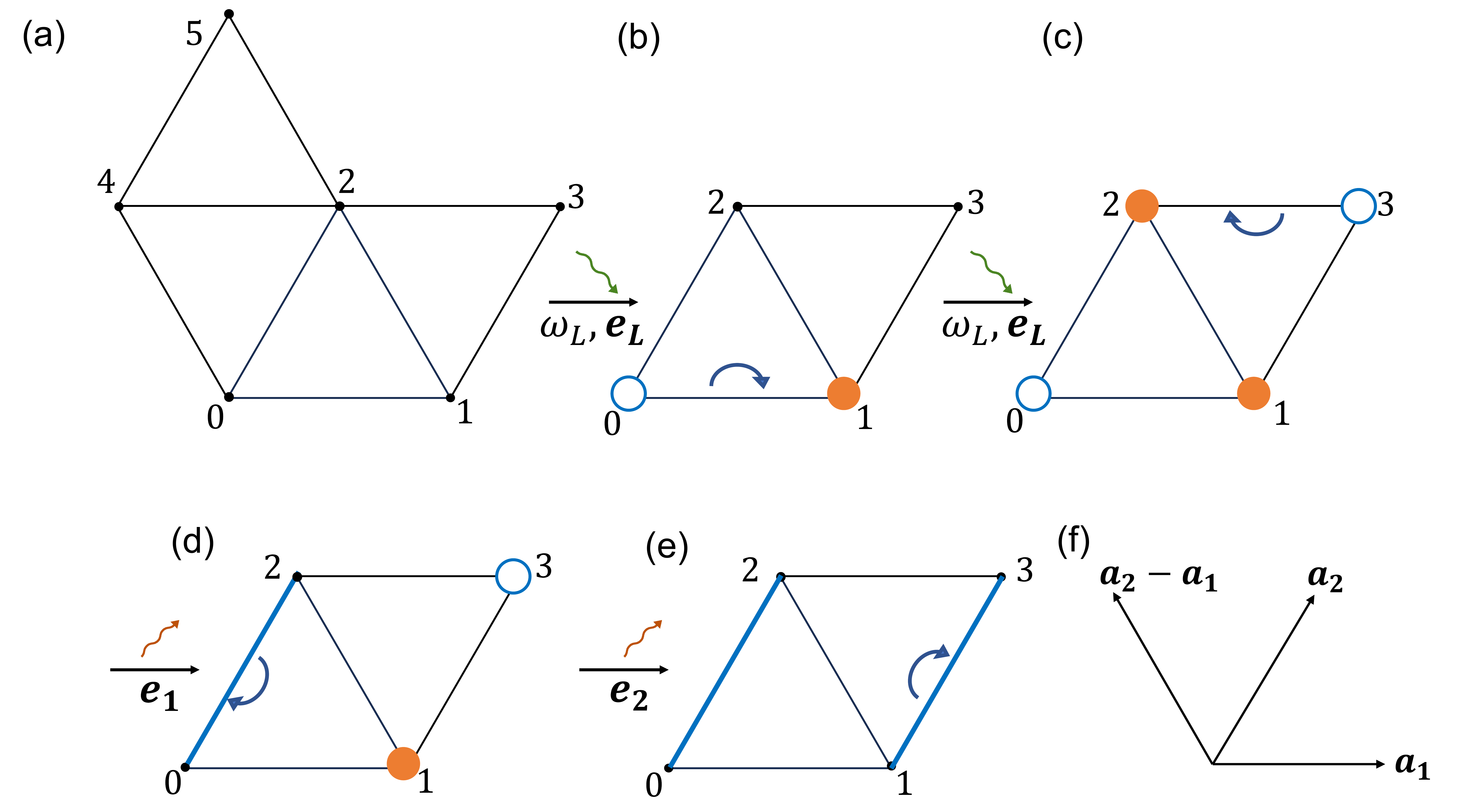}
  \caption{Processes leading to spin chirality matter operator: \textbf{(a)}: We study this motif because the two orientations of equilateral triangles and all six orientations of isosceles obtuse triangles involving the site labeled $0$ can be found in this motif. Any other such triangle can be obtained by translating a triangle from this motif by a lattice vector. A black dot represents a singly occupied electron, i.e., a spin degree of freedom. Blue empty circle is a hole and an orange filled circle is a doubly occupied site (doublon). \textbf{(b)}: Upon absorption of a photon, an electron virtually hops from $0$ to $1$. \textbf{(c)}: Upon absorption of a second photon, an electron hops virtually from 3 to 0. \textbf{(d)}: The doublon-hole pair recombines along the bond from $2$ to $0$ by emitting a photon of polarization $\vb{e}_1$ and frequency around $2\omega_L-U$. \textbf{(e)}: This is similar to the previous step, but the frequency of the second photon is around $U$. \textbf{(f)}: Primitive lattice vectors $\bm{a}_1$ and $\bm{a}_2$ of the triangular lattice. Also shown is the linear combination $\bm{a}_2-\bm{a}_1$. }
  \label{fig:trchirality}
\end{figure}

We now show that the contribution to channel $\mathcal{A}_a$ arises upon combining the current $\hat{C}_2$  with $i\hat{\bm{\mathcal{S}}}_{\bm{\mu}_1,\bm{\mu}_2}(\vb{r},\bm{\mu})\cdot \left[\left(\hat{\vb{S}}_{\vb{r}}-\hat{\vb{S}}_{\vb{r}+\bm{\mu}}\right)/2-i\hat{\vb{S}}_{\vb{r}}\times \hat{\vb{S}}_{\vb{r}+\bm{\mu}}\right]$ from $\hat{B}_1$ (see Eq.~\eqref{eq:BiExp}), thereby yielding a spin operator. Recall that $\hat{\bm{\mathcal{S}}}_{\bm{\mu}_1,\bm{\mu}_2}(\vb{r},\bm{\mu})$ has been defined in Eq.~\eqref{eq:defmathcalS} and is shown schematically in Fig.~\ref{fig:SandC}. The contributing process is shown in Fig.~\ref{fig:sectors}(c) with microscopic counterpart provided in Fig.~\ref{fig:description2}($c'_1$-$c'_6$). The same process for the triangular lattice is shown here in Fig.~\ref{fig:trchirality}. In Eq.~\eqref{eq:BiExp}, tunneling along $\bm{\mu}'_1$ after absorbing a laser photon is shown in Fig.~\ref{fig:trchirality}(b). We can now see from Fig.~\ref{fig:trchirality}(c) that the direction of tunneling induced by the first and second photon absorptions are identical, i.e., $\bm{\mu}'_2=\bm{\mu}'_1$. The same applies to the photon emission process. We note that Eq.~\eqref{eq:BiExp} holds for arbitrarily long-range fermion tunneling. However, for simplicity, in Fig.~\ref{fig:trchirality} and in the expressions below, we restrict to nearest-neighbor tunneling. The term in $\hat{B}_1$ [Eq.~\eqref{eq:BiExp}] which is relevant to Fig.~\ref{fig:trchirality} is (summation over spin indices of fermionic operators is implicit): 
\begin{equation}\label{eq:simpB1}
\begin{aligned}
    &\frac{ig^2_L\tunn^3 g}{(\omega_L -U)^2}\left(\bm{a}_1\cdot \vb{e}_L\right)^2\left(\bm{a}_2\cdot\vb{e}_1\right)\\
    &\times\left(\hat{c}^{\dagger}_{1}\bm{\sigma}\hat{c}_{3}+\hat{c}^{\dagger}_{3}\bm{\sigma}\hat{c}_{1}\right)\cdot \left(\frac{\hat{\vb{S}}_{0}-\hat{\vb{S}}_{2}}{2}-i\vb{S}_{0}\times \vb{S}_{2}\right).
\end{aligned}
\end{equation}
Here, $\bm{a}_1$ and $\bm{a}_2$ are lattice vectors $(1,0)$ and $(1/2,\sqrt{3}/2)$ respectively [See Fig.~\ref{fig:trchirality}(f)]. The subscript for the spins $\hat{\vb{S}}$ and fermionic operator $\hat{c}$ is the site index which runs from $0$ to $5$, the six sites shown in Fig.~\ref{fig:trchirality}. Also, we use the shorthand $\hat{c}^{\dagger}_{1}\bm{\sigma}\hat{c}_{3}$ to denote $\hat{c}^{\dagger}_{1,\alpha}\bm{\sigma}_{\alpha \beta}\hat{c}_{3,\beta}$.

The term from $\hat{C}_2$ [Eq.~\eqref{eq:CiExp}] which is relevant to Fig.~\ref{fig:trchirality} is: 
\begin{equation}\label{eq:simpC2}
    \frac{ig\tunn}{\omega_L-U}\left(\bm{a}_2\cdot\vb{e}_2\right)\left(\hat{c}_{3}^{\dagger}\hat{c}_{1}-\hat{c}_1^{\dagger}\hat{c}_3\right)
\end{equation}
Now, we can see that the electron tunneling in Eq.~\eqref{eq:simpB1} can be combined with that in Eq.~\eqref{eq:simpC2} to give an operator lying purely in the spin sector. To do so, we use the following identity (for combining a pair of fermionic operators into a spin operator) that holds at half-filling (which can be shown using the definition of spin operators in terms of fermions given in Eq.~\eqref{eq:Stofermions}):
\begin{equation}
   \begin{aligned} &\left(\hat{c}^{\dagger}_3\hat{c}_1-\hat{c}^{\dagger}_1\hat{c}_3\right)\left(\hat{c}^{\dagger}_3\bm{\sigma}\hat{c}_1+\hat{c}^{\dagger}_1\bm{\sigma}\hat{c}_3\right)\\
   &\quad =2\left(\hat{\vb{S}}_3-\hat{\vb{S}}_1\right)- 4i \hat{\vb{S}}_3 \times \hat{\vb{S}}_1.
   \end{aligned}
\end{equation}
Using this identity, the contribution from the process shown in Fig.~\ref{fig:trchirality} to $\hat{C}_2\hat{B}_1$ is:
\begin{equation}\label{eq:0132}
\begin{aligned}
    &\frac{4g^2_Lg^2 \tunn^4}{(\omega_L-U)^2} \left(\bm{a}_1\cdot \vb{e}_L\right)^2\left(\bm{a}_2\cdot\vb{e}_1\right)\left(\bm{a}_2\cdot\vb{e}_2\right)\\
    &\times\left(\frac{\hat{\vb{S}}_{2}-\hat{\vb{S}}_{0}}{2}-i\vb{S}_{2}\times \vb{S}_{0}\right)\cdot \left(\frac{\hat{\vb{S}}_{3}-\hat{\vb{S}}_{1}}{2}-i\vb{S}_{3}\times \vb{S}_{1}\right).
\end{aligned}
\end{equation}
Now, for the same set of points $(0,1,2,3)$ [Fig.~\ref{fig:trchirality}], there is an alternative process where the laser-induced tunneling occurs along bonds $02$ and $13$, while the tunneling during photon emission occurs along the bonds $01$ and $23$. The resulting contribution to $\hat{C}_2\hat{B}_1$ is:
\begin{equation}\label{eq:0231}
\begin{aligned}
    &\frac{4g^2_Lg^2 \tunn^4}{(\omega_L-U)^2} \left(\bm{a}_2\cdot \vb{e}_L\right)^2\left(\bm{a}_1\cdot\vb{e}_1\right)\left(\bm{a}_1\cdot\vb{e}_2\right)\\
    &\times\left(\frac{\hat{\vb{S}}_{3}-\hat{\vb{S}}_{2}}{2}-i\vb{S}_{3}\times \vb{S}_{2}\right)\cdot \left(\frac{\hat{\vb{S}}_{1}-\hat{\vb{S}}_{0}}{2}-i\vb{S}_{1}\times \vb{S}_{0}\right).
\end{aligned}
\end{equation}
In addition to the set of points $(0,1,2,3)$ considered above, we have the equivalent of Eq.~\eqref{eq:0132} and Eq.~\eqref{eq:0231} for the set of points $(0,1,2,4)$ and $(0,4,5,2)$ [Fig.~\ref{fig:trchirality}(a)], thereby leading to three pairs of terms per motif (shown in Fig.~\ref{fig:trchirality}(a)). We then add up the contribution of each motif that is obtained by a lattice translation of the motif shown in Fig.~\ref{fig:trchirality}(a). From the resulting expression, we isolate the terms corresponding to scalar spin chirality. We find that the result has a nonzero component along channel $\mathcal{A}_a$, but not $\mathcal{A}_b$. The component along the channel $\mathcal{A}_a$ (Eq.~\eqref{eq:channelAa}) is 
\begin{equation}
    \frac{3\sqrt{3}g_L^2 g^2 \Kconst_1 \Kconst_2 \tunn^4}{2(\omega_L-U)^2}\chi,
\end{equation}
where
\begin{equation}\label{eq:chioperator}
    \begin{aligned}
        \chi\equiv &\sum_{\vb{r}}\left<\left[\left(\ \mysymb{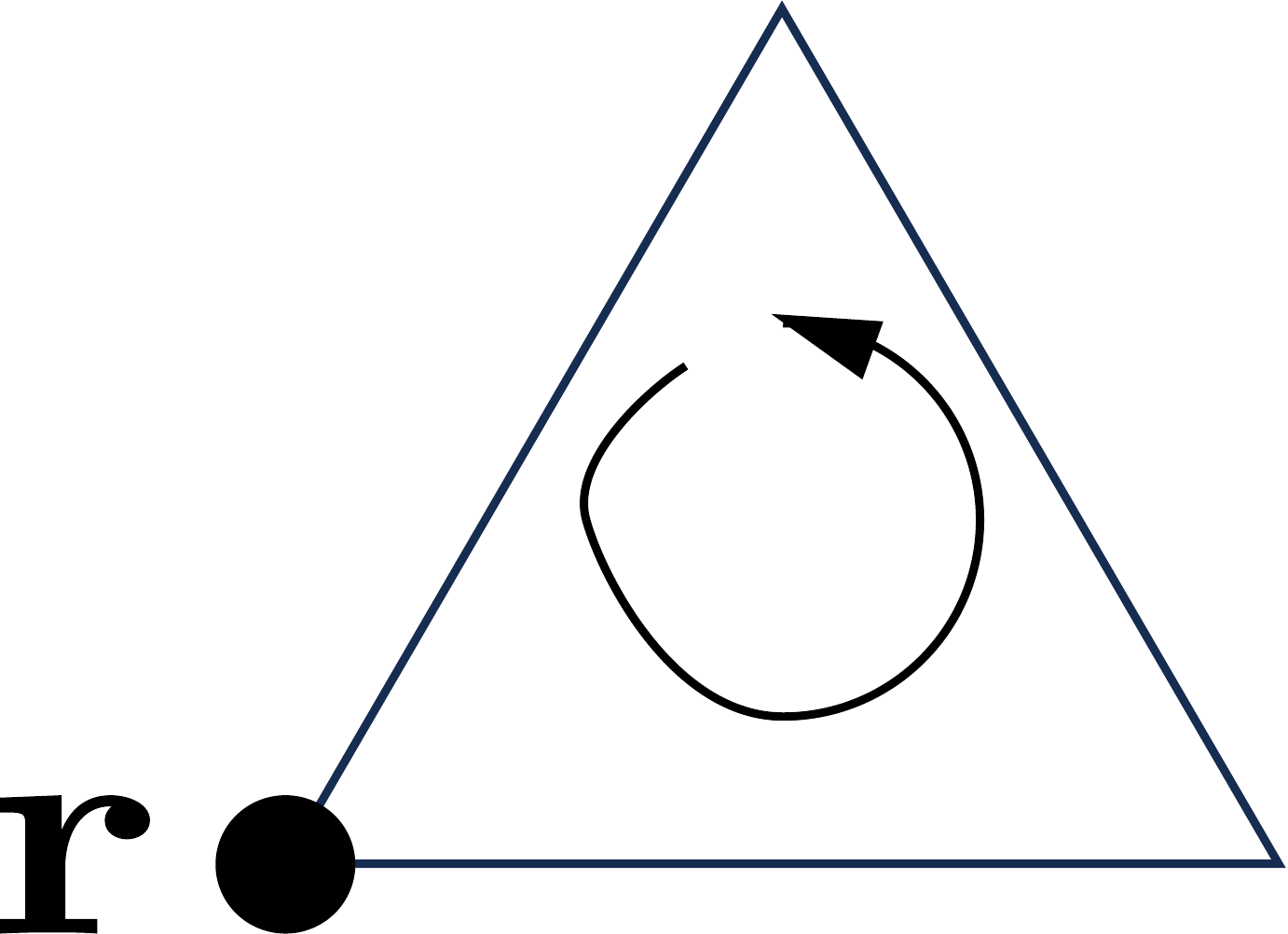}{6.0}+\mysymb{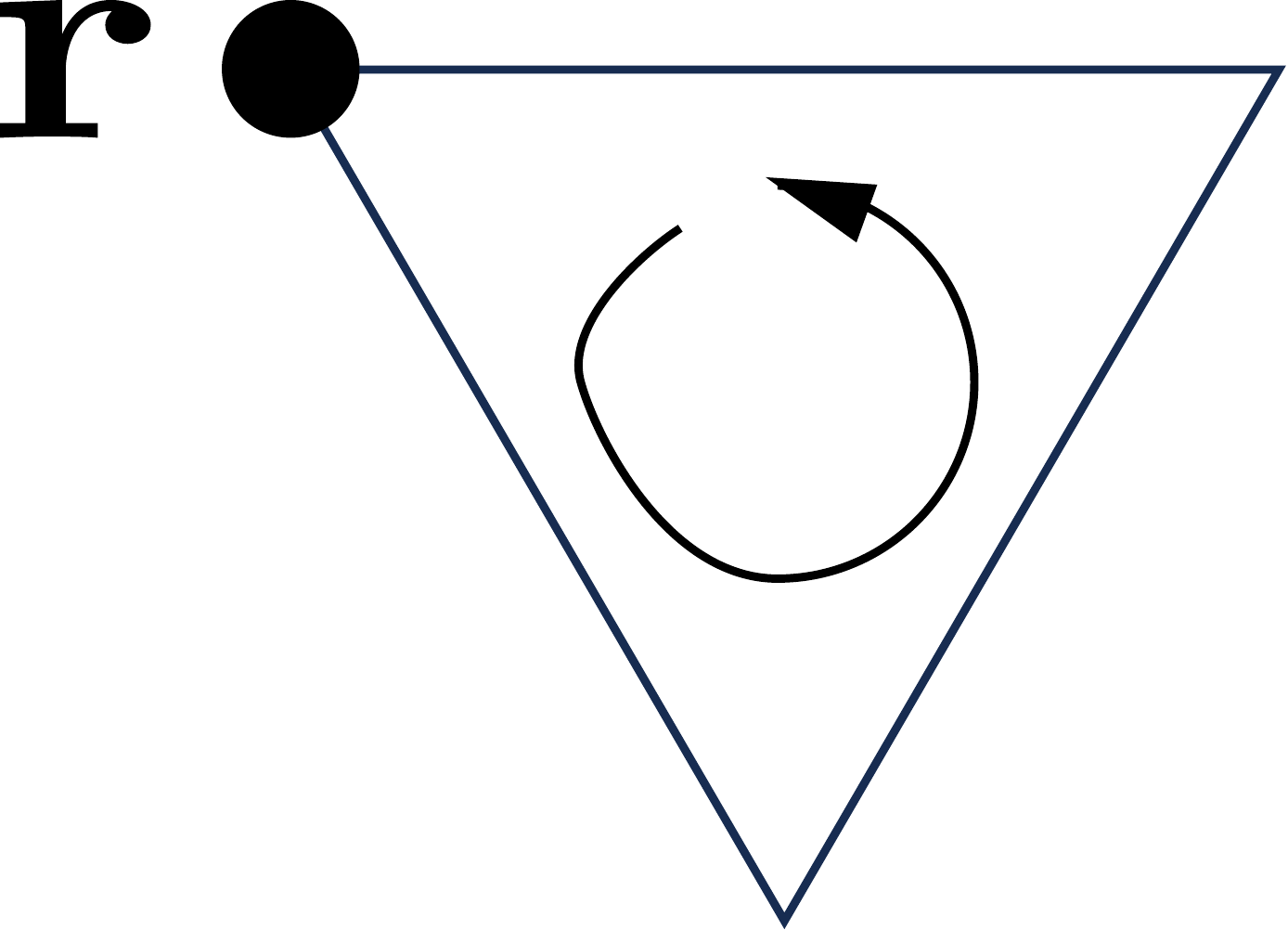}{6.0}\ \right)\right.\right.\\
        &-\frac{1}{3}\left( \ \mysymb{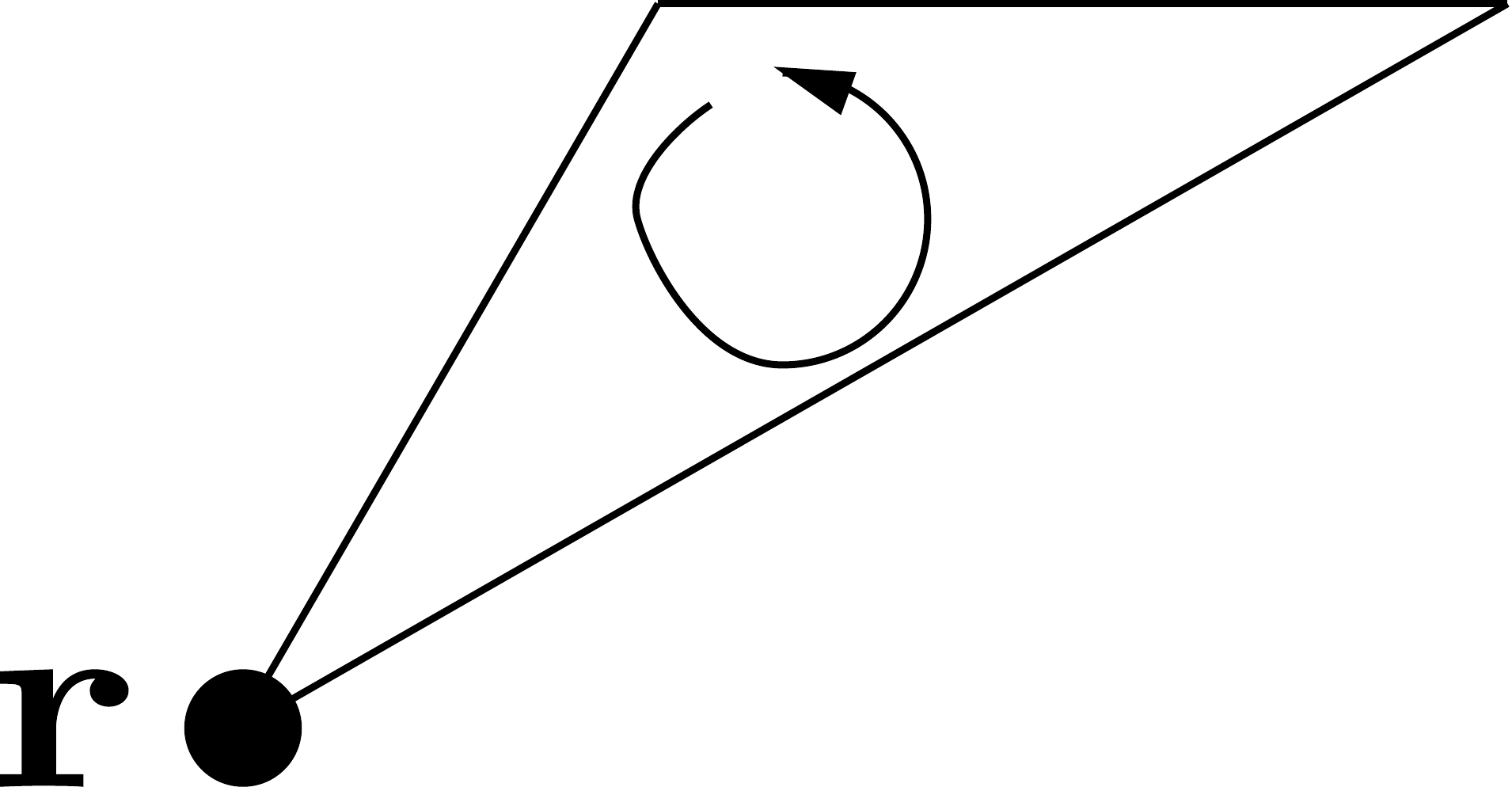}{6.0}+\mysymb{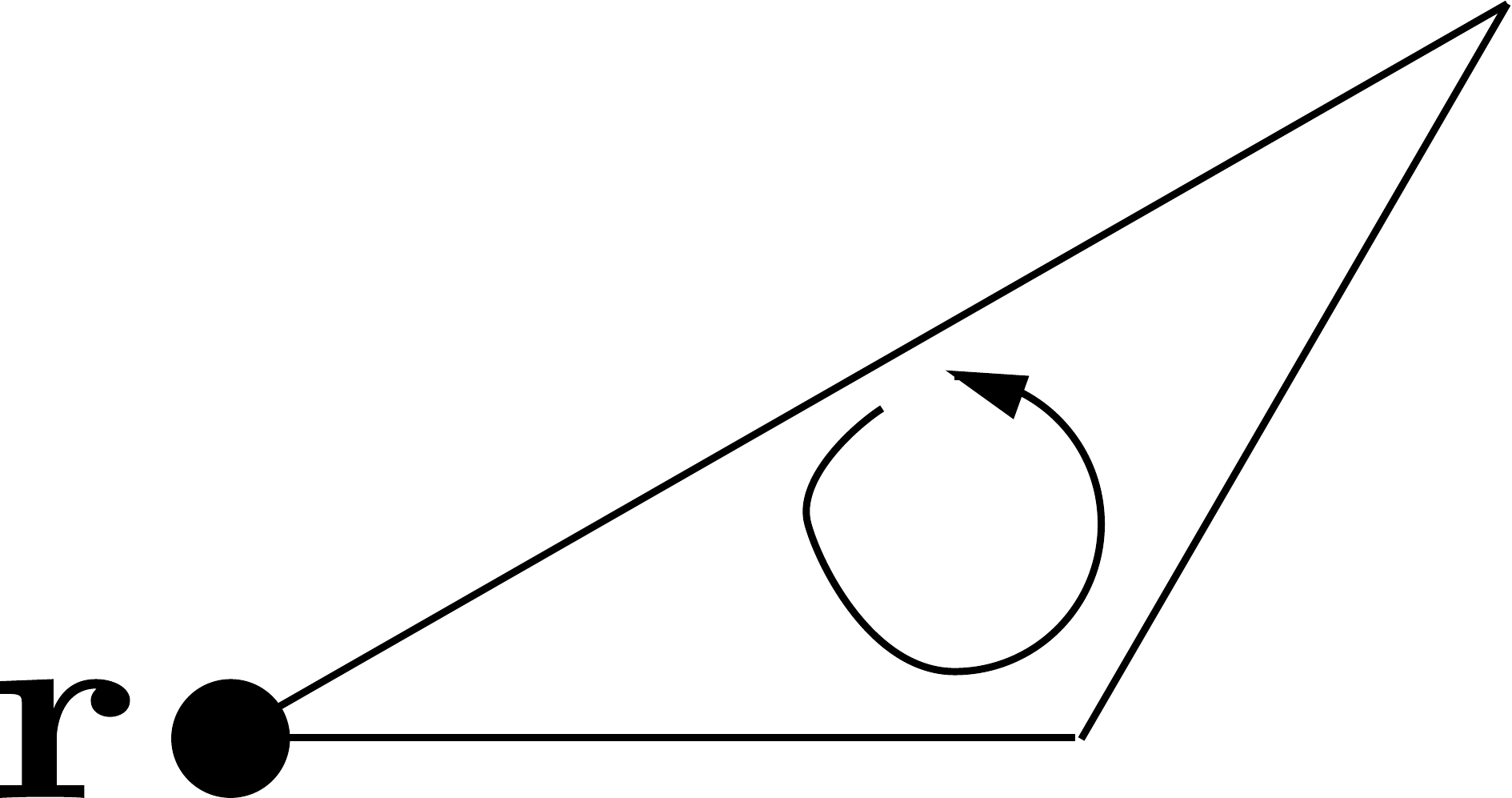}{6.0}+\mysymb{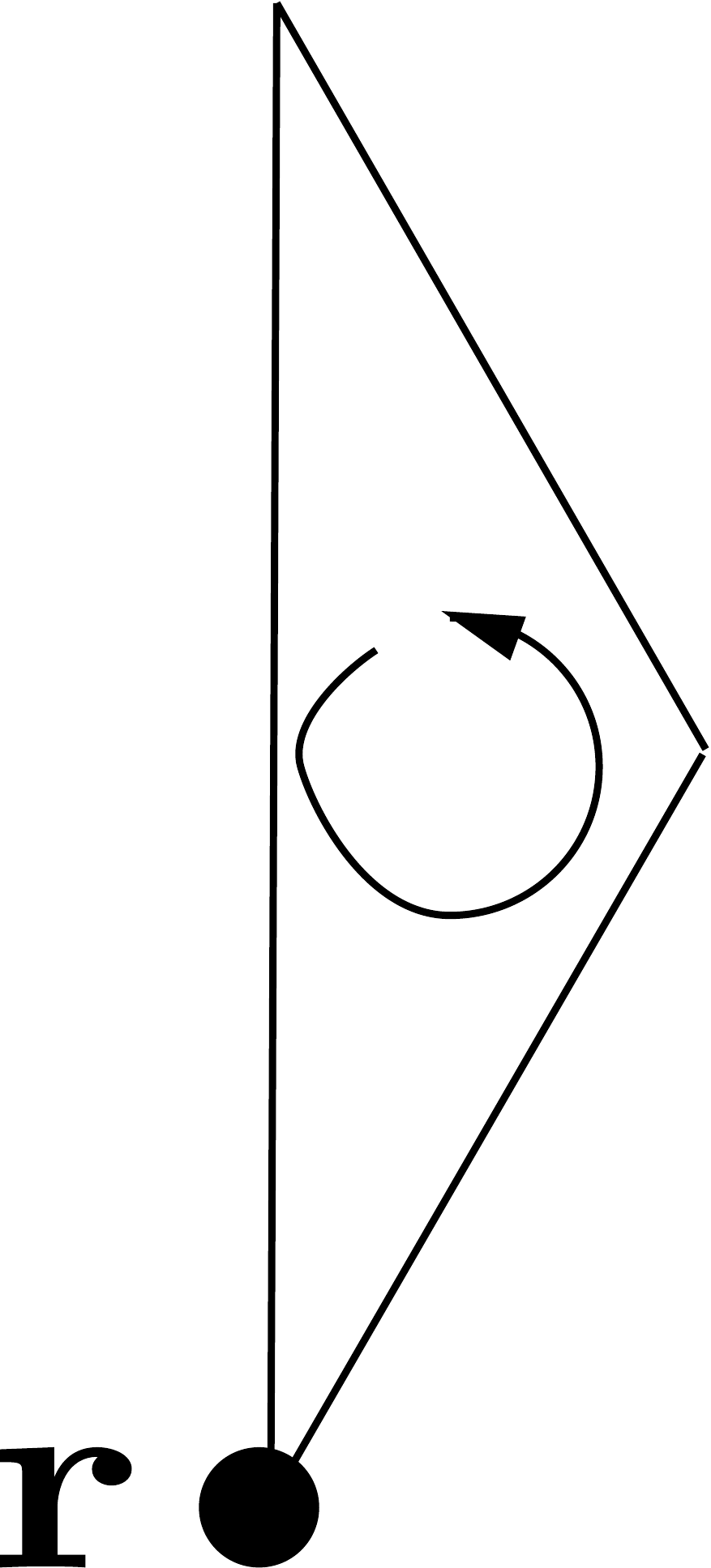}{11.0}+\mysymb{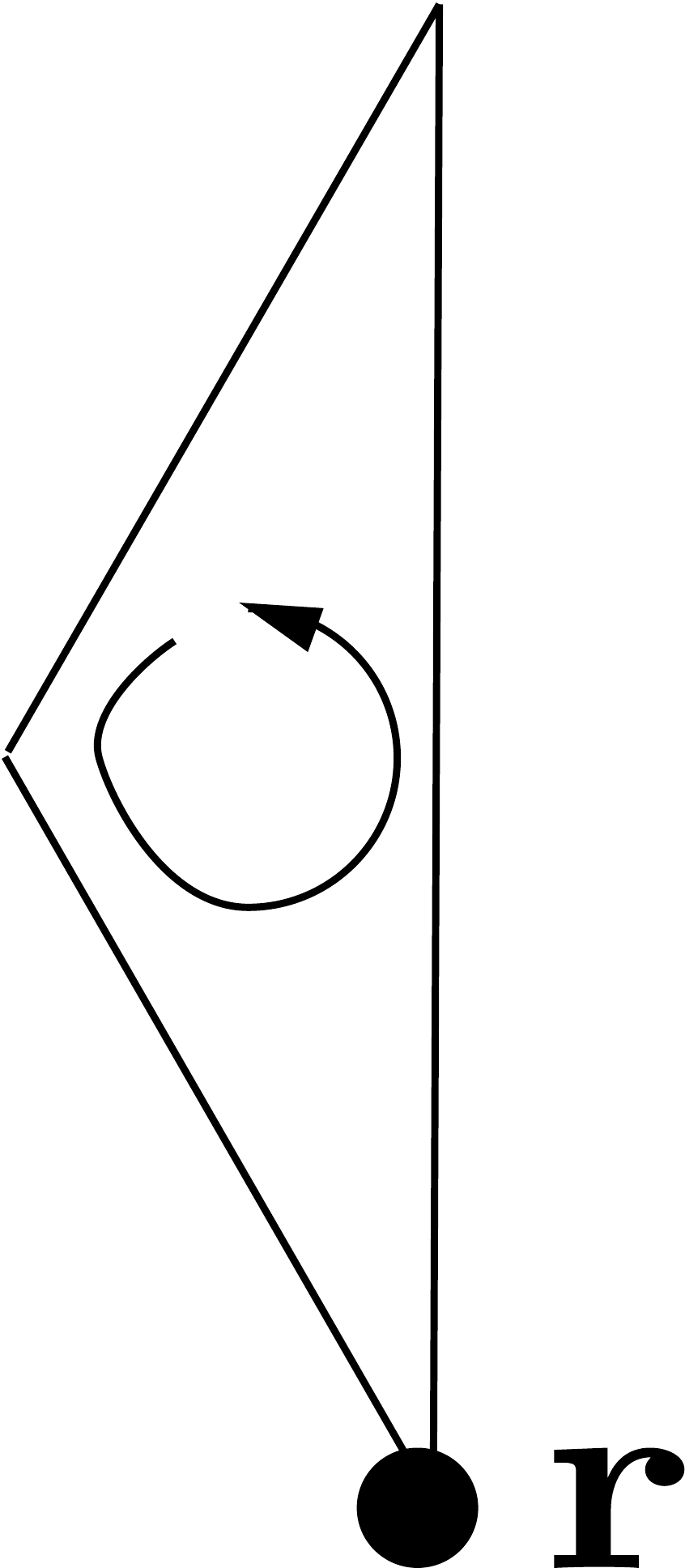}{11.0}\right.\\
        &\quad \quad \quad \left.\left.+\mysymb{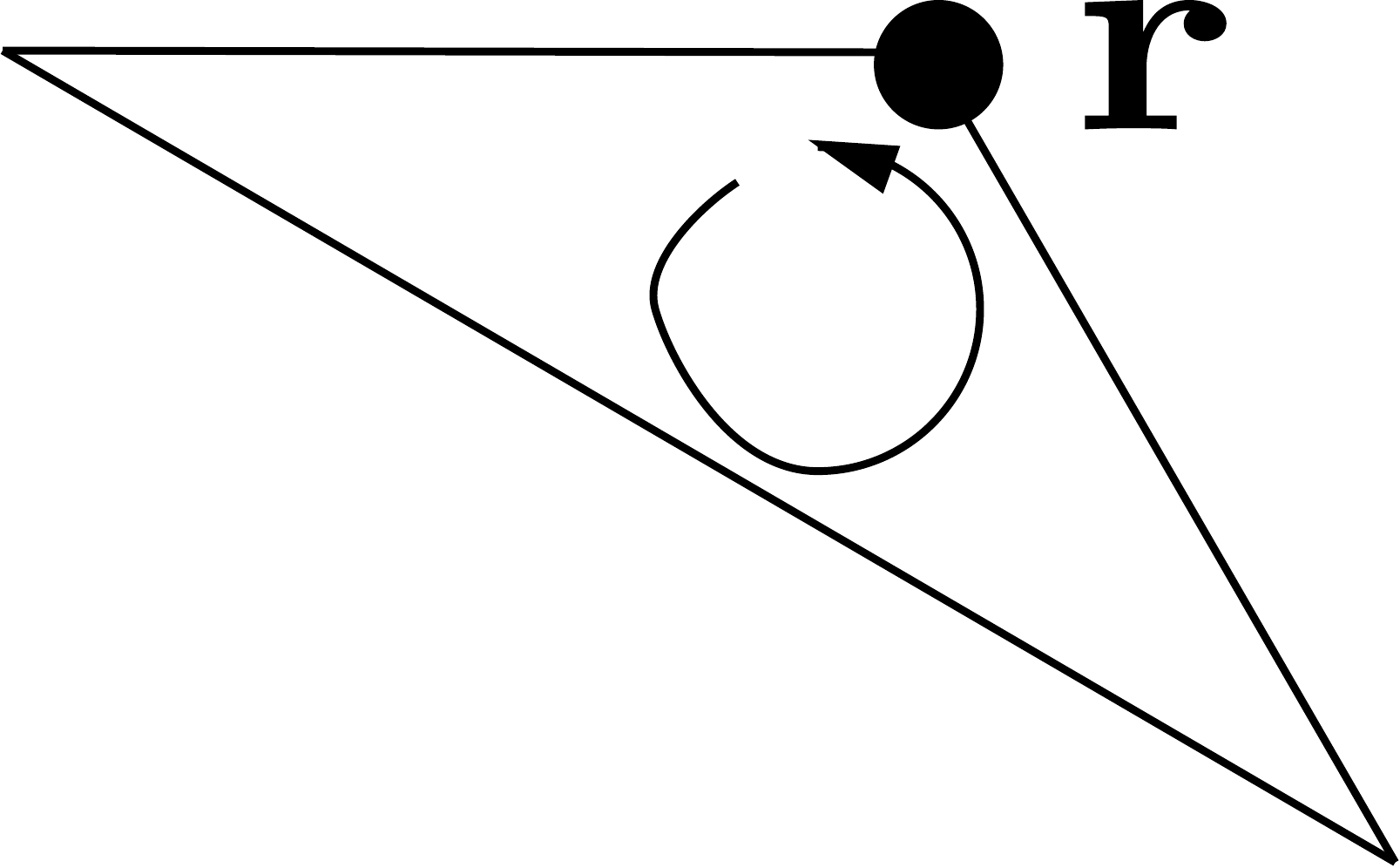}{6.0}+\mysymb{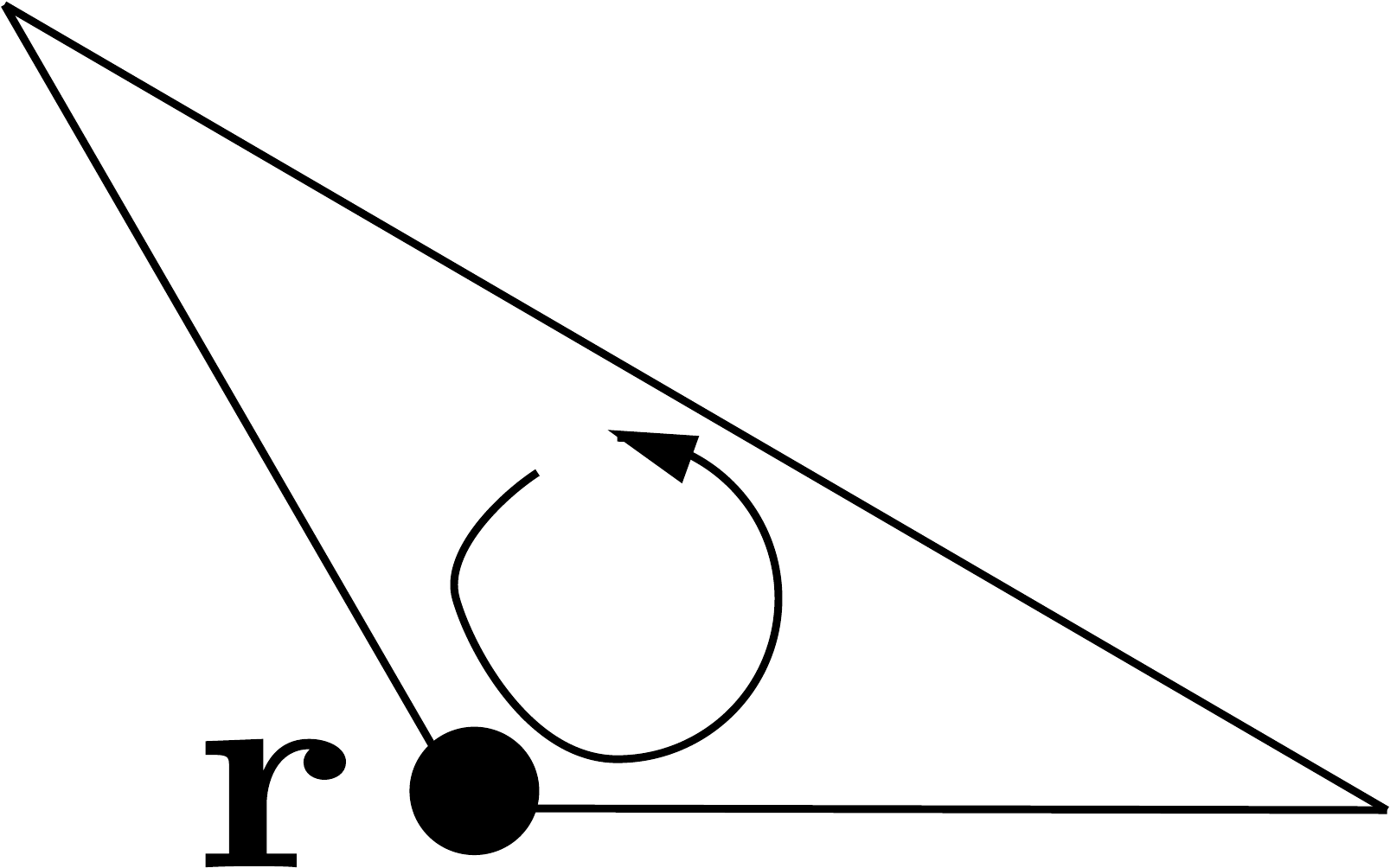}{6.0}\ \Biggr)\right]\right>.
    \end{aligned}
\end{equation}
 Here, we used the notation $\mysymb{trup.pdf}{4.0}\equiv\hat{\vb{S}}_{\vb{r}}\cdot\left(\hat{\vb{S}}_{\vb{r}+\bm{a}_1}\times \hat{\vb{S}}_{\vb{r}+\bm{a}_2}\right)$. Similarly, $\mysymb{tr1.pdf}{3.6}\equiv\hat{\vb{S}}_{\vb{r}}\cdot\left(\hat{\vb{S}}_{\vb{r}+\bm{a}_1+\bm{a}_2}\times \hat{\vb{S}}_{\vb{r}+\bm{a}_2}\right)$, etc. $g_L$ and $g$ are light-matter couplings defined in Sec.~\ref{sec:scales} and $\Kconst_1$ and $\Kconst_2$ are constants related to detector efficiency and photonic density of states, defined in Eq.~\eqref{eq:Kdef}. 

Next, we show that the contribution to the channel $\mathcal{A}_b$ (Eq.~\eqref{eq:channelAb}) arises from combining the current $\hat{C}_2$ with $\hat{\mathcal{K}}_{\bm{\mu}'}(\vb{r},\bm{\mu})\left(4\hat{\vb{S}}_{\vb{r}}\cdot \hat{\vb{S}}_{\vb{r}+\bm{\mu}}-1\right)$, thereby yielding a spin operator. Recall that $\hat{\mathcal{K}}_{\bm{\mu}'}(\vb{r},\bm{\mu})$ (defined in Eq.~\eqref{eq:defKmu} and shown in Fig.~\ref{fig:Kmufig}) is a sum of electron tunnelings along bonds in the direction $\bm{\mu}'$ that also touch the bond $(\vb{r},\bm{\mu})$. One can see that the result is a sum of terms of the following form:
\begin{align}
    &(4\hat{\vb{S}}_{\vb{r}}\cdot \hat{\vb{S}}_{\vb{r}+\bm{\mu}'}-1)(4\hat{\vb{S}}_{\vb{r}}\cdot \hat{\vb{S}}_{\vb{r}+\bm{\mu}}-1)\nonumber\\&\times(\bar{\bm{\mu}}\cdot \vb{e}_L)(\bar{\bm{\mu}}\cdot \vb{e}_1)(\bar{\bm{\mu}'}\cdot \vb{e}_L)(\bar{\bm{\mu}'}\cdot \vb{e}_2)\\
    =&\left[-8i\hat{\vb{S}}_{\vb{r}}\cdot \left(\hat{\vb{S}}_{\vb{r}+\bm{\mu}}\times \hat{\vb{S}}_{\vb{r}+\bm{\mu}'}\right)-4\hat{\vb{S}}_{\vb{r}}\cdot \hat{\vb{S}}_{\vb{r}+\bm{\mu}}-4\hat{\vb{S}}_{\vb{r}}\cdot \hat{\vb{S}}_{\vb{r}+\bm{\mu}'}\right.\nonumber\\&\left. \ +4\hat{\vb{S}}_{\vb{r}+\bm{\mu}}\cdot \hat{\vb{S}}_{\vb{r}+\bm{\mu}'}+1\right](\bar{\bm{\mu}}\cdot \vb{e}_L)(\bar{\bm{\mu}}\cdot \vb{e}_1)(\bar{\bm{\mu}'}\cdot \vb{e}_L)(\bar{\bm{\mu}'}\cdot \vb{e}_2).
\end{align}
The first term in this  equation is a spin chirality operator. Performing a summation over bond directions $\bm{\mu}$ and $\bm{\mu}'$, and lattice sites $\vb{r}$, we find that the resulting component of scalar spin chirality is zero within the channel $\mathcal{A}_a$. In contrast, in channel $\mathcal{A}_b$, we get:
\begin{equation}
    -\frac{3\sqrt{3}\tunn^4 g_L^2 g^2\Kconst_1 \Kconst_2}{(\omega_L-U)^2}\chi,
\end{equation}
Therefore,
\begin{equation}
\begin{aligned}
   \Im \expval{\hat{a}_{d_2}(0)\hat{a}_{d_1}(0)}_{\text{out}}=  &\frac{3\sqrt{3}\tunn^4 g_L^2 g^2 \Kconst_1 \Kconst_2}{2(\omega_L-U)^2} (\mathcal{A}_a-2\mathcal{A}_b)\chi\\
   &+\ldots,
   \end{aligned}
\end{equation}
where ``$\ldots$" includes terms in other symmetry channels which can be filtered out by varying the linear polarizations $\vb{e}_L,$ $\vb{e}_{1}$, $\vb{e}_2$, and taking appropriate linear combinations of the experimental data. We note that the sum of coefficients per unit cell in Eq.~\eqref{eq:chioperator} is zero. {However, we generically do not expect the spin chirality on an equilateral triangle and an obtuse triangle to cancel out, since the two are unrelated by symmetry. Therefore, we generically expect $\Im \expval{\hat{a}_{d_2}(0)\hat{a}_{d_1}(0)}_{\text{out}}$ to be nonzero in a chiral spin liquid. While Eq.~\eqref{eq:chioperator} has been written for nearest-neighbor tunneling, we see from Eq.~\eqref{eq:BiExp} that analogous expressions follow for the case of long-range tunneling as well. The main difference will be the addition of spin chirality operators on larger triangles. These new operators have the same symmetry properties as Eq.~\eqref{eq:chioperator} and are again additional contributions to scalar spin chirality.

In summary, we propose the following experimental protocol to detect scalar spin chirality and identify a chiral spin liquid. (1) The system is driven by a coherent laser beam with a frequency \(\omega_L\) satisfying \(\tunn \ll \abs{\omega_L - U} \ll U\). (2) The phase-sensitive quadrature fluctuation, $\expval{\hat{a}_{d_2}(0)\hat{a}_{d_1}(0)}_\text{out}$ of the filtered sidebands of the scattered light is measured. Within each sector of Fig.~\ref{fig:sectorsbasic}, the filters are assumed to be broadband within their respective frequency windows. Linear combinations of $\expval{\hat{a}_{d_2}(0)\hat{a}_{d_1}(0)}_\text{out}$ at various polarizations are used to extract contributions in the channels \(\mathcal{A}_a\) and \(\mathcal{A}_b\), as defined in Eqs.~(\ref{eq:channelAa}-\ref{eq:channelAb}). A nonzero result directly indicates the presence of scalar spin chirality. 

We note that the generation of entangled photons with frequencies \(2\omega_L - U\) and \(U\) is reminiscent of the four-wave mixing process that naturally occurs in nonlinear media. This process can also be understood as two-mode squeezing of the coherent component of the scattered light \cite{christ2011probing,lemonde2014antibunching,grankin2016inelastic}. The coherent component corresponds to processes where the matter and light are unentangled after the scattering. In contrast, the incoherent components generate excess noise, which can be probed via measurement of the phase-independent part of the quadrature correlation function, i.e., $G^{(1)}(\tau)$.

If we consider $G^{(2)}_{d_1,d_2}$ instead of the squeezing spectrum, the corresponding matter correlator is the absolute value squared of spin chirality. In this section, we focused on detection near $2\omega_L-U$ and $U$. Using $G^{(2)}_{d_1,d_2}$, when both the detectors are centered near frequency $\omega_L$, one can already make interesting predictions for the correlators of $\hat{A}_j$, defined in Eq.~\eqref{eq:AiExp}. This will be the subject of the next two sections.

\section{Contribution of noninteracting magnons to $G^{(2)}$}\label{sec:magnons}
The low-energy Hamiltonian of the Fermi-Hubbard model at half-filling, when projected to the spin sector, yields the Heisenberg model: 
\begin{equation}\label{eq:Heisenberg}
    \hat{H}_{\text{Heisenberg}} = \frac{1}{2}\sum_{\vb{r},\vb{r}'}J_{\vb{r}\vb{r}'}\hat{\vb{S}}_{\vb{r}}\cdot \hat{\vb{S}}_{\vb{r}'},
\end{equation}
where $J_{\vb{r},\vb{r}'}=4\tunn_{\vb{r},\vb{r}'}^2/U$ . 
Depending on the lattice geometry and the values of couplings beyond nearest-neighbor and next-nearest-neighbor couplings, the Heisenberg model is believed to admit a variety of ground states -- both ordered states and spin liquids. If the ground state is magnetically ordered, the excitations are bosonic magnons (which may have a finite lifetime and weak interactions). However, if the ground state is a spin liquid, the excitations can be more exotic, such as anyons with fractional statistics. The goal of this section is to study the contribution to the connected $\mathcal{G}^{(2)}$ from non-interacting magnons. We then determine the conditions under which these contributions can be filtered away, so that any remaining $\mathcal{G}^{(2)}$ signal originates from magnon-magnon interactions or topological contributions.

\subsection{Quadratic magnon Hamiltonian}
Using the standard linearized Holstein-Primakoff
and Bogoliubov transformations \cite{holstein1940field,altland2010condensed},
the Hamiltonian of the Heisenberg model Eq.~\eqref{eq:Heisenberg} can
be written as: 
\begin{equation}\label{eq:magnonham}
    \hat{H}_{\text{low}}=\sum_{\vb{k}}\xi_{\vb{k}} \hat{b}^{\dagger}_{\vb{k}}\hat{b}_{\vb{k}},
\end{equation}
where $\hat{b}_{\vb{k}}^{\dagger}$ is the creation operator of a
magnon carrying momentum $\vb{k}$ and characterized by a dispersion
$\xi_{{\bf k}}$. In spin variables, $\hat{b}_{\vb{k}}^{\dagger}$
generally corresponds to a superposition of $\hat{S}_{\vb{k}}^{+}$
and $\hat{S}_{\vb{k}}^{-}$, where $\hat{S}_{\vb{k}}^{\pm}=\hat{S}_{\vb{k}}^{x}\pm i\hat{S}_{\vb{k}}^{y}$.  

The pure spin sector operators $\hat{A}_{j}$, which
appear in the photonic correlation functions, are defined in Eq.~\eqref{eq:AiExp}.
As outlined in Sec.~\ref{sec:Ramanrev}, $\hat{A}_{j}$ can be decomposed
into irreducible representations of the crystalline point group. In
this section, we assume that this decomposition has been performed
(e.g., by taking linear combinations of experimental data corresponding
to different polarization directions) and that $\hat{A}_{j}$ belongs
to a specific irreducible representation. We proceed to express $\hat{A}_{j}$
in terms of magnon operators. Up to second order in magnon operators,
we have (in the interaction picture): 
\begin{equation}\label{eq:expansionAj}
\begin{aligned}
    \hat{A}_j(-t) =\sum_{\vb{k}; k_x\geq 0}&\left\{\alpha_j(\vb{k})e^{-2i\xi_{\vb{k}}t}\hat{b}^{\dagger}_{\vb{k}}\hat{b}^{\dagger}_{-\vb{k}}\right.\\
    &+(\alpha'_j(\vb{k}))^*\hat{b}_{\vb{k}}\hat{b}_{-\vb{k}}e^{2i\xi_{\vb{k}}t}\\
    &\left.+\beta_j(\vb{k})\left(\hat{b}^{\dagger}_{\vb{k}}\hat{b}_{\vb{k}}+\hat{b}^{\dagger}_{-\vb{k}}\hat{b}_{-\vb{k}}+1\right)\right\},
 \end{aligned}   
\end{equation}
where the coefficients $\alpha_{j}(\vb{k})$, $\alpha'_{j}(\vb{k})$
and $\beta_{j}(\vb{k})$ are determined by the same Bogoliubov transformation
that diagonalizes the free magnon Hamiltonian. Note that earlier,
the subscript $j$ referred to whether the operator $\hat{A}_{j}$
coupled to a photon mode of detector $1$ or $2$. In this section,
the subscript also labels an irreducible representation of the spatial
symmetry group. 

The coefficients $\left\{ \alpha_{j}(\vb{k})\right\} $,
$\left\{ \alpha'_{j}(\vb{k})\right\} $, $\left\{ \beta_{j}(\vb{k})\right\} $,
and their complex conjugates, all transform according to the same
irreducible representation as $\hat{A}_{j}$ does, under spatial rotations
and reflection. Therefore, the following orthogonality conditions
hold: 
\begin{align}\label{eq:orthog1}
    \sum_{\vb{k}}\left(\alpha'_j(\vb{k})\right)^*\alpha_l(\vb{k})&=0, \text{ for } j\neq l,\\
    \label{eq:orthog2}\sum_{\vb{k}}\left(\alpha'_j(\vb{k})\right)\alpha_l(\vb{k})&=0, \text{ for } j\neq l,\\
    \label{eq:orthog3}\sum_{\vb{k}}\left(\beta_j(\vb{k})\right)^*\beta_l(\vb{k})&=0, \text{ for } j\neq l.
\end{align}
\subsection{Contribution of magnons to Raman Scattering, or $G^{(1)}(0)$}
 Let us now assume the spin system is initially prepared in its ground state. Defining the Raman shift $\Omega_j\equiv \omega_L-\omega_j$. Using Eq.~\eqref{eq:G1simp2}, we find that the Raman intensity $G^{(1)}$ detected in channel $j$ for a Lorentzian filter (Eq.~\eqref{eq:Lorentzdef}) is given by 
\begin{equation}
    G^{(1)}_{d_j}(0)=\abs{\Kconst_j}^2\sum_{\vb{k}}\frac{\Gamma^2_j\abs{\alpha_j(\vb{k})}^2}{\Gamma_j^2 +(\Omega_j-2\xi_{\vb{k}})^2}.
\end{equation}
This expression reflects the fact that the photon scattering process can create a magnon pair, and $G^{(1)}(0)$ includes contributions from all magnon pairs with energy $2\xi_{\vb{k}}$ equal to the Raman shift $\Omega_j$ within an uncertainty $\Gamma_j$ set by the frequency filter. It is important to note that the equation involves a discrete sum over momenta, with the number of terms governed by the number of lattice sites \(N\) illuminated by the laser beam.

We now consider the limits $\Gamma_j \to \infty$ (broad filter) and $\Gamma_j \to 0$ (sharp filter). For a broad filter,
\begin{equation}
    G^{(1)}_{d_j}(0)\xrightarrow{\Gamma_j\to \infty} N\alatt^2\abs{\Kconst_j}^2\int_{\text{B.Z.}}\frac{\dd[2]{k}}{(2\pi)^2}\abs{\alpha_j(\vb{k})}^2,
\end{equation}
where B.Z. stands for the first Brillouin zone of the lattice. Next, for a sharp filter,
\begin{equation}
    G^{(1)}_{d_j}(0)\xrightarrow{\Gamma_j\to 0} N\Gamma_j\alatt^2\abs{\Kconst_j}^2\int_{\text{B.Z.}}\frac{\dd[2]{k}}{4\pi}\delta(2\xi_{\vb{k}}-\Omega_j)\abs{\alpha_j(\vb{k})}^2.
\end{equation}
We thus find that $G^{(1)}_{d_j}(0)$ reflects the density of states of magnon pairs (of zero total momentum) at $\Omega_j$, modulated by $\abs{\alpha_j(\vb{k})}^2$. Note that in the sharp filter limit we assume $\Gamma_j$  is still greater than the mean energy spacing due to the finite-size quantization. Thus, $\Gamma_j N$ remains nonzero even in the limit $\Gamma_j \to 0$.
\subsection{Magnon contributions to connected $\mathcal{G}^{(2)}$}
Since the magnon Hamiltonian in Eq.~\eqref{eq:magnonham} is quadratic and the input light is Gaussian, one might na\"ively expect $G^{(2)}_{d_1,d_2}(0)$ to factorize as $G^{(1)}_{d_1}(0)G^{(1)}_{d_2}(0)$. However, we show here that this is not the case. The reason is that the effective coupling to matter via Eq.~\eqref{eq:expansionAj} is nonlinear. This finding is reminiscent of the result of Ref.~\cite{lemonde2014antibunching} on single-mode Gaussian states of light.

We consider the connected part of $\mathcal{G}^{(2)}_{d_1, d_2}(\tau)\equiv G^{(2)}_{d_1,d_2}(\tau)-G^{(1)}_{d_1}(0)G^{(1)}_{d_2}(0)$. For a Lorentzian filter, it is given by:
\begin{equation}\label{eq:spinG2}
    \begin{aligned}
        &\mathcal{G}^{(2)}_{d_1,d_2}(\tau)=\abs{\Kconst_1 \Kconst_2 \Gamma_1 \Gamma_2}^2\iiiint_{0}^{\infty}\dd{t_1}\dd{t_2}\dd{t'_1}\dd{t'_2}\\
        &\times\left\{e^{-\left[\Gamma_1(t_1+t'_1)+\Gamma_2(t_2+t'_2)\right]}e^{i\left[\Omega_1(t_1-t'_1)+\Omega_2(t_2-t'_2)\right]}\right.\\
        & \quad\times \left(\expval{{\bar{\mathbb{T}}\left[\hat{A}^{\dagger}_1(-t'_1)\hat{A}^{\dagger}_2(\tau-t'_2)\right]\mathbb{T}\left[\hat{A}_2(\tau-t_2)\hat{A}_1(-t_1)\right]}}_0\right.\\
        &\left.\left.\quad\quad -\expval{\hat{A}^{\dagger}_1(-t'_1)\hat{A}_1(-t_1)}_{0}\expval{\hat{A}^{\dagger}_2(-t'_2)\hat{A}_2(-t_2)}_{0}\right)\right\}.
    \end{aligned}
\end{equation}
Starting from the ground state of the spin system, the right-most operator (corresponding to the earliest photon emission) first creates a magnon pair as shown in Fig.~\ref{fig:magnonfull}(a), and the photon emitted is red-detuned with respect to the laser frequency. For  $\hat{A}_1$ and $\hat{A}_2$ belonging to the same symmetry channel, $\hat{A}_2$ can annihilate the magnon-pair created by $\hat{A}_1$, leading to the emission of a blue-detuned photon as schematically shown in  Fig.~\ref{fig:magnonfull}(c). For simplicity, we now assume that operators $\hat{A}_1$ and $\hat{A}_2$ are in different symmetry channels. Then, by the orthogonality relations Eq.~(\ref{eq:orthog1}-\ref{eq:orthog3}) the magnon pair created by $\hat{A}_1$ can only be annihilated by $\hat{A}^{\dagger}_1$ (and not by $\hat{A}^{\dagger}_2$). 

Let us now consider the contribution where $\hat{A}_1$ and $\hat{A}_2$ each create a magnon pair of different momenta as shown in Fig.~\ref{fig:magnonfull}(b). Then these  pairs are annihilated by $\hat{A}^{\dagger}_1$ and $\hat{A}^{\dagger}_2$. The term resulting from this process factorizes as $G^{(1)}_{d_1}(0)G^{(1)}_{d_2}(0)$ and thus contributes only to the disconnected part of the full $G^{(2)}$.

We now describe two classes of terms contributing to the connected part $\mathcal{G}^{(2)}_{d_1,d_2}(0)=G^{(2)}_{d_1,d_2}(0)-G^{(1)}_{d_1}(0)G^{(1)}_{d_2}(0)$.
\begin{figure}
  \centering
  \includegraphics[width=0.48\textwidth]{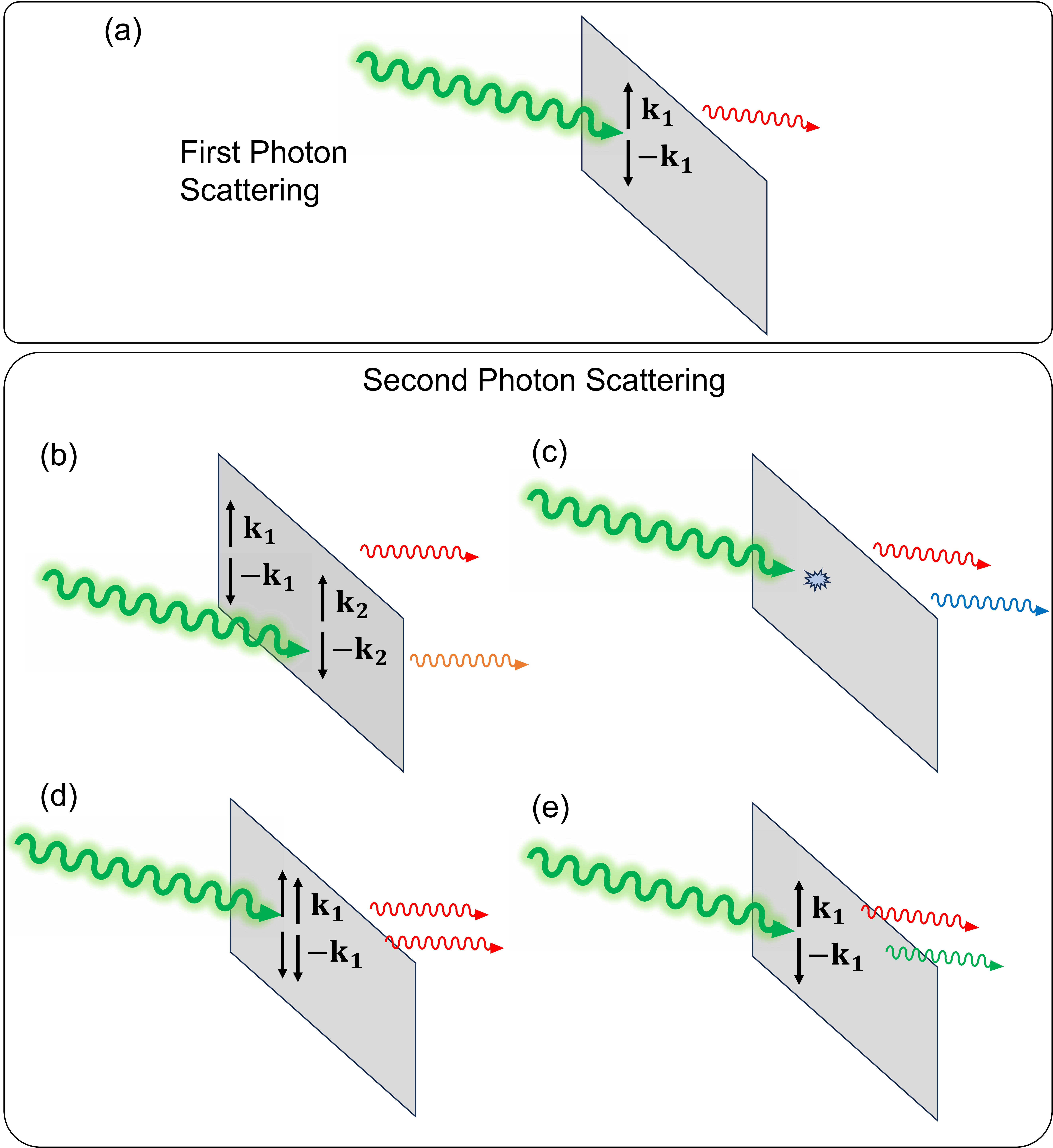}
  \caption{Different magnon processes leading to the scattering of two photons: \textbf{(a)}: Raman scattering of the first photon leads to the creation of a magnon-pair of momenta $\vb{k}_1$ and $-\vb{k}_1$. \textbf{(b)}: The scattering of the second photon could lead to creation of a different magnon-pair of momenta $\vb{k}_2$ and $-\vb{k}_2$. In this case, the second photon is also red-detuned with respect to the laser and generically has a frequency different from the first one. \textbf{(c)}: Scattering of the second photon leads to the annihilation of the magnon-pair created in (a). The photon is blue-detuned in this case. \textbf{(d)}: This is similar to (b), except $\vb{k}_2=\vb{k}_1$. This is treated separately here.} \textbf{(e)}: The second photon is elastically scattered.
  \label{fig:magnonfull}
\end{figure}
\subsubsection{Nonlinear bosonic enhancement}
Now, let us consider the scenario where \(\hat{A}_1\) and \(\hat{A}_2\) create magnon pairs with identical momenta. In this case, the same magnon pair is both created and annihilated twice, as depicted in Fig.~\ref{fig:magnonfull}(d). Using the fact that $\mel{0}{\hat{b}_{-\vb{k}_1}\hat{b}_{\vb{k}_1}\hat{b}_{-\vb{k}_2}\hat{b}_{\vb{k}_2}\hat{b}^{\dagger}_{-\vb{k}_2}\hat{b}^{\dagger}_{\vb{k}_2}\hat{b}^{\dagger}_{-\vb{k}_1}\hat{b}^{\dagger}_{\vb{k}_1}}{0}=4$ for $\vb{k}_1=\vb{k}_2$, and $1$ for $\vb{k}_1\neq \vb{k}_2$, we get the following nonzero contribution to the connected $\mathcal{G}^{(2)}_{d_1,d_2}$:
\begin{equation}\label{eq:magnonbosonicfull} \sum_{\vb{k}}\frac{3\abs{\Kconst_1\Kconst_2}^2\Gamma^2_1\Gamma^2_2\abs{\alpha_1(\vb{k})\alpha_2(\vb{k})}^2}{\left[\Gamma_1^2 +(\Omega_1-2\xi_{\vb{k}})^2\right]\left[\Gamma_2^2 +(\Omega_2-2\xi_{\vb{k}})^2\right]}.
\end{equation}
The discrete summation $\sum_{\vb{k}}$ can be written in the continuum as $N \alatt^2 \int \tfrac{\dd[2]{k}}{(2\pi)^2}$, where $N$ is the number of lattice sites covered by the incoming laser beam. Let us consider the sharp filter case. Here, we will work in the limit where $\Gamma_1\Gamma_2 N$ is nonzero and finite even as one takes $\Gamma_{1,2}\to 0$. In this limit, using the identity $\lim_{ \Gamma\to 0}\tfrac{\Gamma}{\Gamma^2+ x^2}=\pi \delta(x)$,  Eq.~\eqref{eq:magnonbosonicfull} yields:
\begin{equation}\label{eq:bosonicenhresult}
   \begin{aligned}
    &\frac{3\abs{\Kconst_1\Kconst_2}^2(N\Gamma_1\Gamma_2)\alatt^2}{4}\delta(\Omega_1-\Omega_2)\\
    &\quad \quad \quad \quad \quad\times \int \dd[2]{k} \abs{\alpha_1(\vb{k})\alpha_2(\vb{k})}^2 \delta(\Omega_1 - 2\xi_{\vb{k}}). 
    \end{aligned}
\end{equation}
Since this term contains $\delta(\Omega_1-\Omega_2)$, it can be filtered out by choosing $\omega_1 \neq \omega_2$ (and hence $\Omega_1 \neq \Omega_2$). 
\subsubsection{Elastic scattering}
The second contribution to the connected \(\mathcal{G}^{(2)}_{d_1,d_2}\) arises from the following scenario: the first photon is inelastically scattered by the material, resulting in the creation of a magnon pair described by the action of \(\hat{A}_1\) on the magnon vacuum state \(|0\rangle\). The second photon, however, undergoes elastic scattering without generating an additional magnon pair, as illustrated in Fig.~\ref{fig:magnonfull}(e). This process is described by the magnon-conserving terms in \(\hat{A}_2\)  of the form \((\hat{b}^\dagger_{\vb{k}} \hat{b}_{\vb{k}} + \hat{b}^\dagger_{-\vb{k}} \hat{b}_{-\vb{k}} + 1)\).

After a lengthy but straightforward algebraic calculation, we arrive at the following contribution to  Eq.~\eqref{eq:spinG2}, from the integrals over the domains \((t_1 > t_2, \ t'_1 > t'_2)\) and \((t_1 < t_2, \ t'_1 < t'_2)\) :
\begin{equation}\label{eq:magnonordering1}
\begin{aligned}
\sum_{\vb{k}}\Biggl\{&\frac{\abs{\alpha_1(\vb{k})}^2\abs{\beta_2(\vb{k})}^2\Gamma^2_1\Gamma^2_2}{\left[\Gamma^2_1+(2\xi_{\vb{k}}-\Omega_1)^2\right]\left[(\Gamma_1+\Gamma_2)^2+(2\xi_{\vb{k}}-\Omega_1-\Omega_2)^2\right]} \\
&+ \left[1\leftrightarrow 2\right]\Biggr\}\times 9 \abs{\Kconst_1 \Kconst_2}^2
\end{aligned}
\end{equation}
In contrast, the contribution from $(t_1>t_2,\ t'_1<t'_2)$ and $(t_1<t_2 , \ t'_1>t'_2)$ in Eq.~\eqref{eq:spinG2} is given below.
\begin{equation}\label{eq:magnonordering2}
\begin{aligned}
&\sum_{\vb{k}}\Biggl\{\left[\alpha^*_2(\vb{k})\beta^*_1(\vb{k})\beta_2 (\vb{k})\alpha_1 (\vb{k})\frac{\Gamma^2_1\Gamma^2_2e^{-2i\xi_{\vb{k}}\tau}}{\left(\Omega_2 +i \Gamma_2\right)\left(\Omega_1 -i\Gamma_1\right)}\right.\\
    &\quad \times\left(\frac{e^{(i\Omega_2-\Gamma_2) \tau}}{\Omega_1 +\Omega_2 -2\xi_{\vb{k}} +i(\Gamma_1 +\Gamma_2)}+\frac{1-e^{(i\Omega_2-\Gamma_2) \tau}}{\Omega_1-2\xi_{\vb{k}}+i\Gamma_1}\right)\\
    &\left.\quad \times\left(\frac{e^{(-i\Omega_1-\Gamma_1) \tau}}{\Omega_1 +\Omega_2 -2\xi_{\vb{k}} -i(\Gamma_1 +\Gamma_2)}+\frac{1-e^{(-i\Omega_1-\Gamma_1) \tau}}{\Omega_2-2\xi_{\vb{k}}-i\Gamma_2}\right)\right]\\
    &\quad + \text{ c.c.}\Biggr\}\times 9\abs{\Kconst_1\Kconst_2}^2.
    \end{aligned}
\end{equation}
 In Eqs.~\eqref{eq:magnonordering1} and Eq.~\eqref{eq:magnonordering2}, one factor of $\Gamma_1\Gamma_2$ can be combined with $N$ to give a finite constant while converting the sum to an integral, leaving behind the other factor of $\Gamma_1\Gamma_2$ in the integrand. As we take $\Gamma_1 \to 0$, and $\Gamma_2\to 0$, regardless of the order, one can check that the expression in Eq.~\eqref{eq:magnonordering2} goes to 0. On the other hand, for Eq.~\eqref{eq:magnonordering1}, it matters whether one takes $\Gamma_1\to 0$ first, or $\Gamma_2\to 0$. If we first take $\Gamma_2\to 0$ and then $\Gamma_1\to 0$, then the expression in Eq.~\eqref{eq:magnonordering1} becomes:
\begin{equation}\label{eq:gamma2first}
    \begin{aligned}
        \frac{9\abs{\Kconst_1\Kconst_2}^2 \alatt^2 N\Gamma_1 \Gamma_2}{4}\delta(\Omega_1)\int \dd[2]{k}\abs{\alpha_1(\vb{k})\beta_2(\vb{k})}^2\delta(2\xi_{\vb{k}}-\Omega_2).
    \end{aligned}
\end{equation}
On the other hand, if we first take $\Gamma_1\to 0$ and then $\Gamma_2\to 0$, then the expression in Eq.~\eqref{eq:magnonordering1} becomes:
\begin{equation}\label{eq:gamma1first}
    \begin{aligned}
        \frac{9\abs{\Kconst_1\Kconst_2}^2 \alatt^2 N\Gamma_1 \Gamma_2}{4}\delta(\Omega_2)\int \dd[2]{k}\abs{\alpha_2(\vb{k})\beta_1(\vb{k})}^2\delta(2\xi_{\vb{k}}-\Omega_1).
    \end{aligned}
\end{equation}
Eq.~(\ref{eq:gamma2first}-\ref{eq:gamma1first}) correspond to elastic scattering and by choosing $\Omega_1=\omega_L-\omega_1\neq 0$, and $\Omega_2=\omega_L-\omega_2 \neq 0$, this contribution to $\mathcal{G}^{(2)}_{d_1,d_2}$ can be filtered out.

Therefore, if the filter functions are sharp, any contribution to the connected $\mathcal{G}^{(2)}$ from noninteracting bosonic excitations can be eliminated if we: (1) filter out the case when the two magnon pairs created are identical by demanding that the frequencies of the detected photons are different (i.e., impose $\omega_1\neq \omega_2$), and (2) filter out elastic scattering (i.e., impose $\omega_1\neq \omega_L$ and $\omega_2\neq \omega_L$).

Once this filtering is implemented, any remaining contribution to the connected $\mathcal{G}^{(2)}$ should thus arise from interactions or topology.
\section{Application II: Detecting fractional statistics}\label{sec:fracstat}
In this section, we show that if the ground state of the spin sector is topologically ordered, the existence of anyonic excitations with fractional mutual statistics can be detected by measuring photonic correlators of scattered photons. We use the argument from Ref.~\cite{mcginley2024anomalous,mcginley2024signatures}.

We start with an outline of this section. First, in Sec.~\ref{sec:fracstatcondG1}, we consider the conditional $G^{(1)}(\tau)$ [Eq.~\eqref{eq:ConditionalG1}] and observe that it has the same form as a correlator studied in Ref.~\cite{mcginley2024anomalous,mcginley2024signatures}, allowing us to directly use their results. Next, in Sec.~\ref{sec:subfracstat}, we propose another approach using the connected $\mathcal{G}^{(2)}$ where one can explicitly filter out the contributions from topologically trivial magnon excitations discussed in Sec.~\ref{sec:magnons}. The temporal contour of this correlator is slightly different from the one considered in Ref.~\cite{mcginley2024anomalous,mcginley2024signatures}. So we adapt their arguments for our case. We do not repeat their derivation, but summarize their arguments qualitatively in this subsection. We encourage the reader to refer to Ref.~\cite{mcginley2024anomalous,mcginley2024signatures} for a treatment of finite temperature effects and for a discussion on ignoring short-range interactions between anyons.

\subsection{Conditional $G^{(1)}(\tau)$: Signature of existence of fractional statistics}\label{sec:fracstatcondG1}
First, let us consider the conditional $G^{(1)}(\tau)$ defined in Eq.~\eqref{eq:ConditionalG1}, under the following subtraction scheme:
\begin{equation}\label{eq:eqHd1d2}
\begin{aligned}
    &H_{d_1,d_2}(t,\tau)-G^{(1)}_{d_1}(0)G^{(1)}_{d_2}(\tau)\\
    =&\langle\hat{a}^{\dagger}_{d_1}(0)\hat{a}^{\dagger}_{d_2}(t+\tau)\hat{a}_{d_2}(t)\hat{a}_{d_1}(0)\rangle_{\text{out}}+\text{c.c.}\\
    &-\langle\hat{a}^{\dagger}_{d_1}(0)\hat{a}_{d_1}(0)\rangle_{\text{out}} \left(\langle \hat{a}^{\dagger}_{d_2}(t+\tau)\hat{a}_{d_2}(t) \rangle_{\text{out}}+\text{c.c.}\right).
    \end{aligned}
\end{equation}
The experimental scheme to measure this photonic correlator is shown in Fig.~\ref{fig:ConditionalG2}, and explained in detail in Appendix~\ref{app:ConditionalG1}. Let us assume that the filter functions are chosen such that they select the window around $\omega_L$ (as opposed to the sidebands $2\omega_L-U$ and $U$), but is broadband with respect to the finer structure within each sector in Fig.~\ref{fig:sectorsbasic}. Then, using the mapping provided in Table~\ref{tab:dictionary1}, for $t>0$ and $\tau>0$, Eq.~\eqref{eq:eqHd1d2} yields:
\begin{equation}\label{eq:connConditionalG1}
    \begin{aligned}
    \abs{\Kconst_1\Kconst_2}^2&\left[\expval{\hat{A}^{\dagger}_1(0)\hat{A}^{\dagger}_2(t+\tau)\hat{A}_2(t)\hat{A}_{1}(0)}_0+\text{c.c.}\right.\\
    &\left.-\expval{\hat{A}^{\dagger}_1(0)\hat{A}_{1}(0)}_0\left(\expval{\hat{A}^{\dagger}_2(t+\tau)\hat{A}_2(t)}_0+\text{c.c.}\right)\right],
    \end{aligned}
\end{equation}
where the constants $\Kconst_j$ are defined in Eq.~\eqref{eq:Kdef}. 

The detection of a photon by detector $d_j$ corresponds to the application of the operator $\hat{A}_j$, (defined in Eq.~\eqref{eq:AiExp}) on the material state. As long as $\hat{A}_j$ does not commute with the Hamiltonian, $\hat{A}_j$ acting on the ground state will create excitations. If the material is topologically ordered, its excitations can be anyonic. We assume that the material is at zero temperature. Now, if $\ket{I}$ is the ground state, then  $\hat{A}_j \ket{I}$ would generically have an overlap with a state consisting of an anyon pair, unless this is ruled out by symmetry. We will therefore suppose that $\hat{A}_1$ and $\hat{A}_2$ create pairs of anyons whose energy gaps are $\Delta_1$ and $\Delta_2$ respectively. For concreteness, if the ground state is a $\mathbb{Z}_2$ spin liquid, one can think of $\hat{A}_1$ and $\hat{A}_2$ as creating $e$ and $m$ anyon pairs respectively. These pairs get annihilated by $\hat{A}^{\dagger}_1$ and $\hat{A}^{\dagger}_2$ respectively \footnote{Being in different symmetry channels, $\hat{A}^{\dagger}_1$ cannot annihilate anyons created by $\hat{A}_2$ and vice-versa.}. We point out that $\Delta_1$ and $\Delta_2$ are the gaps within the spin sector and are much smaller than the optical gap $U$.

The correlator in Eq.~\eqref{eq:connConditionalG1} is indeed of the same form as the pump-probe susceptibility studied in Ref.~\cite{mcginley2024anomalous,mcginley2024signatures}. Therefore, using the same argument as Ref.~\cite{mcginley2024anomalous,mcginley2024signatures} (which we summarize in Sec.~\ref{sec:subfracstat}), Eq.~\eqref{eq:connConditionalG1} is proportional to $(1-\cos \alpha_{12})\tau^{1/2}$, where $e^{i\alpha_{12}}$ is the braiding phase of an anyon created by $\hat{A}_2$ going around an anyon created by $\hat{A}_1$. This $\tau^{1/2}$ behavior is an experimental signature of the existence of fractional statistics.

One can also look at the following normalized version of the photonic correlator in Eq.~\eqref{eq:eqHd1d2}:
\begin{equation}\label{eq:normalizedcond}
    \bar{H}_{d_1,d_2}(t,\tau)\equiv \frac{H_{d_1,d_2}(t,\tau)-G^{(1)}_{d_1}(0)G^{(1)}_{d_2}(\tau)}{G_{d_2}(\tau)}
\end{equation}

The denominator was shown in Ref.~\cite{mcginley2024anomalous} to go as $1/\tau$. Therefore, $\bar{H}_{d_1,d_2}(t,\tau)\sim \tau^{3/2}$ if there are excitations with fractional mutual statistics. The normalized $\bar{H}_{d_1,d_2}(t,\tau)\sim \tau^{3/2}$. It was shown in Ref.~\cite{mcginley2024anomalous} to be more robust to thermal fluctuations than Eq.~\eqref{eq:eqHd1d2}.

In the next subsection, we show that a similar prediction can be made for the connected $\mathcal{G}^{(2)}_{d_1,d_2}$ correlator instead of Eq.~\eqref{eq:connConditionalG1}. $\mathcal{G}^{(2)}_{d_1,d_2}$ offers additional functionality by enabling filtering out contributions from topologically trivial excitations as discussed in Sec.~\ref{sec:magnons}. Thus, filtering allows us to consider the contribution originating exclusively from the fractional statistics of quasiparticles. 
\subsection{Connected $\mathcal{G}^{(2)}$: Singularity from fractional statistics}\label{sec:subfracstat} 
The spin correlator measured by the connected $\mathcal{G}^{(2)}$ is provided in Eq.~\eqref{eq:spinG2}. It comprises contributions from different worldlines of anyons obeying the constraint that the pair created at time $-t_1$ is annihilated at time $-t'_1$, and similarly the pair created at $\tau-t_2$ is annihilated at $\tau-t'_2$. Among these are those worldlines where one anyon \textit{braids} nontrivially with another. Such paths come with an extra braiding phase. In this section, we show that nontrivial mutual statistics leads to a singularity in the connected $\mathcal{G}^{(2)}$ as well: $\mathcal{G}^{(2)}_{d_1,d_2}(\Omega_1,\Omega_2)\sim\theta(\Omega_1-\Delta_1)\theta(\Omega_2-\Delta_2)\left[ K_2(\Omega_2)(\Omega_1-\Delta_1)^{-3/2}+(1\leftrightarrow 2)\right]$, where $K_j(\Omega_j)$ are system-specific functions (recall that $\Omega_j = \omega_L-\omega_j$). In the limit of sharp frequency filters, we expect the dependence of $G^{(2)}$ on $\tau$ to drop out. Hence, we can set $\tau=0$ from the outset.

We adapt the geometric argument in Ref.~\cite{mcginley2024anomalous,mcginley2024signatures} to show the above result. There is one difference in our case: due to the frequency-filtering, the time at which the first anyon pair is created need not be equal to the time when it is annihilated. 
\begin{figure}
  \centering
  \includegraphics[width=0.48\textwidth]{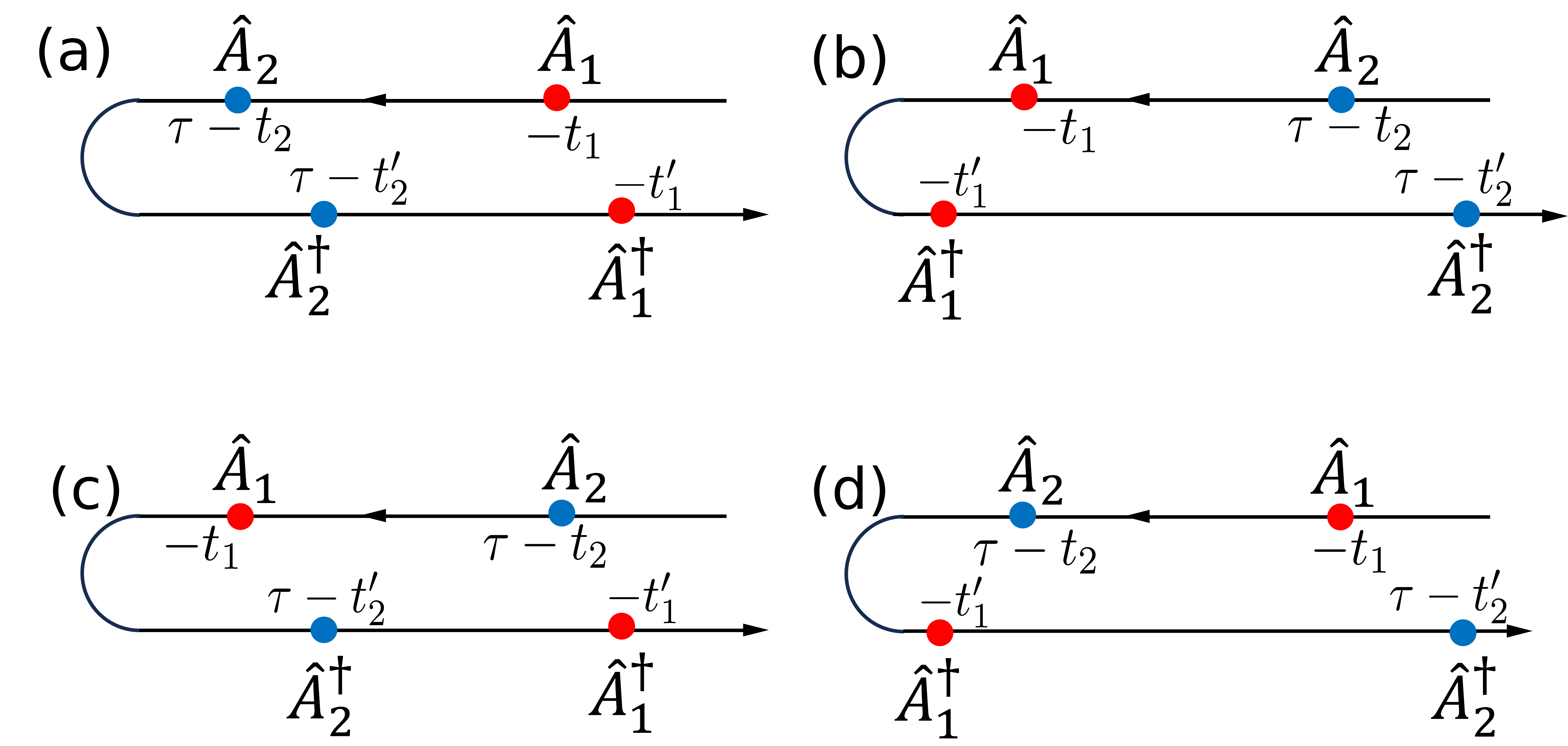}
  \caption{Four operator orderings in Eq.~\eqref{eq:spinG2}: \textbf{(a)}: $\tau-t_2>-t_1$ and $\tau-t'_2>-t'_1$, \textbf{(b)}: $-t_1>\tau-t_2$ and $-t'_1>\tau-t'_2$, \textbf{(c)}: $-t_1>\tau-t_2$ and $\tau-t'_2>-t'_1$, and \textbf{(d)}: $\tau-t_2>-t_1$ and $-t'_1>\tau-t'_2$.}
  \label{fig:allcontours}
\end{figure}

The spin correlator extracted from $\mathcal{G}^{(2)}$, i.e., Eq.~\eqref{eq:spinG2} is a sum of terms with four different operator orderings depending on the ordering within the pairs $(-t_1,-t'_1$), and $(-t_2,-t'_2)$. These orderings are (suppressing the time arguments): $ \expval{\hat{A}^{\dagger}_1\hat{A}^{\dagger}_2\hat{A}_2\hat{A}_1}$, $ \expval{\hat{A}^{\dagger}_2\hat{A}^{\dagger}_1\hat{A}_1\hat{A}_2}$, $\expval{\hat{A}^{\dagger}_1\hat{A}^{\dagger}_2\hat{A}_1\hat{A}_2}$, and $ \expval{\hat{A}^{\dagger}_2\hat{A}^{\dagger}_1\hat{A}_2\hat{A}_1}$ [Fig.~\ref{fig:allcontours} (a), (b), (c), (d) respectively]. 

 First, let us consider the operator ordering in Fig.~\ref{fig:allcontours}(a), i.e., $\expval{\hat{A}^{\dagger}_1(-t'_1)\hat{A}^{\dagger}_2(-t'_2)\hat{A}_2(-t_2)\hat{A}_1(-t_1)}$. In this correlator, first $\hat{A}_1$ and then $\hat{A}_2$ each create anyon pairs at time $-t_1$ and $-t_2$ respectively (shown as red and blue respectively in Fig.~\ref{fig:braidcontour}). The blue pair gets destroyed by $\hat{A}^{\dagger}_2$ at time $-t'_2$, and lastly, at time $-t'_1$, the red anyon pair gets destroyed by $\hat{A}^{\dagger}_1$. If one ignores anyon-anyon interactions, the contributions from worldlines without any braiding get canceled when we look at the connected $\mathcal{G}^{(2)}$ (as shown in Sec.~\ref{sec:magnons}). Furthermore, the argument in Ref.~\cite{mcginley2024anomalous} about the contribution from short-ranged interactions being less singular than the contribution from fractional statistics also applies in our setting. Therefore, we will only study worldlines involving nontrivial braiding of otherwise ``non-interacting" anyons. Each such worldline contributes a topological factor of $(1-\cos \alpha_{12})$ where $e^{i\alpha_{12}}$ is the braiding phase of a blue anyon going around the red one.

\begin{figure}
  \centering
  \includegraphics[width=0.48\textwidth]{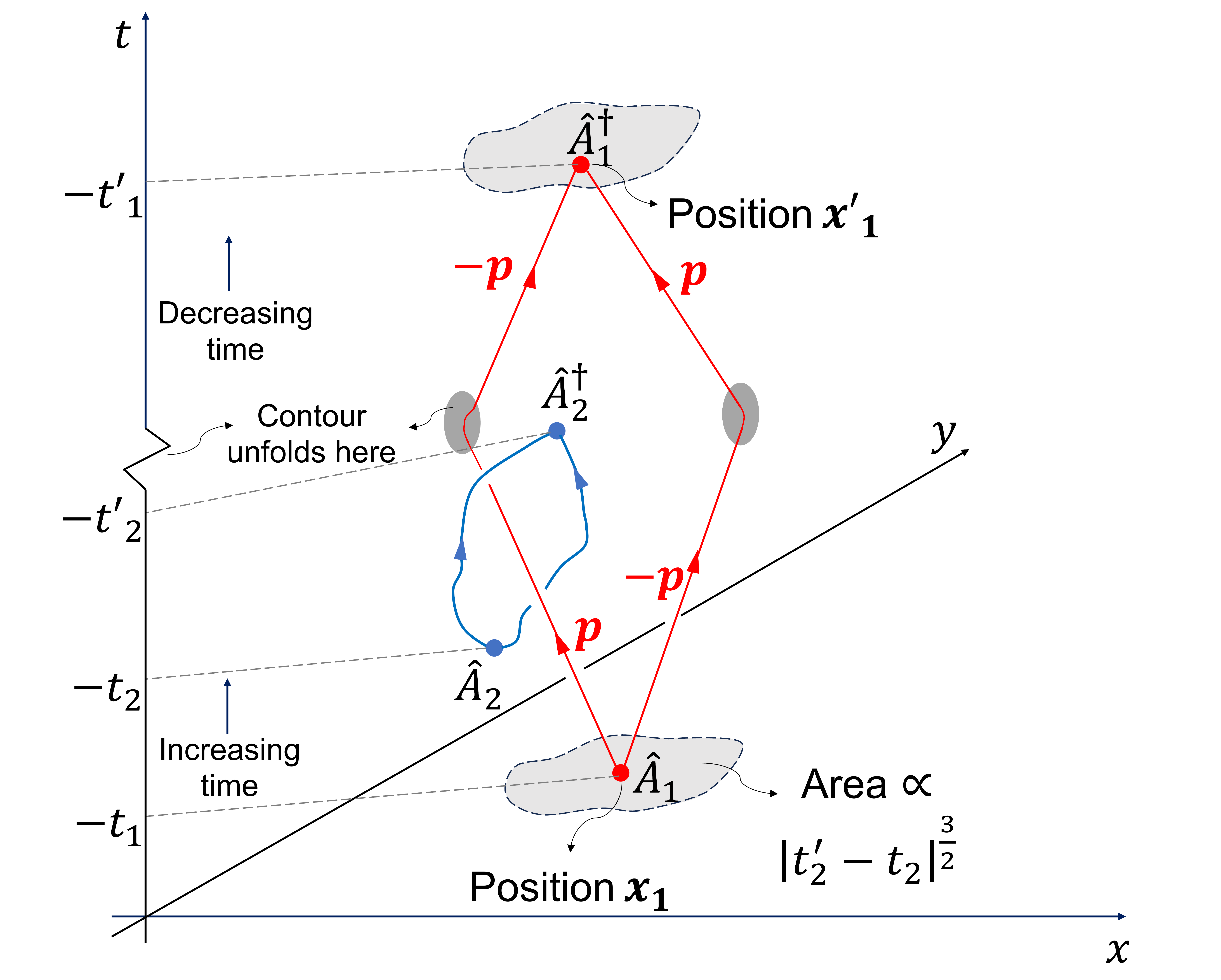}
  \caption{This figure is similar to Fig. 1 of Ref.~\cite{mcginley2024anomalous}, except the backward propagation is unfolded here. Without loss of generality, in this figure, $t_2>t'_2$. In our time axis, from $-t'_2$, onwards, time decreases in the upward direction.  For the operator ordering in Fig.~\ref{fig:allcontours}(a), we depict a history of anyons that contributes to the connected $\mathcal{G}^{(2)}$.  At time $-t_1$, operator $\hat{A}_1$ creates an anyon pair (red) at spatial point $\vb{x}_1$. Since $\hat{A}_1$ is a zero-momentum operator, $\vb{x}_1$ is integrated over the whole area of irradiation. At time $-t_2$, operator $\hat{A}_2$ (blue) creates a second anyon pair that recombines at $-t'_2$ due to wavepacket spreading. Let us consider a fixed worldline loop of the second pair. A path integral over all possible worldlines of the red anyons is dominated by those trajectories where a red anyon ballistically propagates through the blue loop till time $-t'_2$, and then turns around to recombine at time $-t'_1$. For a fixed momentum $\vb{p}$ of the red anyon, it has been shown geometrically that $\vb{x}_1$ can only be from an area $\sim \abs{t'_2-t_2}^{3/2}$~\cite{mcginley2024anomalous}. Since $-t'_1$ and $-t_1$ are not necessarily equal, classically, the red anyons will not recombine if they exactly retrace their path during backward propagation. However, due to quantum wavepacket spreading, there is a nonzero amplitude of recombination, which we conjecture to be independent of $t_2$ and $t'_2$.}
  \label{fig:braidcontour}
\end{figure}

The connected part can be written as a path integral over all trajectories where the blue anyon pair braids around one of the red anyons. Recall that here, $t_1-t_2>0$, and $t'_1-t'_2>0$, but the orderings within the pairs $(t_1, t'_1)$, and $(t_2, t'_2)$ are left unspecified. For the recombination of the second (blue) anyon pair between $t_2$ and $t'_2$, it is not sufficient to consider the ballistic propagation and wavepacket spreading has to be taken into account (see Fig.~\ref{fig:braidcontour}). The amplitude of recombination of the blue pair due to quantum wavepacket spreading is $\sim \abs{t_2-t'_2}^{-d/2}$ where spatial dimension $d=2$.

To study the first (red) anyon pair, following Ref.~\cite{mcginley2024anomalous,mcginley2024signatures}, one can perform a semiclassical analysis in the limit where the time it takes for an anyon pair to recombine is much less than the time difference between creating the first and second anyon pairs, i.e., $\abs{t_1-t_2}\gg \abs{t_2-t'_2}$, $\abs{t'_1-t'_2}\gg\abs{t_2 -t'_2}$, and $\abs{t'_1-t'_2}\gg\abs{t_1 -t'_1}$. In principle, we also need to evaluate correlators outside this limit, since $t_j$ and $t'_j$ are integrated over. But as a first approximation, we obtain the correlator by extrapolating the expressions found in this limit.
 
The path integral is then dominated by those trajectories where one anyon from the red pair propagates ballistically with momentum $\vb{p}$ and passes through the closed worldline of the blue anyon pair~\cite{mcginley2024signatures,mcginley2024anomalous}. At time $-t'_2$, the red anyons turn around to recombine at time $-t'_1$ (here, $\vb{p}$ is integrated over). One such trajectory is shown in Fig.~\ref{fig:braidcontour}.  It was shown in Ref.~\cite{mcginley2024signatures,mcginley2024anomalous} that for a fixed momentum $\vb{p}$ of a red anyon, it can braid through the blue pair only if its starting location $\vb{x}_1$, i.e., the spatial position of creation of the red anyon pair,  belongs to an area that scales as $\sim \abs{t'_2-t_2}^{3/2}F_1(p)$. Computing the function $F_1(p)$ is beyond the scope of this work, and its specific form is not important for our results. Combining all the factors, the overall amplitude for braiding scales as $\sim (1-\cos \alpha_{12})\abs{t_2-t'_2}^{1/2}F_1(p)e^{i\Delta_2(t'_2-t_2)}$.
 
 One caveat in our case is that since $t'_1 \neq t_1$, we need to consider wavepacket spreading even in the red anyon pair to ensure recombination after backward propagation. We conjecture that this recombination amplitude only depends on $(t_1-t'_1)$, and is independent of $t_2$ or $t'_2$. We present a semiclassical argument for this. The center of mass position of the red anyon pair should remain unchanged at the saddle point level, because the blue pair does not impart center of mass momentum to the red pair. If we consider the two red anyons to be moving with equal and opposite momenta $\vb{p}$ and $-\vb{p}$, each momentum $\vb{p}$ contributes an amplitude $e^{i(\tfrac{p^2}{m_1}+\Delta_1)(t'_1-t_1)}$ towards recombination, where $m_1$ is the effective mass of a red anyon. This factor is independent of $t_2$ and $t'_2$. Combining all the above factors, we get 
 \begin{equation}
 \begin{aligned}
     \sim & (1-\cos \alpha_{12})\abs{t_2-t'_2}^{1/2} e^{i\left(\tfrac{p^2}{m_1}+\Delta_1\right)(t'_1-t_1)} F_1(p)\\
     &\times e^{i\Delta_2(t'_2-t_2)}
     \end{aligned}
 \end{equation}
 
  While the geometric argument captures the scaling $\abs{t_2-t'_2}^{1/2}$, it does not capture a phase jump $e^{i\tfrac{\pi}{4}\text{sign}(t_2-t'_2)}$ when $t_2$ and $t'_2$ are exchanged. The phase jump is important to capture the Heaviside step functions $\theta(\Omega_1-\Delta_1)\theta(\Omega_2-\Delta_2)$ in our final result. Fixing the phase requires consideration of the $t_2-t'_2\sim0$ limit, which is beyond the scope of the scaling approach. In order to make progress, we employ an analyticity argument. For this, it is convenient to consider the correlator $\expval{\hat{A}^{\dagger}_1(-t'_1)\hat{A}^{\dagger}_2(-t'_2)\hat{A}_2(-t_2)\hat{A}_1(-t_1)}$ as an analytic continuation of a Euclidean-time correlation function. Therefore, we consider the  analytic continuation defined as:
\begin{equation}
   t_j \to t_j+i\epsilon_j\equiv iu_j \text{ and }t'_j \to t'_j+i\epsilon'_j\equiv i u'_j, 
\end{equation} such that $0<\epsilon_1<\epsilon_2<\epsilon'_2<\epsilon'_1\to 0^+$. This order of limits is required because a zero temperature correlation function $\expval{\hat{P}_1(-iu_n)\hat{P}_2(-iu_{n-1})\ldots \hat{P}_n(-iu_1)}$ is well-defined only if the operators are time-ordered with respect to the arguments $\Re u_n, \Re u_{n-1}, \ldots, \Re u_1$, i.e., if $\Re u_n >\Re u_{n-1} > \ldots > \Re u_{1}$. Here, we use the convention $\hat{P_n}(-iu_n)\equiv e^{u_n\hat{H}_0}\hat{P}_n e^{-u_n\hat{H}_0}$. Now, we know that the correlation function we are computing here is an analytic function of $u_2$ and $u'_2$ when $\Re u'_2>\Re u_2$, i.e., $\epsilon'_2>\epsilon_2$. The factor of $\sim \abs{t_2-t'_2}^{1/2}$ obtained above should thus be replaced by $(u'_2-u_2)^{1/2}$, such that the branch-cut is in the unphysical region where $\Re (u'_2- u_2)<0$. This method is agnostic to the ordering between real times $t_2$ and $t'_2$.  
 \begin{equation}
   \begin{aligned}  &\expval{\hat{A}^{\dagger}_1(-iu'_1)\hat{A}^{\dagger}_2(-iu'_2)\hat{A}_2(-iu_2)\hat{A}_1(-iu_1)}\\
   & \sim   N \alatt^2 (1-\cos \alpha_{12})  e^{-\Delta_1(u'_1-u_1)-\Delta_2(u'_2-u_2)}  \\
   & \quad  \times (u'_2-u_2)^{1/2}\int \frac{\dd[2]p}{(2\pi)^2} e^{-\frac{p^2}{m_1}(u'_1-u_1)}F_1(p).
   \end{aligned}
 \end{equation}
    Analytic continuing back to real-time, while maintaining $\Re (u'_1-u_1)>0$ and $\Re (u'_2-u_2)>0$, we get
 \begin{equation}
   \begin{aligned}  &\expval{\hat{A}^{\dagger}_1(-t'_1)\hat{A}^{\dagger}_2(-t'_2)\hat{A}_2(-t_2)\hat{A}_1(-t_1)}\\
   &\sim N \alatt^2  \abs{t_2-t'_2}^{1/2}e^{i\frac{\pi}{4}\text{sign}(t_2-t'_2)}e^{i\Delta_2(t'_2-t_2+i0^+)}\\
   &\quad \times  (1-\cos \alpha_{12}) \int \frac{\dd[2]p}{(2\pi)^2} e^{i\left(\frac{p^2}{m_1}+\Delta_1\right)(t'_1-t_1+i0^+)}F_1(p),
   \end{aligned}
 \end{equation}
If we insert this expression into Eq.~\eqref{eq:spinG2}, and perform the Fourier transform, we obtain that the contribution of the contour in Fig.~\ref{fig:allcontours}(a) to the connected $\mathcal{G}^{(2)}$ is  $N\theta(\Omega_1-\Delta_1)\theta(\Omega_2-\Delta_2)K_1(\Omega_1)(\Omega_2-\Delta_2)^{-3/2}$. Here, $K_1(\Omega_1)$ is a function obtained by integrating over momenta $\vb{p}$ of the red anyon and performing the Fourier-transform with respect to $t'_1-t_1$. One can now observe that the result for the contour in Fig.~\ref{fig:allcontours}(b) can be obtained by swapping the roles of the red and blue anyons in the above calculation. The resulting contribution is $N\theta(\Omega_1-\Delta_1)\theta(\Omega_2-\Delta_2)K_2(\Omega_2)(\Omega_1-\Delta_1)^{-3/2}$. The physical origin of this singularity in frequency can be traced back to the result in Ref.~\cite{mcginley2024anomalous,mcginley2024signatures} that the amplitude for recombination of one pair of anyons, conditioned on braiding through the other pair of anyons grows as $\abs{t_2-t'_2}^{1/2}$ in time. This growth is in turn due to the fact that the amplitude for one anyon to braid around the other grows with the area enclosed by one anyon loop. We note that the mutual statistics phase $\alpha_{12}$ itself gets absorbed into non-universal factors here, and only the \textit{existence} of fractional statistics can be tested for in this approach.

Now, we consider the contour in Fig.~\ref{fig:allcontours}(c). Here, the blue anyon pair is created before the red pair and is also annihilated before the red pair. In this case, neither anyon can be treated ballistically. If one tries to na\"ively apply the semiclassical argument from Ref.~\cite{mcginley2024anomalous,mcginley2024signatures}, the location of creation of the red anyon pair is drawn from an area scaling as $(u'_2-u_1)(u'_2-u_2)^{1/2}$. This area is smaller than the factor $(u'_2-u_2)^{3/2}$ we found for the contour in Fig.~\ref{fig:allcontours}(a). Thus, the resulting contribution to the connected $\mathcal{G}^{(2)}$ from the ordering 
of Fig.~\ref{fig:allcontours}(c) should also be less singular than that from Fig.~\ref{fig:allcontours}(a). The same applies to the contribution shown in Fig.~\ref{fig:allcontours}(d). Therefore, our final estimate for the singular part of $\mathcal{G}^{(2)}$ is 
\begin{equation}\label{eq:fracstatfinal}
\begin{aligned}
    &\mathcal{G}^{(2)}_{d_1,d_2}(\Omega_1,\Omega_2)\sim N\theta(\Omega_1-\Delta_1)\theta(\Omega_2-\Delta_2)\\
    &\times\left[ K_2(\Omega_2)(\Omega_1-\Delta_1)^{-3/2}+(1\leftrightarrow 2)\right].
    \end{aligned}
\end{equation}
In conclusion, in this section, we demonstrated that the  measurement of correlation functions of the photons scattered off the material enables the detection of the existence of fractional statistics of quasiparticles.  We note that the derivation of Eq.~\eqref{eq:fracstatfinal} involved an assumption that the extent of wavepacket spreading of the first anyon pair is independent of the time at which the second pair was created, as long as the two creation events are sufficiently spaced apart temporally. This assumption should be examined more carefully in future work. 

We also note that in Eq.~\eqref{eq:fracstatfinal}, $\mathcal{G}^{(2)}_{d_1,d_2}(0)$ is proportional to $N$, i.e., the number of sites illuminated by the input laser. Similarly, our results for $\mathcal{G}^{(2)}_{d_1,d_2}$ in Sec.~\ref{sec:magnons}, i.e., Eqs.~(\ref{eq:bosonicenhresult}, \ref{eq:gamma2first}, \ref{eq:gamma1first}) are all proportional to $N$. In contrast, $G^{(1)}_{d_1}(0)G^{(1)}_{d_2}(0)$ is proportional to $N^2$. Hence, the normalized $\mathcal{G}^{(2)}_{d_1,d_2}(0)/G^{(1)}_{d_1}(0)G^{(1)}_{d_2}(0)$ would be proportional to $1/N$. This calls into question whether our prediction in Eq.~\eqref{eq:fracstatfinal} will result in a detectable signal. One way the signal can be detected is if the $(\Omega_j-\Delta_j)^{-3/2}$ singularity makes up for the $1/N$ suppression. Finding whether this is indeed the case requires an explicit microscopic calculation of $\mathcal{G}^{(2)}_{d_1,d_2}$. This is a subject for future work. Another way is if $N$ (proportional to the area of the input light beam) is small enough to overcome the $1/N$ suppression, but large enough for many-body effects to be observed. It is interesting to ask if such a suitable range of $N$ can be achieved using recent advances in near-field spectroscopy~\cite{liu2016nanoscale}.
\section{Potential experimental realization}\label{sec:exprealization}
Our approach is immediately applicable to any correlated material where the system and its optical response can be captured within a Fermi-Hubbard or a generalized Heisenberg model such that the charge gap is in the IR/optical range. Sr$_2$IrO$_4$ is an example of such a transition metal oxide. As shown in Ref.~\cite{kim2008novel}, after taking into account spin-orbit coupling and crystal field splitting, among the ten available $5d$ orbital states, there is a doubly degenerate subspace of effective total angular momentum $J_{\text{eff}}=1/2$. This provides an $SU(2)$-symmetric Hamiltonian in terms of pseudospins. As long as the drive frequency (chosen to be of the order of $U$) is far from gaps to all other orbitals, pseudospin-orbit coupling (and not spin-orbit coupling) can be ignored, and the system is approximately described by Eq.~\eqref{eq:1bhm}. In Sr$_2$IrO$_4$, $U$ is of the order of $100$ meV~\cite{kim2008novel}.

 Furthermore, the expressions Eq.~\eqref{eq:AiExp}, (\ref{eq:BiExp}), (\ref{eq:CiExp}) hold for arbitrarily long-range fermion tunneling. By building on existing literature, our formalism can straightforwardly be extended to systems with spin-orbit coupling~\cite{yang2021non}, and with multiple orbitals per site~\cite{li2020electromagnetic,dmytruk2021gauge}. To do so, there should be no few-photon resonances between the ground state sector and the higher orbitals.

Moreover, our formalism is applicable to any system in which there are two sectors of states separated by an optical gap.  For concreteness, we considered them to be lower Hubbard band (spin sector) and upper Hubbard band (charge sector) (see Fig.~\ref{fig:sectorsbasic}). But our calculation can be adapted to other scenarios. For example, we can modify Fig.~\ref{fig:sectorsbasic} so that the lower and higher energy sectors are valence and conduction bands respectively in a correlated TMD heterobilayer, such as WS$_2$/ WSe$_2$. In this case, a single sector consists of both the upper Hubbard band and lower Hubbard band within the valence/conduction band  \cite{huang2025optical}. In such systems, the emission of photons is strongly susceptible to the nature of the underlying electronic states \cite{upadhyay2024giant}.
For example, it has been observed via photoluminescence (PL) measurements that the diffusion of excitons is affected strongly depending on whether the underlying electronic states are metallic, a Generalized Wigner Crystal or a Mott insulator. In such systems, $U$ is typically a few tens of meV \cite{miao2021strong,lian2024valley,xiong2023correlated,gao2024excitonic,upadhyay2024giant}. Therefore, detection of many-body correlations from PL at temperatures much smaller than $U$ is experimentally accessible. Consequently, it is a promising direction to investigate correlations between emitted photons from such systems. For example, our results from Sec.~\ref{sec:spinchirality} can be relevant to the proposal to realize a chiral spin liquid in TMD heterostructures~\cite{kuhlenkamp2024chiral}.

Furthermore, all photonic correlators in Table~\ref{tab:dictionary1} are either routinely measured in the quantum optics setting or can be measured with existing technology. For example, the phase-sensitive second-order quadrature measurement shown in Fig.~\ref{fig:homodyneG2} is related to two-mode squeezing that has been measured experimentally \cite{boyer2008entangled,de2024characterizing}. $G^{(2)}$ measurements have been used experimentally to study non-classical features of phonons in condensed matter systems \cite{glerean2025ultrafast}.

In optical experiments probing 2D materials, the materials are typically placed on substrates. The laser frequency is far detuned from any resonances within device substrates. Therefore, optical coupling to substrates can be ignored as is done routinely in such experiments  \cite{gao2024excitonic,park2023observation}. 

Now, we address the effects of nonzero temperatures in our work. In Eq.~\eqref{eq:instatedef}, we started with the matter in an arbitrary energy eigenstate $\ket{I}$. As a corollary, our mapping in Table~\ref{tab:dictionary1} holds for the matter starting in any mixed state that is diagonal in the energy eigenbasis, and in particular, for a thermal state. However, we require the temperature to be much smaller than the optical gap $U$, so that we can assume that the matter starts in the spin sector (lower Hubbard band). 
Further, the mapping from phase-sensitive second-order quadrature correlation to scalar spin chirality in Sec.~\ref{sec:spinchirality} is purely at the operator level and is agnostic to temperature and the nature of the ground state in the spin sector.
Regarding Sec.~\ref{sec:fracstat} on the detection of the existence of fractional statistics, at non-zero temperatures, there is a finite thermal population of anyons. The resulting effect on the nonlinear matter correlator was considered in Ref.~\cite{mcginley2024anomalous}. The same analysis applies to our results in Sec.~\ref{sec:fracstat}, i.e., the $\tau^{3/2}$ growth in Eq.~\eqref{eq:normalizedcond}  is cut off by a thermal timescale, and similarly, the $(\Omega_i-\Delta_i)^{-3/2}$ singularity in Sec.~\ref{sec:subfracstat} gets broadened.   
\section{Conclusions and Outlook}\label{sec:conclusions}
In this work, we have proposed a novel probe of many-body states of matter based on the measurement of the quantum statistical properties of scattered light. Furthermore, we demonstrated that it can be used to diagnose exotic properties of strongly correlated electronic systems such as fractional statistics of excitations and scalar spin chirality. 

One important aspect that will be subject to future research concerns the inherent complexity of the electronic band structure. In particular, while this work focused on the single-band Fermi-Hubbard model, many realistic materials, such as cuprates (charge transfer insulators) or insulators based on transition metal dichalcogenides, possess multiple bands. In this case, the specific form of the light-matter interaction, projected to the relevant bands, must account for the orbital structure of the corresponding Bloch wavefunctions.

The formalism developed in this work relies only on the existence of the optical gap and is agnostic to the details of the Hamiltonian describing the low-energy physics. We note that, for a known Hamiltonian, the ground state of a correlated system is not a priori obvious (what symmetries it spontaneously breaks, whether it is a spin liquid, etc.), even if the microscopic Hamiltonian is known.  
It would thus be interesting to investigate the different ground state ansatzes and the corresponding excitations, with the aim of identifying the distinctive signatures in the photon scattering data corresponding to the different many-body states of matter. The extension of these ideas to cold atom systems is also intriguing~\cite{yao2024measuring}.

In this work, we ignored the momentum transfer between electrons and photons. However, in the absence of strong Coulomb binding in photo-excited electron-hole pairs, the dipole approximation can be violated for itinerant electrons, and photon momentum and angular momentum can be imparted to electrons \cite{grass2022two,session2023optical}. Moreover, given the recent development of near-field spectroscopy \cite{liu2016nanoscale}, it is intriguing to explore spatially and spectrally resolved correlations. This question may be especially relevant today in the context of moir\'e materials which come with a much enlarged lattice spacing.

The protocol developed in this work is primarily based on classical driving, and correlations between photons are induced by the nonlinearity of the matter. As a future research direction, it would be of interest to identify scenarios where an additional advantage could be obtained by using input light prepared in squeezed coherent \cite{grankin2021enhancement} or non-Gaussian states.
\begin{acknowledgments}
We thank Ken Burch, Christian Eckhardt, Michele Fava, Pouyan Ghaemi, Lukas Grunwald, Tsung-Sheng Huang, Peter Lunts, Max McGinley, Siddharth Parameswaran, Angel Rubio and Yu-Xin Wang for useful discussions. We acknowledge related preprints that were put forward around the same time as our work. Refs.~\cite{grunwald2024cavity,kass2024many} investigate nonclassical properties of cavity photon modes coupled to a material. 
\end{acknowledgments}

\appendix
\section{Review of $\hat{\mathcal{T}}$-matrix formalism}\label{app:Tmatrixreview}
In this appendix, we review scattering theory using the $\hat{\mathcal{T}}$-matrix formalism. This review is loosely based on Chapter 3 of Ref.~\cite{goldberger2004collision}. The key takeaway from this appendix is Eq.~\eqref{eq:Tmatresrepeat} which serves as a generalization of Fermi's Golden Rule that works to all orders in perturbation theory.
\begin{figure*}[t]
  \centering
  \includegraphics[width=0.99\textwidth]{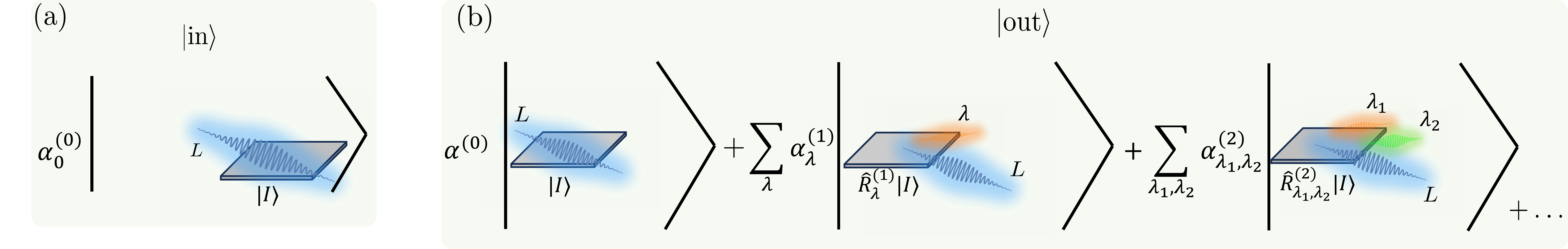}
  \caption{\textbf{Fig.~\ref{fig:schem2} in Interaction Picture}:  Schematic depiction of (a) $\ket{\text{in}}\equiv e^{-i\hat{H}_0T/2}\ket{\Psi(t=-T/2)}$ and (b) $\ket{\text{out}}\equiv e^{i\hat{H}_0T/2}\ket{\Psi(t=T/2)}$. This is a mathematical trick used to bring all wavepackets to where they should have been at $t=0$, according to the noninteracting Hamiltonian $\hat{H}_0$. The states (a) and (b) are respectively obtained by evolving the initial and final states shown in Fig.~\ref{fig:schem2}(a) and \ref{fig:schem2}(b) forward and backward respectively in time till $t=0$. States $\ket{\text{in}}$ and $\ket{\text{out}}$ are identical to Fig.~\ref{fig:schem2}(a) and (b) respectively, except that the light wavepackets have been shifted so as to be in the vicinity of the material. Further, upon doing so, the individual terms may have picked up additional phases $\alpha^{(1)}_\lambda$ etc. (compared to the corresponding terms in Fig.~\ref{fig:schem2})  due to time-evolution.}
  \label{fig:inoutschem}
\end{figure*}

Consider a Hamiltonian
\begin{equation}
    \hat{H}=\hat{H}_0 + \hat{V}.
\end{equation}
For concreteness, one can imagine $\hat{H}_0$ to be the full Hamiltonian of light and matter separately, and $\hat{V}$ is the light-matter interaction. But this formalism is applicable to any quantum scattering problem. At time $-T/2$, we start with a state $\ket{\Psi(-T/2)}$ in the full (light $+$ matter) Hilbert Space. Around $t=0$, light and matter are interacting, and the scattered light is observed at $t=T/2$. The final state is $\ket{\Psi(T/2)}\equiv e^{-i\hat{H}T}\ket{\Psi(-T/2)}$. We are interested in the limit when $T\to \infty$. 
The final state will have several terms with oscillating prefactors of the type $e^{i(E^0_m - E^0_n)T}$ where $E^0_m$ and $E^0_n$ are the energy eigenstates of $\hat{H}_0$. But we are interested in the limit $T\gg 1/(\delta E_{\text{in}})$, ($\delta E_{\text{in}}$ being the uncertainty in energy of the initial state) where terms like $e^{i(E^0_m - E^0_n)T}$ are fast-oscillating and average out to $0$. Therefore, it is useful to have a formalism that directly computes the time-evolved state with such fast-oscillating terms filtered away. This is what the $\T$-matrix formalism does.

We define states $\ket{\text{in}}$ and $\ket{\text{out}}$ as follows:
\begin{align}
    \ket{\Psi(-T/2)}&\equiv e^{-i\hat{H}_0 (-T/2)}\ket{\text{in}}\\
    \bra{\Psi(T/2)}&\equiv \bra{\text{out}} e^{i\hat{H}_0 T/2}.
\end{align}
The purpose of this trick is to allow one to define Heisenberg operators in terms of the eigenstates of $\hat{H}_0$ with respect to time $t=0$, which is a time when the photon wave-packet and the material are already interacting. So $\ket{\text{in}}$ is defined as the initial state evolved forward in time till $t=0$ by the noninteracting Hamiltonian $\hat{H}_0$. In this state, the laser wavepacket spatially overlaps with the material [Fig.~\ref{fig:inoutschem}(a)]. Similarly, $\ket{\text{out}}$ is defined by evolving the final state backward in time till $t=0$ by the noninteracting Hamiltonian $\hat{H}_0$ [Fig.~\ref{fig:inoutschem}(b)]. We suppose the state $\ket{\text{in}}$ is a wave-packet with a narrow spread of energy with respect to $\hat{H}_0$ such that the energy is centred around $E^0_{\textrm{in}}$. For the wavepacket to be far away from the material at $t=-T/2$, it necessarily has a non-zero width in momentum (and therefore energy). Generically, a narrow wavepacket gets wider with time (in real space). Since light is relativistic (i.e., the magnitude of velocity of all the component waves of the wavepacket are equal, so the uncertainty in velocity comes solely in its direction), the velocity of this spreading is maximum in the direction perpendicular to the velocity of the centre of the wavepacket. This spreading velocity has magnitude $\sim c \frac{\sigma_p}{p}$ where $p$ is the mean momentum of the wavepacket and $\sigma_p$ is its uncertainty in momentum. For small enough $\frac{\sigma_p}{p}$, the spreading of the wavepacket fails to catch up with the center itself. Therefore, in the rest of this section, we will ignore wavepacket spreading. (Similar reasoning can also be used to neglect wavepacket spreading in the case of a non-relativistic scatterer.)

Consider the state at time $t=0$:
\begin{equation}
    \ket{\Gamma_-}\equiv e^{-i\hat{H} T/2}e^{i\hat{H}_0 T/2}\ket{\text{in}}.
\end{equation}
Now, let us view $\ket{\Gamma_-}$ as a function of the initial time $-T/2$. We rewrite the above expression by first differentiating with respect to $t'$ (supposing $t'=-T/2$ in the above equation) and then integrating over $t'$ from $-T/2$ to $0$. 
\begin{align}
    \partial_{t'} \ket{\Gamma_-}&=i e^{i \hat{H}t'}\hat{V} e^{-i \hat{H}_0 t'} \ket{\text{in}}\\
    \implies \ket{\Gamma_-} &= \ket{\text{in}} - i \int_{-T/2}^0 \dd t' e^{i \hat{H}t'}\hat{V} e^{-i \hat{H}_0 t'} \ket{\text{in}} \label{eq:ketgamma}
\end{align}
The advantage of rewriting $\ket{\Gamma_-}$ as above is that it makes evident the fact that $\ket{\Gamma_-}$ does not depend on $-T/2$ (as long as $-T/2$ is sufficiently negative). The reason is that $\hat{V}$ acting on $e^{-i \hat{H}_0 t'} \ket{\text{in}}$ returns 0 unless the wavepacket of the photon has some spatial overlap with the material. This observation is the key physics input in the $\hat{\mathcal{T}}$-matrix formalism. In the setting we are imagining, the laser beam comes close to the material only around $t=-t_C<0$ for some time scale $t_C \ll T/2$. Whatever happens before $-t_C$ does not contribute to the above equation. We will soon use this useful fact. 
Let us suppose 
\begin{equation}
    \ket{\text{in}}=\sum_{j}\phi_j^{\text{in}} \ket{\Psi_j^{0}} \text{  and  }\ket{\text{out}}=\sum_{j}\phi_j^{\text{out}}\ket{\Psi^{0}_j}
\end{equation}
where $\ket{\Psi^{0}_j}$ is an eigenstate of $\hat{H}_0$ with energy $E^0_{j}$. We assume that the $\phi_j$'s are narrowly peaked around energy $E_{\text{in},0}$. Also inserting into Eq.~\eqref{eq:ketgamma} a resolution of identity in terms of $\ket{\Psi_J}$ which are eigenstates of $\hat{H}$ with eigenvalue $\mathcal{E}_J$, i.e., $\sum_{J}\ket{\Psi_J}\bra{\Psi_J}$, we get:
\begin{equation}
\begin{aligned}
     &\ket{\Gamma_-}=\ket{\text{in}} - i \sum_j \phi_j^{\text{in}} \sum_J \ket{\Psi_J}\mel{\Psi_J}{\hat{V}}{\Psi^{0}_j}\\ &\quad \quad \quad \quad \quad \quad \times \int_{-T/2}^0 \dd t' e^{i(\mathcal{E}_J - E^0_{j})t'} \\
    =&\ket{\text{in}} + \sum_j \sum_J \ket{\Psi_J} \mel{\Psi_J}{\hat{V}}{\Psi^{0}_j} \phi_j^{\text{in}} \\ & \times \frac{2 \sin^2 \left[\frac{(\mathcal{E}_J - E^0_{j})T/2}{2}\right] + i \sin \left[(\mathcal{E}_J - E^0_{j})T/2\right]}{E^0_{j}-\mathcal{E}_J}\label{eq:oscsum}
    \end{aligned}
\end{equation}
Let us now simplify the following expression from Eq.~\eqref{eq:oscsum}:
\begin{equation}\label{eq:oscpiece}
    \frac{2 \sin^2 ((\mathcal{E}_J - E^0_{j})T/4) + i \sin ((\mathcal{E}_J - E^0_{j})T/2)}{E^0_{j}-\mathcal{E}_J}
\end{equation}
Let us suppose that $T/2 \gg \frac{1}{\delta E^0_{\text{in}}}$, (where $\delta E^0_{\text{in}}$ is the spread in $E^0_{\text{in}}$), i.e., $T/2$ is so large that as $E^0_{j}$ runs through the different eigen-components of the $\ket{\text{in}}$ state, the real part of the numerator of Eq.~\eqref{eq:oscpiece} goes through several cycles of the $\sin^2 ()$ function. Let us consider two limits for $\abs{E^0_{j}-\mathcal{E}_J}$. The first limit is when $\abs{E^0_{j}-\mathcal{E}_J}\gg \delta E^0_{\text{in}}$ (consequently $\abs{E^0_{j}-\mathcal{E}_J}\gg 2/T$). Then the fluctuations in the numerator of Eq.~\eqref{eq:oscpiece} are much stronger than the fluctuations of the denominator. Hence, we can replace the real part of the numerator by its average which is $2\times 1/2$. Thus, the real part of the expression is $1/(E^0_{j}-\mathcal{E}_J)$. But in the opposite limit, when $\abs{E^0_{j}-\mathcal{E}_J}\ll 2/T$, the real part of Eq.~\eqref{eq:oscpiece} tends to $0$. Hence the real part can be approximated by $\mathcal{P}\frac{1}{E^0_{j}-\mathcal{E}_J}$ for large $T/2$. Here, $\mathcal{P}$ stands for principal value and $\mathcal{P}\frac{1}{z}$ is defined as $\lim_{\eta \to 0} \frac{1}{2}\left(\frac{1}{z+i\eta}+\frac{1}{z-i\eta}\right)$. Next, we consider the imaginary part of Eq.~\eqref{eq:oscpiece}, i.e., $-\frac{\sin((\mathcal{E}_J - E^0_{j})T/2)}{\mathcal{E}_J - E^0_{j}}$. As $T/2$ increases, this function becomes sharply peaked around $E^0_{j}- \mathcal{E}_J =0$. As $T/2 \to \infty$, it becomes $-\pi \delta(E^0_{j}-\mathcal{E}_J)$. Thus, in Eq.~\eqref{eq:oscsum}, we can make the following replacement
\begin{equation}
\begin{aligned}
    &\frac{2 \sin^2 \left[\frac{(\mathcal{E}_J - E^0_{j})T/2}{2}\right] + i \sin \left[(\mathcal{E}_J - E^0_{j})T/2\right]}{E^0_{j}-\mathcal{E}_J}\\ \rightarrow & \mathcal{P}\frac{1}{E^0_{j} - \mathcal{E}_{J}} - i\pi \delta (E^0_{j}-\mathcal{E}_J)\\
    =&\lim_{\eta\to 0^+} \frac{1}{E^0_{j}-\mathcal{E}_J +i \eta}.
    \end{aligned}
\end{equation}
This agrees with the intuition provided above that $\ket{\Gamma_-}$ should not depend on $T/2$ as long as it is sufficiently large. Thus, we get:
\begin{equation}
    \ket{\Gamma_-}=\sum_j \phi_j^{\text{in}} \left(\ket{\Psi^{0}_j} +  \frac{1}{E^0_{j} - \hat{H} + i0^+}\hat{V}\ket{\Psi^{0}_j}\right).
\end{equation}
Now, we need to evolve $\ket{\Gamma_-}$ forward in time till $t=T/2$ using $\hat{H}$. This is simple because $\ket{\Psi^{0}_j} +  \frac{1}{E^0_{j} - \hat{H} + i0^+}\hat{V}\ket{\Psi^{0}_j}$ is actually an eigenstate of $\hat{H}$ with eigenvalue $E^0_{j}$. To see this, if we replace $\hat{V}$ with $\hat{H}-\hat{H}_0$, then we get
\begin{align}
    &\ket{\Psi^{0}_j} +  \frac{1}{E^0_{j} - \hat{H} + i0^+}\hat{V}\ket{\Psi^{0}_j}\nonumber \\&=\lim_{\eta \to 0^+}\frac{i\eta}{E^0_{j}-\hat{H}+i\eta} \ket{\Psi^{0}_j} \nonumber \\
    &=\lim_{\eta \to 0^+}\sum_{J}\frac{i\eta}{E^0_{j}-\mathcal{E}_J+i\eta}\ket{\Psi_J}\bra{\Psi_J}\ket{\Psi^{0}_j}.
\end{align}
We see that as we take the limit $\eta \to 0^+$, the only $J$'s that survive are those with $\mathcal{E}_J=E^0_{j}$. Thus we get
\begin{equation}
\begin{aligned}
    &\ket{\Psi^{0}_j} +  \frac{1}{E^0_{j} - \hat{H} + i0^+}\hat{V}\ket{\Psi^{0}_j}\\&=\sum_J \delta_{\mathcal{E}_J=E^0_{j}}\ket{\Psi_J}\bra{\Psi_J}\ket{\Psi^{0}_j}
\end{aligned}
\end{equation}
Therefore,
\begin{equation}
\begin{aligned}
    &e^{-i\hat{H}T/2}\ket{\Gamma_-}\\&=\sum_j \phi_j^{\text{in}}\left\{e^{-iE^0_{j}T/2}\left(\ket{\Psi_j} +  \frac{1}{E^0_{j} - \hat{H} + i0^+}\hat{V}\ket{\Psi_j}\right)\right\}\label{eq:evolved}
    \end{aligned}
\end{equation}
Before proceeding, we make one more formal manipulation (in the style of Dyson equations):
\begin{equation}
\begin{aligned}
    &(E^0_{j}-\hat{H}+i0^+)^{-1}\\
    &=\left(E^0_{j}-\hat{H}_0+i0^+\right)^{-1} \left\{\Id + \hat{V}(E^0_{j}-\hat{H}+i0^+)^{-1}\right\}
\end{aligned}
\end{equation}
Therefore, 
\begin{equation}
\begin{aligned}
    &\frac{1}{E^0_{j} - \hat{H} + i0^+}\hat{V}\\
    &=\frac{1}{E^0_{j}-\hat{H}_0+i0^+} \left\{\hat{V} + \hat{V}\frac{1}{E^0_{j}-\hat{H}+i0^+} \hat{V}\right\}.
    \end{aligned}
\end{equation}
We now define the $\hat{\mathcal{T}}$-matrix as 
\begin{equation}\label{eq:defTmatrix}
\begin{aligned}
    \hat{\mathcal{T}}\equiv & \hat{V} + \hat{V}\frac{1}{E^0_{\text{in}}-\hat{H}+i0^+} \hat{V}\\
    \equiv & \hat{V} + \hat{V}\frac{1}{E^0_{\text{in}}-\hat{H}_0 - \hat{V}+i0^+} \hat{V}
\end{aligned}
\end{equation}
where $E^0_{\text{in}}$ is the energy of the eigenstate of $\hat{H}_0$ appearing in the expansion of $\ket{\text{in}}$ on which $\hat{\mathcal{T}}$ is acting. For example, in the above equation, $\hat{\mathcal{T}}$ is acting on $\ket{\Psi^0_j}$ and hence we should use $E^0_{\text{in}}=E^0_{j}$.  
With this definition at hand, we rewrite Eq.~\eqref{eq:evolved} as
\begin{equation}
\begin{aligned}
    &e^{-i\hat{H}T/2}\ket{\Gamma_-}\\ &=\sum_j \phi_j^{\text{in}}e^{-iE^0_{j}T/2}\left(\ket{\Psi_j^0} +  \frac{1}{E^0_{j} - \hat{H}_0 + i0^+}\hat{\mathcal{T}}\ket{\Psi_j^0}\right).
    \end{aligned}
\end{equation}
Recall that we are interested in calculating $\ket{\text{out}}=e^{i\hat{H}_0 T/2}e^{-i\hat{H} T/2}\ket{\Gamma_-}$ . We have 
\begin{equation}\label{eq:appoutexp}
\begin{aligned}
&\ket{\text{out}}\\&=\sum_{j,k}\ket{\Psi^0_k}\phi_j^{\text{in}}\left\{\delta_{kj}+\frac{e^{-i(E^0_{j}-E^0_{k})T/2}}{(E^0_{j}-E^0_{k})+i0^+}\mel{\Psi^0_k}{\hat{\mathcal{T}}}{\Psi^0_j}\right\}
   \end{aligned}
\end{equation}
Now, by the same argument used before to show that $\ket{\Gamma_-}$ is independent of $-T/2$, the above state, i.e., $\ket{\text{out}}=e^{i\hat{H}_0 T/2}e^{-i\hat{H}T/2}\ket{\Gamma_-}$ should be independent of $T/2$ when $T/2\gg 1/(\delta E^0_{\text{out}})$. Therefore, let us extract the $T/2$-independent piece from the above Eq.~\eqref{eq:appoutexp}.
\begin{equation}
\begin{aligned}
    &\frac{e^{-i(E^0_{j}-E^0_{k})T/2}}{(E^0_{j}-E^0_{k})+i0^+}\\&=e^{-i(E^0_{j}-E^0_{k})T/2}\mathcal{P}\frac{1}{(E^0_{j}-E^0_{k})}-i\pi \delta(E^0_{j}-E^0_{k})\\
    &=\mathcal{P}\frac{\cos\left((E^0_{j}-E^0_{k})T/2\right)}{(E^0_{j}-E^0_{k})} -i\frac{\sin\left((E^0_{j}-E^0_{k})T/2\right)}{(E^0_{j}-E^0_{k})}\\& \quad-i\pi \delta(E^0_{j}-E^0_{k})
\end{aligned}
\end{equation}
In the limit $T/2 \to \infty$, the first term above averages to $0$, and hence does not contribute to the $T$-independent piece. The second term goes to $-i\pi \delta(E^0_{j}-E^0_{k})$. Therefore, in the large $T/2$ limit, 
\begin{equation}
    \frac{e^{-i(E^0_{j}-E^0_{k})T/2}}{(E^0_{j}-E^0_{k})+i0^+} \to -2\pi i \delta(E^0_{j}-E^0_{k}).
\end{equation}
Therefore,
\begin{equation}\label{eq:Tmatres}
\begin{aligned}
    &\ket{\text{out}}\\&=\ket{\text{in}}-\sum_{j,k} 2\pi i \delta(E^0_{j}-E^0_{k})\ket{\Psi^0_k}\bra{\Psi^0_k}\T \ket{\Psi^0_j}\bra{\Psi^0_j}\ket{\text{in}}.
    \end{aligned}
\end{equation}
The above equation is a generalization of Fermi's Golden Rule that works to all orders in $\hat{V}$.
\section{When the incoming laser is modeled as a coherent state instead of a Fock state}\label{app:coherentcase}
In Eq.~\eqref{eq:instatedef} of the main text, we supposed that the radiation part of the $\ket{\text{in}}$ state was in a Fock state (photon-number eigenstate) with $\mathcal{N}_L$ photons in mode $L$. In this appendix, we examine the case when the initial state of the radiation sector is a coherent state. Our purpose is two-fold -- to clarify the definition of the coupling constant $g_L$ and to write an expression for the $\ket{\text{out}}$ state. 

Let us define a coherent state in the radiation sector $\ket{\phi_L}$ as:
\begin{equation}
    \ket{\phi_L}\equiv e^{\phi_L \hat{a}_L^{\dagger}-\phi_L^*\hat{a}_L}\ket{0,\ldots,0}.
\end{equation}
Then the full $\ket{\text{in}}$ state is
\begin{align}
    \ket{\text{in}}&=\ket{I}_M\otimes e^{\phi_L \hat{a}_L^{\dagger}-\phi_L^*\hat{a}_L}\ket{0,\ldots,0}\\
    &=\ket{I}_M\otimes e^{-\frac{\abs{\phi_L}^2}{2}}\sum_{\mathcal{N}_L}\frac{(\phi_L)^{\mathcal{N}_L}}{\sqrt{\mathcal{N}_L!}}\ket{0,\ldots,\mathcal{N}_L,\ldots 0}.
\end{align}
In Eq.~\eqref{eq:tmatentry}, when introducing the $\T$ matrix machinery, we assumed an $\ket{\text{in}}$ state that was an eigenstate of $\hat{H}_0$. When $\ket{\text{in}}$ is not an energy eigenstate, as is the case above, one can decompose it into its energy eigenstates, and for each of them, linearly add up the corresponding $\ket{\text{out}}$ states. 

We argued in the main text that within our approximation, the only processes contributing to $G^{(1)}$ that we keep are those where exactly one photon is absorbed from the laser and one photon is emitted into a different mode. Similarly, the only processes that we keep for $G^{(2)}$ are those where exactly two photons are absorbed from the laser and two are emitted. Therefore, the matrix elements of $\frac{1}{E^0_{\text{in}}-\hat{H}_0}$ are independent of the number of photons $\mathcal{N}_L$ in the initial state. The only dependence of $\ket{\text{out}}$ on the initial state thus comes from the action of $\hat{a}_L$. Therefore, if we make a replacement from a Fock state to a coherent state, we just need to replace $\sqrt{\mathcal{N}_L}$ by $\phi_L$ in expressions for $G^{(1)}$ and $\sqrt{\mathcal{N}_L(\mathcal{N}_L-1)}$ by $\phi_L^2$ in expressions for $G^{(2)}$. For a Fock state, we used $I_L=\frac{\mathcal{N}_L \omega_L c}{\mathcal{V}}$. For a coherent state, we can instead use $I_L=\frac{\abs{\phi_L}^2 \omega_L c}{\mathcal{V}}$. 

Now, we come to the laser-matter coupling. For a Fock state input, in expressions for $G^{(2)}$, we make the identification  
\begin{equation}
    g_L^2 \leftrightarrow \frac{\sqrt{\mathcal{N}_L(\mathcal{N}_L-1)}q_e^2 \alatt^2}{2\varepsilon \mathcal{V}\omega_L}.
\end{equation}
On the other hand, for expressions for $G^{(1)}$, we make the identification
\begin{equation}
    g_L \leftrightarrow \frac{\sqrt{\mathcal{N}_L}q_e \alatt}{\sqrt{2\varepsilon \mathcal{V}\omega_L}}.
\end{equation}
This means that for Fock state input, our definition of $g_L$ is slightly different for $G^{(1)}$ when compared to $G^{(2)}$. But for a coherent state input, the effective laser-matter coupling constants agree.

Now, we are in a position to write an expression for the $\ket{\text{out}}$ state that works for both a Fock state and a coherent state input. We suppose $\ket{\text{in}}=\ket{I}_M \otimes \myket{\psi^{(0)}_L}_R$, as defined in Eq.~\eqref{eq:instatedef} in the main text. Recall from Eq.~\eqref{eq:Tmatresrepeat} and \eqref{eq:tmatentry}, that $\ket{\text{out}}$ can be computed using the $\T$-matrix. In Eq.~\eqref{eq:VGVexp} and Eq.~\eqref{eq:Tsimp}, we simplified the terms of the $\T$-matrix for processes corresponding to absorption and emission of one and two photons respectively. Combining these with the discussion in the above paragraph, we get Eq.~\eqref{eq:outforcoherent} in the main text.
\section{Explicit calculation for matter operator $\hat{R}^{(2)}$ }\label{sec:explicitmicro}
Our goal here is to calculate $\hat{R}^{(2)}_{\lambda_1,\lambda_2}$ defined in Eq.~\eqref{eq:Tsimp} by expanding $\hat{V}_P\hat{\mathbb{G}}_0\hat{V}_P\hat{\mathbb{G}}_0\hat{V}_P\hat{\mathbb{G}}_0\hat{V}_P+\hat{V}_P\hat{\mathbb{G}}_0\hat{V}_D\hat{\mathbb{G}}_0\hat{V}_P+\hat{V}_C\hat{\mathbb{G}}_0\hat{V}_P+\hat{V}_P\hat{\mathbb{G}}_0 \hat{V}_C$, as promised in Sec.~\ref{sec:operators}. Recall that $\hat{\mathbb{G}}_0=\left(E_{\text{in}}-\hat{H}_0\right)^{-1}$. Also, recall that within the dipole approximation, 
\begin{equation}\label{eq:defVP}
    \hat{V}_{P}\approx - \sum_{\bm{\mu}}\left[\hat{\mathcal{J}}_{\bm{\mu}}\bar{\bm{\mu}}\cdot \sum_{\lambda}\frac{g\sqrt{2c}}{\sqrt{\mathcal{V}\omega_{\lambda}}}\left(\vb{e}_{\lambda}\hat{a}_{\lambda}+\vb{e}^*_{\lambda}\hat{a}^{\dagger}_{\lambda}\right)\right],
\end{equation}
where $\hat{\mathcal{J}}_{\bm{\mu}}$ was defined in Eq.~\eqref{eq:defJmu},  $g=\sqrt{\pi\alpha}\alatt$ and we use a convention for summation over $\vb{r}$ and $\bm{\mu}$, so that each bond $(\vb{r},\bm{\mu})$ is counted exactly once (and not double-counted).

Similarly, the diamagnetic term is
\begin{equation}\label{eq:defVD}
    \begin{aligned}
        \hat{V}_{D}\approx& \frac{1}{2} \sum_{(\vb{r},\bm{\mu})}\Biggl\{\left(\tunn_{\vb{r},\vb{r}+\bm{\mu}}\hat{c}^{\dagger}_{\vb{r}+\bm{\mu}, \sigma}\hat{c}_{\vb{r}, \sigma}+\text{h.c.}\right)\\&\quad \times \left[\bar{\bm{\mu}} \cdot \sum_{\lambda}\frac{g\sqrt{2c}}{\sqrt{\mathcal{V}\omega_{\lambda}}}\left(\vb{e}_{\lambda}\hat{a}_{\lambda}+\vb{e}^*_{\lambda}\hat{a}^{\dagger}_{\lambda}\right)\right]^2\Biggr\},
    \end{aligned}
\end{equation}
and the cubic term is
\begin{equation}
    \hat{V}_C\approx \frac{1}{6}\sum_{\bm{\mu}}\hat{\mathcal{J}}_{\bm{\mu}}\left[\bar{\bm{\mu}}\cdot \sum_{\lambda}\frac{g\sqrt{2c}}{\sqrt{\mathcal{V}\omega_{\lambda}}}\left(\vb{e}_{\lambda}\hat{a}_{\lambda}+\vb{e}^*_{\lambda}\hat{a}^{\dagger}_{\lambda}\right)\right]^3.
\end{equation}
 In the $\T$-matrix, each insertion of $\hat{V}_P$ can lead to a photon emission ($\hat{a}^{\dagger}_{\lambda}$) or absorption ($\hat{a}_{\lambda}$). Let us write $\hat{V}_P\equiv \hat{V}_P^{+}+\hat{V}_P^{-}$, where $\hat{V}_P^+$ only consists of photon creation operators and $\hat{V}_P^-$ only consists of photon annihilation operators. Similarly, $\hat{V}_D\equiv \hat{V}_D^{+-}+\hat{V}_D^{++}+\hat{V}_D^{--}$, where $\hat{V}_D^{+-}$ is of the form $\hat{a}^{\dagger}_{\lambda}\hat{a}_{\lambda'}$ and so on. Then we see that Fig.~\ref{fig:sectors}(a) and \ref{fig:sectors}(b) correspond to $\hat{V}^+_P\hat{\mathbb{G}}_0\hat{V}^-_P\hat{\mathbb{G}}_0\hat{V}^+_P\hat{\mathbb{G}}_0\hat{V}^-_P$. Let us denote the contribution from this process to $R^{(2)}_{\lambda_1,\lambda_2}$ as $\hat{R}^{(2)}_{\lambda_1,\lambda_2}\Big|_{a+b}$. Similarly, Fig.~\ref{fig:sectors}(c) corresponds to $\hat{V}^+_P\hat{\mathbb{G}}_0\hat{V}^+_P\hat{\mathbb{G}}_0\hat{V}^-_P\hat{\mathbb{G}}_0\hat{V}^-_P$. Let us denote the contribution from this process as $\hat{R}^{(2)}_{\lambda_1,\lambda_2}\Big|_{c}$. Fig.~\ref{fig:sectors}(d) corresponds to $\hat{V}^+_P\hat{\mathbb{G}}_0\hat{V}^{+-}_D\hat{\mathbb{G}}_0\hat{V}^-_P$ and we denote its contribution to $R^{(2)}_{\lambda_1,\lambda_2}$ as $\hat{R}^{(2)}_{\lambda_1,\lambda_2}\Big|_{d}$. Finally the contribution from the process in Fig.~\ref{fig:cubicmicro} coming from $\hat{V}_P \hat{\mathbb{G}}_0 \hat{V}_C$ and $\hat{V}_C \hat{\mathbb{G}}_0 \hat{V}_P$ are denoted $\hat{R}^{(2)}_{\lambda_1,\lambda_2}\Big|_{e}$ and $\hat{R}^{(2)}_{\lambda_1,\lambda_2}\Big|_{e'}$ respectively. Therefore,
 \begin{equation}
 \begin{aligned}\hat{R}^{(2)}_{\lambda_1,\lambda_2}=&\hat{R}^{(2)}_{\lambda_1,\lambda_2}\Big|_{a+b}+\hat{R}^{(2)}_{\lambda_1,\lambda_2}\Big|_{c}+\hat{R}^{(2)}_{\lambda_1,\lambda_2}\Big|_{d}\\&+\hat{R}^{(2)}_{\lambda_1,\lambda_2}\Big|_{e}+\hat{R}^{(2)}_{\lambda_1,\lambda_2}\Big|_{e'}.
 \end{aligned}
 \end{equation}
\subsection{Processes in Fig.~\ref{fig:sectors}(a) and \ref{fig:sectors}(b)}\label{sec:4a4b}
 These processes are shown pictorially in Fig.~\ref{fig:description2}(a$_1$-a$_6$). Let us first expand out $\hat{V}^+_P\hat{\mathbb{G}}_0\hat{V}^-_P\hat{\mathbb{G}}_0\hat{V}^+_P\hat{\mathbb{G}}_0 \hat{V}^-_P$, keeping the terms that will contribute to $G^{(2)}$:
\begin{equation}\label{eq:Tbigexp1}
\begin{aligned}
    &\hat{V}^+_P\hat{\mathbb{G}}_0\hat{V}^-_P\hat{\mathbb{G}}_0\hat{V}^+_P\hat{\mathbb{G}}_0\hat{V}^-_P\\
    &=g^4\frac{(2c)^2}{\mathcal{V}^2\omega_L}\sum_{\bm{\mu}_1,\bm{\mu}_2,\bm{\mu}'_1,\bm{\mu}'_2}\left(\bar{\bm{\mu}}'_1\cdot \vb{e}_L\right)\left(\bar{\bm{\mu}}'_2\cdot \vb{e}_L\right)\opa^2_L\\
    &\times \sum_{\substack{\lambda_1,\lambda_2\\K,J,K',F}}\Biggl\{\frac{\opa_{\lambda_2}^{\dagger}\opa_{\lambda_1}^{\dagger}}{\sqrt{\omega_{\lambda_1}\omega_{\lambda_2}}}\frac{\left(\bar{\bm{\mu}}_1\cdot\vb{e}^*_{\lambda_1}\right)\left(\bar{\bm{\mu}}_2\cdot\vb{e}^*_{\lambda_2}\right)}{\omega_L - E_{K'I}+i0^+}\\
    &\times \frac{\ketbra{F}\hat{\mathcal{J}}_{\bm{\mu}_2}\ketbra{K}\hat{\mathcal{J}}_{\bm{\mu}'_2}\ketbra{J}\hat{\mathcal{J}}_{\bm{\mu}_1}\ketbra{K'}\hat{\mathcal{J}}_{\bm{\mu}'_1}}{\left(\omega_L - E_{JI}-\omega_{\lambda_1}+i0^+\right)\left(2\omega_L - E_{KI}-\omega_{\lambda_1}+i0^+\right)}\Biggr\}.
\end{aligned}
\end{equation}
See Eq.~\eqref{eq:defJmu} for the definition of $\hat{\mathcal{J}}_{\bm{\mu}}$. Here, we sum over $\ket{J}$, $\ket{K'}$, $\ket{K}$ and $\ket{F}$ that are many-body energy eigenstates of the Fermi-Hubbard model. For convenience, we have defined $E_{JI}\equiv E_J - E_I$. The overall energy-conservation constraint for each $\lambda_1$ and $\lambda_2$, imposed by the $\delta$-function in Eq.~\eqref{eq:G2simp2} is
\begin{equation}\label{eq:energyconserved}
    E_I + 2\omega_L = E_F + \omega_{\lambda_1}+\omega_{\lambda_2}.
\end{equation}
Now consider the three energy dependent factors $\left(2\omega_L - E_{KI}-\omega_{\lambda_1}+i0^+\right)^{-1}$, $\left(\omega_L - E_{JI}-\omega_{\lambda_1}+i0^+\right)^{-1}$ and $\left(\omega_L - E_{K'I}+i0^+\right)^{-1}$. Of these, the first two factors contain $\omega_{\lambda_1}$, a variable that we integrate over, so we cannot estimate them just yet. But we can estimate the third factor -- it is dominated by states $\ket{K'}$ in the single doublon-hole sector. For such states, the factor is of order $1/\left(\omega_L - U\right)$. Within this sector, relative variations in this factor are of order $\tunn/\abs{\omega_L-U}$ that we neglect. We drop contributions from outside this sector because they come with an additional suppression of order  $\abs{U-\omega_L}/U$ (see Fig.~\ref{fig:sectors}). Since Eq.~\eqref{eq:Tbigexp1} now becomes independent of $E_{K'}$, we can replace $\sum_{K'}\ketbra{K'}{K'}$ by the identity operator. Next, to ensure that $\ket{J}$ is in the spin sector, the bond $(\vb{r}'_1,\bm{\mu}'_1)$ along which the first hop occurs should be the same as the bond $(\vb{r}_1,\bm{\mu}_1)$ along which the second hop occurs. Using this fact, we can use the identities \eqref{eq:ferm2spin1} and \eqref{eq:ferm2spin2} given below to simplify the expression in Eq.~\eqref{eq:Tbigexp1}:

\begin{align}
    \hat{c}^\dagger_{\alpha} \hat{c}_{\beta}&= \delta_{\beta \alpha}\frac{\hat{n}}{2} + \left(\hat{\vb{S}}\cdot \bm{\sigma}\right)_{\beta \alpha}\label{eq:ferm2spin1}\\
    \hat{c}_{\alpha} \hat{c}^{\dagger}_{\beta}&= \delta_{\alpha \beta}\left(1-\frac{\hat{n}}{2}\right)-\left(\hat{\vb{S}}\cdot \bm{\sigma}\right)_{\alpha \beta}.\label{eq:ferm2spin2}
\end{align}

We then arrive at the following relation
\begin{equation}\label{eq:ferm2spin3}
\begin{aligned}
    &-\mel{J}{\hat{\mathcal{J}}_{\bm{\mu}}\hat{\mathcal{J}}_{\bm{\mu}'}}{I}\\&\approx\bra{J}\delta_{\bm{\mu},\bm{\mu}'}\sum_{\vb{r}}\abs{\tunn_{\vb{r},\vb{r}+\bm{\mu}}}^2\left(4\hat{\vb{S}}_{\vb{r}}\cdot \hat{\vb{S}}_{\vb{r}+\bm{\mu}}-1\right)\ket{I}\\
    &\text{ if both }\ket{I}\text{ and }\ket{J}\text{ are in the spin sector.}
    \end{aligned}
\end{equation}
Using this in Eq.~\eqref{eq:Tbigexp1}, then symmetrizing the resultant expression between indices $\lambda_1$ and $\lambda_2$, we can read off $\hat{R}^{(2)}_{\lambda_1,\lambda_2}$ [defined in Eq.~\eqref{eq:Tsimp}]. Due to our choice of definition of $\hat{R}^{(2)}_{\lambda_1,\lambda_2}$, all the factors of $\sqrt{2c}/\sqrt{\mathcal{V}\omega_{\lambda}}$ cancel out. We obtain:
\begin{equation}\label{eq:R2ab_bigexp1}
\begin{aligned}
    &\mel{F}{\hat{R}^{(2)}_{\lambda_1,\lambda_2}\Big|_{a+b}}{I}\\
    &=-\frac{g_L^2 g^2}{\omega_L - U}\sum_{(\vb{r}_1,\bm{\mu}_1)}\sum_{\bm{\mu}_2,\bm{\mu}'_2} \sum_{K,J}\Biggl\{\abs{\tunn_{\vb{r}_1,\vb{r}_1+\bm{\mu}_1}}^2\\
    &\times\left[\frac{\left(\bar{\bm{\mu}}'_2\cdot \vb{e}_L\right)\left(\bar{\bm{\mu}}_2\cdot\vb{e}^*_{\lambda_2}\right)}{\omega_{\lambda_1}-\left(2\omega_L - E_{KI}+i0^+\right)} \ \frac{1}{\omega_{\lambda_1}-\left(\omega_L - E_{JI}+i0^+\right)} \right.\\
    &\quad\quad \quad +\Bigl(\lambda_1 \leftrightarrow \lambda_2 \Bigr)\Biggr] \times \bra{F}\hat{\mathcal{J}}_{\bm{\mu}_2}\ketbra{K}
    \hat{\mathcal{J}}_{\bm{\mu}'_2}\ket{J}\\
    &\times \bra{J}\left(4\hat{\vb{S}}_{\vb{r}_1}\cdot \hat{\vb{S}}_{\vb{r}_1 + \bm{\mu}_1}-1\right)\left(\bar{\bm{\mu}}_1\cdot\vb{e}^*_{\lambda_1}\right)\left(\bar{\bm{\mu}}_1\cdot \vb{e}_L\right)\Biggr\}\ket{I}.
\end{aligned}  
\end{equation}
The first term here has two poles, one at $\omega_{\lambda_1}=2\omega_L -E_{KI}+i0^+$, which corresponds to Fig.~\ref{fig:sectors}(a), i.e., the central peak and another at $\omega_{\lambda_1}=\omega_L -E_{JI}+i0^+$, which corresponds to Fig.~\ref{fig:sectors}(b), i.e., the sidebands. The two poles here are reminiscent of the fluorescent triplet of a two-level system studied in Ref.~\cite{dalibard1983correlation}. Our first [Fig.~\ref{fig:sectors}(a)] and second [Fig.~\ref{fig:sectors}(b)] set of poles are analogous to the double Rayleigh process and sidebands respectively of Ref.~\cite{dalibard1983correlation}. The point where the analogy with Ref.~\cite{dalibard1983correlation} breaks is that state $\ket{J}$ in Fig.~\ref{fig:sectors}(a) is generically different from $\ket{I}$, and therefore Fig.~\ref{fig:sectors}(a) is technically not a double Rayleigh process. 

Coming back to our calculation, recall from Eq.~\eqref{eq:G2simp2} that the quantity we are interested in is
Eq.~\eqref{eq:R2ab_bigexp1} multiplied by $\mathcal{F}_{1}(\omega_{\lambda_1})\mathcal{F}_{2}(\omega_{\lambda_2})e^{i\omega_{\lambda_1}\tau} $ and integrated over $\omega_{\lambda_1}$ and $\omega_{\lambda_2}$ with the constraint $2\pi \delta\left(E_{FI}+\omega_{\lambda_1}+\omega_{\lambda_2}-2\omega_L\right)$. In this work, we will assume that the filter function $\mathcal{F}_{i}(\omega)$ is narrow enough to prevent the central peak and sidebands from interfering. In this case, we can expand the above expression around the individual poles of $\omega_{\lambda_1}$ (and similarly, of $\omega_{\lambda_2}$), i.e., 
\begin{equation}
\begin{aligned}
    &\frac{1}{\omega_{\lambda_1}-\left(2\omega_L-E_{KI}+i0^+\right)}\frac{1}{\omega_{\lambda_1}-\left(\omega_L-E_{JI}+i0^+\right)}\\
    \approx&\frac{1}{\omega_L-U}\left[\frac{1}{\omega_{\lambda_1}-\left(\omega_L-E_{JI}+i0^+\right)}\right.\\
    &\left.\quad \quad \quad \quad-\frac{1}{\omega_{\lambda_1}-\left(2\omega_L-E_{KI}+i0^+\right)}\right].
\end{aligned}
\end{equation} 

The upshot is that we only need to look at Eq.~\eqref{eq:R2ab_bigexp1} around the two poles, corresponding to either the central peak or the sidebands. With this understanding, we can write $\hat{R}^{(2)}_{\lambda_1,\lambda_2}\Big|_{a+b}$ as a sum of two terms: $\hat{R}^{(2)}_{\lambda_1,\lambda_2}\Big|_{a}$ and $\hat{R}^{(2)}_{\lambda_1,\lambda_2}\Big|_{b}$, near the first and second pole respectively. Let us look at them one by one.
\subsubsection{Process in Fig.~\ref{fig:sectors}(a)}\label{sec:4aonly}
Since $\hat{R}^{(2)}_{\lambda_1,\lambda_2}\Big|_{a}$ is evaluated near $\omega_{\lambda_1}= \omega_L -E_{JI}$, the factor $\left[\omega_{\lambda_1}-(2\omega_L - E_{KI}+i0^+)\right]^{-1}$ becomes $\left[E_{KI}  - \omega_L\right]^{-1}$. This can be approximated as $(U-\omega_L)^{-1}$. Then, the dependence on $E_{K}$ drops out and we can replace $\sum_K \ketbra{K}$ by the identity operator. Then, following the same reasoning explained in the paragraph below Eq.~\eqref{eq:energyconserved}, we can simplify the fermionic terms into a spin singlet projection operator. Doing so, we get
\begin{equation}\label{eq:R2aexp}
\begin{aligned}
    &\mel{F}{\hat{R}^{(2)}_{\lambda_1,\lambda_2}\Big|_{a}}{I}=\frac{-g_L^2 g^2}{\left(\omega_L - U\right)^2}\sum_{\substack{(\vb{r}_1,\bm{\mu}_1)\\(\vb{r}_2,\bm{\mu}_2)}}\abs{\tunn_{\vb{r}_2,\vb{r}_2+\bm{\mu}_2}}^2\abs{\tunn_{\vb{r}_1,\vb{r}_1+\bm{\mu}_1}}^2\\
    &\times \sum_{J}\left\{\left\{\frac{1}{\omega_{\lambda_1}-\left(\omega_L - E_{JI}+i0^+\right)} \right.\right.\\
    &\times \bra{F}\left(4\hat{\vb{S}}_{\vb{r}_2}\cdot \hat{\vb{S}}_{\vb{r}_2 + \bm{\mu}_2}-1\right)\left(\bar{\bm{\mu}}_2\cdot \vb{e}_L\right)\left(\bar{\bm{\mu}}_2\cdot\vb{e}^*_{\lambda_2}\right)\ket{J}\\
     &\times \bra{J}\left(4\hat{\vb{S}}_{\vb{r}_1}\cdot \hat{\vb{S}}_{\vb{r}_1 + \bm{\mu}_1}-1\right)\left(\bar{\bm{\mu}}_1\cdot \vb{e}_L\right)\left(\bar{\bm{\mu}}_1\cdot\vb{e}^*_{\lambda_1}\right)\biggr\}\ket{I}\\
     & \quad \quad + \left\{\lambda_1 \leftrightarrow \lambda_2\right\}\biggl\},
\end{aligned}
\end{equation}
where $\ket{J}$ lies in the spin sector. ($\ket{I}$ and $\ket{F}$ lie in the spin sector, as usual.) As anticipated in Sec.~\ref{sec:processes}, the process in Fig.~\ref{fig:sectors}(a) involves operators entirely in the spin sector.
\subsubsection{Process in Fig.~\ref{fig:sectors}(b)}\label{sec:4bonly}
Along similar lines, one can simplify $\hat{R}^{(2)}_{\lambda_1,\lambda_2}\Big|_{b}$, which is evaluated near $\omega_{\lambda_1}=2\omega_L -E_{KI}$. As we would expect, this time, it is not possible to rewrite all the fermionic operators in terms of spins. Instead, we get
\begin{equation}\label{eq:R2bexp}
    \begin{aligned}
           &\mel{F}{\hat{R}^{(2)}_{\lambda_1,\lambda_2}\Big|_{b}}{I}=\sum_{K}\sum_{(\vb{r}_1,\bm{\mu}_1)}\left\{ -\frac{g_L^2 g^2\abs{\tunn_{\vb{r}_1,\vb{r}_1+\bm{\mu}_1}}^2}{\left(\omega_L - U\right)^2} \right.\\
    &\times \frac{\mel{F}{\sum_{\bm{\mu}_2}\left(\bar{\bm{\mu}}_2\cdot\vb{e}^*_{\lambda_2}\right)\hat{\mathcal{J}}_{\bm{\mu}_2}}{K}\bra{K}\sum_{\bm{\mu}'_2}\left(\bar{\bm{\mu}}'_2\cdot \vb{e}_L\right)\hat{\mathcal{J}}_{\bm{\mu}'_2}}{\omega_{\lambda_1}-\left(2\omega_L - E_{KI}+i0^+\right)} \\
    &\times \left(4\hat{\vb{S}}_{\vb{r}_1}\cdot \hat{\vb{S}}_{\vb{r}_1 + \bm{\mu}_1}-1\right)\left(\bar{\bm{\mu}}_1\cdot \vb{e}_L\right)\left(\bar{\bm{\mu}}_1\cdot\vb{e}^*_{\lambda_1}\right)\Biggr\}\ket{I}\\
    & \quad \quad +\left\{\lambda_1 \leftrightarrow \lambda_2\right\},
    \end{aligned}
\end{equation}
where $\ket{K}$ is an energy eigenstate in the single doublon-hole sector. ($\ket{I}$ and $\ket{F}$ lie in the spin sector, as usual.) Therefore, the process in Fig.~\ref{fig:sectors}(b) necessarily includes operators in the charge sector.
\subsection{Process in Fig.~\ref{fig:sectors}(c)}\label{sec:4c}
We show this process pictorially in Fig.~\ref{fig:description2}(c$_1$-$c'_6$). This process, whose energy level schematic is in Fig.~\ref{fig:sectors}(c), arises from the term $\hat{V}^+_P\hat{\mathbb{G}}_0\hat{V}^+_P\hat{\mathbb{G}}_0\hat{V}^-_P\hat{\mathbb{G}}_0\hat{V}^-_P$.  
Expanding it out,
\begin{equation}\label{eq:Tbigexp2}
\begin{aligned}
    &\hat{V}^+_P\hat{\mathbb{G}}_0\hat{V}^+_P\hat{\mathbb{G}}_0\hat{V}^-_P\hat{\mathbb{G}}_0\hat{V}^-_P
    \\&=g^4\sum_{\bm{\mu}_1,\bm{\mu}'_1,\bm{\mu}_2,\bm{\mu}'_2}\left(\bar{\bm{\mu}}'_1\cdot \vb{e}_L\right)\left(\bar{\bm{\mu}}'_2\cdot \vb{e}_L\right)\opa^2_L\\
    &\times \sum_{\substack{\lambda_1,\lambda_2\\K,\tilde{J},K',F}}\Biggl\{\frac{\opa_{\lambda_2}^{\dagger}\opa_{\lambda_1}^{\dagger}\left(\bar{\bm{\mu}}_1\cdot\vb{e}^*_{\lambda_1}\right)\left(\bar{\bm{\mu}}_2\cdot\vb{e}^*_{\lambda_2}\right)}{\omega_L-E_{K'I}+i0^+}\\
    &\times \frac{\ketbra{F}\hat{\mathcal{J}}_{\bm{\mu}_2}\ketbra{K}\hat{\mathcal{J}}_{\bm{\mu}_1}\ketbra{\tilde{J}}\hat{\mathcal{J}}_{\bm{\mu}'_2}\ketbra{K'}\hat{\mathcal{J}}_{\bm{\mu}'_1}}{\left(2\omega_L - E_{\tilde{J}I}+i0^+\right)\left(2\omega_L - E_{KI}-\omega_{\lambda_1}+i0^+\right)}\Biggr\},
\end{aligned}
\end{equation}
where $\ket{K'}$ and $\ket{K}$ are in the single doublon-hole sector, $\ket{\tilde{J}}$ is in the two doublon-hole sector, and $\ket{I}$ and $\ket{F}$ are in the spin sector. Like before, $\left(\omega_L - E_{K'I}\right)^{-1}$ can be approximated as $\left(\omega_L -U\right)^{-1}$. Also, $\left(2\omega_L - E_{\tilde{J}I}\right)^{-1}$ can be approximated as $\left(2\omega_L -2U\right)^{-1}$. Then the dependence on both $E_{K'}$ and $E_{\tilde{J}}$ drop out. Thus, 
\begin{equation}\label{eq:R2cexp}
    \begin{aligned}
           &\mel{F}{\hat{R}^{(2)}_{\lambda_1,\lambda_2}\Big|_{c}}{I}=\frac{i g_L^2 g^2}{2\left(\omega_L - U\right)^2} \\
    &\times \left\{\sum_K\left[\frac{\bra{F}\sum_{\bm{\mu}_2}\left(\bar{\bm{\mu}}_2\cdot\vb{e}^*_{\lambda_2}\right)\hat{\mathcal{J}}_{\bm{\mu}_2}\ketbra{K}}{\omega_{\lambda_1}-\left(2\omega_L - E_{KI}+i0^+\right)}\right.\right. \\
    &\quad \times \sum_{\vb{r}_1,\bm{\mu}_1,\sigma_1}\left(\bar{\bm{\mu}}_1\cdot\vb{e}^*_{\lambda_1}\right)\left(\tunn_{\vb{r}_1,\vb{r}_1+\bm{\mu}_1}\hat{c}^{\dagger}_{\vb{r}_1+\bm{\mu}_1,\sigma_1}\hat{c}_{\vb{r}_1\sigma_1}-\text{h.c.}\right)\\
    &\quad \times  \tilde{\sum}_{\tilde{J}}\ketbra{\tilde{J}}\\&\quad \times \sum_{\substack{\vb{r}'_2,\bm{\mu}'_2,\\\sigma'_2}}\left(\bar{\bm{\mu}}'_2\cdot \vb{e}_L\right)\left(\tunn_{\vb{r}'_2,\vb{r}'_2+\bm{\mu}'_2}\hat{c}^{\dagger}_{\vb{r}'_2+\bm{\mu}'_2,\sigma'_2}\hat{c}_{\vb{r}'_2\sigma'_2}-\text{h.c.}\right)\\
    &  \times\sum_{\substack{(\vb{r}'_1,\bm{\mu}'_1),\\\sigma'_1}}\left(\bar{\bm{\mu}}'_1\cdot \vb{e}_L\right)\left(\tunn_{\vb{r}'_1,\vb{r}'_1+\bm{\mu}'_1}\hat{c}^{\dagger}_{\vb{r}'_1+\bm{\mu}'_1,\sigma'_1}\hat{c}_{\vb{r}'_1\sigma'_1}-\text{h.c.}\right)\ket{I}\Biggr]\\
    &\quad +\left[\lambda_1 \leftrightarrow \lambda_2\right]\Biggr\},
    \end{aligned}
\end{equation}
where $\tilde{\sum}_{\tilde{J}}\ketbra{\tilde{J}}$ is a projector onto the sector with two doublons and two holes.
Consider the first three hops -- the first one along bond $(\vb{r}'_1,\bm{\mu}'_1)$, the second one along bond $(\vb{r}'_2,\bm{\mu}'_2)$, and the third along $(\vb{r}_1,\bm{\mu}_1)$. At the end of the second hop, there are two doublons and two holes. At the end of the third hop, there is one doublon and one hole. This can only happen in the two qualitatively distinct ways shown in Fig.~\ref{fig:2doublon1} and Fig.~\ref{fig:2doublon2}. 
\begin{figure*}[ht]
  \centering
  \includegraphics[width=0.78\textwidth]{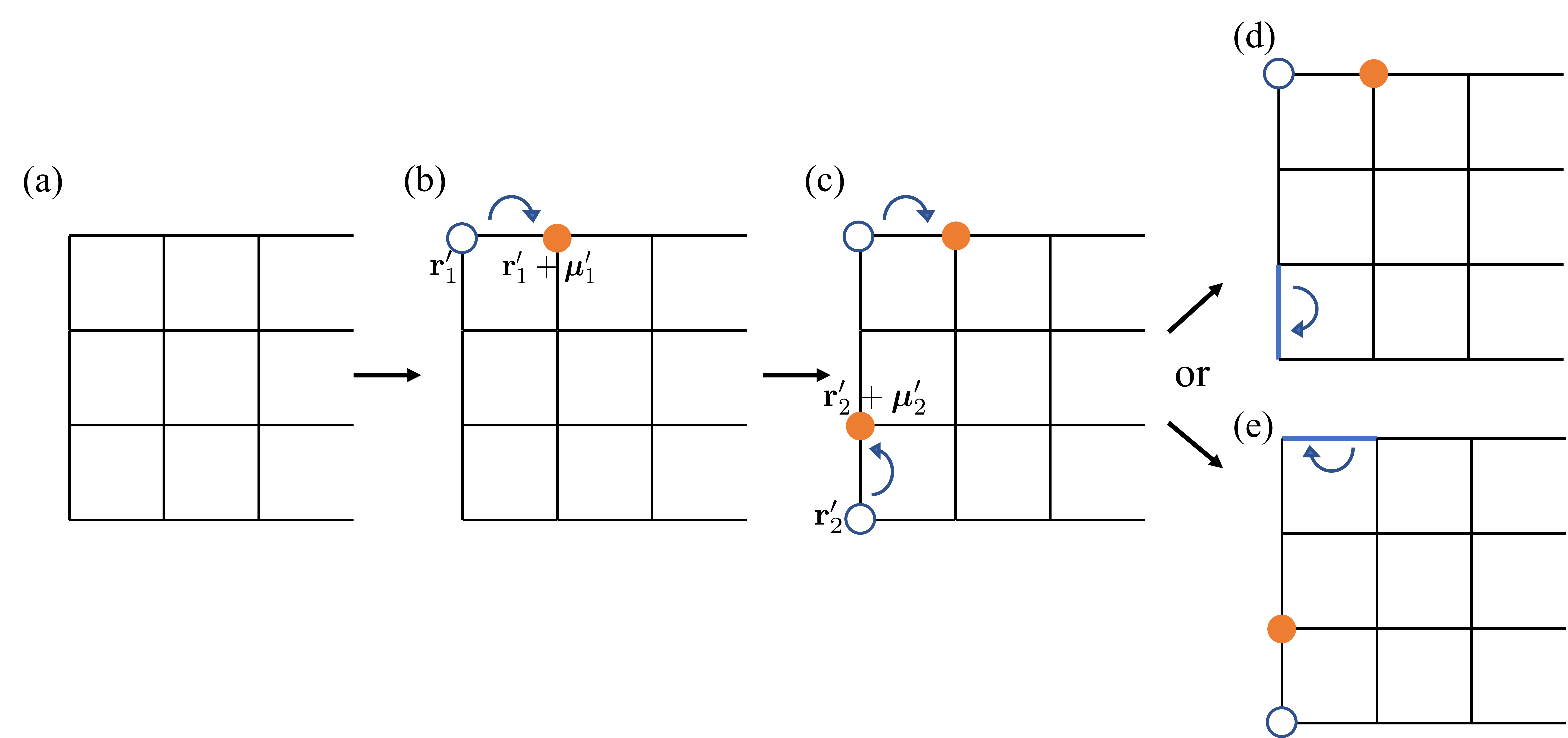}
  \caption{First class of processes contributing to Fig.~\ref{fig:sectors}(c). For figures in this paper, an empty circle denotes a hole at the lattice site and a filled circle denotes a doublon. The absence of any circle denotes a spin (whose state is left unspecified). We use a curved blue arrow to denote an electron hopping from the tail to the head of the arrow. The configuration shown in each figure is the \textit{result} of such a hop shown by the arrow on the \textit{same} figure. Here, we show a square lattice for concreteness. But our results hold for any lattice. We suppose $\bm{\mu}'_1$ and $\bm{\mu}'_2$ are in the $x$ and $y$ directions respectively. \textbf{(a)}: One starts with a spin state. \textbf{(b)}: Through a photon absorption, an electron hops from $\vb{r}'_1$ to $\vb{r}'_1+\bm{\mu}'_1$. \textbf{(c)}: Through a photon absorption, an electron hops from $\vb{r}'_2$ to $\vb{r}'_2+\bm{\mu}'_2$. At this point, there are two doublon-hole pairs as shown. Now there are two choices of doublon-hole pairs to annihilate via a photon emission -- either \textbf{(d)}: the one created second, or \textbf{(e)}: the one created first.}
  \label{fig:2doublon1}
\end{figure*}
\begin{figure*}[ht]
  \centering
  \includegraphics[width=0.78\textwidth]{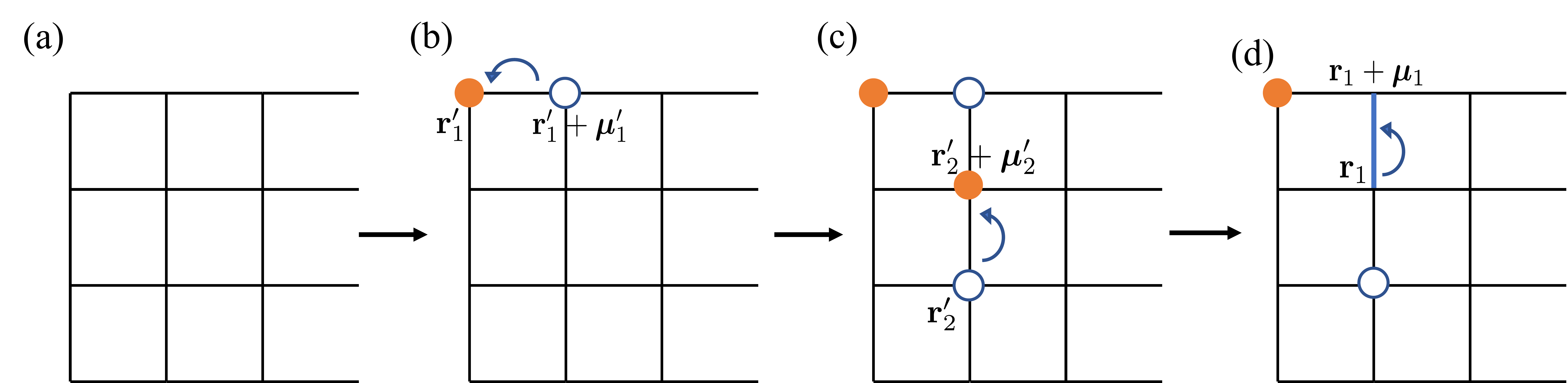}
  \caption{Second class of processes contributing to Fig.~\ref{fig:sectors}(c). Here, the doublon-hole pair that is annihilated differs from either of the two pairs that were created, but is instead made of one hole and one doublon from each pair. For this to be possible, the two bonds along which the doublon-hole pairs were created should be connected to each other by another bond.   \textbf{(a)}: Spin state. \textbf{(b)}: Creation of first doublon-hole pair. \textbf{(c)}: Creation of second doublon-hole pair. \textbf{(d)}: Annihilation of a doublon-hole pair.  }
  \label{fig:2doublon2}
\end{figure*}
First, in Fig.~\ref{fig:2doublon1}, the third hop annihilates the doublon-hole pair created in either the first hop or the second hop.

In contrast, in the process shown in  Fig.~\ref{fig:2doublon2}, the two doublon-hole pairs created by the first two hops are on neighbouring bonds. The third hop then annihilates a doublon-hole pair, not along either of the previous two bonds, but instead along the bond connecting the first two hops. 

Let us analyze the process in Fig.~\ref{fig:2doublon1} first. Here, the bond $(\vb{r}_1,\bm{\mu}_1)$ is identical to either $(\vb{r}'_1,\bm{\mu}'_1)$ [Fig.~\ref{fig:2doublon1}(e)] or $(\vb{r}'_2,\bm{\mu}'_2)$ [Fig.~\ref{fig:2doublon1} (d)]. Let us go with the former first. In this case, the bond $(\vb{r}'_2,\bm{\mu}'_2)$ corresponding to the second hop should share no site in common with the bond $(\vb{r}_1,\bm{\mu}_1)$. Therefore, we can replace $i\sum_{\vb{r}'_2,\sigma'_2}\left(\tunn_{\vb{r}'_2,\vb{r}'_2+\bm{\mu}'_2}\hat{c}^{\dagger}_{\vb{r}'_2+ \bm{\mu}'_2,\sigma'_2}\hat{c}_{\vb{r}'_2\sigma'_2}-\text{h.c.}\right)$
by
\begin{equation*}
\mathcal{J}_{\bm{\mu'}_2}-i\sum_{\vb{r}'_2,\sigma'_2}\eta^{(\vb{r}'_2,\bm{\mu}'_2)}_{(\vb{r}_1,\bm{\mu}_1)}\left(\tunn_{\vb{r}'_2,\vb{r}'_2+\bm{\mu}'_2}\hat{c}^{\dagger}_{\vb{r}'_2+ \bm{\mu}'_2,\sigma'_2}\hat{c}_{\vb{r}'_2\sigma'_2}-\text{h.c.}\right), 
\end{equation*}
where the symbol $\eta_{(\vb{r}_1,\bm{\mu}_1)}^{(\vb{r}'_2,\bm{\mu}'_2)}$ is a function of two bonds $(\vb{r}_1,\bm{\mu}_1)$ and $(\vb{r}'_2,\bm{\mu}'_2)$, and was defined in Eq~\eqref{eq:defeta}. Once we enforce this constraint, we can drop the projector to the two doublon-hole sector, $\sum_{\Tilde{J}}\ketbra{\tilde{J}}$ in Eq.~\eqref{eq:R2cexp}, since the projector is enforced automatically. Now, we can commute the bond $(\vb{r}_1,\bm{\mu}_1)$ to the right in Eq.~\eqref{eq:R2cexp} through the bond $(\vb{r}'_2,\bm{\mu}'_2)$. Then the resulting expression has the currents through two identical bonds next to each other, and we replace it with $\abs{\tunn_{\vb{r}_1,\vb{r}_1+\bm{\mu}_1}}^2 \left(\bar{\bm{\mu}}_1\cdot \vb{e}_L\right)\left(\bar{\bm{\mu}}_1\cdot\vb{e}^*_{\lambda_1}\right)\left(4\hat{\vb{S}}_{\vb{r}_1}\cdot \hat{\vb{S}}_{\vb{r}_1+\bm{\mu}_1}-1\right)$. It is easy to see that the second option [Fig.~\ref{fig:2doublon1} (d)] gives the same result. Let us denote the sum of contributions shown in Fig.~\ref{fig:2doublon1} (d) and Fig.~\ref{fig:2doublon1} (e) by $\mel{F}{\hat{R}^{(2)}_{\lambda_1,\lambda_2}\Big|_{c_1}}{I}$. It thus equals 
\begin{equation}\label{eq:R2csimp}
         \begin{aligned}
           &\mel{F}{\hat{R}^{(2)}_{\lambda_1,\lambda_2}\Big|_{c_1}}{I}=\frac{g_L^2 g^2}{\left(\omega_L - U\right)^2} \\
    &\times\left\{ \sum_K\Biggl\{\frac{\bra{F}\sum_{\bm{\mu}_2}\left(\bar{\bm{\mu}}_2\cdot\vb{e}^*_{\lambda_2}\right)\hat{\mathcal{J}}_{\bm{\mu}_2}\ketbra{K}}{\omega_{\lambda_1}-\left(2\omega_L - E_{KI}+i0^+\right)}\right. \\
    &\times \sum_{\vb{r}_1,\bm{\mu}_1}\Biggl[\sum_{\bm{\mu}'_2}\left(\bar{\bm{\mu}}'_2\cdot \vb{e}_L\right)\biggl(\mathcal{J}_{\bm{\mu'}_2}-\hat{\mathcal{K}}_{\bm{\mu}'_2}\left(\vb{r}_1,\bm{\mu}_1\right)\biggr)\abs{\tunn_{\vb{r}_1,\vb{r}_1+\bm{\mu}_1}}^2\\
    & \quad  \times  \left(4\hat{\vb{S}}_{\vb{r}_1}\cdot \hat{\vb{S}}_{\vb{r}_1 + \bm{\mu}_1}-1\right)\left(\bar{\bm{\mu}}_1\cdot \vb{e}_L\right)\left(\bar{\bm{\mu}}_1\cdot\vb{e}^*_{\lambda_1}\right)\Biggr]\Biggr\}\ket{I}\\
    &\quad + \left\{\lambda_1 \leftrightarrow \lambda_2\right\}\Biggr\},
    \end{aligned}
 \end{equation}
 where $\hat{\mathcal{K}}_{\bm{\mu}'_2}\left(\vb{r}_1,\bm{\mu}_1\right)$ is a Hermitian local operator supported near the bond $\left(\vb{r}_1,\bm{\mu}_1\right)$ in a way that depends on the direction $\bm{\mu}'_2$. It was defined in Eq.~\eqref{eq:defKmu} and shown pictorially in Fig.~\ref{fig:Kmufig}.
 
If we now re-examine Eq.~\eqref{eq:R2bexp}, we find that it actually gets cancelled by part of Eq.~\eqref{eq:R2csimp}. Therefore, 
\begin{equation}\label{eq:R2bandc1}
\begin{aligned}
    &\mel{F}{\hat{R}^{(2)}_{\lambda_1,\lambda_2}\Big|_{b}+\hat{R}^{(2)}_{\lambda_1,\lambda_2}\Big|_{c_1}}{I}=\frac{-g^2_Lg^2}{\left(\omega_L - U\right)^2}\\
    &\times\left\{ \sum_{K} \Biggl\{\frac{\sum_{\bm{\mu}_2}\left(\bar{\bm{\mu}}_2\cdot\vb{e}^*_{\lambda_2}\right)\bra{F}\hat{\mathcal{J}}_{\bm{\mu}_2}\ketbra{K}}{\omega_{\lambda_1}-\left(2\omega_L - E_{KI}+i0^+\right)}\right.\\
    &\times\sum_{\vb{r}_1,\bm{\mu}_1}\biggl[\sum_{\bm{\mu}'_2}\left(\bar{\bm{\mu}}'_2\cdot \vb{e}_L\right)\hat{\mathcal{K}}_{\bm{\mu}'_2}\left(\vb{r}_1,\bm{\mu}_1\right)\left(4\hat{\vb{S}}_{\vb{r}_1}\cdot \hat{\vb{S}}_{\vb{r}_1 + \bm{\mu}_1}-1\right)\\
    &  \  \times \abs{\tunn_{\vb{r}_1,\vb{r}_1+\bm{\mu}_1}}^2\left(\bar{\bm{\mu}}_1\cdot \vb{e}_L\right)\left(\bar{\bm{\mu}}_1\cdot\vb{e}^*_{\lambda_1}\right)\biggr]\Biggr\}\ket{I}+ \left\{\lambda_1 \leftrightarrow \lambda_2\right\}\Biggr\}.
\end{aligned}
\end{equation}
Now, let us consider the contribution from the process shown in Fig.~\ref{fig:2doublon2} to Eq.~\eqref{eq:R2cexp}, that we will denote by $\mel{F}{\hat{R}^{(2)}_{\lambda_1,\lambda_2}\Big|_{c_2}}{I}$. In this process, the bonds $(\vb{r}'_2,\bm{\mu}'_2)$, $(\vb{r}_1,\bm{\mu}_1)$ and $(\vb{r}'_1,\bm{\mu}'_1)$ form a train, when put together successively (see Fig.~\ref{fig:2doublon2}(d)). The result of this process is a back-and-forth hopping of an electron across the bond $(\vb{r}_1,\bm{\mu}_1)$ as well as the transport of an electron from $\vb{r}'_2$ to $\vb{r}'_1$ or vice-versa. Thus, the resulting operator only involves spin (and is not charged) at sites $\vb{r}_1$ and $\vb{r}_1+\bm{\mu}_1$, but involves charged operators at sites adjacent to the bond $(\vb{r}_1,\bm{\mu}_1)$. Simplifying Eq.~\eqref{eq:R2cexp} for this process, we get 
\begin{equation}\label{eq:R2c2simp}
    \begin{aligned}
           &\mel{F}{\hat{R}^{(2)}_{\lambda_1,\lambda_2}\Big|_{c_2}}{I}=-\frac{g_L^2g^2}{\left(\omega_L - U\right)^2} \\
    &\times\left\{ \sum_{K} \Biggl\{\frac{\sum_{\bm{\mu}_2}\left(\bar{\bm{\mu}}_2\cdot\vb{e}^*_{\lambda_2}\right)\bra{F}\hat{\mathcal{J}}_{\bm{\mu}_2}\ket{K}\bra{K}}{\omega_{\lambda_1}-\left(2\omega_L - E_{KI}+i0^+\right)}\right. \\
    &\times \sum_{(\vb{r}_1,\bm{\mu}_1),\bm{\mu}'_1,\bm{\mu}'_2}\Biggl[\left(\bar{\bm{\mu}}_1\cdot\vb{e}^*_{\lambda_1}\right)\left(\bar{\bm{\mu}}'_1\cdot \vb{e}_L\right)\left(\bar{\bm{\mu}}'_2\cdot \vb{e}_L\right)\\
    &\  \times\Biggl(i\hat{\bm{\mathcal{S}}}_{\bm{\mu}'_1,\bm{\mu}'_2}(\vb{r}_1,\bm{\mu}_1)\cdot\biggl(\frac{\hat{\vb{S}}_{\vb{r}_1}-\hat{\vb{S}}_{\vb{r}_1+\bm{\mu}_1}}{2}-i\hat{\vb{S}}_{\vb{r}_1}\times \hat{\vb{S}}_{\vb{r}_1+\bm{\mu}_1}\biggr) \\
    &\quad  +\hat{\mathcal{C}}_{\bm{\mu}'_1,\bm{\mu}'_2}(\vb{r}_1,\bm{\mu}_1)\left(\hat{\vb{S}}_{\vb{r}_1}\cdot \hat{\vb{S}}_{\vb{r}_1 + \bm{\mu}_1}-\frac{1}{4}\right)\Biggl)\Biggr]\Biggr\}\ket{I}\\
    &+ \left\{\lambda_1 \leftrightarrow \lambda_2\right\}\Biggr\},
    \end{aligned}
\end{equation}
where $\hat{\bm{\mathcal{S}}}_{\bm{\mu}'_1,\bm{\mu}'_2}(\vb{r}_1,\bm{\mu}_1)$ and $\hat{\mathcal{C}}_{\bm{\mu}'_1,\bm{\mu}'_2}(\vb{r}_1,\bm{\mu}_1)$ are operators that result in tunneling of charge from one site adjacent to the bond $(\vb{r}_1, \bm{\mu}_1)$ to another adjacent to the same bond, for example, from the empty circle to the filled circle in Fig.~\ref{fig:2doublon2}(d). These operators were defined in Eq.~\eqref{eq:defmathcalS} and Eq.~\eqref{eq:defmathcalC}, and shown pictorially in Fig.~\ref{fig:SandC}.

Both of the above operators are symmetric under exchanging the bond directions $\bm{\mu}'_1$ with $\bm{\mu}'_2$. Under reversing the orientation of the bond $(\vb{r}_1,\bm{\mu}_1)$, i.e., by replacing it with $(\vb{r}_1+\bm{\mu}_1,-\bm{\mu}_1)$, the operator $\hat{\bm{\mathcal{S}}}_{\bm{\mu}'_1,\bm{\mu}'_2}(\vb{r}_1,\bm{\mu}_1)$ remains invariant, while $\hat{\mathcal{C}}_{\bm{\mu}'_1,\bm{\mu}'_2}(\vb{r}_1,\bm{\mu}_1)$ flips sign, as is needed for consistency of Eq.~\eqref{eq:R2c2simp}. Also, note that $\hat{\bm{\mathcal{S}}}_{\bm{\mu}'_1,\bm{\mu}'_2}(\vb{r}_1,\bm{\mu}_1)$ and $\hat{\mathcal{C}}_{\bm{\mu}'_1,\bm{\mu}'_2}(\vb{r}_1,\bm{\mu}_1)$ transform as spin triplet and spin singlet respectively under spin rotation.
\subsection{Process in Fig.~\ref{fig:sectors}(d): Diamagnetic term}\label{sec:4d}
\begin{figure}[h]
  \centering
  \includegraphics[width=0.48\textwidth]{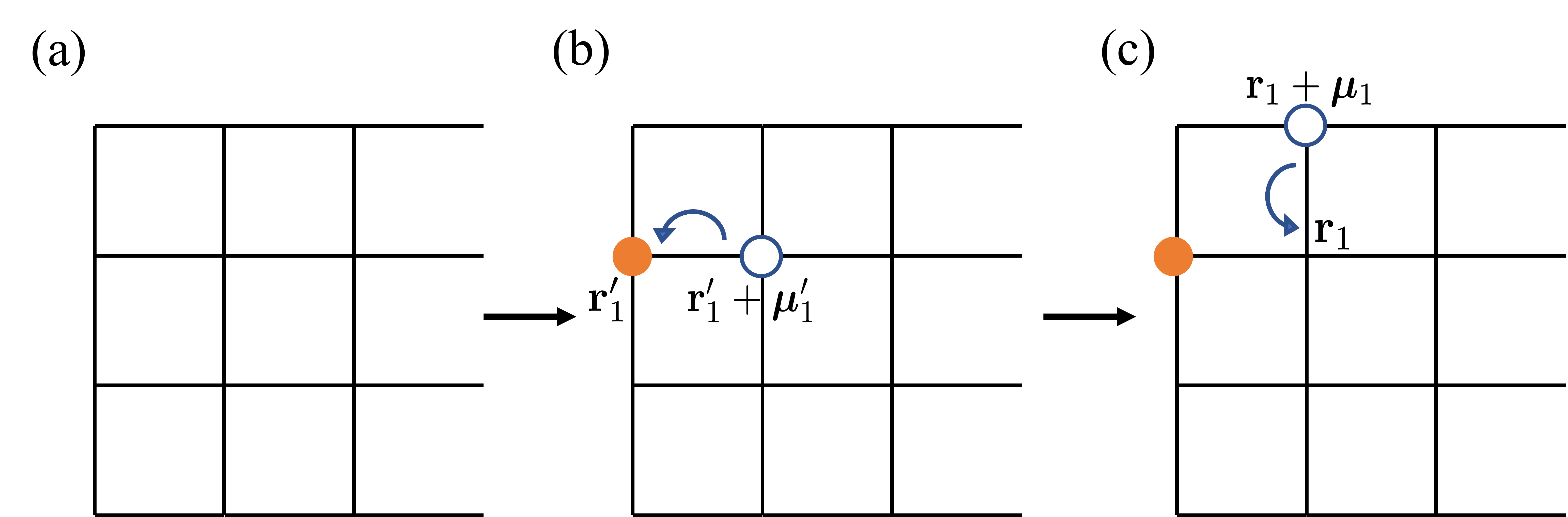}
  \caption{\textbf{(a)}: Spin state. \textbf{(b)}: First doublon-hole pair is created via the paramagnetic term. \textbf{(c)}: A hole moves via the diamagnetic term.}
  \label{fig:diamagnetic1}
\end{figure}
Let us now look at $\hat{R}^{(2)}_{\lambda_1,\lambda_2}\Big|_{d}$, i.e., the contribution from $\hat{V}^+_P\hat{\mathbb{G}}_0\hat{V}^{+-}_D\hat{\mathbb{G}}_0\hat{V}^-_P$, in other words, the process shown in Fig.~\ref{fig:sectors}(d). This process involves the diamagnetic term, and is illustrated pictorially in Fig.~\ref{fig:description2}(d$_1$-d$_5$). We have,
\begin{equation}\label{eq:diamag1}
 \begin{aligned}
    &\bra{F}\hat{R}^{(2)}_{\lambda_1,\lambda_2}\Big|_{d}\ket{I}=ig_L^2 g^2\\
     &\times \sum_{K,K'}\Biggl\{ \sum_{\bm{\mu}_2}\left(\bar{\bm{\mu}}_2\cdot\vb{e}^*_{\lambda_2}\right)\bra{F}\hat{\mathcal{J}}_{\bm{\mu}_2}\ketbra{K}\\
     &\times\sum_{\substack{\vb{r}_1,\bm{\mu}_1,\sigma_1\\ \vb{r}'_1,\bm{\mu}'_1,\sigma'_1}}\frac{\left(\bar{\bm{\mu}}'_1\cdot \vb{e}_L\right)\left(\bar{\bm{\mu}}_1\cdot \vb{e}_L\right)\left(\bar{\bm{\mu}}_1\cdot\vb{e}^*_{\lambda_1}\right)}{\left(\omega_L - E_{K'I}+i0^+\right)\left(2\omega_L - E_{KI}-\omega_{\lambda_1}+i0^+\right) }\\
    &\times \left(\tunn_{\vb{r}_1,\vb{r}_1+\bm{\mu}_1}\hat{c}^{\dagger}_{\vb{r}_1+\bm{\mu}_1,\sigma_1}\hat{c}_{\vb{r}_1,\sigma_1}+\text{h.c.}\right)\ketbra{K'}\\
    &\times\left(\tunn_{\vb{r}'_1,\vb{r}'_1+\bm{\mu}'_1}\hat{c}^{\dagger}_{\vb{r}'_1+\bm{\mu}'_1,\sigma'_1}\hat{c}_{\vb{r}'_1,\sigma'_1}-\text{h.c.}\right)\ket{I}+\biggl[\lambda_1 \leftrightarrow \lambda_2\biggr]\Biggr\}.
 \end{aligned} 
\end{equation}
Here, $\ket{K'}$ and $\ket{K}$ are in the single doublon-hole sector, and $\ket{I}$ and $\ket{F}$ are in the spin sector. Like before, $\left(\omega_L - E_{K'I}\right)^{-1}$ can be approximated as $\left(\omega_L -U\right)^{-1}$. The pattern of electron hopping in this process is depicted in Fig.~\ref{fig:diamagnetic1}. 

To ensure that the state $\ket{K}$ remains in the single doublon-hole subspace, the process leading from Fig.~\ref{fig:diamagnetic1}(b) to (c) should be just a hopping of a doublon or a hole, and should not result in the formation of an additional doublon-hole pair. Therefore, the bonds $(\vb{r}_1,\bm{\mu}_1)$ and $(\vb{r}'_1,\bm{\mu}'_1)$ should have exactly one site in common. Therefore, either $\vb{r}'_1=\vb{r}_1$ or $\vb{r}'_1=\vb{r}_1-\bm{\mu}_1$. This allows us to simplify Eq.~\eqref{eq:diamag1} to 
\begin{align}
        &\bra{F}\hat{R}^{(2)}_{\lambda_1,\lambda_2}\Big|_{d}\ket{I}=\frac{-g_L^2 g^2}{\omega_L - U}\nonumber\\
         &\times\left\{ \sum_{K}\Biggl\{ \sum_{\bm{\mu}_2}\frac{\left(\bar{\bm{\mu}}_2\cdot\vb{e}^*_{\lambda_2}\right)\bra{F}\hat{\mathcal{J}}_{\bm{\mu}_2}\ketbra{K}}{\omega_{\lambda_1}-(2\omega_L - E_{KI}+i0^+)}\right.\nonumber \\
         & \quad \quad \quad\times\sum_{\vb{r},\bm{\mu}',\bm{\mu}}\sum_{\substack{s, s' =\pm  1\\s\bm{\mu} \neq s' \bm{\mu}'}}\biggl[\left(s'\bar{\bm{\mu}}'\cdot \vb{e}_L\right)\left(s\bar{\bm{\mu}}\cdot \vb{e}_L\right)\vb{e}^*_{\lambda_1}\cdot\nonumber\\
         & \quad \quad \quad\times \Bigl[ \left(s\bar{\bm{\mu}}-s'\bar{\bm{\mu}}'\right)\hat{\bm{\mathcal{J}}}^S_{\vb{r},s\bm{\mu},s'\bm{\mu}'}\cdot\hat{\vb{S}}_{\vb{r}}\nonumber\\
         & \quad \quad \quad \quad+\frac{i}{2}\left(s\bar{\bm{\mu}}+s'\bar{\bm{\mu}}'\right)\hat{\mathcal{H}}_{{\vb{r}},s\bm{\mu},s'\bm{\mu}'} \Bigr]\biggr]\Biggr\}+ \left\{\lambda_1 \leftrightarrow \lambda_2\right\}\Biggr\}
\end{align}
where $\hat{\mathcal{H}}_{\vb{r},s\bm{\mu},s'\bm{\mu}'}$ and $ \hat{\bm{\mathcal{J}}}^S_{\vb{r},s\bm{\mu},s'\bm{\mu}'}$ were defined in Eq.~\eqref{eq:defHrr} and \eqref{eq:defJSrr} respectively.
\subsection{Contribution from the cubic term}\label{sec:cubicappendix}
\begin{figure}
    \includegraphics[width=0.48\textwidth]{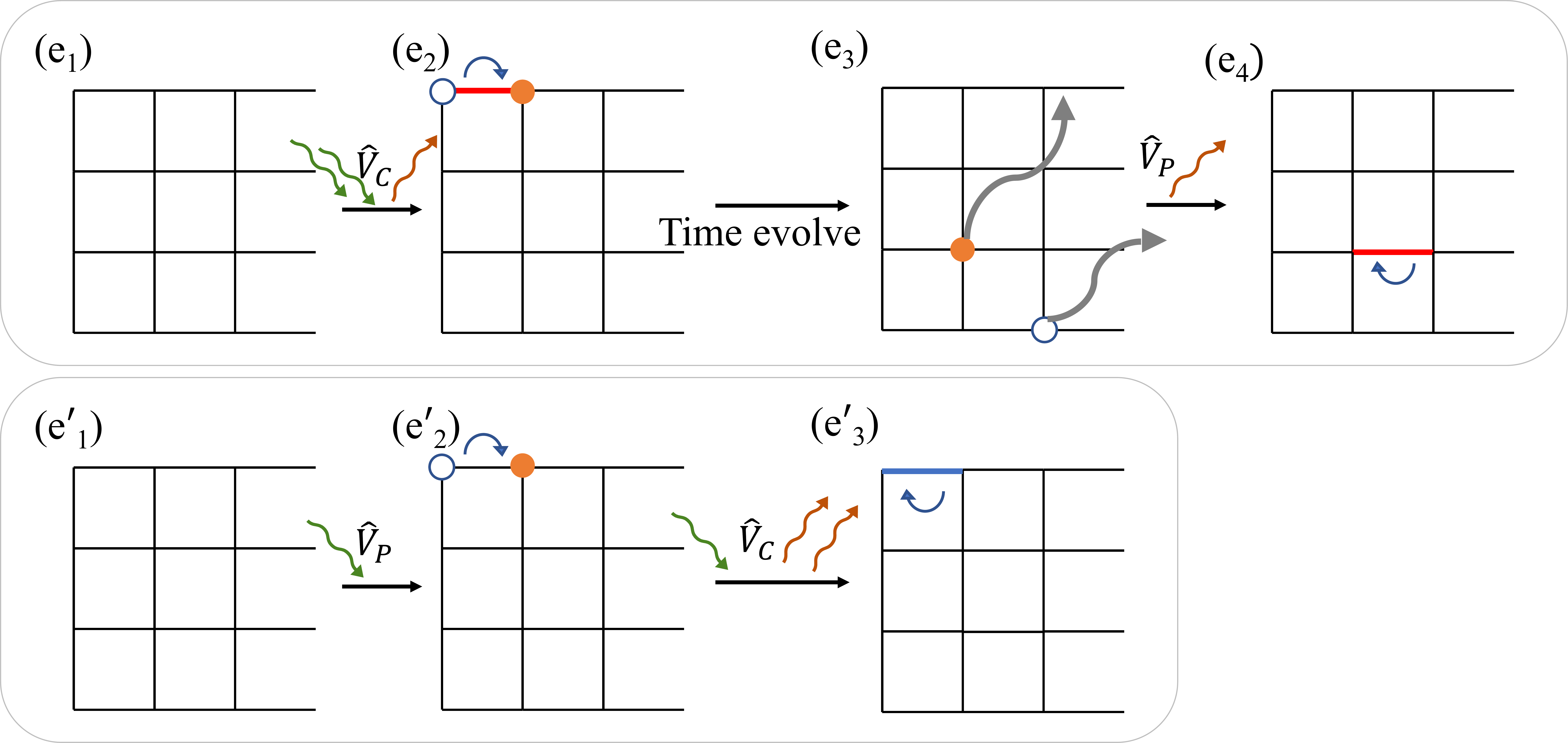}
    \caption{Microscopic processes involving the cubic term $\hat{V}_C$. This Figure is a continuation of Fig.~\ref{fig:description2}, and hence we use (e$_j$) and ($\text{e}'_j$) as the subfigure indices. (\textbf{e$_1$-e$_4$}): Via the $\hat{A}^3$ term, absorption of two photons followed by emission of one results in an electron tunneling across a bond. Then the doublon-hole pair recombines to emit the second photon. This process couples to the charge sector because (\textbf{e$'_1$-e$'_4$}): A photon is absorbed via the paramagnetic term leading to off-resonant electron tunneling. Then, two photons are absorbed and one photon is emitted via the $\hat{A}^3$ term resulting in the electron tunneling back.}
    \label{fig:cubicmicro}
\end{figure}

Finally, we calculate the contribution from the cubic term $\hat{V}_C$ to the $\T$-matrix: $\hat{V}_P\hat{\mathbb{G}}_0\hat{V}_C+\hat{V}_C\hat{\mathbb{G}}_0\hat{V}_P$ (shown in Fig.~\ref{fig:cubicmicro}). First, the contribution from $\hat{V}_P\hat{\mathbb{G}}_0\hat{V}_C$ is:
\begin{equation}  
\begin{aligned}
&\bra{F}\hat{R}^{(2)}_{\lambda_1,\lambda_2}\Big|_{e}\ket{I}=\frac{-g_L^2 g^2}{2}\\
&\times \sum_{\bm{\mu}_1,\bm{\mu}_2}\Biggl[\left(\bm{\mu}_2\cdot \vb{e}_{L}\right)^2\left(\bm{\mu}_1\cdot \vb{e}^*_{\lambda_1}\right)\left(\bm{\mu}_2\cdot \vb{e}^*_{\lambda_2}\right)\\
& \quad \quad \quad \quad \left.\times\frac{\bra{F}\hat{\mathcal{J}}_{\bm{\mu}_2}\ket{K}\bra{K}\hat{\mathcal{J}}_{\bm{\mu}_1}\ket{I}}{2\omega_L-E_{KI}-\omega_{\lambda_1}+i0^+} +[\lambda_1 \leftrightarrow \lambda_2]\right].
\end{aligned}
\end{equation}
Next, the contribution from $\hat{V}_C\hat{\mathbb{G}}_0\hat{V}_P$ is:
\begin{equation}\label{eq:R2corected}  
\begin{aligned}
&\bra{F}\hat{R}^{(2)}_{\lambda_1,\lambda_2}\Big|_{e'}\ket{I}=\frac{g_L^2 g^2}{\omega_L-U}\\
&\times \sum_{(\vb{r},\bm{\mu})}\left(\bar{\bm{\mu}}\cdot \vb{e}_{L}\right)^2\left(\bar{\bm{\mu}}\cdot \vb{e}^*_{\lambda_1}\right)\left(\bar{\bm{\mu}}\cdot \vb{e}^*_{\lambda_2}\right)\abs{\tunn_{\vb{r},\vb{r}+\bm{\mu}}}^2\\
& \quad \quad \quad \quad \times \left(4\hat{\vb{S}}_{\vb{r}}\cdot\hat{\vb{S}}_{\vb{r}+\bm{\mu}}-1 \right). 
\end{aligned}
\end{equation}
\subsection{Consolidating all contributions}
After all the simplifications in the previous subsection, we are now ready to write an expression for the total two-photon amplitude 
\begin{equation}\label{eq:R2general}
\begin{aligned}
    \mel{F}{\hat{R}^{(2)}_{\lambda_1,\lambda_2}}{I}=&\bra{F}\hat{R}^{(2)}_{\lambda_1,\lambda_2}\Big|_{a}\ket{I}+\bra{F}\hat{R}^{(2)}_{\lambda_1,\lambda_2}\Big|_{b}\ket{I}\\+&\bra{F}\hat{R}^{(2)}_{\lambda_1,\lambda_2}\Big|_{c}\ket{I}+\bra{F}\hat{R}^{(2)}_{\lambda_1,\lambda_2}\Big|_{d}\ket{I}.\\
    &\bra{F}\hat{R}^{(2)}_{\lambda_1,\lambda_2}\Big|_{e}\ket{I}+\bra{F}\hat{R}^{(2)}_{\lambda_1,\lambda_2}\Big|_{e'}\ket{I}.
\end{aligned}
\end{equation}
The result leads to Eq.~\eqref{eq:R2final} in the main text. We discuss below the corrections to Eq.~\eqref{eq:R2final} due to Eq.~\eqref{eq:R2corected}.

\begin{equation}\label{eq:R2finaltimecubic}
    \begin{aligned}
        &\mel{F}{\hat{R}^{(2)}_{\lambda_1,\lambda_2}}{I}\\
        &=-i\int_{-\infty}^{\infty}\dd{t}e^{-i(\omega_{\lambda_1}-\omega_L)t}\mel{F}{\mathbb{T}\left[\hat{A}_2(0) \hat{A}_1(-t)\right]}{I}\\
        &-i\int_{-\infty}^{\infty} \dd{t}\bra{F}\left[\theta(t)e^{-i\left(\omega_{\lambda_1}- 2\omega_L\right) t}\hat{C}_2(0)\hat{B}_1(-t)\right.\\
        &\left. \quad \quad \quad \quad \quad \quad +\theta(-t)e^{-i\omega_{\lambda_1} t} \hat{C}_1(-t)\hat{B}_2(0)\right]\ket{I}\\
        &+\frac{1}{\omega_L-U} \bra{F}\hat{D}_{12}\ket{I},
    \end{aligned}
\end{equation}
where $\hat{D}_{12}$ is given by:
\begin{equation}
\begin{aligned}
    \hat{D}_{12}=&\sum_{(\vb{r},\bm{\mu)}}\left(\vb{e}_L\cdot \bar{\bm{\mu}}\right)^2 \left(\vb{e}^*_1\cdot \bar{\bm{\mu}}\right)\left(\vb{e}^*_2\cdot \bar{\bm{\mu}}\right)\abs{\tunn_{\vb{r},\vb{r}+\bm{\mu}}}^2 \\ &\quad \quad \quad \times\left(4\hat{\vb{S}}_{\vb{r}}\cdot\hat{\vb{S}}_{\vb{r}+\bm{\mu}} -1\right).
    \end{aligned}
\end{equation}
This contribution arises from the $\hat{A}_{\vb{r},\vb{r}'}^3$ term in light-matter interactions depicted in Fig.~\ref{fig:cubicmicro}(e$'_1$-e$'_4$). $\hat{D}_{12}$ is a pure spin operator, and is hence gauge-invariant. Thus, $\hat{D}_{12}$ is not needed to maintain gauge-invariance of the remaining terms in Eq.~\eqref{eq:R2finaltime}.  We now note that the dependence of $\hat{D}_{12}$ on polarizations is fine-tuned, and hence by choosing appropriate polarization channels, the contribution from $\hat{D}_{12}$ can be eliminated. Therefore, in this work, we drop $\hat{D}_{12}$ since it does not affect our main results. Developing a careful understanding of $\hat{D}_{12}$ is a direction for future work. 

Thus, Eq.~\eqref{eq:R2final}, combined with the definitions in Eq.~\eqref{eq:AiExp},\eqref{eq:CiExp} and \eqref{eq:BiExp}, gives the amplitude to find two photons -- one in mode $\lambda_1$ and another in mode $\lambda_2$, entirely in terms of the matrix elements of matter operators between initial matter eigenstate $\ket{I}$ and final matter eigenstate $\ket{F}$. Now, we would like to get rid of the explicit dependence on the intermediate states $\ket{J}$ and $\ket{K}$ in Eq.~\eqref{eq:R2final}. To do so, we use the following identity (for some $\omega$ and $\omega_0$): $\frac{1}{\omega-\omega_0-i0^+}=i\int_{-\infty}^{\infty}\dd{t}  \theta(t) e^{-i(\omega-\omega_0-i0^+)t}$, where $\theta(t)$ is a step function that is 1 for $t>0$ and $0$ otherwise. By using the constraint $\omega_{\lambda_2}=2\omega_L-\omega_{\lambda_1}-E_{FI}$, Eq.~\eqref{eq:R2final} can be viewed as a function of only one frequency $\omega_{\lambda_1}$. We then use the above identity to trade the denominators in Eq.~\eqref{eq:R2final} in favor of the $t$-dependent phases. Now, these phases can be absorbed into the Heisenberg evolution of the operators $\hat{A}_j$ and $\hat{B}_j$ (defined as $\hat{A}_j(t)=e^{i\hat{H}_0 t}\hat{A}_j e^{-i\hat{H}_0 t}$). This allows us to rewrite Eq.~\eqref{eq:R2final} as Eq.~\eqref{eq:R2finaltime} in the main text.
\section{Details of measuring phase-sensitive quadrature correlations and Conditional $G^{(1)}$}\label{app:homodyne}
\subsection{Time-averaging phase-sensitive quadrature measurements}
In Table~\ref{tab:dictionary1} of the main text, we presented phase-sensitive quadrature correlations $\expval{\opa_{d_j}(0)}e^{i\theta}+\text{c.c.}$ and $\expval{\opa_{d_2}(\tau)\opa_{d_1}(0)}e^{i\theta}+\text{c.c.}$ In Sec.~\ref{sec:photonicdefs} and Fig.~\ref{fig:homodyneG2} of the main text, we described a setup to measure these. In this Appendix, we address the $t$-dependence of Eq.~\eqref{eq:quadraturetrick} and Eq.~\eqref{eq:homodynetrick2} of the Main Text. Our goal here is to extract a $t$-independent piece from them, because fast-oscillating quantities cannot be robustly measured experimentally. Further, we should justify why the matter correlators in Table~\ref{tab:dictionary1} do not depend on $t$. We do so by time-averaging Eq.~\eqref{eq:quadraturetrick} and Eq.~\eqref{eq:homodynetrick2} over a long time. First, let us consider Eq.~\eqref{eq:quadraturetrick}.
\begin{equation}\label{eq:quadraturetaverage}
\begin{aligned}
    &\lim_{T\to \infty}\frac{1}{T}\int\limits_{-T/2}\limits^{T/2}\dd{t}\left[G^{(1)}_{d_j;+}(\theta;0;t)-G^{(1)}_{d_j;-}(\theta;0;t)\right]\\
    =&\lim_{T\to \infty}\frac{1}{T}\int\limits_{-T/2}\limits^{T/2}\dd{t}\sqrt{\frac{I_{\text{L.O.}}}{2}}\left(\expval{\hat{a}_{d_j}(t)}_{\text{out}}e^{i(\omega_L t + \theta)}+ \text{c.c.}\right).
\end{aligned}
\end{equation}
 From the definition of $\hat{a}_{d_j}$ in Eq.~\eqref{eq:addef} of the main text, $\hat{a}_{d_j}(t)e^{i\omega_L t}$ is a sum of terms of the form $e^{i(\omega_L -\omega_{\vb{k}})t}$, for each mode of momentum $\vb{k}$. The long-time average will select only those modes satisfying $\omega_{\vb{k}}=\omega_L$, i.e., elastic scattering. This implies that the energy of the material remains unchanged. Independent of this discussion, we concluded in Eq.~\ref{eq:X1simp} of Sec.~\ref{sec:photoncorrsimp} that in the (leading order) process contributing to $\expval{\hat{a}_{d_j}(0)}_{\text{out}}$, the matter sector remains in the same state $\ket{I}$ after scattering, and the scattered photon has frequency $\omega_L$. In this Appendix, we obtain the same condition from time-averaging. Therefore, combining the two discussions, we conclude that to evaluate the time-average in Eq.~\ref{eq:quadraturetaverage}, one can simply set $t=0$, thus justifying why we used $t=0$ in the Main Text [Eq.~\eqref{eq:X1def}] for the quadrature measurement.

 Next, let us consider the time average of Eq.~\eqref{eq:homodynetrick2} to obtain the phase-sensitive second order quadrature correlator. The relevant integral over $t$ here is $\lim_{T\to \infty}\frac{1}{T}\int_{-T/2}^{T/2}\dd{t}e^{2i\omega_L t}\expval{\hat{a}_{d_2}(t+\tau)\hat{a}_{d_1}(t)}_{\text{out}}$. Once again, using Eq.~\eqref{eq:addef}, we see that $e^{2i\omega_L t}\hat{a}_{d_2}(t+\tau)\hat{a}_{d_1}(t)$ is a sum of terms of the form $e^{i(2\omega_L-\omega_{\vb{k}_1}-\omega_{\vb{k}_2})t}$. The long-time average selects modes satisfying $2\omega_L=\omega_{\vb{k}_1}+\omega_{\vb{k}_2}$, therefore implying that the energy of the matter state remains unchanged after scattering. This is once again consistent with our independent conclusion in Eq.~\eqref{eq:X2plussimp} of Sec.~\ref{sec:photoncorrsimp}, where we considered $\expval{\hat{a}_{d_2}(\tau)\hat{a}_{d_1}(0)}_{\text{out}}$ and showed that the matter state remains unchanged in processes relevant to this correlator. Therefore, the result of setting $t=0$, (as in Sec.~\ref{sec:photoncorrsimp}) is identical to performing a long-time average over $t$, as in Eq.~\eqref{eq:homodynetrick2}.

 Finally, we note that by setting $\theta_A=\pi/4$ and $\theta_B=-\pi/4$ in Eq.~\eqref{eq:homodynetrick2}, one can measure $\Im \expval{\hat{a}_{d_2}(0)\hat{a}_{d_1}(0)}_{\text{out}}$, as desired in Sec.~\ref{sec:spinchirality}.
 \subsection{Measuring Conditional $G^{(1)}$}\label{app:ConditionalG1}
 In this section, we provide the details for the experimental scheme shown in Fig.~\ref{fig:ConditionalG2} of the Main Text to measure $H_{d_1,d_2}(t,\tau)\equiv \langle\hat{a}^{\dagger}_{d_1}(0)\hat{a}^{\dagger}_{d_2}(t+\tau)\hat{a}_{d_2}(t)\hat{a}_{d_1}(0)\rangle_\text{out}+\text{c.c.}$. One conditions on a photon detection at detector $d_1$ (shown in orange in Fig.~\ref{fig:ConditionalG2}) at time $0$. The second photon is given a time delay $t$ with respect to the first, and is passed through a beamsplitter. One arm of the beam splitter is given an additional time delay $\tau$ with respect to the other, and they are mixed again using a second beamsplitter. The detectors at the two output arms of the second beamsplitters are labeled $d_2;+$ and $d_2;-$. We have:
 \begin{align}
     \hat{a}_{d_2;+}(t+\tau)&=\frac{1}{\sqrt{2}}\left[\hat{a}_{d_2}(t)+\hat{a}_{d_2}(t+\tau)\right],\\
     \hat{a}_{d_2;-}(t+\tau)&=\frac{1}{\sqrt{2}}\left[\hat{a}_{d_2}(t)-\hat{a}_{d_2}(t+\tau)\right].
 \end{align}
The conditional measurement to be made here is a $G^{(2)}(t+\tau)$ measurement between the pair of detectors $(d_1,d_2;+)$ and $(d_1,d_2;-)$. Let us consider:
\begin{equation}
\begin{aligned}
    &G^{(2)}_{d_1,d_2;+}(t+\tau)-G^{(2)}_{d_1,d_2;-}(t+\tau)\\
    =&\expval{\hat{a}^{\dagger}_{d_1}(0)\hat{a}^{\dagger}_{d_2;+}(t+\tau)\hat{a}_{d_2;+}(t+\tau)\hat{a}_{d_1}(0)}_{\text{out}}\\
    &\quad -\expval{\hat{a}^{\dagger}_{d_1}(0)\hat{a}^{\dagger}_{d_2;-}(t+\tau)\hat{a}_{d_2;-}(t+\tau)\hat{a}_{d_1}(0)}_{\text{out}}\\
    &=\expval{\hat{a}^{\dagger}_{d_1}(0)\hat{a}^{\dagger}_{d_2}(t+\tau)\hat{a}_{d_2}(t)\hat{a}_{d_1}(0)}_{\text{out}}+\text{ c.c.}
\end{aligned}
\end{equation}
Therefore, the above difference measures the desired correlator. 
\bibliography{g2references}

\end{document}